\documentclass[fp,twocolumn]{jpsj3}

\usepackage{txfonts}

\usepackage[usenames]{color}

\newcommand{\bras}[1]{\langle#1\rvert}
\newcommand{\kets}[1]{\lvert#1\rangle}

\newcommand{\mean}[1]{\left<#1\right>}
\newcommand{\means}[1]{\langle#1\rangle}

\title{Hunting Majorana Fermions in Kitaev Magnets}

\author{Yukitoshi Motome$^1$\thanks{motome@ap.t.u-tokyo.ac.jp} and Joji Nasu$^2$}
\inst{$^1$Department of Applied Physics, University of Tokyo, Bunkyo, Tokyo 113-8656, Japan \\
$^2$Department of Physics, Yokohama National University, Hodogaya, Yokohama 240-8501, Japan}

\abst{
A Majorana fermion is a fermionic particle that is its own antiparticle.
Since the theoretical discovery by Ettore Majorana in 1937, the exotic particle has long been searched in particle physics.
In the last few decades, however, it has attracted renewed interest in condensed matter physics, where it can be realized as an elementary excitation (quasiparticle) in quantum states of matter, such as the fractional quantum Hall states and topological superconductors.
In this review, we discuss another platform for Majorana fermions, the quantum spin liquid.
The quantum spin liquid is a bizarre quantum phase of insulating magnets, firstly proposed by Philip Anderson in 1973, in which interacting magnetic moments remain disordered down to the lowest temperature under strong quantum fluctuations.
They are characterized by topological entanglement and fractional excitations, whose possible application to topological quantum computation is recently discussed intensively.
As a prime candidate for such exotic states, we here focus on the Kitaev magnets, a subgroup of the spin-orbit Mott insulators, which have been a subject of intense research initiated by the seminal works by Alexei Kitaev in 2006 and by George Jackeli and Giniyat Khaliullin in 2009. 
After a brief overview of the Kitaev model and the fractionalization of spins in the exact ground state, we review recent explosive development in this rapidly growing field, with a focus on numerical solutions of the Kitaev model at finite temperatures and the comparison with experiments.
The key concept is {\it thermal fractionalization} --- two types of fractional excitations manifest themselves at largely different temperatures.
This leads to distinct thermodynamics and spin dynamics in a variety of experimentally measurable quantities.
We discuss such peculiar behaviors as the signatures of fractional quasiparticles, in careful comparison with the available experimental data for the candidate materials of the Kitaev magnets.
Our review gives an overview of the current status of the identification of Majorana fermions in the Kitaev magnets, which would serve as a basis for further experimental and theoretical studies toward the manipulation of the exotic particles for topological quantum computation.
}

\begin{document}
\maketitle

\section{Introduction}
\label{sec:introduction}

Majorana fermions are charge-neutral spin-1/2 particles that are their own antiparticles.
They were theoretically discovered by Ettore Majorana in 1937 in a real solution for the Dirac equation~\cite{Majorana1937}.
The Majorana fermions are distinguished from the ordinary fermions in the complex solution, called the Dirac fermions.
The Dirac fermions are not their own antiparticles, and can be described by the annihilation and creation operators, $f$ and $f^\dagger$, respectively.
Two Majorana operators are defined by using $f$ and $f^\dagger$ as
\begin{align}
\gamma_1 = \frac{f - f^\dagger}{i}, \ \ \
\gamma_2 = f + f^\dagger.
\label{eq:f_g}
\end{align}
The definitions immediately yield that their creation and annihilation are equivalent:
\begin{align}
\gamma_i^\dagger = \gamma_i,
\label{eq:g_rel1}
\end{align}
and they satisfy the anticommutation relation
\begin{align}
\{\gamma_i, \gamma_j\} = 2 \delta_{ij},
\label{eq:g_rel2}
\end{align}
where $\delta_{ij}$ is the Kronecker delta ($i,j = 1,2$).
Equation~(\ref{eq:f_g}) indicates that the occupied and unoccupied states of the Dirac fermion can be described by a pair of Majorana fermions.
This means that one Majorana fermion carries half degrees of freedom of one Dirac fermion.

Since the intriguing proposal by Ettore Majorana, the physical example of the exotic particles has long been sought in particle physics.
Within the standard model, all the fermionic particles are the Dirac fermions, except for the neutrino.
Thus, the neutrino has long been studied as a prime candidate for the Majorana fermion, but its nature is not settled yet~\cite{Doi1985,Mohapatra2006,Akhmedov2015}.
Another candidates have been discussed for superpartners in the supersymmetry model, but no evidence was established to date.

In the last few decades, the Majorana fermions have attracted renewed interest by their possible realization in condensed matter physics~\cite{Wilczek2009}.
In this case, they appear not as elementary particles but as elementary excitations (quasiparticles) in quantum states of matter.
In general, quantum many-body states under electron correlations can host emergent quasiparticles, which have distinct nature from the constituent electrons.
In some cases, the elementary excitations are described by more than one types of quasiparticles, which  looks like the electrons are fractionalized into several particles.
This is called {\it fractionalization}.
For instance, in the two-dimensional (2D) fractional quantum Hall states, the elementary charge $-e$ is fractionalized into fractional charges, e.g., $e/3$, and as a result, the elementary excitations of the system are described by emergent quasiparticles called anyons that do not obey either Dirac-Fermi or Bose-Einstein statistics.

In the context of the fractionalization, the emergence of Majorana fermions has been discussed for several different quantum states, such as the edge modes in the $\nu=5/2$ fractional quantum Hall state~\cite{Moore1991,Stomer1999,Read2000,Nayak2008,Jain2015}, the zero modes in $p$-wave superconductors\cite{Read2000,DasSarma2006}, and the bound states in topological superconductors~\cite{Jackiw1981,Fu2008,Sato2016,Sato2017}.
Since these Majorana fermions originate from the fractionalization of fundamental particles, i.e., electrons, they acquire topological entanglement and intrinsically nonlocal nature.
Owing to the unusual properties, the emergent Majorana fermions have drawn a great attention for the possible application to topological quantum computation~\cite{Kitaev2003,Freedman2003}.

In this review, we focus on another realization of Majorana fermions in insulating magnets called Mott insulators.
In these systems, electrons are spatially localized due to strong electron correlations, and hence, the charge degree of freedom is inactive.
Instead, what can be fractionalized here is the spin degree of freedom. 
Such a possibility of spin fractionalization has been discussed to take place in the quantum spin liquid (QSL), which is a quantum disordered state in the Mott insulators, firstly proposed by Philip Anderson in 1973~\cite{Anderson1973}.
In the QSL, any conventional symmetry breaking is precluded by strong quantum fluctuations, and the localized spins remain disordered but quantum entangled.
Several types of QSLs have been predicted depending on the symmetry of the system, and they host their own fractional quasiparticles~\cite{Wen2004,Savary2017}.
For instance, in the so-called $Z_2$ QSLs, the spin excitations are supposed to be fractionalized into two types of elementary excitations, spinons and visons; the spinons are charge-neutral spin-1/2 particlelike excitations, while the visons are topological excitations defined by their stringlike traces~\cite{Read1989,Wen1991}.

Most of such arguments, however, lack rigorous grounds, as there are less well-defined QSLs in more than one dimension. 
Thus far, tremendous efforts have been made for geometrically-frustrated antiferromagnets in two and three dimensions, but there are few examples where the ground state is strictly shown to be a QSL~\cite{Balents2010,Lacroix2011,Diep2013,Zhou2017}. 
A main difficulty lies in the lack of suitable theoretical methods: 
Any approximate theories may miss the essential aspects of the quantum entanglement in QSLs, and numerical methods require extremely high precisions to select out the true ground state from a macroscopic number of quasi-degenerate states under strong frustration. 
Thus, it has remained a big challenge to identify fractional spin excitations in QSLs.

The situation has been changed dramatically over the past decade through two breakthroughs.
One is the proposal of the exactly-solvable model in the seminal paper by Alexei Kitaev in 2006~\cite{Kitaev2006}, which is now called the Kitaev model.
The model is a spin-1/2 model defined on a 2D honeycomb structure with bond-dependent interactions.
The ground state is exactly obtained to be a QSL, in which the spin excitations are fractionalized into two types of quasiparticle excitations: itinerant spinon-like excitations, which are described by the Majorana fermions, and localized ones that constitute vison-like excitations.
The other breakthrough was brought by G. Jackeli and G. Khaliullin in 2009~\cite{Khaliullin2005,Jackeli2009}.
They pointed out that the Kitaev model can be materialized in a class of the Mott insulating magnets with  the strong spin-orbit coupling.
Stimulated by their argument, several materials have been nominated as the candidates for the Kitaev QSL, such as iridium oxides $A_2$IrO$_3$ ($A$=Li and Na) and a ruthenium trichloride $\alpha$-RuCl$_3$.
These two breakthroughs have driven intense research for the Kitaev QSL from both theoretical and experimental viewpoints.

In the present article, we give an overview of the recent progress in this rapidly growing field.
Several review articles are already available for the Kitaev QSL and its candidates~\cite{Nussinov2015,Trebst2017preprint,Winter2017a,Hermanns2018,Knolle2019a,Takagi2019,Janssen2019}.
Here we particularly focus on the finite-temperature ($T$) aspects of the fractional excitations, which are relevant to identify them in the candidate materials.
Since the exact solution of the Kitaev model is limited to the ground state, the authors and their collaborators have developed several numerical techniques to study the finite-$T$ properties~\cite{Nasu2014,Nasu2015,Yoshitake2016,Yoshitake2017a,Yoshitake2017b,Mishchenko2017}, and calculated the experimental observables, such as the specific heat and entropy, static spin-spin correlations~\cite{Nasu2015}, magnetic susceptibility, inelastic neutron scattering spectra, spin-lattice relaxation rate in the nuclear magnetic resonance (NMR)~\cite{Yoshitake2016,Yoshitake2017a,Yoshitake2017b}, Raman scattering spectra~\cite{Nasu2016}, and longitudinal and transverse components of the thermal conductivity~\cite{Nasu2017a}.
Through the detailed comparison of the theoretical results with experimental data, signatures of the fractional excitations have been accumulated for the Kitaev candidate materials.
We will discuss in detail such comparisons in this review.

The structure of this article is as follows.
In Sec.~\ref{sec:Model}, we introduce the Kitaev model and the fractional excitations derived from the exact solution for the QSL ground state.
After introducing the Hamiltonian in Sec.~\ref{sec:Hamiltonian}, we briefly discuss the origin of the peculiar bond-dependent interaction in the Kitaev model in Sec.~\ref{sec:Jackeli-Khaliullin}.
In Sec.~\ref{sec:Majrep}, we describe a Majorana representation of the spin operators, which is different from the original one introduced by Kitaev but useful for numerical techniques developed for finite-$T$ calculations.
After an overview of the exact QSL ground state and the fractional excitations in Sec.~\ref{sec:groundstate} and \ref{sec:frac}, respectively, we discuss the effects of finite $T$, an external magnetic field, and other exchange interactions in Sec.~\ref{sec:finiteT}, \ref{sec:field}, and \ref{sec:other_exchanges}, respectively.
These additional effects on the Kitaev QSL are schematically summarized in the potential phase diagrams in Sec.~\ref{sec:schematic_phase_diagram}.

In Sec.~\ref{sec:Tfrac}, we discuss one of the distinct aspects in the thermodynamics of the Kitaev model, which we call thermal fractionalization.
In Sec.~\ref{sec:2D_honeycomb}, as the prototypical behaviors, two successive crossovers are discussed for the Kitaev model on the 2D honeycomb structure.
Then, a peculiar phase transition with time-reversal symmetry breaking is overviewed for a 2D triangle-honeycomb structure in Sec.~\ref{sec:2D_CSL}.
In Sec.~\ref{sec:3D}, we showcase several unconventional phase transitions found for three-dimensional (3D) extensions of the Kitaev model, which can be regarded as gas-liquid-solid transitions in terms of the spin degree of freedom.
We also briefly discuss spontaneous breaking of time-reversal symmetry in the 3D cases.
These crossovers and phase transitions are summarized in Sec.~\ref{sec:phase_diagram}.

In Sec.~\ref{sec:candidates}, we introduce several candidate materials for the Kitaev QSL.
We discuss the fundamental aspects of quasi-2D iridium oxides in Sec.~\ref{sec:2D_iridates}, a ruthenium trichloride in Sec.~\ref{sec:RuCl3}, and 3D iridium oxides in Sec.~\ref{sec:3D_iridates}.

In Sec.~\ref{sec:theory_exp}, we compare theoretical results for the Kitaev model with experimental data for the candidate materials, focusing on the quasi-2D materials.
We discuss the two successive crossovers in the specific heat and entropy in Sec.~\ref{sec:Cv_S}, the saturation of static spin correlations measured from optical probe in Sec.~\ref{sec:SS}, and peculiar $T$ dependence of the magnetic susceptibility in Sec.~\ref{sec:chi}.
Then, we turn to the signatures of the fractional excitations in the spin dynamics: the dynamical spin structure factor measured in inelastic neutron scattering in Sec.~\ref{sec:Sqw} and the NMR relaxation rate in Sec.~\ref{sec:1/T1}.
From the comparison, we discuss the dichotomy between static and dynamical spin correlations as a signature of the thermal fractionalization.
More direct signatures of fermionic excitations are discussed for the thermal conductivity in Sec.~\ref{sec:kappa_xx} and the Raman scattering in Sec.~\ref{sec:Raman}; in the latter, the unusual fermionic nature is clearly identified in a wide-$T$ range.
Finally, in Sec.~\ref{sec:kappa_xy}, a direct evidence of the Majorana nature and the topological state is discussed for the thermal Hall conductivity.
Section~\ref{sec:summary} is devoted to the summary and perspectives.
In Appendix, we describe the details of the Majorana-based numerical techniques.

\section{Kitaev model and Majorana fermions}
\label{sec:Model}

\subsection{Hamiltonian}
\label{sec:Hamiltonian}

\begin{figure}[t]
 \begin{center}
  \includegraphics[width=0.8\columnwidth,clip]{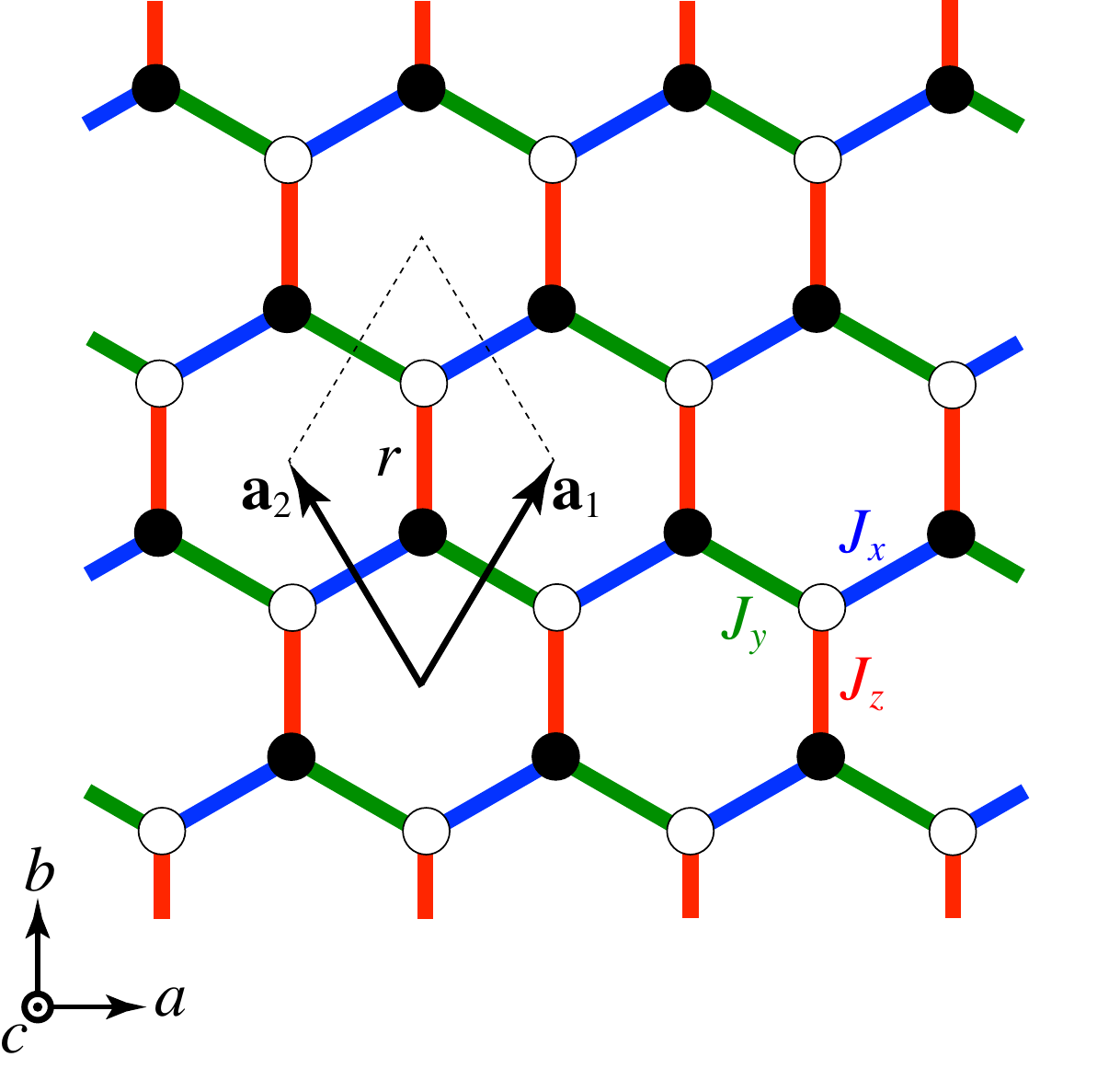}
  \caption{
(Color online) 
Schematic picture of the Kitaev model defined on a honeycomb structure with three kinds of interactions $J_x$, $J_y$, and $J_z$ on the $x$, $y$, and $z$ bonds, respectively.
$\textbf{a}_1$ and $\textbf{a}_2$ are the primitive translation vectors and $r$ labels the unit cell including the $z$ bond.
The Cartesian coordinate axes $(a,b,c)$ are also shown.
}
  \label{fig:lattice}
 \end{center}
\end{figure}

The Kitaev model is a quantum spin model with localized spin-1/2 magnetic moments with bond-dependent anisotropic interactions~\cite{Kitaev2006}.
The model was originally introduced on a 2D honeycomb structure, while it can be extended to any tricoordinate structures in any spatial dimensions (some examples will be shown in Sec.~\ref{sec:Tfrac}).
We mostly focus on the honeycomb case in this review.
The exchange interactions are all Ising type, but the spin component depends on the three types of nearest-neighbor (NN) bonds on the tricoordinate structure.
The Hamiltonian is given by
\begin{align}
{\cal H} = - \sum_{\mu=x,y,z} J_\mu \sum_{\langle i,j \rangle_\mu} S_i^\mu S_j^\mu,
\label{eq:H}
\end{align}
where $J_\mu$ is the exchange coupling constant on the $\mu$ bonds and $S_i^\mu$ is the $\mu$ component of the spin-1/2 operator at site $i$; the sum of $\langle i,j \rangle_\mu$ is taken for NN spin pairs on the $\mu$ bonds.
A schematic picture of the model is shown in Fig.~\ref{fig:lattice}.

As it is impossible to optimize all the bond energies simultaneously, the bond-dependent anisotropic interactions lead to severe frustration despite the absence of geometrical frustration in the lattice structure.
Indeed, the classical counterpart of the Kitaev model, where the spins are regarded as the classical vectors, has an infinite numbers of energetically degenerate ground states~\cite{Baskaran2008}.
In the quantum case, however, this macroscopic classical degeneracy is lifted and a QSL ground state is realized as described in Sec.~\ref{sec:Majrep} and \ref{sec:groundstate}.

\subsection{Origin of bond-dependent anisotropic interactions}
\label{sec:Jackeli-Khaliullin}

\begin{figure}[t]
 \begin{center}
  \includegraphics[width=0.95\columnwidth,clip]{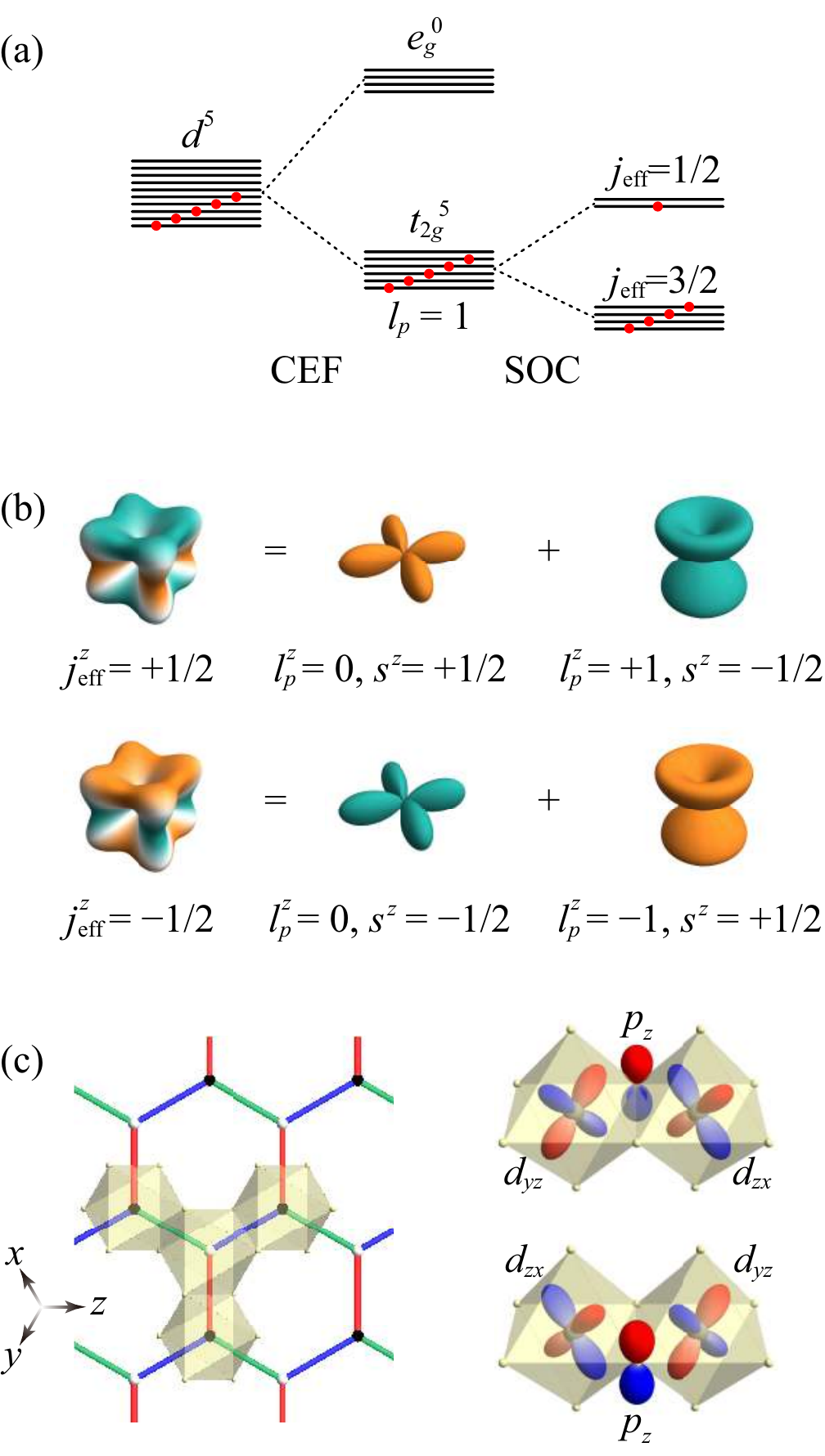}
  \caption{
(Color online) 
(a) Energy scheme of the atomic $d$-orbital states occupied by five electrons in the presence of the cubic crystalline electric field (CEF) and the spin-orbit coupling (SOC).
(b) Pictorial representation of the $j_{\rm eff} =1/2$ Kramers doublet in Eq.~(\ref{eq:jeff=1/2}).
(c) Schematic picture of the lattice structure with edge-sharing ligand octahedra (left), and two kinds of the exchange processes by the indirect $d$-$p$-$d$ hoppings yielding the Kitaev coupling (right).
The coordinate axes $(x,y,z)$, which point from the center to the corners of an (ideal) octahedron, are shown in the left panel. 
They are common to the spin axes set by the SOC in the corresponding Kitaev model. 
The objects with blue and red ellipsoids in the right panel represent $d$ and $p$ orbitals.
}
  \label{fig:ligand}
 \end{center}
\end{figure}

The peculiar form of the interactions in Eq.~(\ref{eq:H}), which is often called the Kitaev coupling, can be realized in a class of Mott insulators with strong spin-orbit coupling.
This intriguing possibility was theoretically pointed out by Jackeli and Khaliullin~\cite{Jackeli2009}, following the pioneering work by Khaliullin~\cite{Khaliullin2005}.
They pointed out two requisites for the Kitaev coupling:
(i) localized magnetic moments arising from spin-orbital entanglement, each of which carries an effective angular momentum $j_{\rm eff}=1/2$, and
(ii) quantum interference between the exchange processes by indirect hoppings of the localized electrons via ligands.

It was argued that the requisite (i) is satisfied in the low-spin $d^5$ configuration under the cubic crystalline electric field and the strong spin-orbit coupling.
This is schematically shown in Fig.~\ref{fig:ligand}(a).
The tenfold degenerate states (including spin) for the $d$-orbital manifold are split by the cubic crystalline electric field into the low-energy sixfold $t_{2g}$ manifold ($d_{xy}$, $d_{yz}$, and $d_{zx}$) and the high-energy fourfold $e_g$ manifold ($d_{3z^2-r^2}$ and $d_{x^2-y^2}$).
Five $d$ electrons occupy the $t_{2g}$ states in the low-spin state, as shown in the middle panel of Fig.~\ref{fig:ligand}(a).
The $t_{2g}$ manifold is isomorphic to the $p$-orbital states; the angular momentum for the $t_{2g}$ manifold is effectively described by $\textbf{l}_{t_{2g}}= -\textbf{l}_p$, where $\textbf{l}_p$ is the $l=1$ angular momentum operator obeying the commutation relations.
The bases are explicitly written as
\begin{align}
&\kets{l_{t_{2g}}^z = 0 } = \kets{l_p^z = 0} = \kets{ d_{xy}}, \\
&\kets{ l_{t_{2g}}^z = \pm 1 } = \kets{l_p^z = \mp 1} = \frac{1}{\sqrt2}\kets{ d_{zx}} \pm i \kets{ d_{yz}}.
\end{align}
When the angular momentum $l_p=1$ is coupled with the spin angular momentum $s=1/2$ by the spin-orbit coupling (SOC), the $t_{2g}$ manifold is further split into the low-energy $j_{\rm eff}=3/2$ quartet and the high-energy $j_{\rm eff}=1/2$ doublet.
Thus, the low-spin $d^5$ state ends up with the one-hole state in the $j_{\rm eff}=1/2$ doublet, as shown in the right panel of Fig.~\ref{fig:ligand}(a).
The $j_{\rm eff}=1/2$ doublet comprises a time-reversal Kramers pair, which is described by
\begin{align}
\Big\lvert j_{\rm eff}^z=\pm \frac12 \Big\rangle =
\sqrt{\frac13} \Big\lvert l_p^z=0, s^z=\pm \frac12 \Big\rangle 
-\sqrt{\frac23} \Big\lvert l_p^z=\pm 1, s^z=\mp \frac12 \Big\rangle.
\label{eq:jeff=1/2}
\end{align}
The schematic pictures are shown in Fig.~\ref{fig:ligand}(b).
The $g$ factor of the $j_{\rm eff}=1/2$ doublet is isotropic and negative ($\simeq -2$), whose sign is opposite to that of the anomalous $g$-factor of the electron spin due to the orbital contribution~\cite{Abragam1970}.

On the other hand, the requisite (ii) is satisfied in an edge-sharing network of the ligand octahedra with the $d^5$ cations in the centers, as shown in the left panel of Fig.~\ref{fig:ligand}(c).
In this geometry, there are two different paths for the indirect $d$-$p$-$d$ hopping via two ligands shared by the edge-sharing octahedra, as shown in the right panel of Fig.~\ref{fig:ligand}(c).
The exchange processes by the two paths cause the quantum interference, which cancels out the isotropic Heisenberg exchange interactions and makes the higher-order Kitaev coupling the leading contribution.
The Kitaev coupling has a contribution from the Hund's-rule coupling in the exchange process, and therefore, it is expected to be ferromagnetic (FM).

The two requisites are approximately satisfied, e.g., in the spin-orbit Mott insulators with Ir$^{4+}$ and Ru$^{3+}$ ions.
Indeed, some iridium and ruthenium compounds, such as $A_2$IrO$_3$ ($A$=Na and Li) and $\alpha$-RuCl$_3$ have been intensively studied as the candidates for the model in Eq.~(\ref{eq:H}); see Sec.~\ref{sec:candidates} for more details.
In these compounds, however, other exchange couplings such as the isotropic Heisenberg ones are also present due to the deviation from the ideal situation.
Effects of such other interactions will be discussed in Sec.~\ref{sec:other_exchanges}.

Recently, the Kitaev coupling was also predicted for other systems.
One is the systems with the high-spin $d^7$ configuration, such as Co$^{2+}$ ions~\cite{Liu2018,Sano2018,Yan2019,Yao2019preprint,Zhong2019preprint}.
In this case, while the $j_{\rm eff}=1/2$ moments arise from a different energy scheme from that in the low-spin $d^5$ case, the underlying mechanism for the exchange processes is basically common, and hence, the Kitaev coupling is FM.
Another candidates are explored for $f$-electron compounds~\cite{Li2017,Rau2018,Jang2019,Luo2019preprint,Xing2019preprint}.
In particular, for the $f^1$ electron configuration, an antiferromagnetic (AFM) Kitaev coupling ($J_\mu <0$) was theoretically predicted, in contrast to the $d^5$ and $d^7$ cases~\cite{Jang2019}.
The sign change is brought by the different atomic energy scheme and the different shapes of the $f$ orbitals.
We will return to this point in Sec.~\ref{sec:field}.

\subsection{Majorana representation}
\label{sec:Majrep}

In the seminal paper, Kitaev showed that the ground state of the model in Eq.~(\ref{eq:H}) is exactly obtained by introducing a Majorana representation of the spin operators~\cite{Kitaev2006}.
In the exact solution, each spin-1/2 operator is represented by four Majorana fermion operators.
Later, another Majorana representation was introduced, which gives the same exact solution~\cite{Chen2007,Feng2007,Chen2008}.
In this case, the spin-1/2 operator is represented by two Majorana fermions.
In this article, we briefly review the latter Majorana representation, as it is used in the numerical simulations in the later sections.
The advantage of the latter is in the size of the Hilbert space.
The former Kitaev's representation doubles the Hilbert space and requires a projection to the original subspace to obtain physical results.
It is not straightforward to deal with the projection in the numerical methods~\cite{note1}.
On the other hand, such a projection is not necessary in the latter representation, as the size of the Hilbert space is retained.

\begin{figure}[t]
 \begin{center}
  \includegraphics[width=0.8\columnwidth,clip]{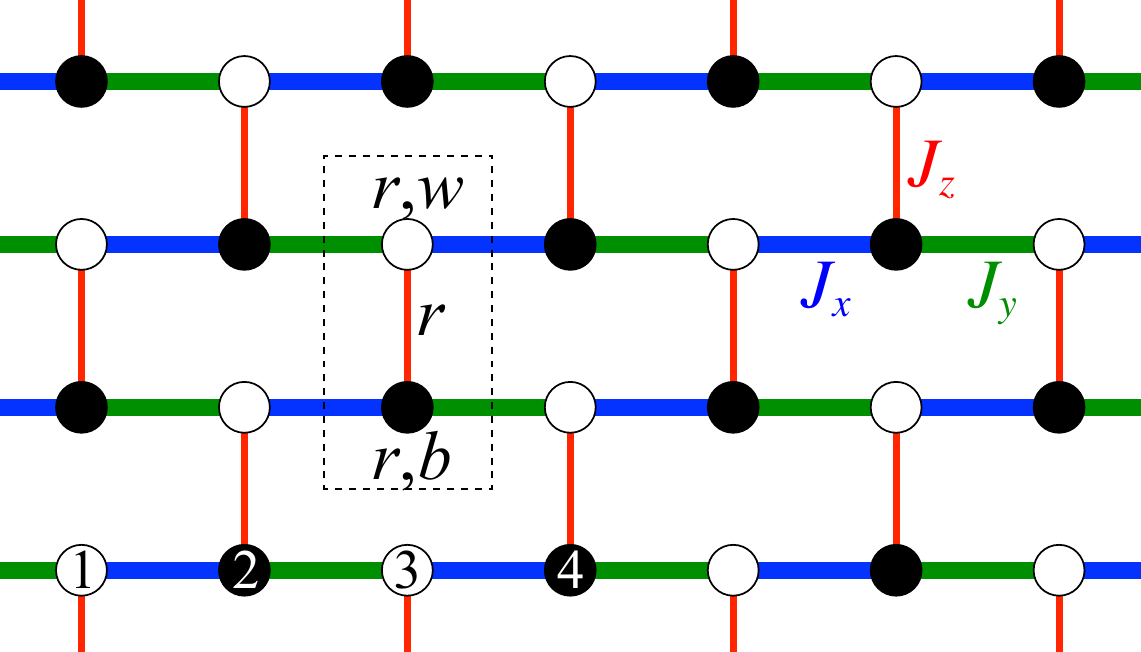}
  \caption{
(Color online) 
Schematic picture of the one-dimensional chains consisting of the $x$ and $y$ bonds, which are shown by the thick blue and green lines, respectively.
The honeycomb structure in Fig.~\ref{fig:lattice} is deformed into a brick-wall structure.
The dotted square represents the unit cell including the $r$th $z$ bond; the two sites are denoted as $r,b$ and $r,w$ [see Eqs.~(\ref{eq:H_spinless}) and (\ref{eq:H_Maj})]. 
In the Jordan-Wigner transformation in Eq.~(\ref{eq:JWtrsf}), the sites are numbered from the bottom left as partly shown in the figure.
}
  \label{fig:lattice-jw}
 \end{center}
\end{figure}

In the Majorana representation, we first apply the Jordan-Wigner transformation to the model in Eq.~(\ref{eq:H}), by regarding the system as a one-dimensional chain composed of the $x$ and $y$ bonds; see Fig.~\ref{fig:lattice-jw}.
In the Jordan-Wigner transformation, the spin operators are rewritten by spinless fermion operators as
\begin{align}
S_i^+ = (S_i^-)^\dagger = S_i^x + i S_i^y=\prod_{i'=1}^{i-1}(1-2n_{i'}) a_i^\dagger, \ \
S_i^z = n_i - \frac12,
\label{eq:JWtrsf}
\end{align}
where $a_i^\dagger$ and $a_i$ are the creation and annihilation operators for the spinless fermions, respectively, and $n_i = a_i^\dagger a_i$ is the number operator;
$a_i^\dagger$ and $a_i$ satisfy the anticommutation relations as
\begin{align}
\{a_i^\dagger, a_j\} = \delta_{ij}, \ \
\{a_i^\dagger, a_j^\dagger\} = 0, \ \
\{a_i, a_j\} = 0,
\end{align}
where $\delta_{ij}$ is the Kronecker delta.
Then, by considering that the honeycomb structure is bipartite, the Hamiltonian in Eq.~(\ref{eq:H}) is transformed into
\begin{align}
{\cal H} &= \frac{J_x}{4} \sum_{\langle r',w;r,b\rangle_x} (a_{r',w}-a_{r',w}^\dagger)(a_{r,b}+a_{r,b}^\dagger) \nonumber \\
&- \frac{J_y}{4} \sum_{\langle r,b;r',w\rangle_y} (a_{r,b}+a_{r,b}^\dagger)(a_{r',w}-a_{r',w}^\dagger) \nonumber \\
&- \frac{J_z}{4} \sum_r (2n_{r,b}-1)(2n_{r,w}-1),
\label{eq:H_spinless}
\end{align}
where the subscripts $b$ and $w$ label the two sublattices in the $r$th unit cell with one $z$ bond (see Figs.~\ref{fig:lattice} and \ref{fig:lattice-jw}); $r,b$ and $r',w$ in the sums in the first and second terms are taken for all NN pairs on the $x$ and $y$ bonds, colored by blue and green in Fig.~\ref{fig:lattice-jw}, respectively.
Note that the so-called boundary terms appear in the Jordan-Wigner transformation for the systems under periodic boundary conditions.
The boundary terms are nonlocal and hard to treat in the numerical simulations.
One way to avoid this is to consider the systems under open boundary conditions.
Another is just to neglect the boundary terms; their contributions are expected to be negligible in the thermodynamic limit.

Next, we replace the spinless fermion operators by Majorana fermion operators.
This is done by
\begin{align}
& \gamma_{r,w} = \frac{a_{r,w} - a_{r,w}^\dagger}{i}, \ \ \
\bar{\gamma}_{r,w} = a_{r,w} + a_{r,w}^\dagger,
\label{eq:gamma_w} \\
& \gamma_{r,b} = a_{r,b} + a_{r,b}^\dagger, \ \ \
\bar{\gamma}_{r,b} = \frac{a_{r,b} - a_{r,b}^\dagger}{i},
\label{eq:gamma_b}
\end{align}
where $\gamma$ and $\bar{\gamma}$ are the Majorana fermion operators.
These are the same as Eq.~(\ref{eq:f_g}).
By using Eqs.~(\ref{eq:gamma_w}) and (\ref{eq:gamma_b}), Eq.~(\ref{eq:H_spinless}) is rewritten into
\begin{align}
{\cal H} &=  \frac{i J_x}{4} \sum_{\langle r',w;r,b\rangle_x} \gamma_{r',w} \gamma_{r,b}
- \frac{i J_y}{4} \sum_{\langle r,b;r',w\rangle_y} \gamma_{r,b} \gamma_{r',w} \nonumber \\
&- \frac{i J_z}{4} \sum_r\eta_r \gamma_{r,b} \gamma_{r,w},
\label{eq:H_Maj}
\end{align}
where $\eta_r$ in the last term is defined on the $z$ bond as
\begin{align}
\eta_r = i \bar{\gamma}_{r,b} \bar{\gamma}_{r,w}.
\label{eq:eta_r}
\end{align}

\begin{figure}[t]
 \begin{center}
  \includegraphics[width=0.9\columnwidth,clip]{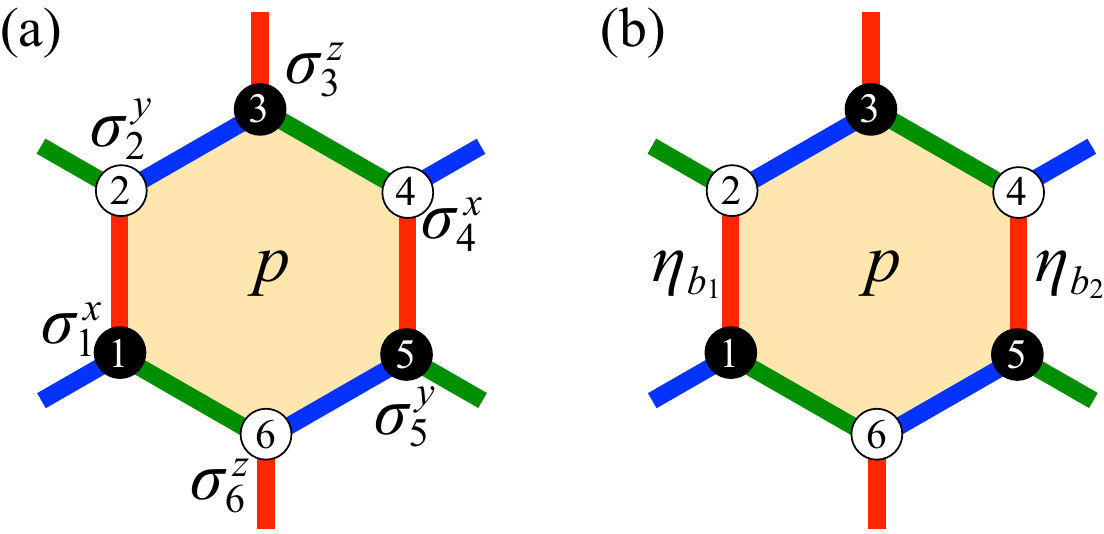}
  \caption{
(Color online) 
Representations of the $Z_2$ flux $W_p$ for a plaquette $p$ by using (a) the spin operators $\sigma_j^{\tilde{\mu}}$ at the six vertices [Eq.~(\ref{eq:Wp1})] and (b) the $Z_2$ variables $\eta_r$ on the two $z$ bonds [Eq.~(\ref{eq:Wp2})].
}
  \label{fig:hexagon}
 \end{center}
\end{figure}

The bond variable $\eta_r$ in Eq.~(\ref{eq:eta_r}) satisfies the following relations:
\begin{align}
&[{\cal H}, \eta_r] = 0, \ \ \eta_r^2 =1 \ \ \mbox{for all $r$}, \\
&[\eta_r, \eta_{r'}]=0. 
\end{align}
This means that each $\eta_r$ is a constant of motion and takes $\pm 1$.
Thus, $\{\eta_r \}$ are $Z_2$ conserved quantities.
It is worth noting that they are related with another conserved quantities called the $Z_2$ fluxes denoted by $W_p$, which were introduced in the paper by Kitaev~\cite{Kitaev2006}.
$W_p$ is defined for each hexagonal plaquette on the honeycomb structure as
\begin{align}
W_p = \prod_{j \in p} \sigma_j^{\bar{\mu}},
\label{eq:Wp1}
\end{align}
where the product is taken for the six sites on the plaquette $p$ in the clockwise manner [see Fig.~\ref{fig:hexagon}(a)]; $\bar{\mu}$ is the index for the bond connected to the site $i$ which is not included in the sides of $p$, and $\sigma_i^\mu$ is the $\mu$th component of the Pauli matrices ($S_j^\mu = \frac{\hbar}{2}\sigma_j^\mu$, where $\hbar$ is the reduced Planck constant and taken to be unity hereafter).
By using the algebra of the Pauli matrices and the equations above, $W_p$ is also expressed as
\begin{align}
W_p = \prod_{r\in p} \eta_r,
\label{eq:Wp2}
\end{align}
where the product is taken for the two $z$ bonds belonging to the hexagonal plaquette $p$ [see Fig.~\ref{fig:hexagon}(b)].

\subsection{Quantum spin liquid ground state}
\label{sec:groundstate}

\begin{figure*}[t]
 \begin{center}
  \includegraphics[width=1.6\columnwidth,clip]{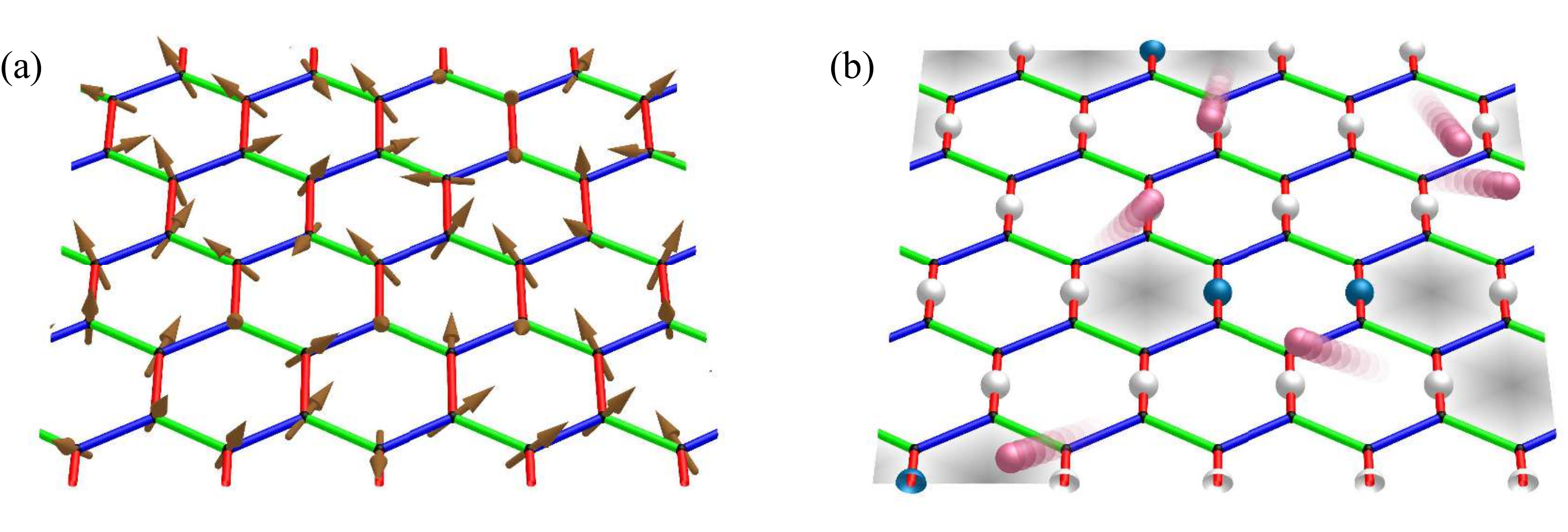}
  \caption{
(Color online) 
Schematic figures of the Kitaev model in (a) the spin representation in Eq.~(\ref{eq:H}) and (b) the Majorana representation in Eq.~(\ref{eq:H_Maj}).
The arrows in (a) represent the spins $\mathbf{S}_i$.
In (b), the itinerant Majorana fermions $\gamma_i$ are represented by the pink spheres, and the localized $Z_2$ variables $\eta_r$ taking $+1$ ($-1$) are by the white (blue) spheres.
The gray hexagons in (b) stand for the excited fluxes with $W_p=-1$.
}
  \label{fig:majorana}
 \end{center}
\end{figure*}

The Majorana representation of the Hamiltonian in Eq.~(\ref{eq:H_Maj}) shows that the original spin model in Eq.~(\ref{eq:H}) is mapped to the system with itinerant Majorana fermions $\{\gamma_j \}$ coupled with the $Z_2$ conserved variables $\{\eta_r \}$ or the $Z_2$ fluxes $\{W_p \}$ via Eq.~(\ref{eq:Wp2}).
The situation is schematically shown in Fig.~\ref{fig:majorana}.
The Hamiltonian is in a bilinear form of $\{\gamma_j \}$, namely, there is no quantum interactions between the Majorana fermions $\{\gamma_i \}$; they interact only with the $Z_2$ variables $\{\eta_r \}$.
This means that the Hamiltonian can be written in a block diagonalized form classified by different configurations of $\{\eta_r \}$ or $\{W_p \}$ as follows. 
The total Hamiltonian matrix with the dimension $2^N$ is decomposed into a direct sum of the sectors specified by $\{W_p \}$ configurations. 
The number of $\{W_p\}$ configurations is $2^{N/2}$. 
The block Hamiltonian in each sector has thus the dimension $2^N / 2^{N/2} = 2^{N/2}$, and it is represented by a $N\times N$ bilinear form of Majorana operators with hopping matrix elements including $\{W_p\}$ as $c$-numbers. 
This decomposition enables one to find the ground state, in principle, by comparing the energies in all the sectors, as the energy in each sector is easily obtained for the noninteracting fermion problem.

For this problem, a mathematical proof, called Lieb's theorem, offers the exact solution for the lowest energy state~\cite{Lieb1994}.
This theorem tells the flux configuration which gives the lowest energy state in the systems with mirror symmetry with respect to the plane not including the lattice sites.
In the present model on the honeycomb structure, we can apply this theorem to the cases when at least two of three $J_\mu$ are equal.
The exact ground state for these symmetric cases is given in the sector with all $W_p=+1$, which is called the flux-free state.
On the other hand, Lieb's theorem does not apply to the cases with generic $J_\mu$.
Nevertheless, by comparing the energies for different configurations of $\{W_p \}$, the flux-free state is deduced to be the ground state in the entire parameter space of $J_\mu$~\cite{Kitaev2006}.

\begin{figure}[t]
 \begin{center}
  \includegraphics[width=\columnwidth,clip]{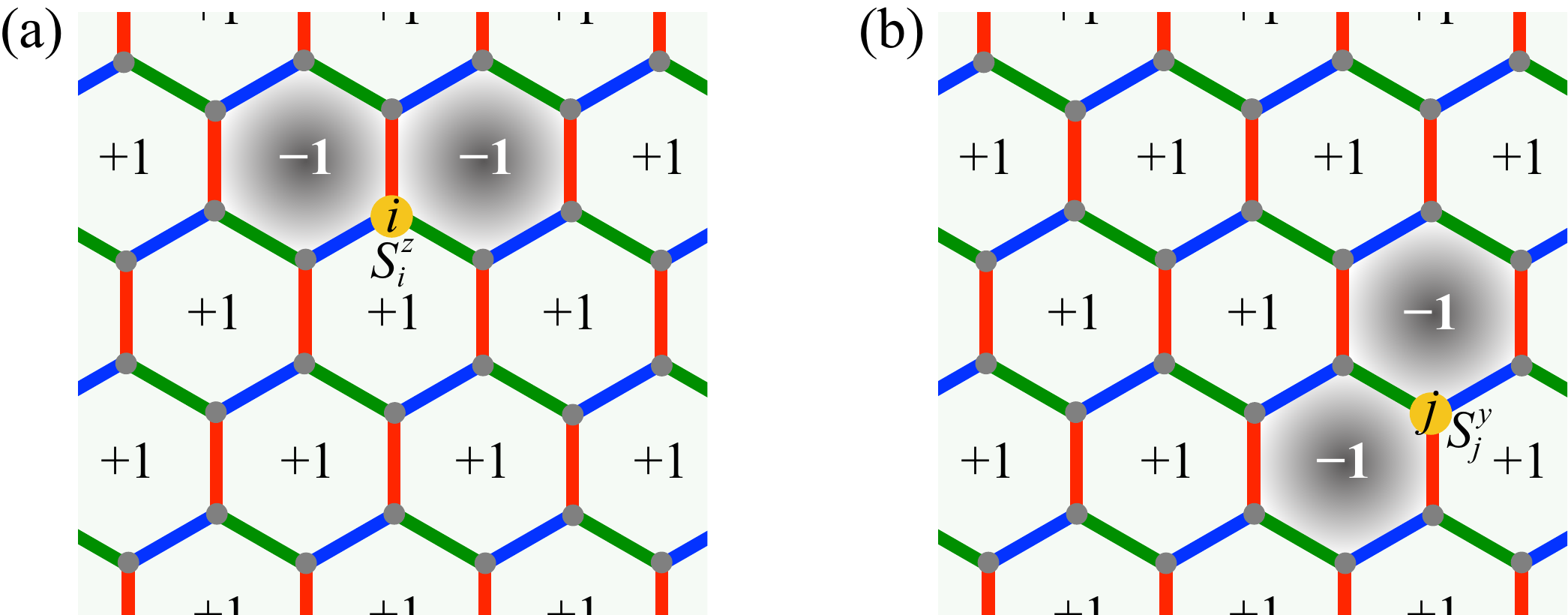}
  \caption{
(Color online) 
Configurations $\{W_p\}$ for the states (a) $S_i^z\kets{\Psi}$ and (b) $S_j^y\kets{\Psi}$, where $\kets{\Psi}$ represents the flux-free state.
$S_i^\mu$ flips two $W_p$ on both sides of the $\mu$ bond connected to the site $i$.
As the states with different $\{W_p\}$ are orthogonal to each other, $\bras{\Psi}S_i^z S_j^y\kets{\Psi}=0$.
}
  \label{fig:flip_flux}
 \end{center}
\end{figure}

The flux-free state is a QSL.
This was explicitly shown by calculating the spin correlations~\cite{Baskaran2007}.
The spin correlations have nonzero values only for the $\mu$ components on the NN $\mu$ bonds as well as the same sites, namely,
\begin{align}
\langle S_i^\mu S_j^\nu \rangle \neq 0 \ \ {\rm only \ for} \ \ \mu=\nu \ \ {\rm and} \ \ i,j \in \langle i,j \rangle_\mu.
\label{eq:SS}
\end{align}
All other further-neighbor correlations vanish.
This is concluded from the fact that a spin operator $S_i^\mu$ flips two neighboring $W_p$ sandwiching the $\mu$ bond including the site $i$; only the combinations of $S_i^\mu$ and $S_j^\nu$ satisfying the condition in Eq.~(\ref{eq:SS}) conserve the flux-free configuration of $W_p$ (see Fig.~\ref{fig:flip_flux}).
Thus, the spin correlations are extremely short-ranged in the flux-free state.
This means that the flux-free ground state does not break any symmetry of the system, and hence, it is a rare realization of the exact QSL in more than one dimension.

Note that Eq.~(\ref{eq:SS}) holds for arbitrary flux configurations.
This suggests that further-neighbor spin correlations beyond the NN sites are always zero even at finite $T$ where fluxes with $W_p=-1$ are thermally excited.
This is indeed confirmed by numerical studies introduced in Sec.~\ref{sec:finiteT} and Appendix.

\subsection{Fractional excitations}
\label{sec:frac}

For the flux-free ground state, there are two types of excitations.
One is the excitations in terms of the itinerant Majorana fermions $\{\gamma_j \}$, and the other is for the $Z_2$ fluxes $\{W_p \}$.
These are quasiparticle excitations arising from the fractionalization of the spin degree of freedom.

\begin{figure*}[t]
 \begin{center}
  \includegraphics[width=2\columnwidth,clip]{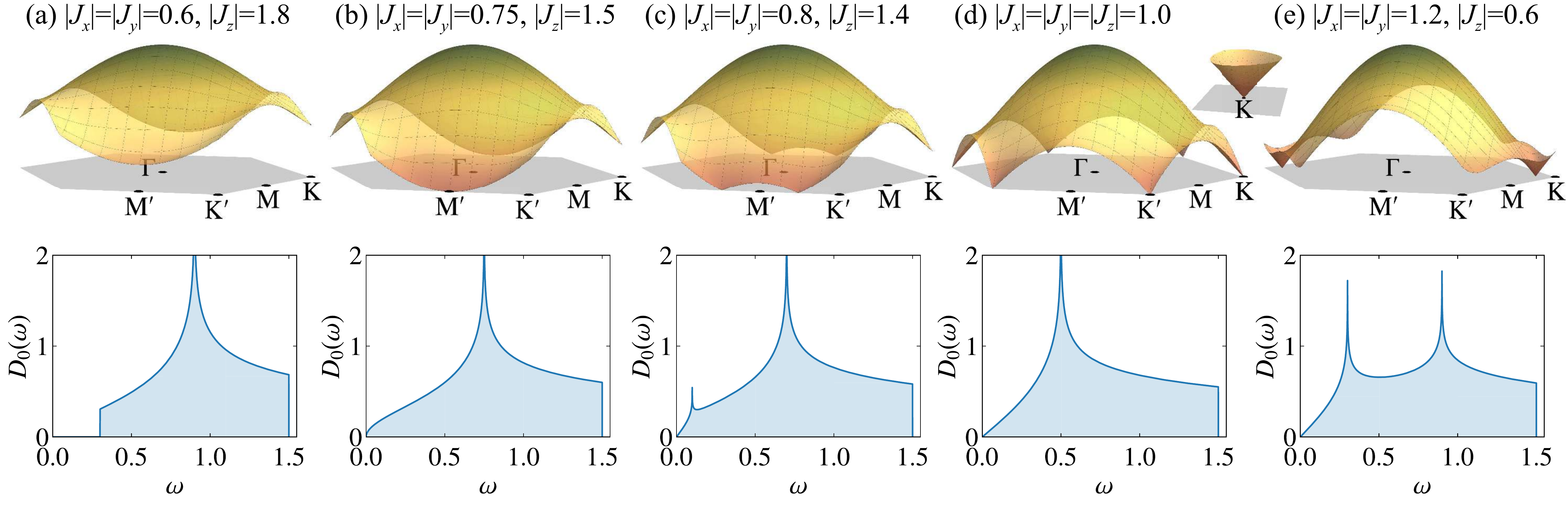}
  \caption{
(Color online) 
Dispersion relations of the complex fermion band in the first Brillouin zone and the density of states $D_0(\omega)$ for the flux-free state at several sets of the exchange parameters with $|J_x|+|J_y|+|J_z|=3$.
The inset of (d) shows the extended plot of the Dirac-like linear dispersion around the K point.
}
  \label{fig:band}
 \end{center}
\end{figure*}

The former excited states are constructed by exciting complex fermions $\{f_k^\dagger\}$, which are comprised as linear combinations of Majorana fermions $\{\gamma_j\}$ with complex amplitudes (see Appendix~\ref{sec:quantum-monte-carlo}). 
They are noninteracting fermions traversing on the honeycomb structure with the NN hopping.
The dispersion relation is given by~\cite{Kitaev2006}
\begin{align}
E(\mathbf{k}) = \left| \varepsilon(\mathbf{k}) \right|,
\label{eq:disp}
\end{align}
where
\begin{align}
\varepsilon(\mathbf{k}) = \frac{1}{2} \left\{J_x \exp(i \mathbf{k}\cdot\mathbf{a}_1)
+ J_y \exp(i \mathbf{k}\cdot\mathbf{a}_2) + J_z \right\}.
\end{align}
Here, $\mathbf{a}_1 = (\frac12, \frac{\sqrt{3}}{2})$ and $\mathbf{a}_2 = (-\frac12, \frac{\sqrt{3}}{2})$ are the primitive translation vectors (see Fig.~\ref{fig:lattice}), whose lengths are taken to be unity.
The dispersion relation in Eq.~(\ref{eq:disp}) is depicted in Fig.~\ref{fig:band} for several sets of the parameters $J_x$, $J_y$, and $J_z$.
In the isotropic case with $J_x=J_y=J_z$, $E(\mathbf{k})$ becomes gapless at the point nodes located at the corners of the Brillouin zone (K and K' points), as shown in the upper panel of Fig.~\ref{fig:band}(d).
Near the gapless nodal points, $E(\mathbf{k})$ has a linear dispersion, similar to the Dirac nodes in the dispersion of $\pi$ electrons in graphene, as shown in the inset of Fig.~\ref{fig:band}(d).
This leads to the $\omega$-linear dependence of the density of states (DOS) in the low-energy limit, as shown in the lower panel of Fig.~\ref{fig:band}(d).
The gapless nature is retained for small anisotropy in $J_x$, $J_y$, and $J_z$, despite a shift of the nodal points; see Figs.~\ref{fig:band}(c) and \ref{fig:band}(e).
The two nodal points approach each other while increasing the anisotropy, and finally merge at some point, as exemplified for $|J_z| = |J_x| + |J_y|$ in Fig.~\ref{fig:band}(b).
With a further increase of the anisotropy, $E(\mathbf{k})$ is gapped in the entire Brillouin zone, as exemplified in Fig.~\ref{fig:band}(a).

\begin{figure*}[t]
 \begin{center}
  \includegraphics[width=1.9\columnwidth,clip]{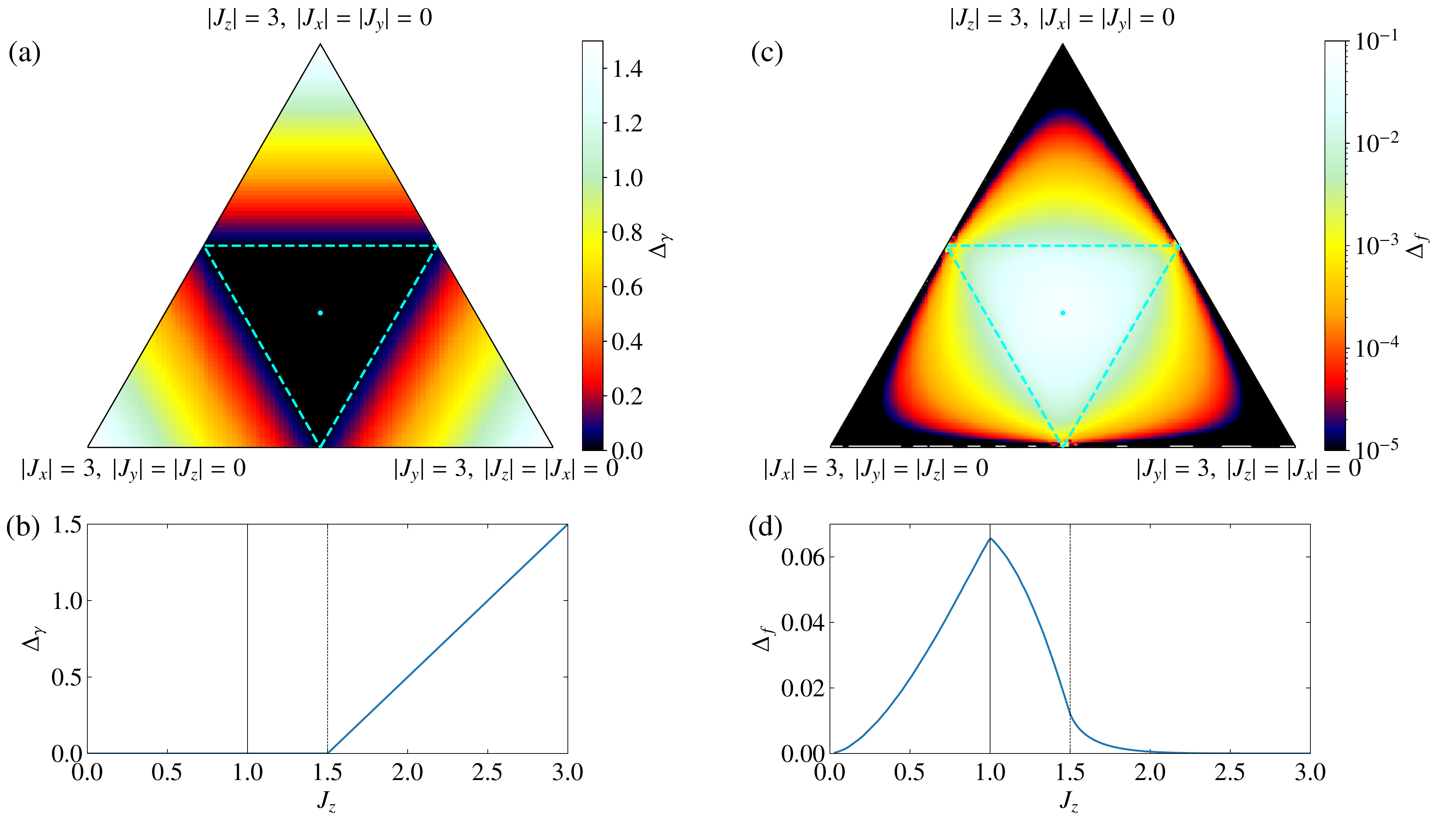}
  \caption{
(Color online) 
(a) Excitation gap in the itinerant fermion band in the flux-free ground state, $\Delta_\gamma$, on the plane of $|J_x|+|J_y|+|J_z|=3$.
The cyan dot at the center stands for the isotropic point and the dotted lines represent the
boundaries between the gapless and gapped phases.
(b) $J_z$ dependence of $\Delta_\gamma$ with $J_x=J_y=(3-J_z)/2$ corresponding to the cut along the vertical line through the isotropic point in (a).
(c) and (d) Corresponding plots for the flux gap $\Delta_f$, which is defined by the lowest energy change by flipping two neighboring $W_p$.
See the phase diagram in Fig.~5 in Ref.~\citen{Kitaev2006}.
}
  \label{fig:aniso}
 \end{center}
\end{figure*}

The magnitude of the excitation gap $\Delta_\gamma$ is plotted in the entire parameter space in Fig.~\ref{fig:aniso}(a). 
$\Delta_\gamma$ is zero in the center triangle defined by the conditions $|J_x| \leq |J_y| + |J_z|$, $|J_y| \leq |J_z| + |J_x|$, and $|J_z| \leq |J_x| + |J_y|$ (dashed lines in the figure).
Meanwhile, $\Delta_\gamma$ becomes nonzero in the other three outer triangles and increases as increasing the anisotropy in the Kitaev coupling; the contours are parallel to the gapless-gapped boundaries.
The $J_z$ dependence of the gap is shown in Fig.~\ref{fig:aniso}(b) along the center vertical line in Fig.~\ref{fig:aniso}(a) [$J_x=J_y=(3-J_z)/2$], indicating that $\Delta_\gamma$ increases linearly with $J_z$ in the gapped phase for $J_z>1.5$.
Thus, there are two different phases with respect to the excitations of the itinerant Majorana fermions:
the gapless phase including the isotropic point and the gapped one including the anisotropic limits.

On the other hand,  the other types of excitations are generated by flipping $W_p$ from the flux-free ground state~\cite{Kitaev2006}.
It turns out that they are always gapped and dispersionless reflecting the localized nature of $W_p$.
The lowest-energy excited state is given by a pair flip of neighboring two $W_p$.
The excitation gap $\Delta_f$ is plotted on the $J_x$-$J_y$-$J_z$ phase diagram in Fig.~\ref{fig:aniso}(c).
The gap is nonzero in the entire parameter space, except for the anisotropic limits at the three corners of the phase diagram; it remains small in the gapped phases in Fig.~\ref{fig:aniso}(a) but becomes large rapidly in the gapless phase.
As shown in Fig.~\ref{fig:aniso}(d),  along the center vertical line in Fig.~\ref{fig:aniso}(c), $\Delta_f$ becomes maximum at the isotropic point with $J_x=J_y=J_z$.

Thus, the two different types of the fractional excitations have distinct excitation spectra.
The fermionic excitations from the itinerant Majorana fermions are dispersive and become both gapless and gapped depending on the anisotropy in the exchange coupling constants.
Meanwhile, the $Z_2$ flux excitations are always gapped with a flat dispersion.
The energy scales are also largely different for these two excitations;
the bandwidth for the former is roughly set by the sum of three $J_\mu$, while the excitation gap for the latter is much smaller by more than one order of magnitude.
This large difference in the energy scales affects the thermodynamics and the spin dynamics in a peculiar fashion, as discussed in the later sections.

\subsection{Effect of finite temperature}
\label{sec:finiteT}

The exact solution and related arguments above are limited to zero temperature ($T=0$).
At finite $T$, the $Z_2$ flux excitations are generated by thermal fluctuations, and the exact solution is no longer available.
As discussed in Sec.~\ref{sec:groundstate}, however, the model in Eq.~(\ref{eq:H_Maj}) is defined by noninteracting fermions coupled with thermally-fluctuating $Z_2$ variables $\{ \eta_r \}$.
As $\{ \eta_r \}$ are regarded as classical variables taking $\pm 1$, the situation is similar to the Falicov-Kimball model~\cite{Falicov1969} and the double-exchange model with Ising spins~\cite{Zener1951,Motome2001}.
This enables us to study the finite-$T$ properties by developing numerical techniques similar to those used for such fermion models.
The authors and their collaborators have developed the quantum Monte Carlo (QMC) method free from the negative sign problem~\cite{Nasu2014,Nasu2015,Mishchenko2017} and the cluster dynamical mean-field theory (CDMFT)~\cite{Yoshitake2016,Yoshitake2017a}.
These Majorana-based techniques are efficient to compute thermodynamic quantities, but they cannot be applied to the quantities not commuting with $\{\eta_r\}$, e.g., dynamical spin correlations.
To overcome this difficulty, the authors and their collaborators have also developed the continuous-time QMC (CTQMC) method based on the Majorana representation~\cite{Yoshitake2016,Yoshitake2017a,Yoshitake2017b}.
The details of each method are presented in Appendix.

As will be described in detail in Sec.~\ref{sec:Tfrac}, an interesting finding at finite $T$ is that the two distinct fractional excitations manifest themselves clearly in the thermodynamic behavior of the system.
Specifically, in the 2D honeycomb case, the two largely different energy scales lead to two crossovers at largely different temperatures.
One appears at $T=T_H$ in the order of the characteristic energy scale of the itinerant Majorana fermions [more precisely, the center of mass (COM) of the DOS for the fermion band; see Sec.~\ref{sec:Tfrac}], and the other takes place at $T=T_L$ in the order of the excitation gap in terms of the localized $Z_2$ fluxes.
These two characteristic temperatures show up in many observables, not only thermodynamic quantities,
but also the spin dynamics, as discussed in the later sections.

\subsection{Effect of a magnetic field}
\label{sec:field}

\begin{figure}[t]
 \begin{center}
  \includegraphics[width=0.7\columnwidth,clip]{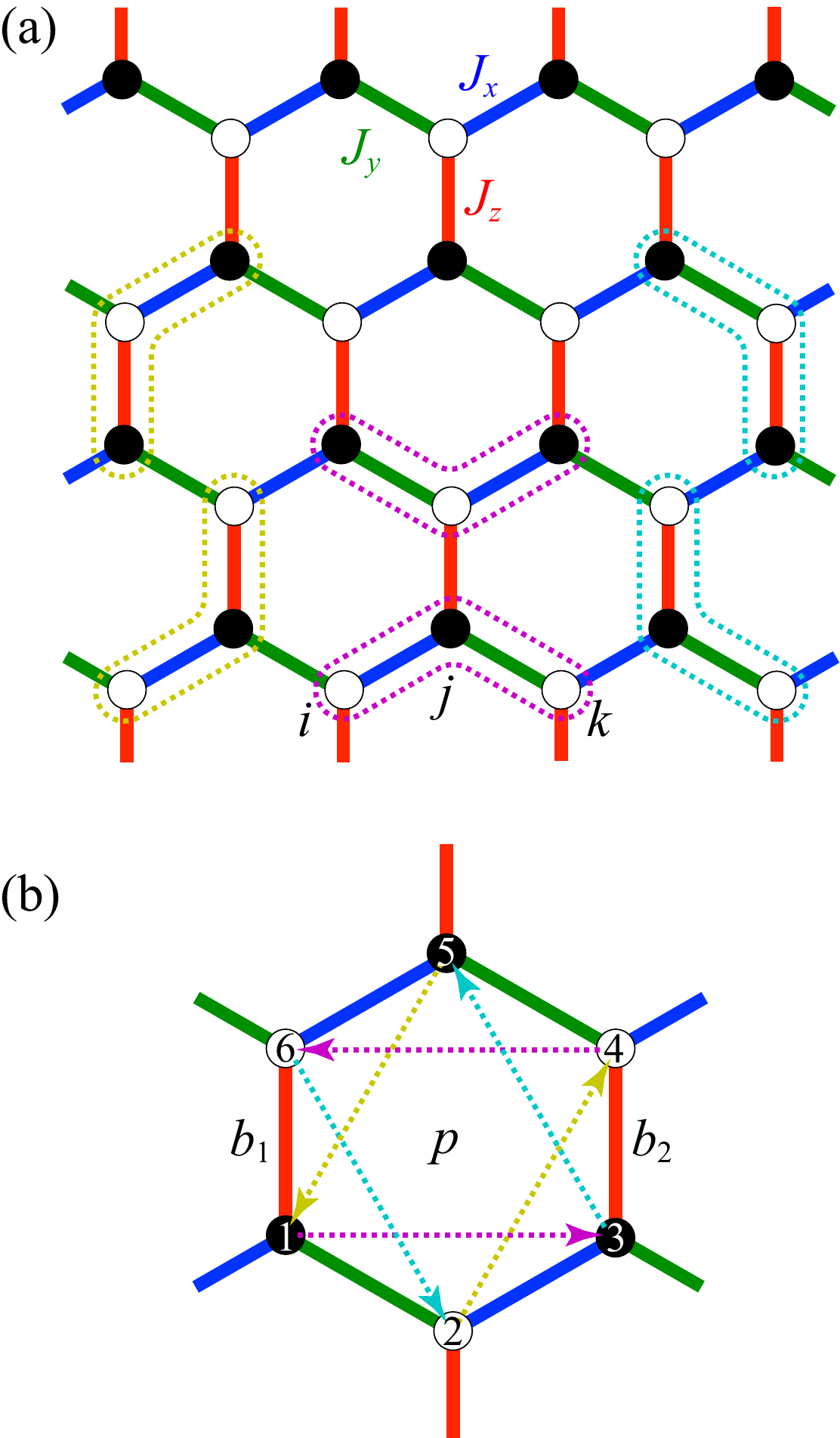}
  \caption{
(Color online) 
(a) Six kinds of neighboring three sites $\{ i,j,k \}$ in Eq.~(\ref{eq:H'_spin}).
(b) Second-neighbor hoppings of Majorana fermions in Eq.~(\ref{eq:H'_Maj}).
The color of the arrows indicates the corresponding type of the three-site terms in (a).
}
  \label{fig:lattice_mag}
 \end{center}
\end{figure}

Let us return to the flux-free ground state and discuss the effect of an external magnetic field at $T=0$.
The Zeeman coupling to the magnetic field, $- \mathbf{h} \cdot \sum_i \mathbf{S}_i$, spoils the exact solvability, because it makes the flux operators $W_p$ in Eq.~(\ref{eq:Wp1}) and $\eta_r$ in Eq.~(\ref{eq:eta_r}) no longer conserved.
(Note that the sign of the $g$ factor is opposite to that for electron spins, as discussed in Sec.~\ref{sec:Jackeli-Khaliullin}.) 
Nonetheless, Kitaev suggested an interesting possibility by using the perturbation theory with respect to the field strength~\cite{Kitaev2006}.
In the perturbation theory, the lowest-order relevant term is in the third order of $\mathbf{h}$ as
\begin{align}
{\cal H}' = - \tilde{h} \sum_{\{i,j,k \}} S_i^x S_j^y S_k^z \propto - \frac{h_x h_y h_z}{J^2} \sum_{\{i,j,k \}} S_i^x S_j^y S_k^z,
\label{eq:H'_spin}
\end{align}
where $\mathbf{h} = (h_x,h_y,h_z)$ and the Kitaev couplings are set to be isotropic, $J_x=J_y=J_z=J$, for simplicity; here, all the intermediate states are assumed to have an excitation energy of $J$.
The sum of $\{ i,j,k \}$ is taken for neighboring three sites [see Fig.~\ref{fig:lattice_mag}(a)].
Note that $\{W_p\}$ and $\{\eta_r\}$ remain conserved within the perturbation theory since the flux configurations are identical between the initial and final states by definition.

\begin{figure}[t]
 \begin{center}
  \includegraphics[width=\columnwidth,clip]{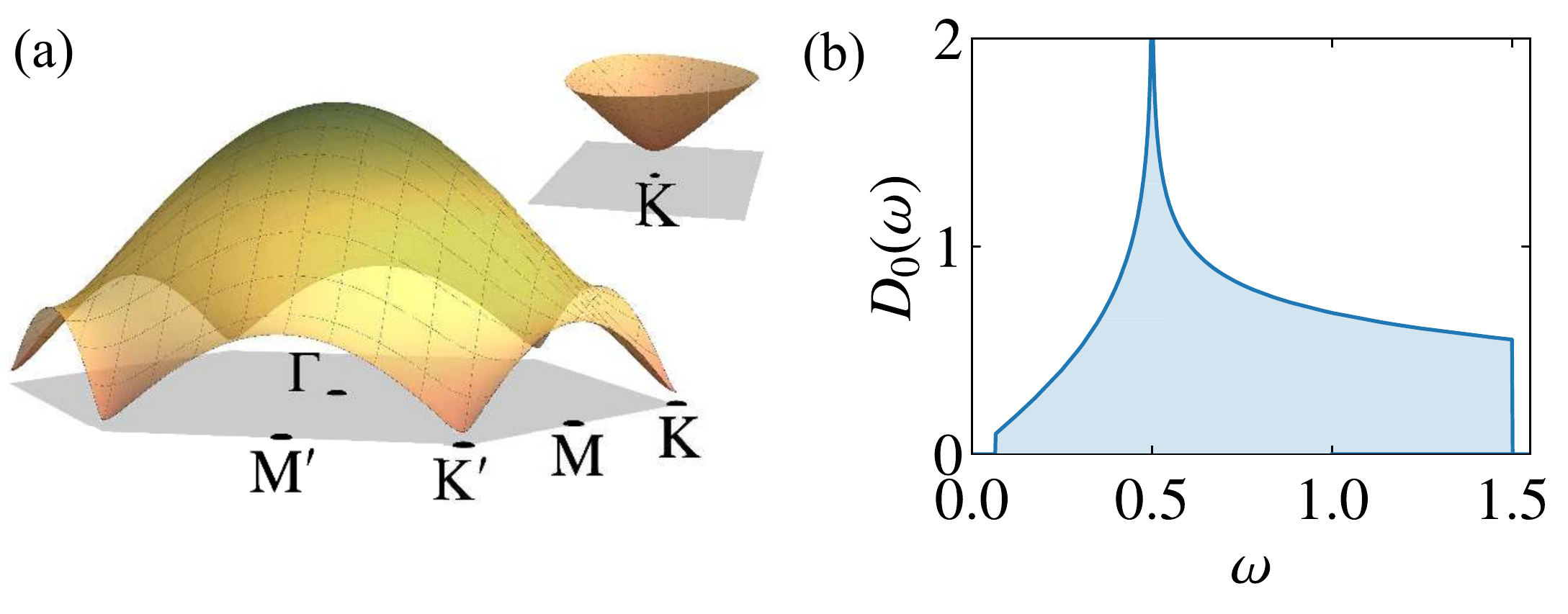}
  \caption{
(Color online) 
(a) Dispersion relation of the complex fermion band in the first Brillouin zone for the isotropic case $J_x=J_y=J_z=J$ with the effective magnetic field $\tilde{h}=0.05J$.
The inset shows the extended plot of the gapped dispersion around the K point.
(b) Corresponding DOS.
}
  \label{fig:band_mag}
 \end{center}
\end{figure}

\begin{figure}[t]
 \begin{center}
  \includegraphics[width=0.85\columnwidth,clip]{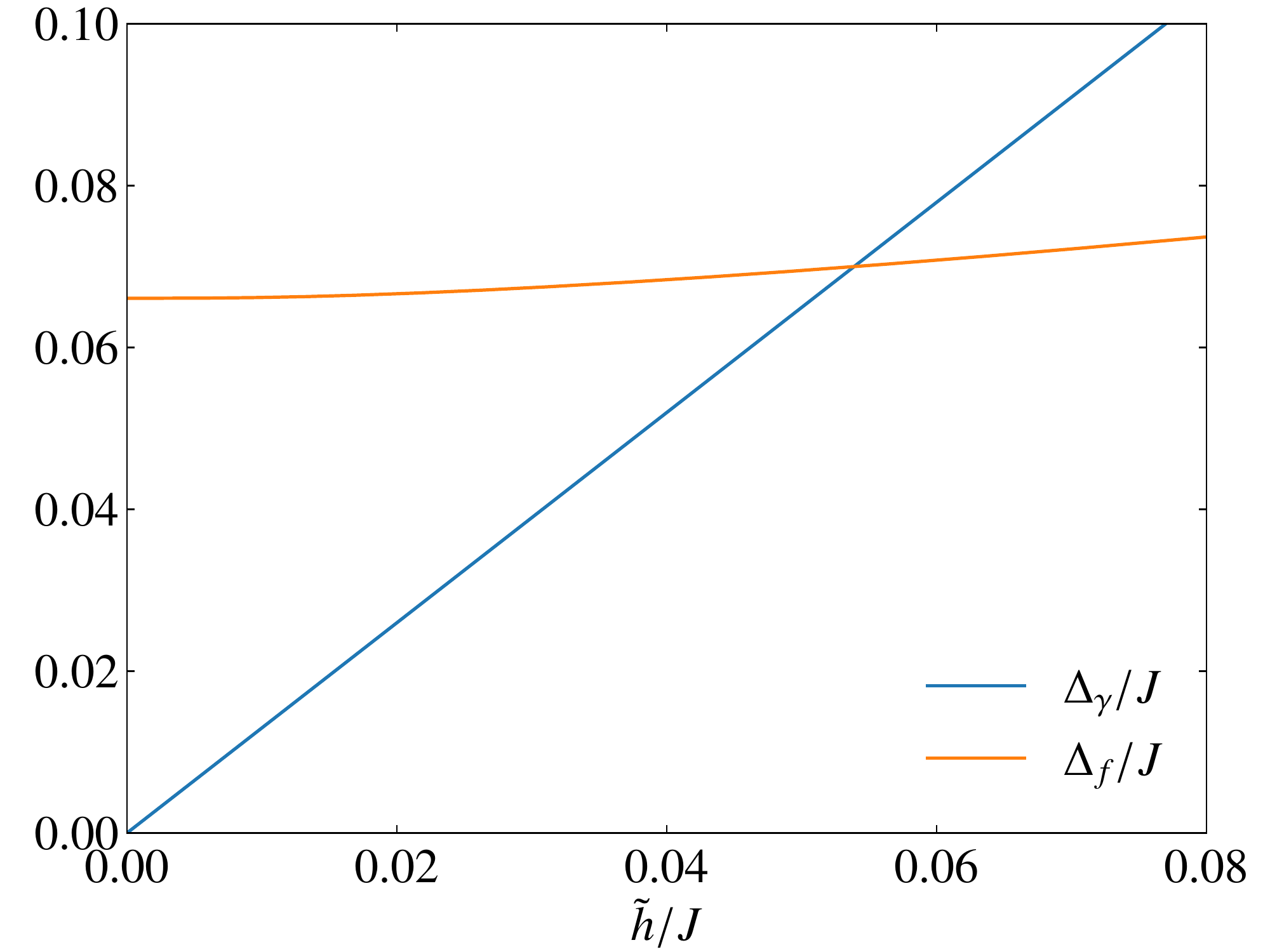}
  \caption{
(Color online) 
Gaps in the itinerant fermion band, $\Delta_\gamma$, and for the flux excitation, $\Delta_f$, as functions of the effective magnetic field $\tilde{h}$ in the isotropic case $J_x=J_y=J_z=J$.
The same plot is found in the Supplemental Material for Ref.~\citen{Nasu2017a}.
}
  \label{fig:eff_mag}
 \end{center}
\end{figure}

By using the Majorana representation in Sec.~\ref{sec:Majrep}, Eq.~(\ref{eq:H'_spin}) is written in the form
\begin{align}
{\cal H}' = - \frac{i \tilde{h}}{8} \sum_p &(\gamma_{p_1}\gamma_{p_3} + \eta_{b_2}\gamma_{p_3}\gamma_{p_5} + \eta_{b_1}\gamma_{p_5}\gamma_{p_1} \nonumber\\
&+ \gamma_{p_4}\gamma_{p_6} + \eta_{b_1}\gamma_{p_6}\gamma_{p_2} + \eta_{b_2}\gamma_{p_2}\gamma_{p_4}),
\label{eq:H'_Maj}
\end{align}
where the sites $p_1$--$p_6$ and the bonds $b_1$ and $b_2$ are defined for the plaquette $p$ as shown in Fig.~\ref{fig:lattice_mag}(b)~\cite{Nasu2017a}.
Equation~(\ref{eq:H'_Maj}) shows that the weak magnetic field induces the complex second-neighbor hopping of the itinerant Majorana fermions coupled with the $Z_2$ bond variables $\{\eta_r \}$.
This modulates the dispersion relation from Eq.~(\ref{eq:disp}) to~\cite{Kitaev2006}
\begin{align}
E(\mathbf{k}) = \pm \sqrt{|\varepsilon(\mathbf{k})|^2 + \Delta(\mathbf{k})^2},
\end{align}
where
\begin{align}
 \Delta(\mathbf{k}) = \frac{\tilde{h}}{2} \left\{ -\sin(\mathbf{k} \cdot \mathbf{a}_1) + \sin(\mathbf{k} \cdot \mathbf{a}_2) +\sin[\mathbf{k} \cdot (\mathbf{a}_1-\mathbf{a}_2)] \right\}.
\end{align}
Thus, while the fermionic excitation in the isotropic case with $J_x=J_y=J_z$ has the gapless nodal points at the K and K' points [see Fig.~\ref{fig:band}(d)], the magnetic field opens a gap proportional to $\tilde{h} \propto h_x h_y h_z$ as $\Delta_\gamma=\frac34\sqrt{3}\tilde{h}$ (see Fig.~\ref{fig:band_mag}).
On the other hand, the flux gap $\Delta_f$ is almost independent of $\tilde{h}$.
These behaviors are plotted in Fig.~\ref{fig:eff_mag}.

\begin{figure}[t]
 \begin{center}
  \includegraphics[width=0.7\columnwidth,clip]{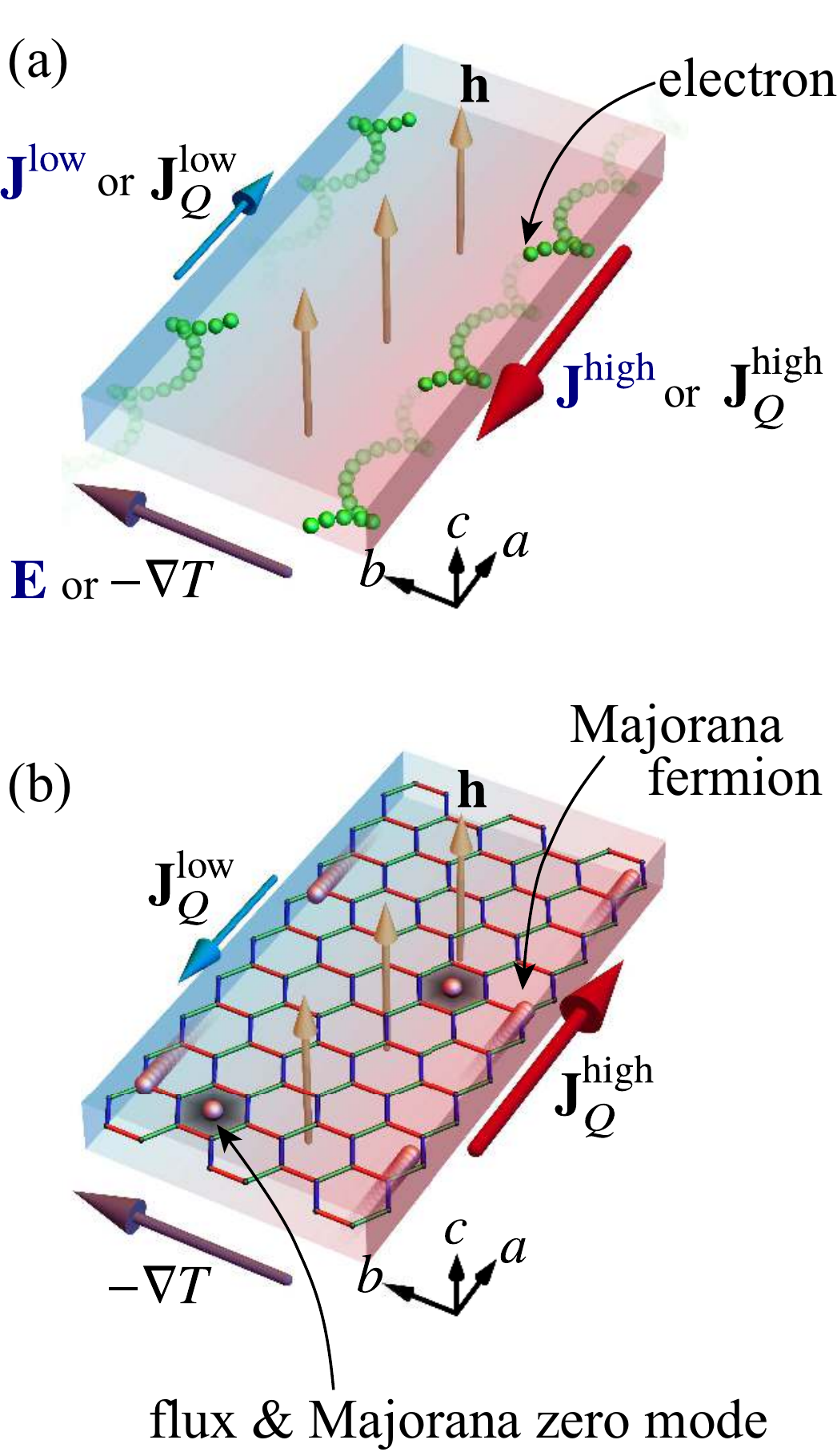}
  \caption{
(Color online) 
Schematic pictures of (a) the integer quantum Hall effect and (b) its Majorana counterpart expected for the Kitaev model under the magnetic field.
In (a), under the magnetic field $\textbf{h}$ perpendicular to the sample plane, an electric field $\textbf{E}$ (thermal gradient $-\nabla T$) causes unbalance in the edge electric (thermal) currents $\textbf{J}^{\rm high}$ ($\textbf{J}_Q^{\rm high}$) and $\textbf{J}^{\rm low}$ ($\textbf{J}_Q^{\rm low}$), which leads to the quantized (thermal) Hall effect.
In contrast, in (b), the Majorana fermions do not carry electric charge, and hence, edge electric currents do not appear under the electric field; however, edge thermal currents can appear under a thermal gradient.
In this case, the magnetic field is not necessarily perpendicular to the plane; any direction, even in the plane, leads to the thermal Hall effect.
In this situation, each excited flux is associated with a Majorana zero mode as schematically shown in (b), which behaves as a nonabelian anyon (see Sec.~\ref{sec:summary}).
}
  \label{fig:hall}
 \end{center}
\end{figure}

Interestingly, the model in the presence of the second-neighbor hopping in Eq.~(\ref{eq:H'_Maj}) is formally equivalent to a Majorana fermion version of the model for the spontaneous quantum Hall effect proposed by F.~Duncan~Haldane~\cite{Haldane1988}.
This equivalence shows that the gapped fermion band in the magnetic field is topologically nontrivial.
The gapped state is a Majorana Chern insulator with the Chern number $C=\pm 1$ (the sign is set by that of $h_x h_y h_z$~\cite{Kitaev2006}).
Thus, similar to other topologically-nontrivial insulators with nonzero Chern numbers, the gapped topological state in the weak magnetic field is predicted to possess gapless chiral edge modes~\cite{Kitaev2006}.
In contrast to the integer quantum Hall states, such chiral edge currents cannot be detected by electromagnetic measurements, as the Majorana fermions do not carry any electric charges;
however, they could be observed by heat measurements (see Fig.~\ref{fig:hall}).
There are two distinct features in this thermal Hall effect by the Majorana fermions. 
One is that the thermal Hall conductivity divided by $T$ is predicted to be quantized at half of that for the integer quantum Hall state~\cite{Kitaev2006}. 
This is because each Majorana fermion carries half degrees of freedom of an electron, as mentioned in Sec.~\ref{sec:introduction}. 
The other feature is that the half-quantized thermal Hall effect can be induced by the magnetic field in any direction, even in-plane directions.
This is because the chiral Majorana edge currents are induced by the Zeeman effect enhanced on the spins near the edges (see Sec.~III in Supplemental Material in Ref.~\citen{Nasu2017a}), in contrast to the electric edge currents from skipping orbits by the Lorentz force. 
This interesting phenomenon specific to the Majorana fermions will be discussed in Sec.~\ref{sec:kappa_xy}.

Beyond the perturbation theory, any rigorous argument is not available thus far.
Nonetheless, many numerical studies have been performed to clarify the effect of the magnetic field at $T=0$.
One of the earliest studies was done by the density matrix renormalization group for the Kitaev-Heisenberg model (see Sec~\ref{sec:other_exchanges})~\cite{Jiang2011}.
The Kitaev coupling was assumed to be isotropic and FM ($J_x=J_y=J_z=J>0$), and the magnetic field was applied along the [111] direction with the strength $h$.
The results indicate that the topologically-nontrivial QSL state predicted by the perturbation theory survives up to the critical field $h_c \simeq 0.018J$, and turns into a topologically-trivial forced FM state above $h_c$.
This has been confirmed, e.g., by the exact diagonalization and other density matrix renormalization group calculations~\cite{Yadav2016,Zhu2018,Gohlke2018,Hickey2019,Gordon2019}.

Recently, considerable attention has been drawn to the case with AFM Kitaev couplings.
While the perturbation theory above is common to the FM and AFM cases, different aspects appear between the two cases when going beyond the perturbation.
The most intriguing aspect is the possibility of another topological QSL in the intermediate-field region~\cite{Zhu2018,Gohlke2018,Nasu2018,Liang2018,Hickey2019,Ronquillo2019,Patel2019}.
It was argued that the AFM Kitaev model undergoes successive phase transitions from the low-field QSL connected to the topological QSL in the perturbed region to another topological QSL, and to the forced FM state, while increasing the field.
Although candidate materials with the AFM Kitaev couplings have not been identified thus far, this interesting possibility has attracted much interest.
Note that recently there are several theoretical proposals for material realization of the AFM Kitaev couplings, for instance, by using $f$ electrons~\cite{Jang2019} and polar asymmetry perpendicular to the honeycomb plane~\cite{Sugita2019preprint}.

Finite-$T$ calculations under a magnetic field are more difficult.
For instance, we cannot apply the sign-free Majorana-based QMC method, since it assumes the conservation of $\{W_p\}$ and $\{\eta_r\}$.
Nonetheless, one can study finite-$T$ properties of the Hamiltonian with the effective magnetic field in Eqs.~(\ref{eq:H'_spin}) and (\ref{eq:H'_Maj}) derived from the perturbation, by using the sign-free Majorana-based QMC method.
Such applications will be discussed in Sec.~\ref{sec:kappa_xy}.
In addition, a CTQMC technique has recently been developed and applied to the region where the negative sign problem is not severe, as discussed in Sec.~\ref{sec:Cv_S}, \ref{sec:Sqw}, and \ref{sec:1/T1}~\cite{Yoshitake2019preprint}.

\subsection{Effect of other exchange interactions}
\label{sec:other_exchanges}

As briefly mentioned in Sec.~\ref{sec:Jackeli-Khaliullin}, in reality, there exist other types of the exchange couplings.
A generic Hamiltonian proposed for realistic compounds is given by 
\begin{align}
{\cal H}_{\rm generic} = \sum_{\langle i,j \rangle} \mathbf{S}_i^{\rm T} \hat{J}_{\mu_{ij}} \mathbf{S}_j,
\end{align}
where $\mu_{ij}$ denotes the type of $ij$ bond, and the $3 \times 3$ matrix $\hat{J}_{\mu_{ij}}$ is parametrized, e.g., for the $z$ bond as
\begin{align}
\hat{J}_z
=
\left(
\begin{array}{ccc}
J_{\rm Heis}			&\Gamma	&\Gamma'\\
\Gamma	&J_{\rm Heis} 			&\Gamma'\\
\Gamma'	&\Gamma'	&J_{\rm Heis}+J_z
\end{array}
\right).
\end{align}
Here, $J_{\rm Heis}$ is the coupling constant for the isotropic Heisenberg interaction, and $\Gamma$ and $\Gamma'$ are for the symmetric off-diagonal interactions. 
$\hat{J}_x$ and $\hat{J}_y$ are obtained by $C_3$ rotations. 

\begin{figure}[t]
 \begin{center}
  \includegraphics[width=0.9\columnwidth,clip]{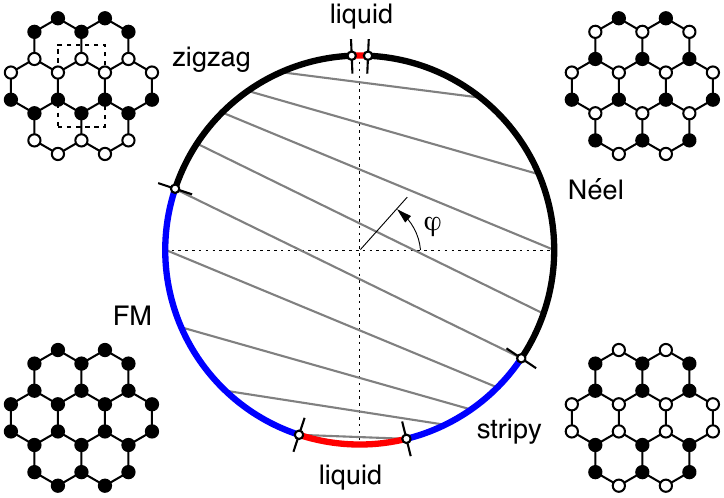}
  \caption{
(Color online) 
Phase diagram of the Kitaev-Heisenberg model with $J/J_{\rm Heis}=2\tan\varphi$ obtained by the exact diagonalization of the 24-site cluster, where $J$ and $J_{\rm Heis}$ are the Kitaev and Heisenberg exchange constants, respectively.
Reprinted with permission from Ref.~\citen{Chaloupka2013} $\copyright$ (2013) the American Physical Society.
}
  \label{fig:jackeli}
 \end{center}
\end{figure}

In the early stage of the research of the Kitaev QSL, the case with $\Gamma=\Gamma'=0$ has been intensively studied.
The model is called the Kitaev-Heisenberg model.
Figure~\ref{fig:jackeli} displays the ground-state phase diagram obtained by the exact diagonalization~\cite{Chaloupka2010,Chaloupka2013}.
In this case, the model exhibits at least four magnetically-ordered phases in addition to two regions of the Kitaev QSLs: FM, zigzag, N\'eel, and stripy phases.
An important finding in this phase diagram is that the Kitaev QSL is found in narrow but finite parameter windows with nonzero $J_{\rm Heis}$ in both FM and AFM Kitaev cases.
The result suggests that the Kitaev QSL is not a singular property limited to the pure Kitaev model but survives against additional exchange couplings.
This has encouraged material exploration for the Kitaev QSL.

Through such experimental exploration as well as computational studies of the spin Hamiltonians on the basis of first-principles calculations, it has been realized that beside the Heisenberg interaction, the symmetric off-diagonal interaction $\Gamma$ plays a role. 
Indeed, $\Gamma$ can be larger than $J_{\rm Heis}$ from the perturbative argument~\cite{Rau2014}. 
Thus, the model including $\Gamma$ has also been studied~\cite{Rau2014,Rusnacko2019}, for which the ground-state phase diagram becomes richer.
The effect of $\Gamma'$ was also studied~\cite{Rusnacko2019}.

From the materials perspectives, the crucial question is how these other exchange interactions affect the QSL behavior in the exact solution for the Kitaev model.
Unfortunately, in most of the candidate materials found thus far, the lowest-$T$ state shows a magnetic order, such as the zigzag type and an incommensurate noncollinear type (see Sec.~\ref{sec:candidates}).
An exception was recently found for H$_3$LiIr$_2$O$_6$~\cite{Kitagawa2018}.
The absence of long-range ordering in this material was discussed on the basis of the stacking manner of the honeycomb layers~\cite{Slagle2018}, the role of the hydrogens~\cite{Yadav2018,Li2018}, and the relevance of disorder in the exchange interactions~\cite{Knolle2019b}, but it remains unclear how to reconcile the sharp NMR lines observed down to the lowest $T$~\cite{Kitagawa2018} to these scenarios.

\subsection{Schematic phase diagram}
\label{sec:schematic_phase_diagram}

\begin{figure*}[t]
 \begin{center}
  \includegraphics[width=2\columnwidth,clip]{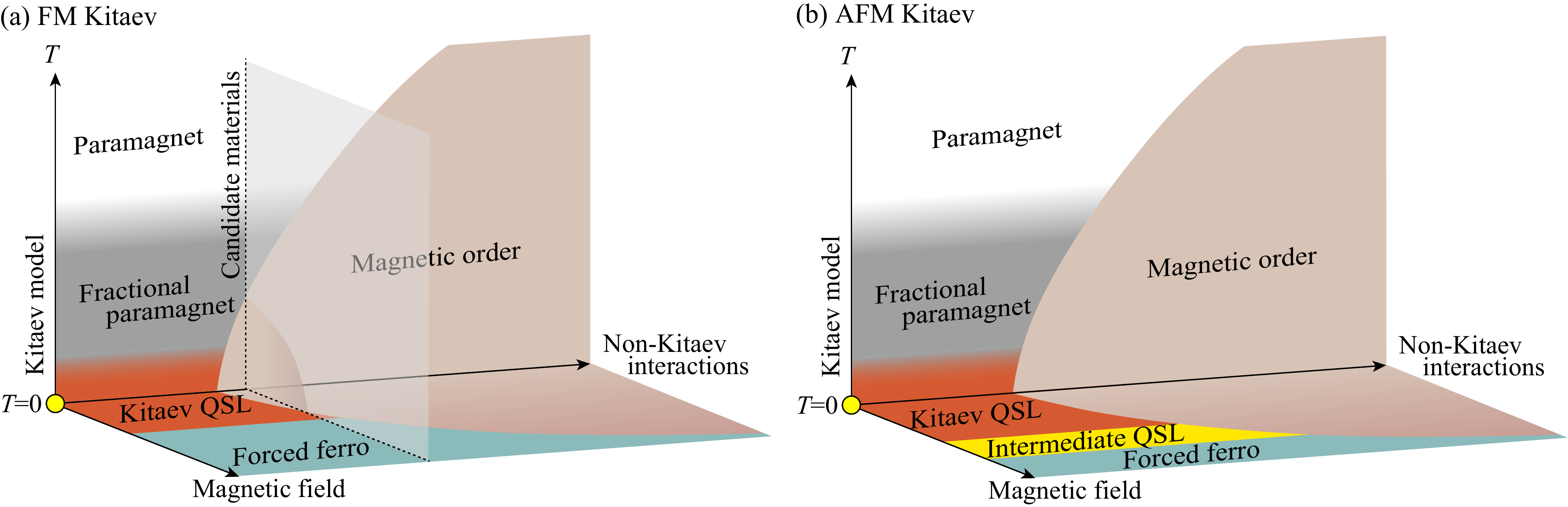}
  \caption{
(Color online) 
Schematic phase diagrams while changing temperature $T$, magnetic field, and non-Kitaev interactions for the cases with (a) FM and (b) AFM Kitaev couplings. 
The yellow circle at the origin represents the exact QSL ground state for the Kitaev model. 
}
  \label{fig:sketch_phase}
 \end{center}
\end{figure*}

Figure~\ref{fig:sketch_phase} summarizes the arguments in the previous sections into the schematic phase diagrams.
The phase diagrams are displayed for both cases with the FM and AFM Kitaev coupling, in the parameter space of temperature $T$, external magnetic field $h$, and other non-Kitaev interactions; the origin corresponds to the QSL state found in the exact solution for the Kitaev model.

Let us first discuss the FM case shown in Fig.~\ref{fig:sketch_phase}(a), which is believed to be relevant to most of the existing candidate materials.
As discussed in Sec.~\ref{sec:other_exchanges}, the Kitaev QSL state survives in a finite region in the ground state against the non-Kitaev interactions.
Above the threshold, the ground state exhibits some magnetic ordering whose spin structure depends on the detailed forms of the non-Kitaev interactions.
The magnetic order is expected to survive at finite $T$ due to the spin anisotropy as well as the three dimensionality, and the critical temperature $T_c$ will rise as the non-Kitaev interactions increase.

On the other hand, while raising $T$ in the QSL region below the threshold, the system undergoes two crossovers as briefly mentioned in Sec.~\ref{sec:finiteT} (the details will be discussed in Sec.~\ref{sec:Tfrac}).
Considering the realistic value of $J \sim 200$-$300$~K for $A_2$IrO$_3$~\cite{Foyevtsova2013,Katukuri2014,Yamaji2014,Winter2016} and $J \sim 100$-$200$~K for $\alpha$-RuCl$_3$~\cite{Winter2016,Sandilands2015,Banerjee2016,Yadav2016,Do2017}, the high-$T$ crossover takes place at $T_H \sim 80$-$110$~K and $\sim 40$-$80$~K, respectively.
The temperature scales are significantly higher than $T_c$ for these compounds, $T_c \sim 15$~K and $\sim 7$~K, respectively.
On the other hand, $T_L \sim 1$-$2$~K and $\sim 0.5$-$1$~K are lower than $T_N$.
Thus, we believe that the candidate materials are located at the vertical dashed line in Fig.~\ref{fig:sketch_phase}(a).
If this is the case, there is a considerable $T$ window between $T_H$ and $T_c$, where one can expect unconventional behavior arising from the fractionalization; this will be discussed in detail in Sec.~\ref{sec:Tfrac}.

When applying the external magnetic field, as discussed in Sec.~\ref{sec:field}, the QSL survives up to a nonzero field strength, but it is taken over by the forced FM state in the larger field region.
An interesting question is whether the QSL behavior can be captured in the candidate materials after the magnetic order is suppressed by the magnetic field.
We depict Fig.~\ref{fig:sketch_phase}(a) so that there is a narrow but nonzero window for such field-induced QSL.
This intriguing possibility has attracted upsurge interest in $\alpha$-RuCl$_3$, as will be described in Sec.~\ref{sec:Sqw}, \ref{sec:1/T1}, and \ref{sec:kappa_xy}.

Figure~\ref{fig:sketch_phase}(b) represents the corresponding phase diagram for the AFM Kitaev case.
The overall structure is similar to the FM case in Fig.~\ref{fig:sketch_phase}(a), but there is a qualitative difference in the behavior in the magnetic field.
As described in Sec.~\ref{sec:field}, in the AFM Kitaev case, the system appears to exhibit two successive phase transitions including the intermediate QSL phase~\cite{Zhu2018,Gohlke2018,Nasu2018,Liang2018,Hickey2019,Ronquillo2019,Patel2019}.
Note that the scale of the magnetic field is almost ten times larger compared to the FM case (this is also indicated by the large difference in the magnitude of the magnetic susceptibility in Sec.~\ref{sec:chi}).
Although no realistic compounds with the AFM Kitaev coupling are at hand thus far, the peculiar phase diagram is worth investigating and will stimulate further material exploration.

\section{Thermal fractionalization}
\label{sec:Tfrac}

In this section, we discuss a distinguished thermodynamic property of the Kitaev model, which we call {\it thermal fractionalization}~\cite{Nasu2015}.
As discussed in Sec.~\ref{sec:frac}, the exact QSL ground state hosts two types of quasiparticles, itinerant Majorana fermions and localized $Z_2$ fluxes, which have largely separated energy scales.
The two energy scales show up in the thermodynamic behavior as two characteristic temperatures.
The higher characteristic temperature $T_H$ is related with the itinerant Majorana fermions, which is roughly set by the COM of the fermion DOS (see Fig.~\ref{fig:band}).
At $T\simeq T_H$, the system exhibits a crossover irrespective of the spatial dimensions as well as the details of the model.
Meanwhile, the lower one $T_L$ is related with the localized $Z_2$ fluxes, which is roughly set by the $Z_2$ flux gap [see Fig.~\ref{fig:aniso}(b)].
In contrast to the universal crossover at $T_H$, the behavior at $T\simeq T_L$ depends on the nature of the localized $Z_2$ flux excitations in each system; it can be either a crossover or a phase transition.
Thus, the Kitaev model, in general, exhibits three distinct states:
a conventional paramagnetic (PM) state for $T \gtrsim T_H$,
an unconventional PM state for $T_L \lesssim T \lesssim T_H$, and
the (asymptotic) QSL state for $T \lesssim T_L$.
We call the intermediate $T$ region the fractional PM state, where the thermal fractionalization makes the system distinct from the conventional PM state.

We discuss these intriguing behaviors by the thermal fractionalization in this section.
They have been unveiled by the recently-developed numerical methods based on the Majorana representation of the Kitaev model at zero field.
In Sec.~\ref{sec:2D_honeycomb}, we present the results for the 2D Kitaev model on the honeycomb structure, which provides a canonical example of two successive crossovers at $T_H$ and $T_L$.
We also discuss a variant of the Kitaev model in two dimensions in Sec.~\ref{sec:2D_CSL}, which exhibits a phase transition to a chiral spin liquid (CSL), instead of the low-$T$ crossover at $T_L$.
In Sec.~\ref{sec:3D}, we present the results for the Kitaev models defined on several 3D tricoordinate lattices, in which various types of the phase transitions take place between three states of matter in terms of the spin degree of freedom.
Finally, in Sec.~\ref{sec:phase_diagram}, we summarize the phase diagrams for the crossovers and phase transitions found in the 2D and 3D Kitaev models.

\subsection{Successive crossovers in the 2D honeycomb case}
\label{sec:2D_honeycomb}

\subsubsection{Crossovers caused by thermal fractionalization}
\label{sec:crossovers}

\begin{figure}[t]
 \begin{center}
  \includegraphics[width=0.9\columnwidth,clip]{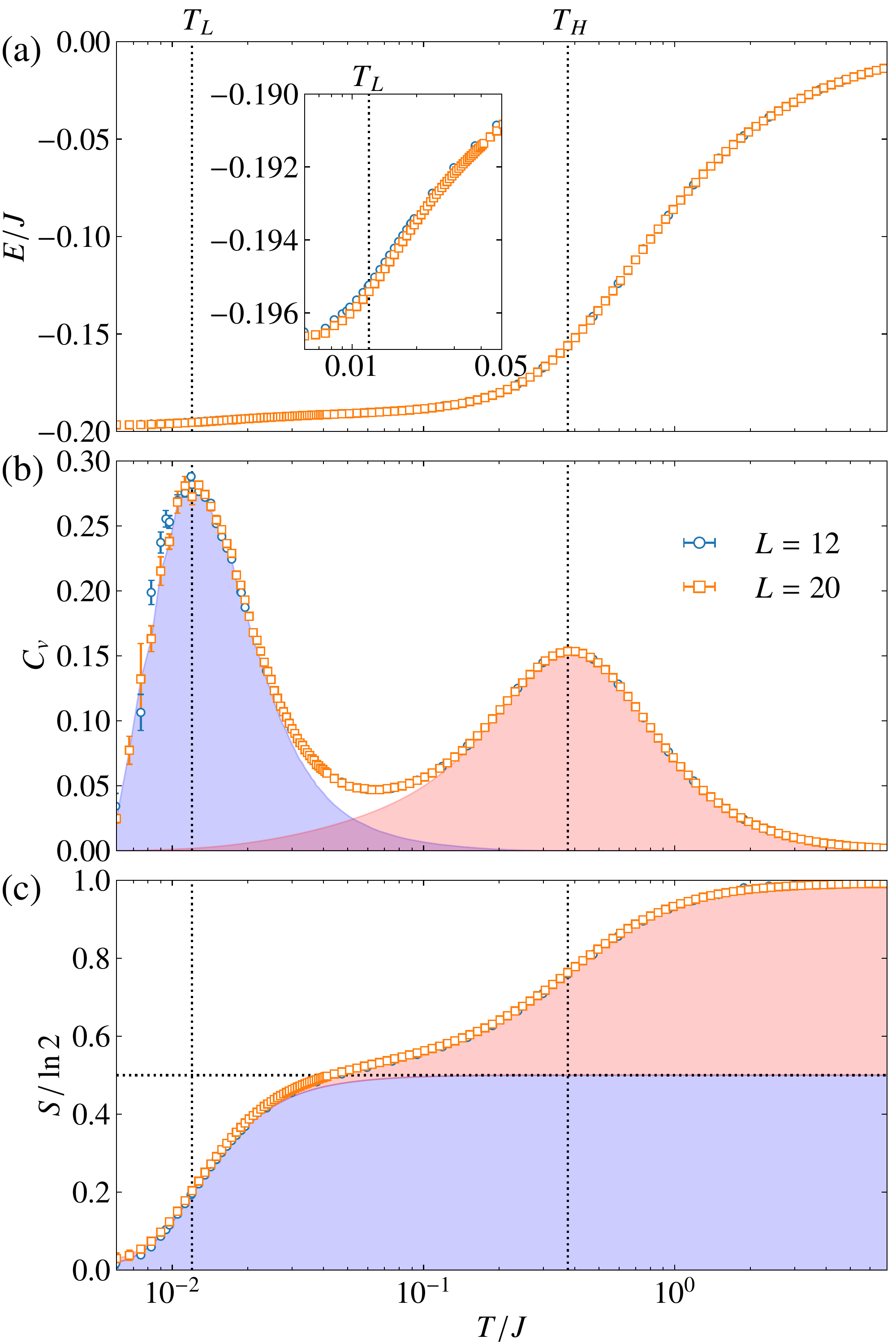}
  \caption{
(Color online) 
$T$ dependences of (a) the internal energy $E$, (b) the specific heat $C_v$, and (c) the entropy $S$ per site for the honeycomb Kitaev model with isotropic coupling $J_x=J_y=J_z=J$.
The data are obtained by the Majorana-based QMC simulations for the clusters with $N=2L^2$ spins ($L=12$ and $20$).
The vertical dotted lines represent $T_L$ and $T_H$.
The inset in (a) is an extended plot around $T_L$.
The reddish and bluish shades in (b) and (c) show the contributions from the itinerant Majorana fermions and the localized $Z_2$ fluxes, respectively. 
The horizontal dotted line in (c) represents $\frac{1}{2}\ln 2$.
The data for $L=12$ were taken from Ref.~\citen{Nasu2015}, and the data for the specific heat for $L=20$ were taken from Ref.~\citen{Nasu2017a}.
The data for $L=20$ in (a) and (c) as well as the decomposition into the two types of fractional quasiparticles are newly added in (b) and (c).
}
  \label{fig:ene}
 \end{center}
\end{figure}

Let us begin with the original Kitaev model defined on the honeycomb structure.
Figure~\ref{fig:ene} shows the $T$ dependences of the internal energy $E$, specific heat $C_v$, and entropy $S$ per site for the isotropic Kitaev coupling $J_x=J_y=J_z=J$~\cite{Nasu2015} (the results are common to the FM and AFM Kitaev couplings).
The calculations were performed by using the QMC simulations based on the Majorana representation for the clusters with $N=2L^2$ spins (see Appendix~\ref{sec:quantum-monte-carlo}).
As shown in Fig.~\ref{fig:ene}(a) and its inset, the internal energy $E$ decreases rapidly at two temperatures, $T_H \simeq 0.375J$ and $T_L \simeq 0.012J$, while the decrease at $T_L$ is much smaller than that at $T_H$.
Correspondingly, the specific heat $C_v$ exhibits two peaks as shown in Fig.~\ref{fig:ene}(b), both of which show no significant system-size dependence, indicating that these are crossovers.
Interestingly, as plotted in Fig.~\ref{fig:ene}(c), the entropy $S$ is released successively by half $\ln 2$ at each crossover.
This peculiar behavior is considered to originate from the thermal fractionalization in which the original spin degree of freedom carrying the entropy of $\ln 2$ is fractionalized into the two types of quasiparticles each carrying the entropy of half $\ln 2$.
This is confirmed by the decomposition of $C_v$ and $S$ into the contributions from the itinerant Majorana fermions and the localized $Z_2$ fluxes [see Eqs.~(\ref{eq:Cv_Maj}) and (\ref{eq:Cv_flux}) in Appendix~\ref{sec:quantum-monte-carlo}], as shown in Figs.~\ref{fig:ene}(b) and \ref{fig:ene}(c).

\begin{figure}[t]
 \begin{center}
  \includegraphics[width=0.9\columnwidth,clip]{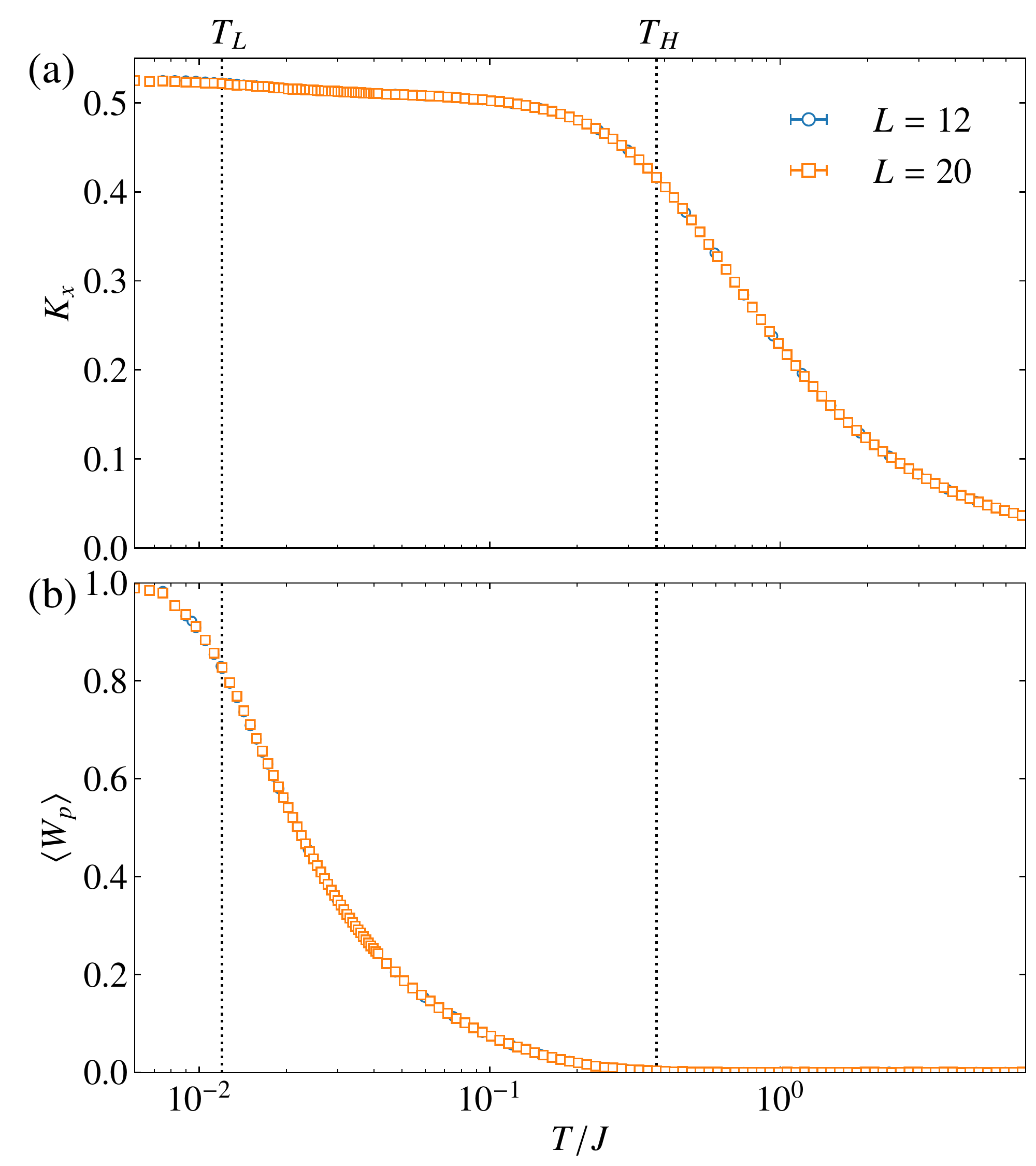}
  \caption{
(Color online) 
$T$ dependences of (a) the measure of the kinetic energy of itinerant Majorana fermions, $K_x$, and (b) the thermal average of the $Z_2$ flux, $\langle W_p \rangle$.
Note that $K_x = -\frac23 E = 4\langle S_i^x S_j^x \rangle$ for the isotropic case.
The data for $L=12$ were taken from Ref.~\citen{Nasu2015}.
The data for $L=20$ are newly added .
}
  \label{fig:corr}
 \end{center}
\end{figure}

The role of the two fractional quasiparticles in the two crossovers is shown in more explicit way by calculating the quantities associated with each quasiparticle.
Figure~\ref{fig:corr}(a) plots the measure of the kinetic energy of the itinerant Majorana fermions,
$K_x = -i \langle \gamma_i \gamma_j \rangle_x$, where the thermal average $\langle \cdots \rangle_x$ is calculated on the $x$ bond.
Note that this quantity is related with the internal energy as $E = -\frac32 K_x$ in the isotropic case.
Also, it is equivalent to the spin correlation on the $x$ bonds, $4\langle S_i^x S_j^x \rangle_x$.
The result indicates that the measure of the Majorana kinetic energy increases rapidly around $T=T_H$, and does not change largely in the lower-$T$ region.
This suggests that the Fermi degeneracy of the complex fermions composed of the itinerant Majorana fermions sets in at $T\simeq T_H$.
On the other hand, Fig.~\ref{fig:corr}(b) displays the thermal average of the $Z_2$ flux, $\langle W_p \rangle$.
While it becomes nonzero from high $T$ around $T_H$, it grows rapidly around $T=T_L$ and approaches $\langle W_p \rangle = 1$ (the value in the flux-free ground state) below $T_L$.

These results clearly show that the crossover at $T_H$ is caused by the itinerant Majorana fermions, and that at $T_L$ is by the localized $Z_2$ fluxes.
The former corresponds to the Fermi degeneracy of the complex fermions composed of the itinerant Majorana fermions, and the latter to the asymptotic freezing of the $Z_2$ fluxes into the flux-free state.
Thus, these two crossovers are manifestations of the thermal fractionalization in thermodynamics.
While decreasing $T$, the fractionalization of the spin degree of freedom sets in around $T_H$ with the entropy release of half $\ln 2$ by the Fermi degeneracy, and the system enters into an unconventional PM state, dubbed the fractional PM state, below $T \simeq T_H$.
In the fractional PM region, the $Z_2$ fluxes remain disordered as the states with flipped $W_p$ are thermally excited beyond the flux gap.
By approaching $T_L$ with a further decrease of $T$, however, the thermal excitations of the $Z_2$ fluxes are suppressed, and the system crosses over into the asymptotic QSL state below $T \simeq T_L$ with the entropy release of the rest half $\ln 2$ by the freezing of $W_p$.
The picture of the successive crossovers will be further discussed in Sec.~\ref{sec:phase_diagram}.

\subsubsection{Crossovers temperature scales}
\label{sec:Tscales}

What determines the values of the two crossover temperatures $T_H$ and $T_L$?
From the above arguments, it is naturally expected that $T_H$ is set by the Fermi degeneracy temperature, which is roughly given by the COM of the fermion DOS, and that $T_L$ is set by half of the gap for the lowest excitation of the $Z_2$ fluxes (the flux gap is defined for a two-flux excitation); see Sec.~\ref{sec:frac}.
In the isotropic case with $J_x=J_y=J_z=J$, the COM of the fermion DOS is at $\simeq 0.762J$ and the half of the flux gap is $\simeq 0.0328J$.
Note that the COM of the fermion DOS is less sensitive to the flux sector, but here we use the value for the disordered flux configuration, which we call $G_\gamma$, corresponding to the high-$T$ limit, as the fluxes are almost disordered near $T_H$ as shown in Fig.~\ref{fig:corr}(b).
Considering that the specific heat peak in the two-level system with a gap of unity appears at $T = \zeta \simeq 0.417$, we note that the numbers $0.762J \times \zeta \simeq 0.318J$ and $0.0328J \times \zeta \simeq 0.0136J$ are very close to $T_H \simeq 0.375J$ and $T_L \simeq 0.012J$, respectively, which confirms the above expectation.

\begin{figure}[t]
 \begin{center}
  \includegraphics[width=0.95\columnwidth,clip]{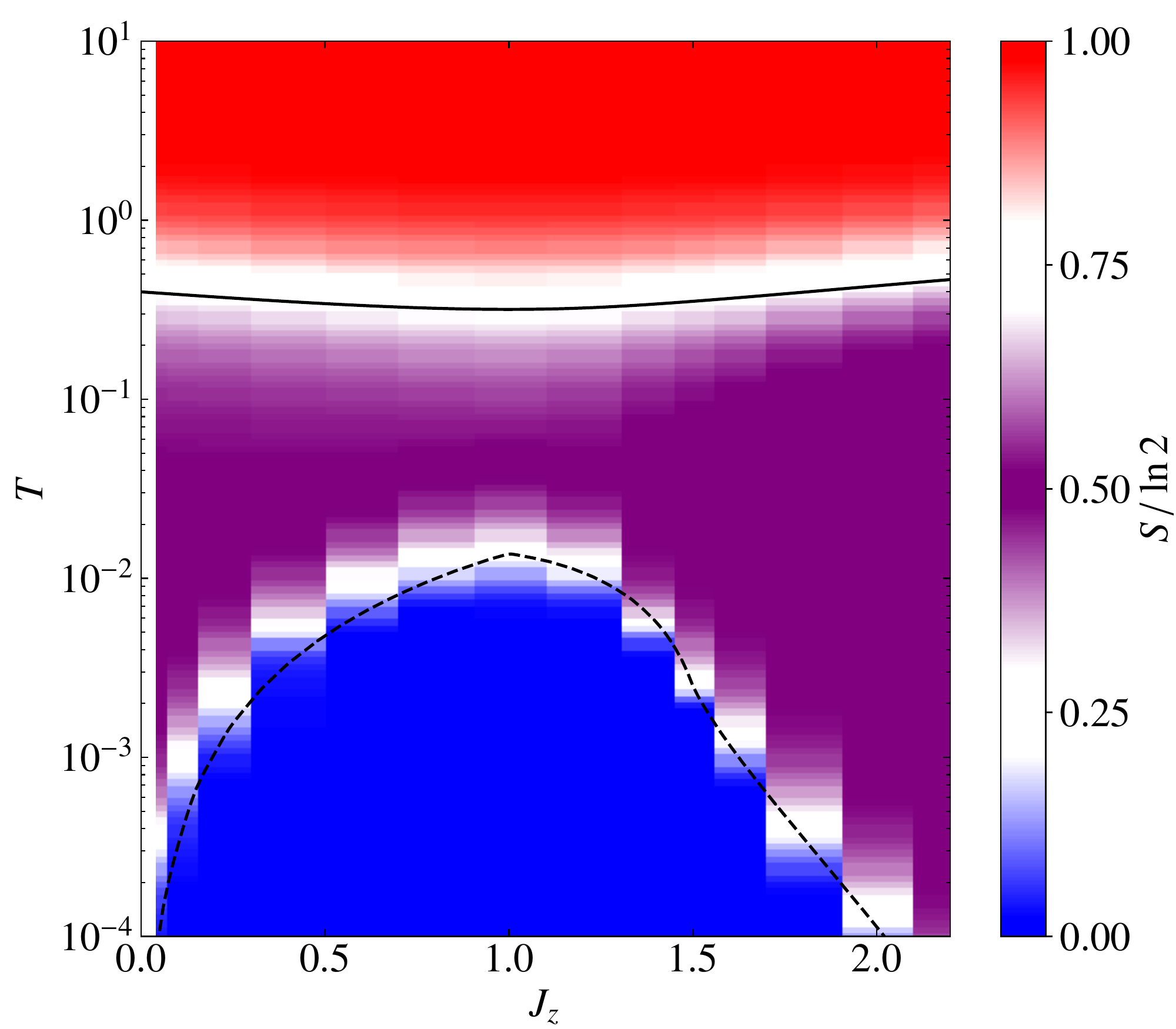}
  \caption{
(Color online) 
Contour plot of the entropy per site normalized by $\ln 2$ as functions of $T$ and $J_z$ with $J_x=J_y=(3-J_z)/2$ for the $L=12$ cluster.
The solid and dashed curves represent $\zeta G_\gamma$ and $\frac12 \zeta \Delta_f$, respectively, where $G_\gamma$ and $\Delta_f$ are the COM of the fermion DOS for the disordered flux configuration corresponding to the high-$T$ limit and the flux gap for the flux-free ground state, respectively; $\zeta\simeq 0.417$ is the peak temperature of the specific heat in the two-level system with a gap of unity.
The factor 1/2 in $\frac12 \zeta \Delta_f$ is introduced since $\Delta_f$ means the energy cost to excite two neighboring fluxes.
The data of the entropy were taken from Ref.~\citen{Nasu2015}.
The two curves are newly added.
}
  \label{fig:s_aniso}
 \end{center}
\end{figure}

\begin{figure}[t]
 \begin{center}
  \includegraphics[width=0.95\columnwidth,clip]{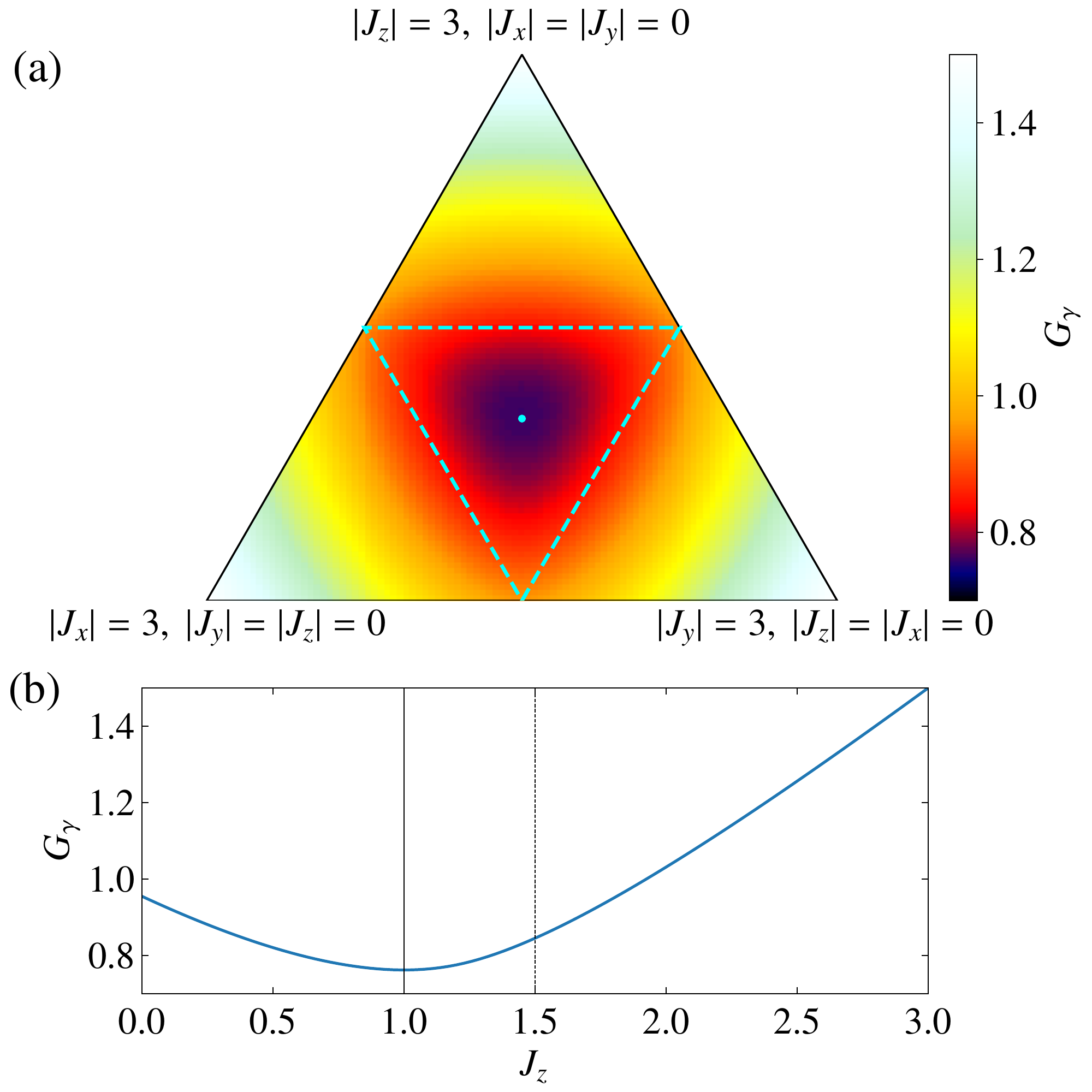}
  \caption{
(Color online) 
(a) COM of the fermion DOS for the disordered flux configuration, $G_\gamma$, on the plane of $|J_x|+|J_y|+|J_z|=3$.
(b) $J_z$ dependence of $G_\gamma$ with $J_x=J_y=(3-J_z)/2$ corresponding to the cut along the vertical line through the isotropic point in (a).
}
  \label{fig:aniso3}
 \end{center}
\end{figure}

We can further examine these correspondences by varying the anisotropy of the Kitaev coupling.
Figure~\ref{fig:s_aniso} shows the contour plot of the entropy per site, $S$, normalized by $\ln 2$ while changing $J_z$ with $J_x=J_y$ and $J_x+J_y+J_z=3$.
The two white regions with $S/\ln 2 \simeq 0.75$ and $\simeq 0.25$ roughly corresponds to $T_H$ and $T_L$, respectively.
As shown in the figure, $T_H$ does not show a drastic change against $J_z$, whereas $T_L$ does: $T_L$ has a peak around the isotropic point with $J_z=J$ and rapidly decreases by increasing the anisotropy with both $J_z \to 0$ and $J_z \to 3$.
For comparison, we plot the effective activation temperatures defined by the COM of the fermion DOS for the disorder flux configuration, $\zeta G_\gamma$, and the $Z_2$ flux gap, $\frac12 \zeta \Delta_f$, by the solid and dashed curves in Fig.~\ref{fig:s_aniso}.
The former does not change so much for $J_z$ similar to $T_H$ (see Fig.~\ref{fig:aniso3}), while the latter depends largely on $J_z$ similar to $T_L$ [see Figs.~\ref{fig:aniso}(c) and ~\ref{fig:aniso}(d)];
in the entire range of $J_z$, $\zeta G_\gamma$ and $\frac12 \zeta \Delta_f$ coincide well with $T_H$ and $T_L$, respectively.
The results further confirm the correspondences of $T_H$ and $T_L$ to the energy scales of the itinerant Majorana fermions and the localized $Z_2$ fluxes.

\subsubsection{Majorana metal}
\label{sec:Majorana_metal}

\begin{figure*}[t]
 \begin{center}
  \includegraphics[width=1.8\columnwidth,clip]{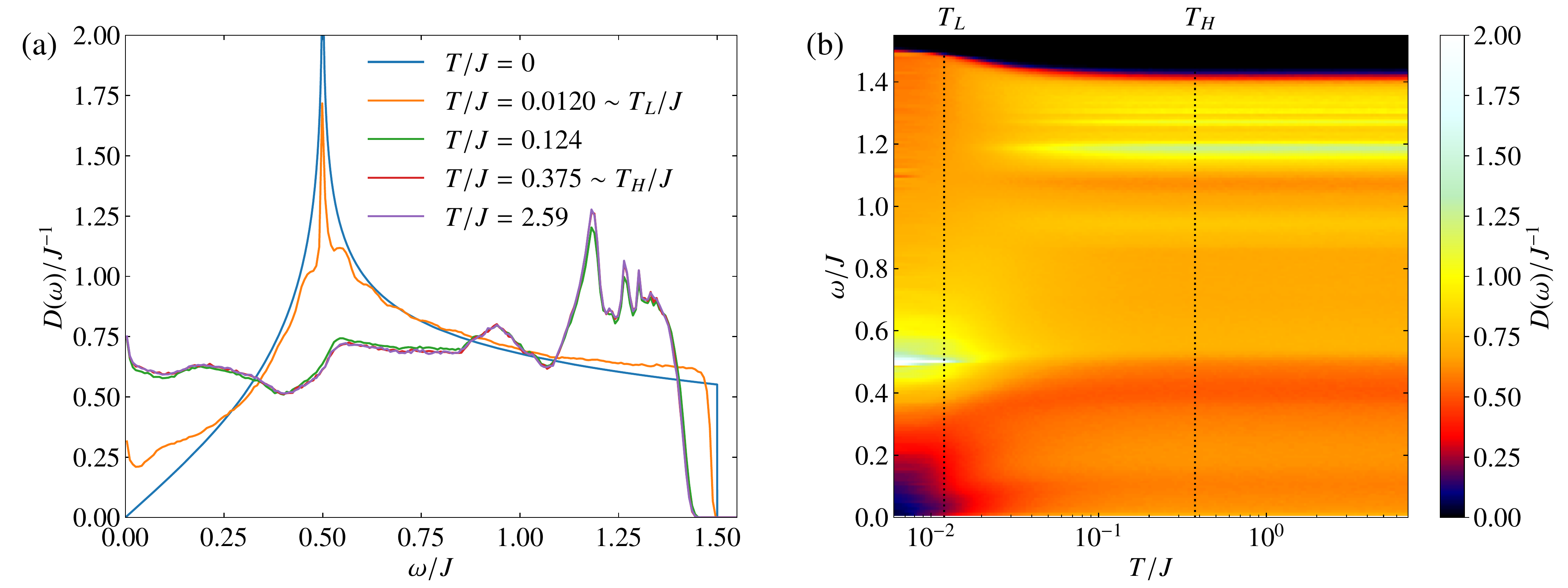}
  \caption{
(Color online) 
(a) $T$ dependence of the fermion DOS $D(\omega)$ for the $L=12$ cluster.
See also Fig.~3(a) in Ref.~\citen{Nasu2015}.
(b) The contour plot as a function of $T$ and $\omega$.
}
  \label{fig:dosmap}
 \end{center}
\end{figure*}

\begin{figure}[t]
 \begin{center}
  \includegraphics[width=0.9\columnwidth,clip]{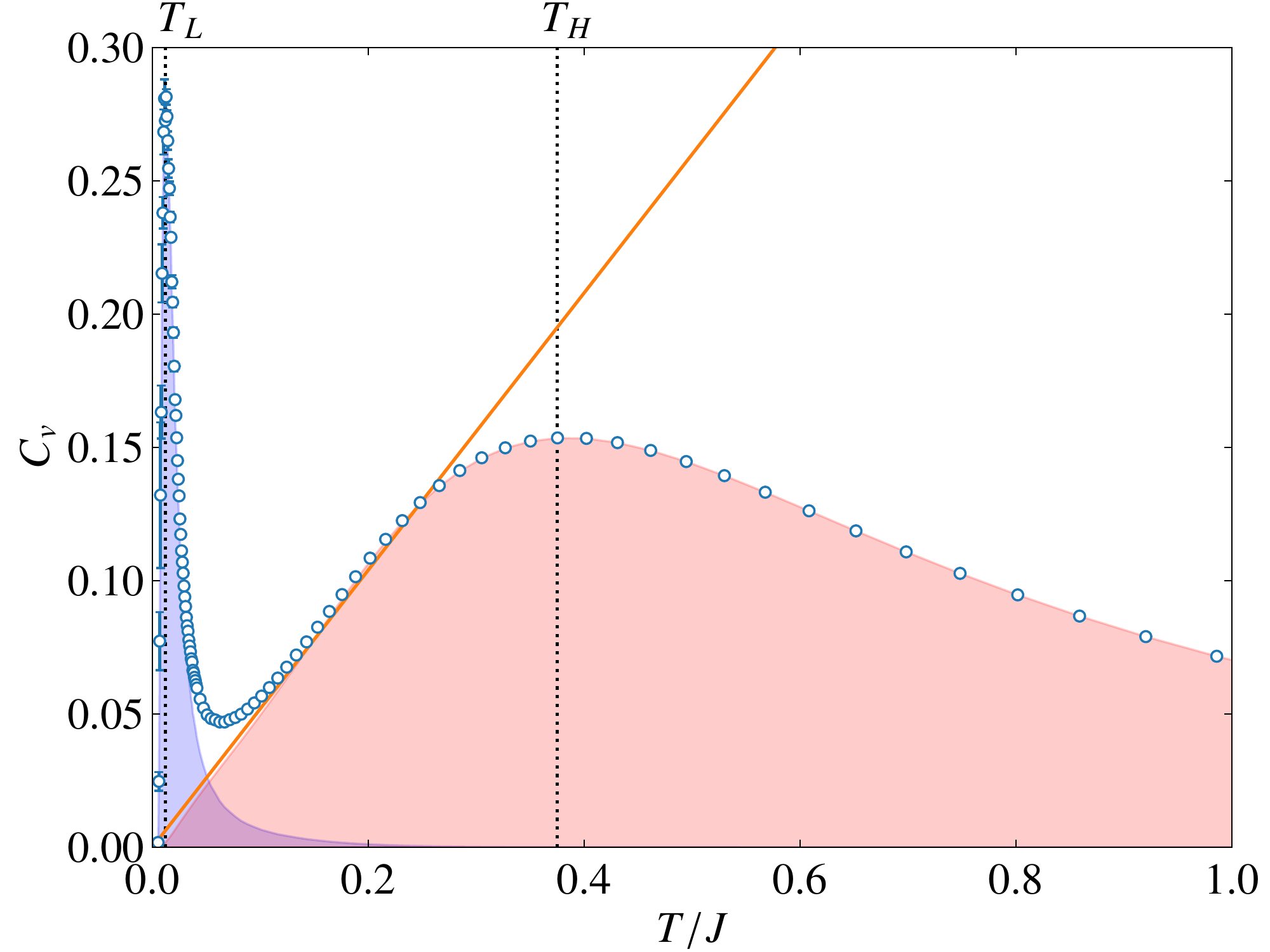}
  \caption{
(Color online) 
Plot of the data in Fig.~\ref{fig:ene}(b) in the $T$-linear scale.
The orange solid line is a $T$-linear function to fit the data between $T_L$ and $T_H$.
A similar plot for a slightly anisotropic case is found in Ref.~\citen{Nasu2015}.
}
  \label{fig:cvlinear}
 \end{center}
\end{figure}

Let us discuss the excitation spectrum of the itinerant Majorana fermions while changing $T$.
Figure~\ref{fig:dosmap} shows the $T$ dependence of the fermion DOS. 
While the overall structure of the DOS below $T_L$ is similar to that in the flux-free ground state as shown in Fig.~\ref{fig:dosmap}(b), the DOS rapidly changes its form above $T_L$.
In particular, in the low-energy region, the energy-linear behavior at the lower band edge is quickly smeared out and the DOS at zero energy becomes nonzero, as shown in Fig.~\ref{fig:dosmap}(a)~\cite{Nasu2015,note2}.
This is due to the thermal excitations of the $Z_2$ fluxes, which disturb the Dirac-like linear dispersion in the flux-free ground state.
The nonzero DOS at the band bottom indicates that the fractional PM state above $T_L$ is regarded as a ``Majorana metal'', in analogy with the 2D conventional metal that has nonzero DOS at the band edges.
Needless to say, the present system is an insulator with localized magnetic moments, and hence, the particles traversing the system are not electrons but the Majorana fermions.
This is why we call the unconventional state the Majorana metal.
An interesting consequence of this Majorana metallic state is observed in the specific heat $C_v$.
As shown in Fig.~\ref{fig:cvlinear}, $C_v$ shows $T$-linear dependence in the $T$ window between $T_L$ and $T_H$, reflecting the ``metallic'' nature of the system~\cite{Nasu2015}.
The itinerant quasiparticles also contribute heat conductions, as will be discussed in Sec.~\ref{sec:kappa_xx} and \ref{sec:kappa_xy}.

\subsection{Phase transitions to 2D chiral spin liquids}
\label{sec:2D_CSL}

\begin{figure}[t]
 \begin{center}
  \includegraphics[width=0.7\columnwidth,clip]{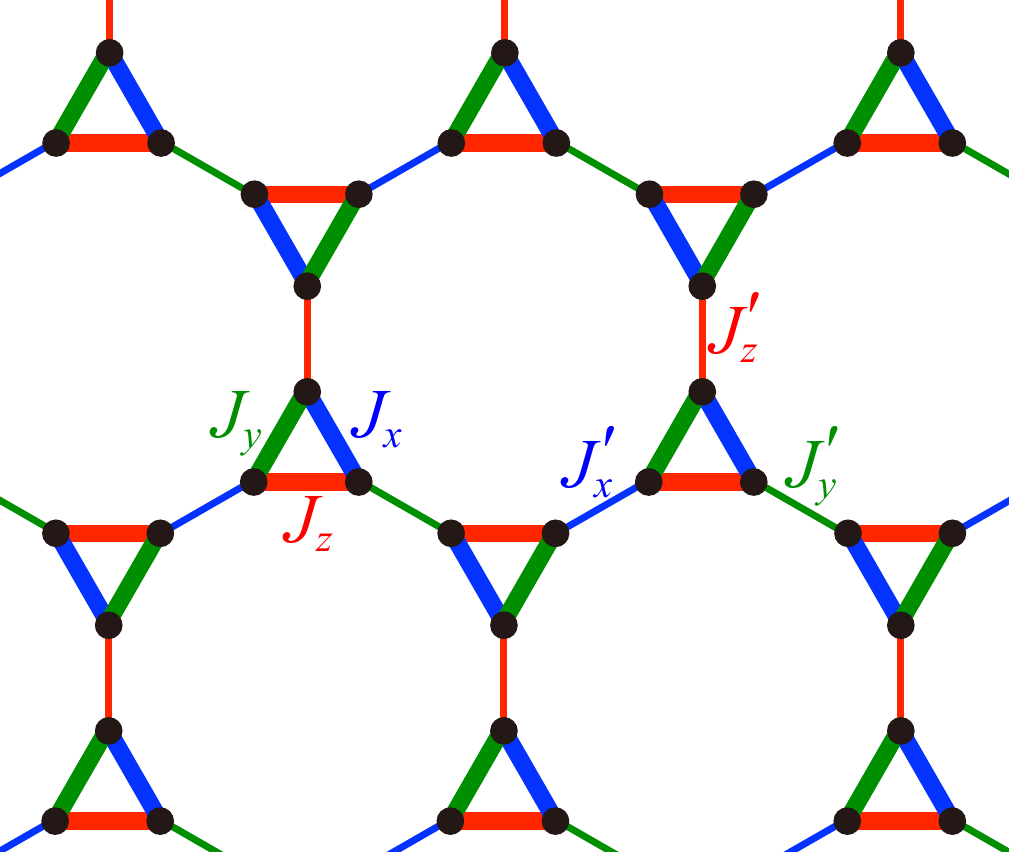}
  \caption{
(Color online) 
Schematic picture of the triangle-honeycomb structure.
The two sets of Kitaev couplings are also shown.
}
  \label{fig:lattice-yk}
 \end{center}
\end{figure}

Let us turn to a variant of the Kitaev model in two dimensions, which is defined on a modified lattice structure, called the triangle-honeycomb structure.
The structure is obtained by replacing all the vertices of the honeycomb structure by triangles, as shown in Fig.~\ref{fig:lattice-yk}.
The Kitaev model can be extended straightforwardly to this tricoordinate lattice structure, but one can define two different sets of the Kitaev coupling, ($J_x, J_y, J_z$) and ($J'_x, J'_y, J'_z$), for the two types of NN bonds, intra-triangle and inter-triangle ones, respectively (see Fig.~\ref{fig:lattice-yk})~\cite{Yao2007}.
In the following, we consider the case with $J_x=J_y=J_z=J$ and $J'_x=J'_y=J'_z=J'$.

The most important difference from the honeycomb model is that the lattice structure includes the elementary loops with odd number of sites.
As pointed out in the seminal paper by Kitaev~\cite{Kitaev2006}, the Kitaev model defined on the lattices with such odd cycles may break time-reversal symmetry spontaneously, as the flux operator defined on an odd-cycle plaquette describes a time-reversal pair.
Indeed, H.~Yao and S.~A.~Kivelson showed that the ground state of the triangle-honeycomb Kitaev model becomes a CSL with spontaneous breaking of time-reversal symmetry~\cite{Yao2007}.
Interestingly, there are two different CSLs: topologically-nontrivial one for $J'/J<\sqrt3$ and topologically-trivial one for $J'/J>\sqrt3$.
For the topologically nontrivial (trivial) CSL, the flux excitations obey non-Abelian (Abelian) statistics. 
The topologically nontrivial phase is characterized by a nonzero Chern number in the band structure, and exhibits a chiral Majorana edge state under open boundary conditions.
The topological nature was also explained by the fact that the low-energy effective model in the limit of $J'/J\to 0$ has a similar form to the effective Hamiltonian for the honeycomb model in a magnetic field with the three-spin term in Eq.~(\ref{eq:H'_spin})~\cite{Dusuel2008}.

\begin{figure}[t]
 \begin{center}
  \includegraphics[width=\columnwidth,clip]{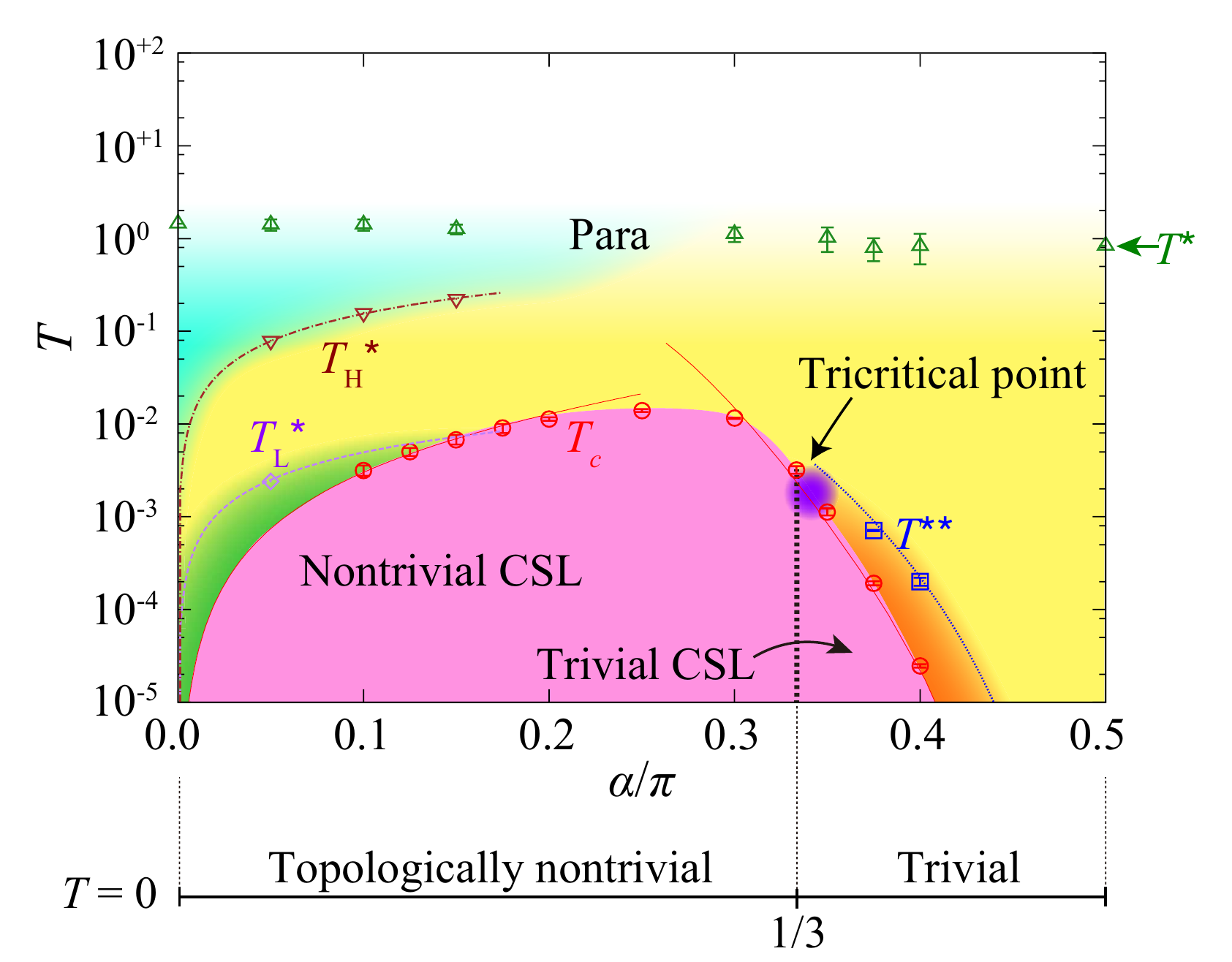}
  \caption{
(Color online) 
Finite-$T$ phase diagram of the triangle-honeycomb Kitaev model with $(J,J')=4(\cos\alpha, \sin\alpha)$ obtained by the Majorana-based QMC simulations.
Reprinted with permission from Ref.~\citen{Nasu2015b} $\copyright$ (2015) the American Physical Society.
}
  \label{fig:phase-yk}
 \end{center}
\end{figure}

Thermodynamic properties of this model were studied by using the Majorana-based QMC simulations~\cite{Nasu2015b}.
In contrast to the honeycomb case in Sec.~\ref{sec:2D_honeycomb}, the model exhibits a finite-$T$ phase transition instead of the crossover at $T_L$.
This is due to the spontaneous breaking of time-reversal symmetry by the freezing of the $Z_2$ fluxes; while the freezing does not break any symmetry in the honeycomb case, it breaks time-reversal symmetry in the triangle-honeycomb case because of the odd-cycle plaquettes on the triangles.
Interestingly, the transition was found to be continuous in the topologically-nontrivial region for $J'/J \lesssim \sqrt3$, and the estimated critical exponents are close to those of the 2D Ising universality class, but discontinuous in the topologically-trivial region for $J'/J \gtrsim \sqrt3$.
This suggests the existence of the tricritical point in between, while the precise location and the nature are not fully identified.
The obtained phase diagram is presented in Fig.~\ref{fig:phase-yk}.

\begin{figure}[t]
 \begin{center}
  \includegraphics[width=0.95\columnwidth,clip]{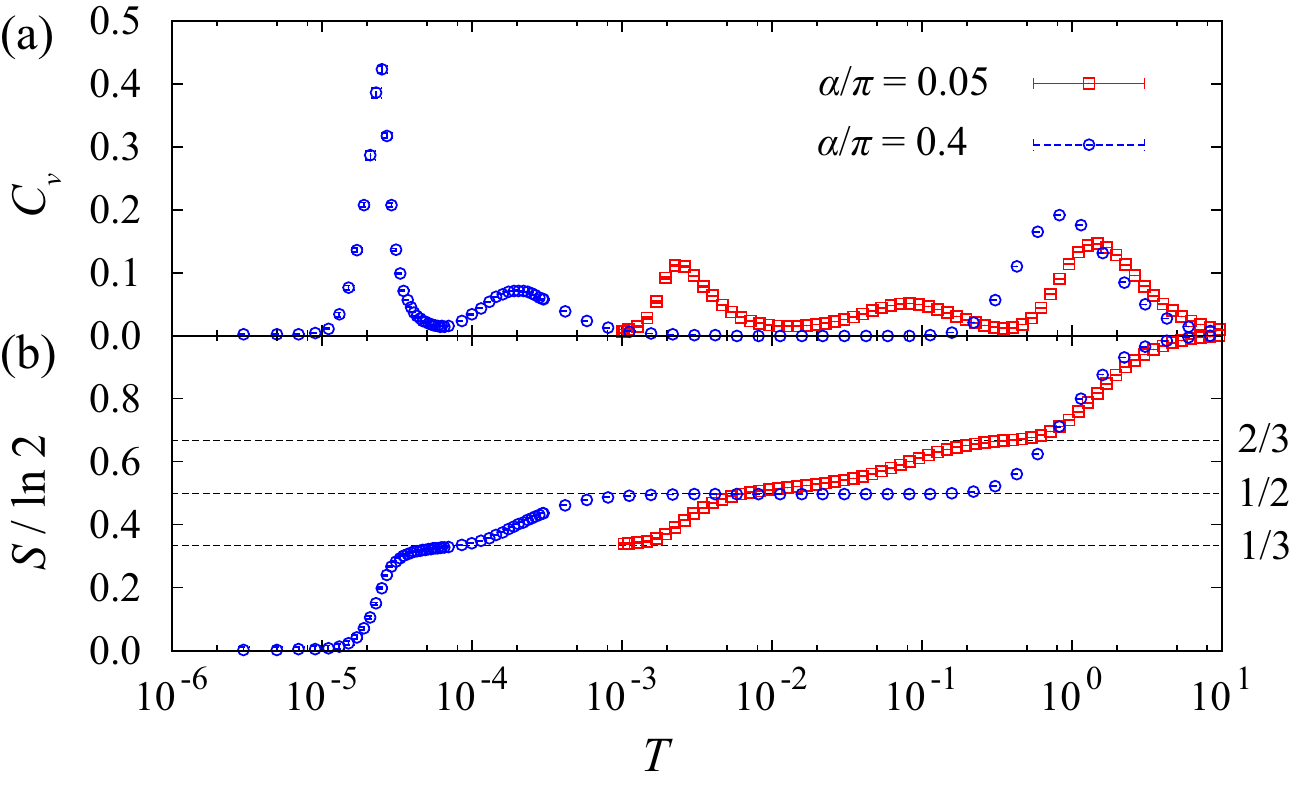}
  \caption{
(Color online) 
$T$ dependences of (a) the specific heat and (b) the entropy per site in the triangle-honeycomb Kitaev model with $(J,J')=4(\cos\alpha, \sin\alpha)$.
The horizontal dotted lines in (b) denote $1/3$, $1/2$, and $2/3$ of $\ln 2$. 
Reprinted with permission from Ref.~\citen{Nasu2015b} $\copyright$ (2015) the American Physical Society.
}
  \label{fig:CvS-yk}
 \end{center}
\end{figure}

Then, what happens to the high-$T$ crossover at $T_H$ found in the honeycomb case?
It was shown that while the crossover takes place also in the triangle-honeycomb case, the amount of entropy released in the crossover can be different from the honeycomb case depending on the parameter $J'/J$~\cite{Nasu2015b}.
In the topologically-trivial region for $J'/J \gtrsim \sqrt3$, the entropy release is the same as in the honeycomb case, half $\ln 2$.
But in this case, the system exhibits another crossover, where the entropy of $\frac16 \ln 2$ corresponding to fluxes on the dodecagons is released.
Finally, the rest of entropy $\frac13 \ln 2$ corresponding to fluxes on the triangles is released at the phase transition to the CSL.
The typical $T$ dependences of the specific heat $C_v$ and the entropy $S$ per site are shown in Fig.~\ref{fig:CvS-yk}.
On the other hand, in the topologically-nontrivial region for $J'/J \lesssim \sqrt3$, the entropy release at the high-$T$ crossover is $\frac13 \ln 2$; the remaining entropy $\frac23 \ln 2$ corresponds to fourfold degeneracy in each triangle in the isolated triangle limit ($J'/J\to 0$).
In this case, the system exhibits two additional crossovers, at each of which the entropy of $\frac16 \ln 2$ is released; see the typical behavior in Fig.~\ref{fig:CvS-yk}.
As mentioned before, in the isolated triangle limit, the system is effectively described by the honeycomb Kitaev model in a weak magnetic field, and hence, these two crossovers correspond to $T_L$ and $T_H$ in the honeycomb Kitaev model.
Note that the lowest-$T$ crossover appear to merge into the phase transition for $J'/J \gtrsim 0.1$ (see Fig.~\ref{fig:phase-yk}).

Thus, the complicated behaviors are found in the high-$T$ crossovers depending on two types of the Kitaev coupling, $J$ and $J'$.
Nonetheless, the important point is that the highest-$T$ crossover occurs at the temperature almost independent of $J'/J$ ($T^*$ in Fig.~\ref{fig:phase-yk}).
This originates from the Fermi degeneracy of the complex fermions composed of the itinerant Majorana fermions, similar to the honeycomb case in Sec.~\ref{sec:2D_honeycomb}.
Hence, the comparison between the honeycomb and triangle-honeycomb cases implies that the high-$T$ crossover arising from the itinerant Majorana fermions is commonly seen in the variants of the Kitaev model, while the low-$T$ one from the localized $Z_2$ fluxes may appear differently depending on the nature of the $Z_2$ flux in each system.
We will further examine this conjecture in several examples in three dimensions in Sec.~\ref{sec:3D}.

\subsection{Phase transitions in three dimensions}
\label{sec:3D}

In this section, we discuss the thermodynamic behaviors in some variants of the Kitaev model in three dimensions.
In Sec.~\ref{sec:loop_proliferation}, we present the results for the 3D Kitaev model on the so-called hyperhoneycomb structure.
In this model, the low-$T$ crossover at $T_L$ in the 2D honeycomb case in Sec.~\ref{sec:2D_honeycomb} is replaced by a phase transition as in the triangle-honeycomb case in Sec.~\ref{sec:2D_CSL}, but the low-$T$ phase is not a CSL but the Kitaev QSL in this case.
We discuss the origin of the phase transition to the QSL on the basis of the distinct nature of the $Z_2$ flux excitations in three dimensions.
In Sec.~\ref{sec:gas-liq-solid}, we present the results for the model which exhibits a phase transition to a conventional magnetically-ordered phase in addition to that to the QSL.
Finally, we discuss the phase transitions to 3D CSLs on a lattice structure with odd-cycle plaquettes, dubbed the hypernonagon lattice in Sec.~\ref{sec:3D_CSL}.

\subsubsection{Phase transition by loop proliferation: gas-liquid transition}
\label{sec:loop_proliferation}

\begin{figure}[t]
 \begin{center}
  \includegraphics[width=0.65\columnwidth,clip]{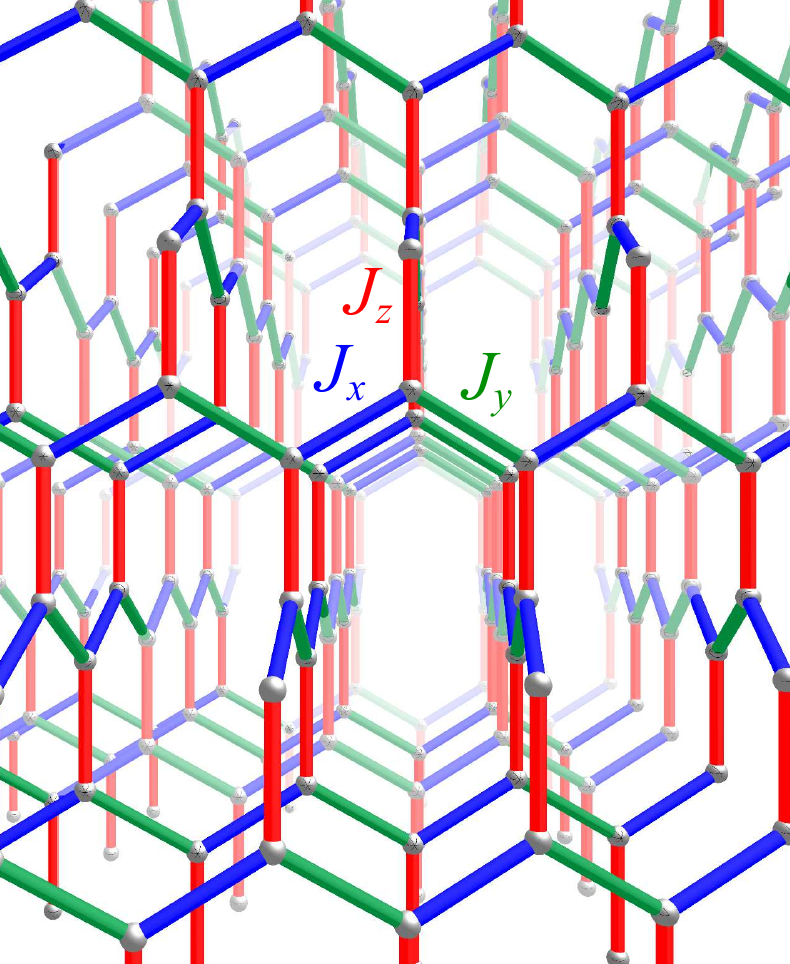}
  \caption{
(Color online) 
Schematic picture of the hyperhoneycomb structure where the 3D Kitaev model is defined.
}
  \label{fig:lattice-hh}
 \end{center}
\end{figure}

S.~Mandal and N.~Surendran discussed a variant of the Kitaev model on a 3D lattice structure~\cite{Mandal2009}, which was later called the hyperhoneycomb structure shown in Fig.~\ref{fig:lattice-hh}.
The lattice is in a series of extensions of the honeycomb structure to three dimensions~\cite{Modic2014}, and surprisingly, it is realized in a candidate material $\beta$-Li$_2$IrO$_3$~\cite{Takayama2015} (see Sec.~\ref{sec:3D_iridates} for the details).
Mandal and Surendran showed that the hyperhoneycomb Kitaev model retains the exact solvability and the ground state offers a 3D exact QSL.
They also argued the peculiar nature of the $Z_2$ flux excitations, which we will discuss below.

\begin{figure}[t]
 \begin{center}
  \includegraphics[width=0.9\columnwidth,clip]{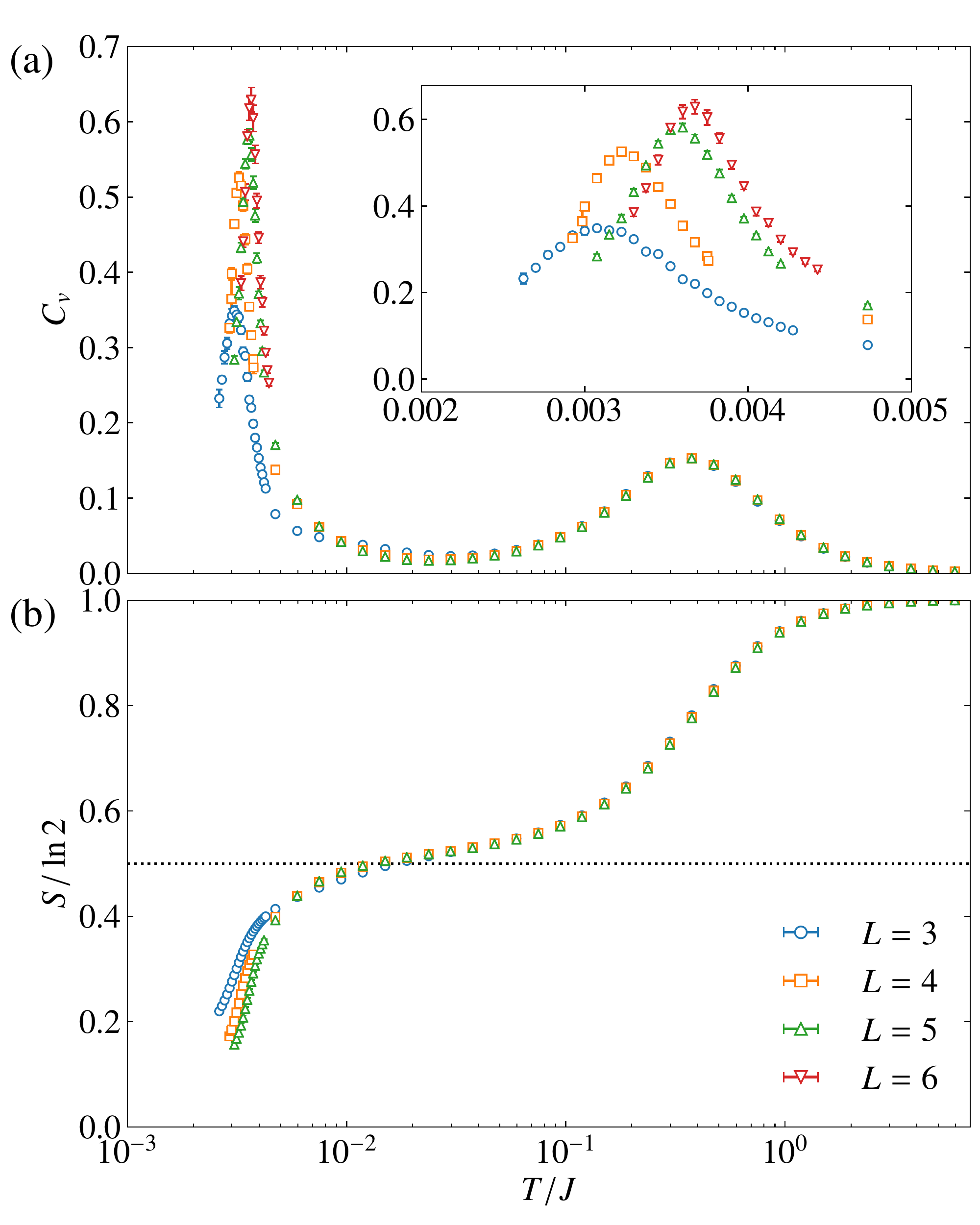}
  \caption{
(Color online) 
$T$ dependences of (a) the specific heat and (b) the entropy per site for the 3D Kitaev model on the hyperhoneycomb lattice with isotropic coupling $J_x=J_y=J_z=J$.
The data are obtained by the Majorana-based QMC simulations for the clusters with $N=4L^3$ spins ($L=3$-$6$).
The extended plot around the low-$T$ peak is shown in the inset of (a).
The horizontal dotted line in (b) represents $\frac12 \ln 2$.
The data are taken from Ref.~\citen{Nasu2014}.
}
  \label{fig:3d-cv}
 \end{center}
\end{figure}

Finite-$T$ behavior of this 3D model was studied by the Majorana-based QMC simulations~\cite{Nasu2014}.
$T$ dependences of the specific heat and entropy are presented in Fig.~\ref{fig:3d-cv} for the isotropic case with $J_x=J_y=J_z=J$.
Although the overall behaviors are similar to those for the 2D honeycomb case in Figs.~\ref{fig:ene}(b) and \ref{fig:ene}(c), clear differences appear at low $T$; while the low-$T$ peak in the specific heat does not depend on the system size in the 2D case, the height becomes higher and the width gets narrower in the present 3D case as increasing the system size [see also the inset of Fig.~\ref{fig:3d-cv}(a)].
This signals a phase transition instead of the crossover.
A similar phase transition was also found by larger-scale simulations for the effective model in the anisotropic limit of the Kitaev coupling, called the Kitaev toric code~\cite{Nasu2014b}.
In the present case, however, in contrast to the transition to the CSL in Sec.~\ref{sec:2D_CSL}, the freezing of the $Z_2$ fluxes does not break time-reversal symmetry, as the lattice structure does not include odd cycles.
Then, what happens in this finite-$T$ phase transition?

The phase transition is caused by a change of topological nature in the excitations of the $Z_2$ fluxes~\cite{Nasu2014,Nasu2014b}.
In the 3D case, the localized $Z_2$ fluxes cannot be flipped independently because of the local constraint arising from the lattice geometry~\cite{Mandal2009}.
Any 3D lattices have closed volumes composed of several plaquettes.
For any closed volume, the product of the $Z_2$ flux operators $W_p$ becomes an identity because of the algebra of the Pauli matrices~\cite{Mandal2009}.
This gives the local constraint that does not allow to flip the $Z_2$ fluxes independently: The excitations are only allowed in a form of closed loops composed of flipped $W_p$.
This is in contrast with the 2D cases where there is no local constraint (there is a global constraint $\prod_p W_p = 1$, but it does not affect thermodynamics).

What happens in the 3D case is as follows.
While raising $T$ from the flux-free QSL ground state, the localized $Z_2$ fluxes are thermally excited in the form of closed loops. 
At low $T$, the loop lengths are short compared to the system size.
With a further increase of $T$, however, excitation loops with their lengths comparable to the system size are proliferated at some $T$ because of the entropic gain, which leads to the topological transition.
The critical temperature $T_c$ is set by the loop tension arising from the excitation energy proportional to the loop length~\cite{Kimchi2014}.
Thus, the finite-$T$ phase transition in this 3D Kitaev model is caused by the loop proliferation.
The picture of this topological transition will be further discussed in Sec.~\ref{sec:phase_diagram}.

The phase transition takes place between the high-$T$ PM state and the low-$T$ QSL state.
The former is regarded as ``gas'' in terms of the spin degree of freedom, while the latter is regarded as ``liquid'', both of which preserve the symmetry of the system.
Therefore, the phase transition is regarded as a ``gas-liquid'' transition in the spin degree of freedom.
In contrast to the conventional gas-liquid transition, which is discontinuous in general, the numerical results in Fig.~\ref{fig:3d-cv} do not find any discontinuity.
The analysis of the effective model in the anisotropic limit concludes that the phase transition is continuous and belongs to the inverted 3D Ising universality class; 
the confined loops are favored in the low-$T$ (high-$T$) phase in the 3D toric code (Ising model).
Note that the closed loops in the 3D Ising model are composed of interacting spins, which appear in the high-$T$ expansion and contribute to the partition function. 
The order parameter of this peculiar transition is not described by any local quantities but it can be identified by a global quantity called the Wilson loop, which is given by the product of all $W_p$ on the plane defined by a given loop~\cite{Nasu2014}. 
Note that the Wlison loop measures the parity of the total number of the excited $W_p$ lines penetrating the plane.
Thus, this phase transition caused by the loop proliferation evades from the conventional Landau-Ginzburg-Wilson theory for the continuous phase transitions.

A similar phase transition was found also for another 3D Kitaev model defined on the so-called hyperoctagon lattice~\cite{Mishchenko2017}.
The origin of the phase transition is common.
This suggests that the loop proliferation works as a common mechanism for the gas-liquid phase transition in 3D Kitaev models.
The comparative study between the hyperhoneycomb and hyperoctagon cases confirmed the correlation between $T_c$ and the loop tension~\cite{Mishchenko2017}.

\subsubsection{Three states of matter: Gas-liquid-solid transition}
\label{sec:gas-liq-solid}

\begin{figure}[t]
 \begin{center}
  \includegraphics[width=0.9\columnwidth,clip]{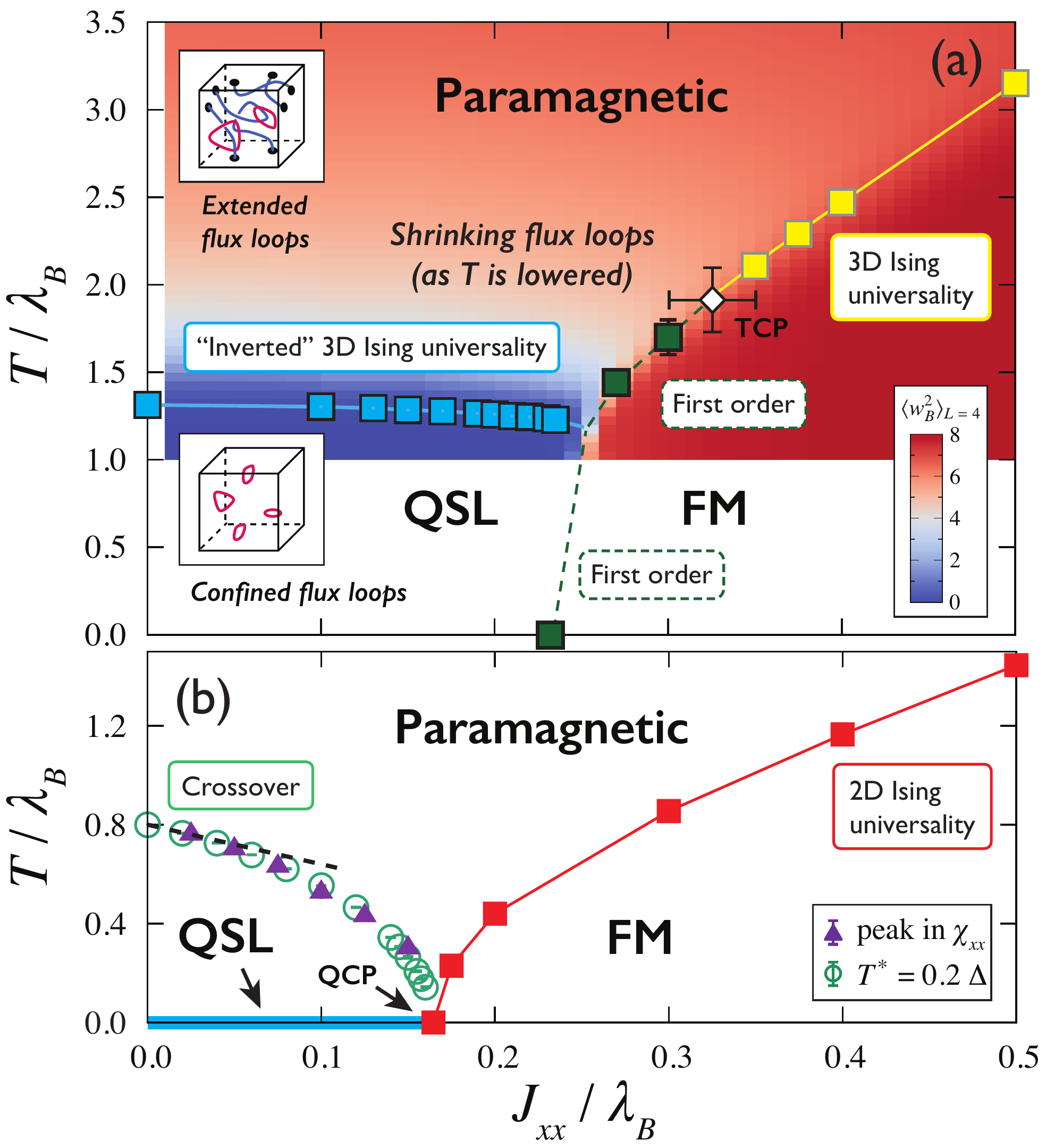}
  \caption{
(Color online) 
Finite-$T$ phase diagrams for (a) the 3D and (b) 2D Kitaev toric code with the FM Ising interaction $J_{xx}$.
$\lambda_B$ is the coupling constant in the toric code.
TCP in (a) and QCP in (b) denote the tricritical point and the quantum critical point, respectively.
Reprinted with permission from Ref.~\citen{Kamiya2015} $\copyright$ (2015) the American Physical Society.
}
  \label{fig:3d-toric-phase}
 \end{center}
\end{figure}

Stimulated by the finding of the gas-liquid phase transition in the spin degree of freedom, the phase transitions for three states of matter, gas, liquid, and solid, were investigated for the Kitaev toric code with additional ferromagnetic Ising interaction~\cite{Kamiya2015}.
In this model, while increasing the Ising interaction, the QSL ground state is taken over by a FM ordered state, which is regarded as ``solid''.
Hence, one can expect the phase transitions between the three states of matter.
Figure~\ref{fig:3d-toric-phase}(a) shows the phase diagram obtained by extensive QMC simulations (in this case, not the Majorana-based QMC but the continuous-time world-line QMC in the original spin representation)~\cite{Kamiya2015}.
The result indicates that the gas-liquid transition described in Sec.~\ref{sec:loop_proliferation} survives against the FM Ising interaction with a slight decrease of the critical temperature, but at some point, it changes into a phase transition between the high-$T$ PM state and the low-$T$ FM state, which is a gas-solid transition.
The first-order transition line between the QSL and FM phases extends from $T=0$ to the tricritical points on the gas-liquid and gas-solid transition lines (the former is not identified within the numerical precision).
In the PM state near the bifurcation of the phase boundaries, an interesting proximity effect was found in the flux loop excitations~\cite{Kamiya2015}.

Similar study was conducted also for the 2D case~\cite{Kamiya2015}.
The result is shown in Fig.~\ref{fig:3d-toric-phase}(b).
In contrast to the 3D case in Fig.~\ref{fig:3d-toric-phase}(a), the QSL phase is limited to zero $T$, while there is a crossover at finite $T$. 
The crossover $T$ decreases as the Ising interaction increases, and finally goes to zero at the quantum critical point.
For larger Ising interactions, the FM state evolves with continuous growth of $T_c$.
The phase transition at $T_c$ is continuous and belongs to the 2D Ising universality class.
Thus, the phase transitions for three states of matter in the spin degree of freedom look qualitatively different between the 3D and 2D cases, owing to the distinct nature of the $Z_2$ flux excitations.

The above study of three states of matter has been limited to the toric code corresponding to the anisotropic limit of the Kitaev coupling.
The issue in a more realistic parameter region remains for future study, which is potentially relevant to understanding of the properties of 3D candidate materials for the Kitaev model (see Sec.~\ref{sec:3D_iridates}).
The 2D case is also worth investigating~\cite{Mandal2011}; indeed, in a weakly anisotropic case, an interesting liquid-liquid phase transition between the Kitaev QSL and a spin-nematic quantum paramagnet was found before entering the FM ordered state~\cite{Nasu2017b}.

\subsubsection{Phase transitions to 3D chiral spin liquids}
\label{sec:3D_CSL}

In Sec.~\ref{sec:2D_CSL}, we discussed finite-$T$ phase transitions to 2D CSLs with spontaneous breaking of time-reversal symmetry.
Similar transitions in three dimensions were studied for the 3D Kitaev model defined on the lattice structure with odd cycles, dubbed the hypernonagon structure~\cite{Kato2017,Mishchenko2019preprint}.
In the 3D case, there is an interesting possibility of successive phase transitions, since the 3D Kitaev models can exhibit a topological transition by the loop proliferation discussed in Sec.~\ref{sec:loop_proliferation}, in addition to the spontaneous time-reversal symmetry breaking.
Such a possibility was studied for two anisotropic limits of the Kitaev coupling in the hypernonagon Kitaev model~\cite{Kato2017}.
The numerical results indicate that the system exhibits a single discontinuous phase transition with simultaneous occurrence of the loop proliferation and time-reversal breaking.
Interestingly, however, the low-$T$ CSL state is not a flux-free state but shows a nonuniform spatial order of the $Z_2$ fluxes.
The study was extended to other parameter regions apart from the anisotropic limits, and at least five distinct phases with different nonuniform flux orders were discovered~\cite{Mishchenko2019preprint}.

Most of the studies of CSLs thus far have been limited to two dimensions since the pioneering work by V. Kalmeyer and R. B. Laughlin~\cite{Kalmeyer1987}.
The above results offer the examples of 3D CSLs that allow detailed studies of their nature and the phase transitions owing to the exact solvability of the Kitaev model.
Further development on this interesting issue will be expected by using the extensions of the Kitaev model.

\subsection{Phase diagram}
\label{sec:phase_diagram}

\begin{figure}[t]
 \begin{center}
  \includegraphics[width=\columnwidth,clip]{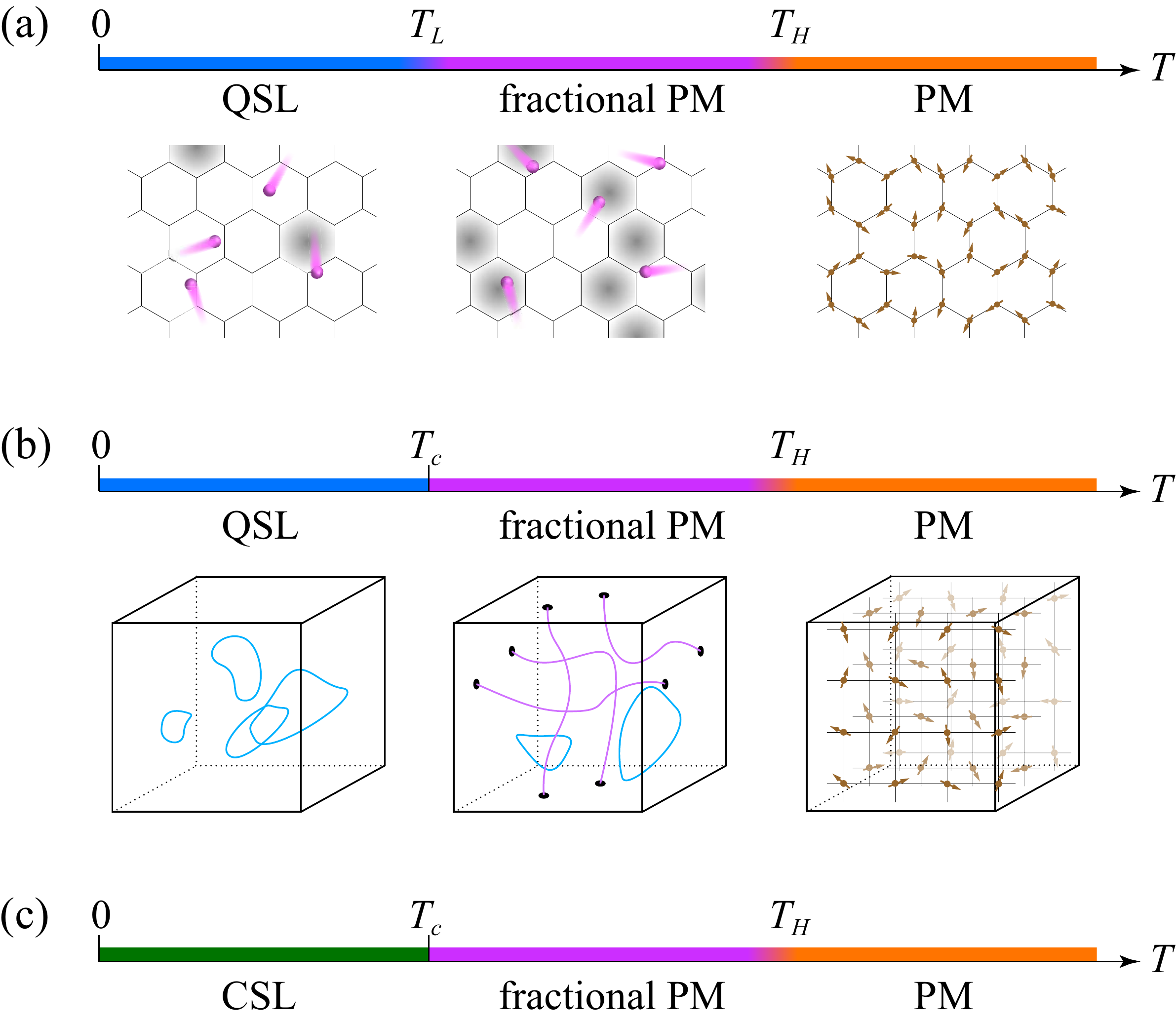}
  \caption{
(Color online) 
Schematic finite-$T$ phase diagrams of the Kitaev models for
(a) the 2D cases like the honeycomb case in Sec.~\ref{sec:2D_honeycomb},
(b) the 3D cases like the hyperhoneycomb case in Sec.~\ref{sec:loop_proliferation}, and
(c) the 2D and 3D cases like the triangle-honeycomb case in Sec.~\ref{sec:2D_CSL} and the hypernonagon case in Sec.~\ref{sec:3D_CSL}, respectively.
In (a), the systems exhibits three states separated by two crossovers at $T_L$ and $T_H$: the asymptotic QSL for $T\lesssim T_L$, the fractional PM for $T_L\lesssim T\lesssim T_H$, and the conventional PM for $T\gtrsim T_H$.
The lower panels show the schematic pictures of the three states.
The magenta spheres, the gray hexagons, and the arrows represent the itinerant Majorana fermions, the flipped localized $Z_2$ fluxes, and the spins, respectively (see Fig.~\ref{fig:majorana}).
In (b), the systems undergo a ``gas-liquid" phase transition at $T_c$ from the low-$T$ QSL to the fractional PM and a crossover at $T_H$ to the conventional PM.
In the schematic picture in the lower panels, the cyan and purple lines represent short and extended loops composed of the flipped localized $Z_2$ fluxes, respectively.
In (c), the phase transition at $T_c$ occurs between the low-$T$ CSL and the fractional PM states.
}
  \label{fig:2d3d-phase}
 \end{center}
\end{figure}

As a brief summary of Sec.~\ref{sec:Tfrac}, we show schematic phase diagrams at finite $T$ for the Kitaev models in both two and three dimensions.
Figure~\ref{fig:2d3d-phase}(a) displays the 2D honeycomb case~\cite{Nasu2015}, which will be common to other 2D cases without odd cycles in the lattice structure.
In this case, the system exhibits two crossovers at $T=T_H$ and $T_L$.
The former temperature scale is set by the COM of the itinerant fermion DOS, and the latter by the excitation gap for the localized $Z_2$ fluxes.
The finite-$T$ state is separated into three by these two crossovers: the conventional PM for $T\gtrsim T_H$, the fractional PM for $T_L\lesssim T \lesssim T_H$, and the asymptotic QSL for $T\lesssim T_L$.
The schematic picture of each region is shown in the lower panels of Fig.~\ref{fig:2d3d-phase}(a).

Meanwhile, Fig.~\ref{fig:2d3d-phase}(b) displays the 3D counterpart, inferred from the results for the 3D hyperhoneycomb~\cite{Nasu2014,Nasu2014b} and hyperoctagon cases~\cite{Mishchenko2017}.
In this case, while the high-$T$ crossover at $T_H$ remains in a similar manner, the low-$T$ one is replaced by the phase transition of gas-liquid type caused by the loop proliferation.
The difference arises from the distinct nature of the localized $Z_2$ flux excitations.
The schematic picture in terms of the excited loops is shown in the lower panels of Fig.~\ref{fig:2d3d-phase}(b).

Figure~\ref{fig:2d3d-phase}(c) presents the schematic phase diagram in the presence of odd cycles in the lattice structure.
In this case, the system can show a finite-$T$ phase transition to a CSL with spontaneous breaking of time-reversal symmetry.
The transition is a single phase transition from the fractional PM state to the CSL for the 2D triangle-honeycomb~\cite{Nasu2015b} and the 3D hypernonagon cases~\cite{Kato2017,Mishchenko2019preprint}.
It can be split into multiple transitions in the latter as discussed in Sec.~\ref{sec:3D_CSL}, but such behavior has not been found thus far.

\section{Material candidates}
\label{sec:candidates}

In this section, we briefly overview candidates for materialization of the Kitaev QSL.
We here focus on some of $4d$- and $5d$-electron compounds.
Readers who are interested in more details including other candidates are referred to other review articles~\cite{Trebst2017preprint,Winter2017a,Takagi2019}.

\subsection{Quasi-2D iridates}
\label{sec:2D_iridates}

As discussed in Sec.~\ref{sec:Jackeli-Khaliullin}, Jackeli and Khaliullin pointed out two requisites for materialization of the Kitaev coupling.
They nominated $A_2B$O$_3$-type layered compounds as a good candidate.
Following this proposal, J. Chaloupka and his coworkers have pointed out that this is indeed the case for the quasi-2D honeycomb iridium oxides, Na$_2$IrO$_3$ and $\alpha$-Li$_2$IrO$_3$~\cite{Chaloupka2010}.
These two compounds have a common quasi-2D lattice structure with the honeycomb layers composed of edge-sharing IrO$_6$ octahedra~\cite{Singh2010,Singh2012,Freund2016}, as shown in Fig.~\ref{fig:ligand}(c); the crystal symmetry belongs to space group $C2/m$.
(We put the prefix $\alpha$ only for Li$_2$IrO$_3$ since it has polymorphs as introduced in Sec.~\ref{sec:3D_iridates}.)
In these compounds, the formal valence of the Ir ions is $4+$, and hence, the outermost $5d$ shell is partially occupied by five electrons.
As described in Sec.~\ref{sec:Jackeli-Khaliullin}, this leads to the low-spin $5d^5$ state under the cubic crystalline electric field, and furthermore, comprises the $j_{\rm eff}=1/2$ pseudospin with the influence of the strong spin-orbit coupling [see Figs.~\ref{fig:ligand}(a) and \ref{fig:ligand}(b)].
The importance of both spin-orbit coupling and Coulomb interaction and the formation of the $j_{\rm eff}=1/2$ state have been confirmed by spectroscopic measurements~\cite{Comin2012,Sohn2013}. 
The pseudospins are expected to interact with each other via the Kitaev coupling through the perturbation processes in the edge-sharing geometry [see Fig.~\ref{fig:ligand}(c)].
The predominant Kitaev coupling was experimentally confirmed for Na$_2$IrO$_3$ by using diffuse X-ray scattering~\cite{Chun2015} and torque magnetometry~\cite{Das2019}.
It was supported also by theoretical estimates based on first-principles calculations~\cite{Foyevtsova2013,Katukuri2014,Yamaji2014,Winter2016}.

Despite the presence of the predominant Kitaev coupling, these candidates do not show QSL behavior in the low-$T$ limit; instead, they undergo a phase transition to a magnetically-ordered phase at low $T$.
Na$_2$IrO$_3$ exhibits a zigzag-type AFM ordering at the critical temperature $T_N\simeq 15$~K~\cite{Singh2010,Liu2011,Ye2012}, while $\alpha$-Li$_2$IrO$_3$ exhibits an incommensurate spiral ordering at almost the same $T$~\cite{Singh2012,Freund2016,Williams2016}.
The magnetic orders are considered to be induced by non-Kitaev couplings in the honeycomb layer as well as interlayer couplings, which are weaker than the Kitaev coupling.
Thus, it is widely believed that the compounds are proximate to the Kitaev QSL, whereas the low-$T$ properties are hindered by the parasitic magnetic orders.

There have been several efforts to realize the Kitaev QSL by suppressing the non-Kitaev interactions.
Theoretically, it was proposed that a thin film~\cite{Yamaji2014} and a heterostructure~\cite{Winter2016} might be helpful for this purpose.
Also, experimentally, the chemical substitutions of $A$-site ions locating between the honeycomb layers were attempted, and $A'_3$LiIr$_2$O$_6$ with $A'$=Ag, Cu, and H were synthesized~\cite{Todorova2011,Roudebush2016,Kitagawa2018}.
These compounds have a different stacking manner from Na$_2$IrO$_3$ and $\alpha$-Li$_2$IrO$_3$.
Among them, H$_3$LiIr$_2$O$_6$ is intriguing since it does not show any magnetic ordering down to the lowest $T$~\cite{Kitagawa2018}, while disorder effects have been argued, as discussed in the end of Sec.~\ref{sec:other_exchanges}.
In addition, Cu$_2$IrO$_3$ was recently nominated as a candidate, but in this case also the effect of chemical disorder was pointed out~\cite{Abramchuk2017,Choi2019,Kenny2019,Takahashi2019}.

\subsection{$\alpha$-RuCl$_3$}
\label{sec:RuCl3}

Another candidate is a Ru trichloride $\alpha$-RuCl$_3$, which was firstly pointed out in Ref.~\citen{Plumb2014}.
This compound has a similar quasi-2D layered honeycomb structure with edge-sharing RuCl$_6$ octahedra, but the crystal symmetry is controversial among $P3_112$, $C2/m$, and $\bar{R}3$ depending on the samples~\cite{Fletcher1963,Fletcher1967,Brodersen1968,Kubota2015,Johnson2015,Cao2016,Do2017}.
This might be related with the fact that the honeycomb layers are weakly coupled with each other via the van der Waals interaction~\cite{Kim2016}.
The formal valence of the Ru ions is $3+$, and hence, the $4d^5$ electron configuration offers a playground for the Kitaev coupling similar to the iridium oxides in the previous section.
The formation of the $j_{\rm eff}=1/2$ state was confirmed, e.g, by the spectroscopic measurements with the help of first-principles calculations~\cite{Plumb2014,Koitzsch2016,Sinn2016,Yadav2016,Winter2016}.

Unfortunately, this compound also exhibits a magnetic order of zigzag type at low $T$~\cite{Sears2015,Johnson2015,Cao2016}.
The critical temperature is, however, scattered between $T_N\simeq 6.5$~K and $\simeq 14$~K depending on the samples.
It is believed that the samples with stacking faults show rather high $T_N$; the lowest $T_N=6.5$~K was reported for a single crystal with $\bar{R}3$ symmetry~\cite{Do2017}.

One of the advantages in $\alpha$-RuCl$_3$ is the feasibility of inelastic neutron scattering, which is a powerful tool to probe spin dynamics (note that Ir is a neutron absorber).
Recently, several measurements have been done in a wide range of energy and wave vector.
The results will be discussed in comparison with theoretical results for the Kitaev model in Sec.~\ref{sec:Sqw}.

Another advantage is that the zigzag magnetic order in $\alpha$-RuCl$_3$ can be suppressed by an external magnetic field of $\sim 8$~T applied within the $ab$ plane~\cite{Kubota2015,Johnson2015}.
This opens an interesting possibility to realize QSL behavior in the field-induced PM region.
We will discuss the recent development on this issue in Sec.~\ref{sec:Sqw}, \ref{sec:1/T1}, and \ref{sec:kappa_xy}.

Last but not least, $\alpha$-RuCl$_3$ has a unique aspect owing to the fact that this compound is a van der Waals material:
The weak interlayer coupling allows to fabricate the samples in a thin film form~\cite{Weber2016,Ziatdinov2016,Gronke2018,Zhou2019a}.
More recently, interesting electronic properties were observed for heterostructures between a thin film of $\alpha$-RuCl$_3$ and graphene~\cite{Zhou2019b,Mashhadi2019,Biswas2019preprint,Gerber2019preprint}.
Such fabrication of thin films and heterostructures will stimulate further studies on interesting physics arising from the potential fractional excitations in this Kitaev candidate magnet.

\subsection{3D iridates}
\label{sec:3D_iridates}

\begin{figure}[t]
 \begin{center}
  \includegraphics[width=0.9\columnwidth,clip]{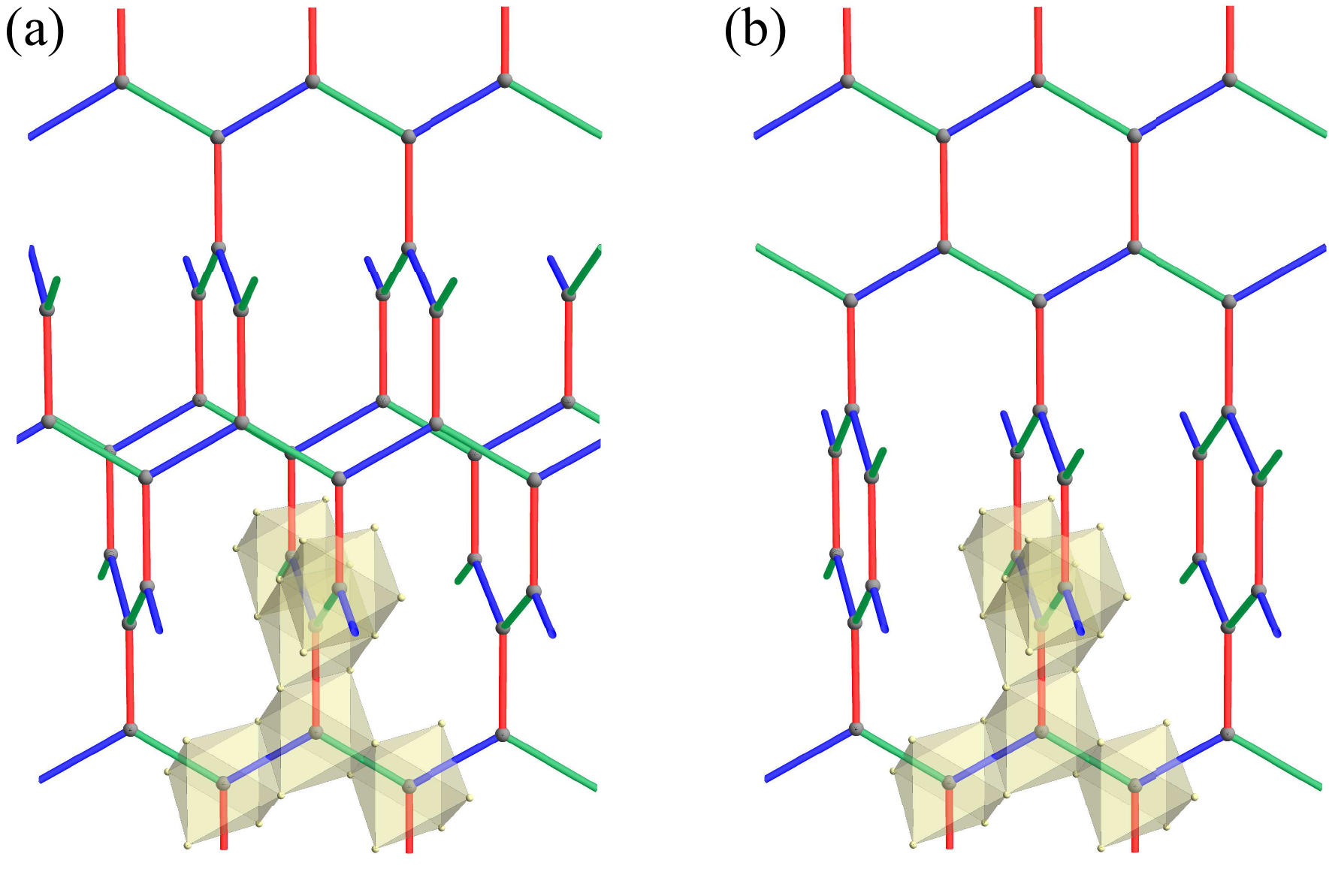}
  \caption{
(Color online) 
Schematic pictures of (a) the hyper- and (b) stripy-honeycomb structures with edge-sharing octahedra, which are realized in $\beta$- and $\gamma$-Li$_2$IrO$_3$, respectively.
}
  \label{fig:lattice-hh-sh}
 \end{center}
\end{figure}

Finally, we introduce two polymorphs of Li$_2$IrO$_3$: $\beta$-Li$_2$IrO$_3$ and $\gamma$-Li$_2$IrO$_3$.
These two compounds have 3D networks of the edge-sharing IrO$_6$ octahedra, instead of the quasi-2D layered one in $\alpha$-Li$_2$IrO$_3$.
$\beta$-Li$_2$IrO$_3$ has the so-called hyperhoneycomb structure with space group $Fddd$ [Fig.~\ref{fig:lattice-hh-sh}(a); see also Fig.~\ref{fig:lattice-hh}]~\cite{Takayama2015}, and $\gamma$-Li$_2$IrO$_3$ has the stripy-honeycomb structure with space group $Cccm$ [Fig.~\ref{fig:lattice-hh-sh}(b)]~\cite{Modic2014}.
Both structures belong to a series of the harmonic honeycomb structures~\cite{Modic2014}.
In both cases, the local coordination is common to $\alpha$-Li$_2$IrO$_3$, and the Ir ions comprise tricoordinate lattices, for which the Kitaev model can be extended in a straightforward manner.
Thus, these polymorphs have attracted attention as candidates for the 3D Kitaev QSL discussed in Sec.~\ref{sec:loop_proliferation}~\cite{Kim2015a,Katukuri2016}.
However, they show spiral magnetic ordering at rather high temperature $T_N\sim 40$~K~\cite{Takayama2015,Biffin2014a,Modic2014,Biffin2014b}.
Interestingly, the magnetic orders can be suppressed by applying relatively small magnetic fields~\cite{Ruiz2017,Modic2017} as well as external pressure~\cite{Breznay2017,Takayama2019}.

\section{Comparative study between theory and experiment}
\label{sec:theory_exp}

In this section, we discuss the signatures of thermal fractionalization in the Kitaev QSL through the comparison between theory and experiment.
On the theoretical side, we concentrate on the Kitaev model in Eq.~(\ref{eq:H}) defined on the honeycomb structure, neglecting other additional interactions discussed in Sec.~\ref{sec:other_exchanges}, as it allows to obtain reliable results by well-controlled numerical techniques.
All the following results are for the isotropic Kitaev coupling $J_x=J_y=J_z=J$.
Meanwhile, on the experimental side, we present the data for three candidates: the honeycomb iridium oxides, Na$_2$IrO$_3$ and $\alpha$-Li$_2$IrO$_3$, and the Ru trichloride $\alpha$-RuCl$_3$.

\subsection{Specific heat and entropy}
\label{sec:Cv_S}

\begin{figure}[t]
\begin{center}
\includegraphics[width=0.9\columnwidth,clip]{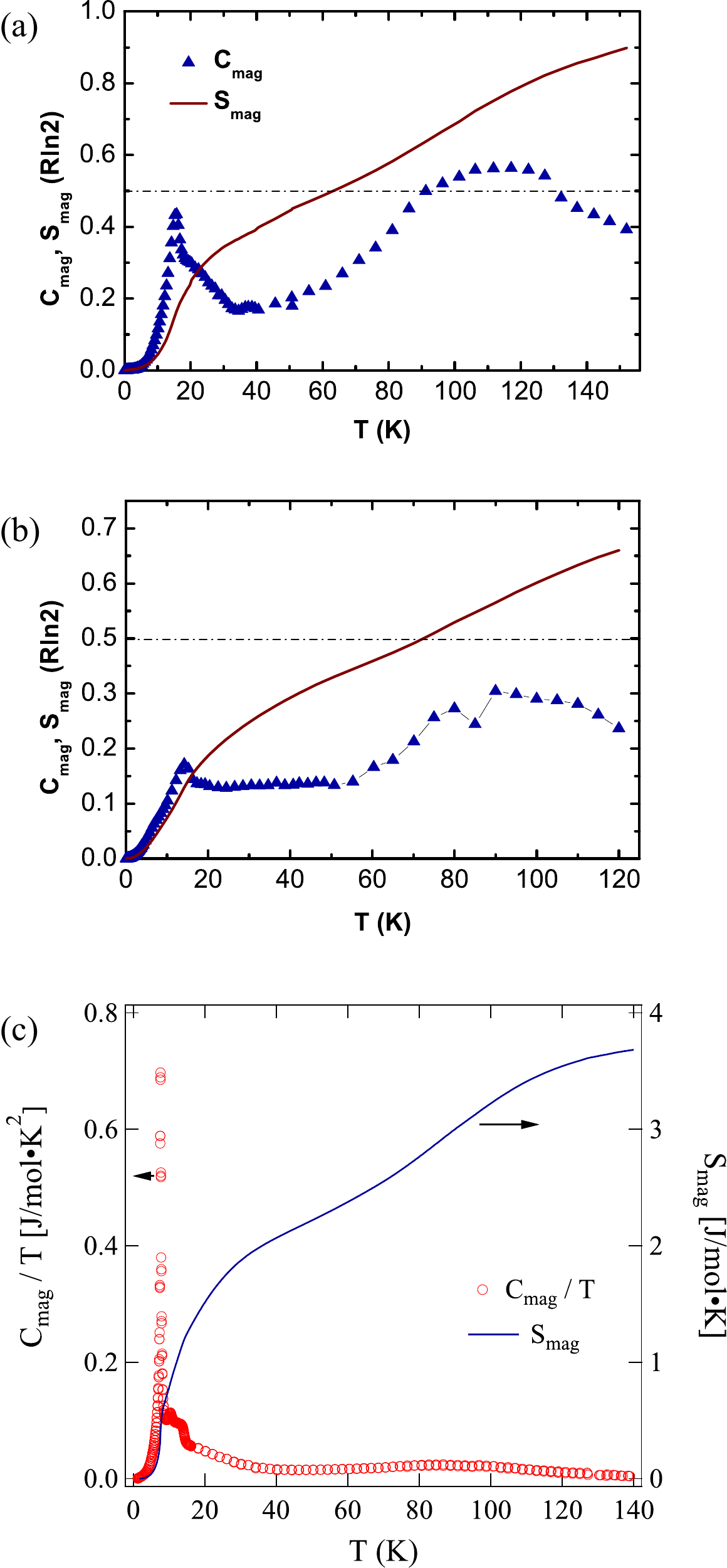}
\caption{
(Color online) 
$T$ dependences of the specific heat ($C_{\rm mag}$) and entropy ($S_{\rm mag}$) for (a) Na$_2$IrO$_3$, (b) $\alpha$-Li$_2$Ir$_3$, and (c) $\alpha$-RuCl$_3$.
In (c), the specific heat divided by $T$ is plotted.
The magnetic contributions are extracted by subtracting the data for the nonmagnetic compounds, (a) Na$_2$SnO$_3$, (b) Li$_2$SnO$_3$, and (c) ScCl$_3$.
The figures (a) and (b) are reprinted with permission from Ref.~\citen{Mehlawat2017} $\copyright$ (2017) the American Physical Society. 
The figure (c) is reprinted with permission from Ref.~\citen{Kubota2015} $\copyright$ (2015) the American Physical Society.
}
\label{fig:CandS}
\end{center}
\end{figure}

Let us first begin with the comparison for the specific heat and entropy.
Figure~\ref{fig:CandS}(a) displays the experimental data for a candidate material for the Kitaev model, Na$_2$IrO$_3$~\cite{Mehlawat2017}.
The specific heat exhibits a broad peak around 110~K, in addition to a sharp anomaly at $T_N\simeq 15$~K associated with the magnetic ordering.
The entropy is released corresponding to the high-$T$ broad peak, and shows an interesting $T$ dependence with inflection points; the decrease becomes slow around 60~K, where the entropy is roughly half $R \ln 2$ ($R$ is the gas constant).
With a further decrease of $T$, the entropy is continuously released, and finally, decreases rapidly at the magnetic phase transition at $T_N\simeq 15$~K.
Qualitatively similar behaviors were observed for the related compound $\alpha$-Li$_2$IrO$_3$~\cite{Mehlawat2017} and another candidate $\alpha$-RuCl$_3$~\cite{Kubota2015}, as shown in Figs.~\ref{fig:CandS}(b) and \ref{fig:CandS}(c), respectively.
In addition, in a recent study for $\alpha$-RuCl$_3$~\cite{Do2017}, $T$-linear behavior of the specific heat was reported in the intermediate $T$ region, as suggested for the Majorana metal in Sec.~\ref{sec:Majorana_metal}.

At first glance, these experimental data look similar to the theoretical results for the Kitaev model presented in Sec.~\ref{sec:crossovers}, except for the sharp anomaly at the magnetic transition temperature.
Then, it is natural to ask whether the similarities provide experimental evidence for the thermal fractionalization arising from the Kitaev QSL.
The answer is that although they look consistent with theory, it is difficult to admit them as strong evidence.
On one hand, the broad peak in the specific heat at high $T$ is in fact commonly seen in frustrated magnets; the suppression of magnetic ordering by the frustration leaves development of short-range spin correlations, which gives rise to the entropy release in the high-$T$ region.
This is also the case in the Kitaev model: As shown in Fig.~\ref{fig:corr}(a), the crossover at $T=T_H$ is related with the growth of NN spin correlations.
Hence, the broad peak of the specific heat alone cannot be evidence of the thermal fractionalization.
On the other hand, the approximately half $R \ln 2$ entropy at the shoulderlike feature also looks consistent with the theoretical result, but this is again not conclusive, considering that in general it is not easy to precisely estimate the lattice contributions in experiments.
Also, theoretically, it is difficult to predict how non-Kitaev interactions, which are inevitably present in real materials, affect the behavior of the entropy at low $T$~\cite{Yamaji2016,Suzuki2018}.

Then, what could be evidence in these thermodynamic quantities?
A specific feature to the Kitaev QSL is the low-$T$ crossover at $T=T_L$ by the freezing of the localized $Z_2$ fluxes.
Unfortunately, in the candidate materials shown above, $T_L \simeq 0.012J$ is considered to be around 1~K, which is lower than the critical temperatures.
Thus, the interesting behavior associated with the $Z_2$ fluxes, if any, is hindered by the parasitic magnetic ordering caused by non-Kitaev interactions.
A potential route to unveil the crossover behavior is to suppress the magnetic order by applying an external magnetic field, as discussed in Sec.~\ref{sec:field}.
Such an experiment was indeed performed for $\alpha$-RuCl$_3$, and a peak was observed in the region where the magnetic order is suppressed by the magnetic field~\cite{Sears2017,Wolter2017,Widmann2019}.
Meanwhile, the specific heat in the magnetic field was recently calculated for the Kitaev model by using a newly-developed CTQMC method~\cite{Yoshitake2019preprint};
a similar peak was obtained in the high-field region, while the data at low $T$ and low field are lacked because of the negative sign problem.
Further detailed comparison is necessary to identify the signature of the $Z_2$ fluxes.

\subsection{Spin correlation}
\label{sec:SS}

\begin{figure}[t]
\begin{center}
\includegraphics[width=0.7\columnwidth,clip]{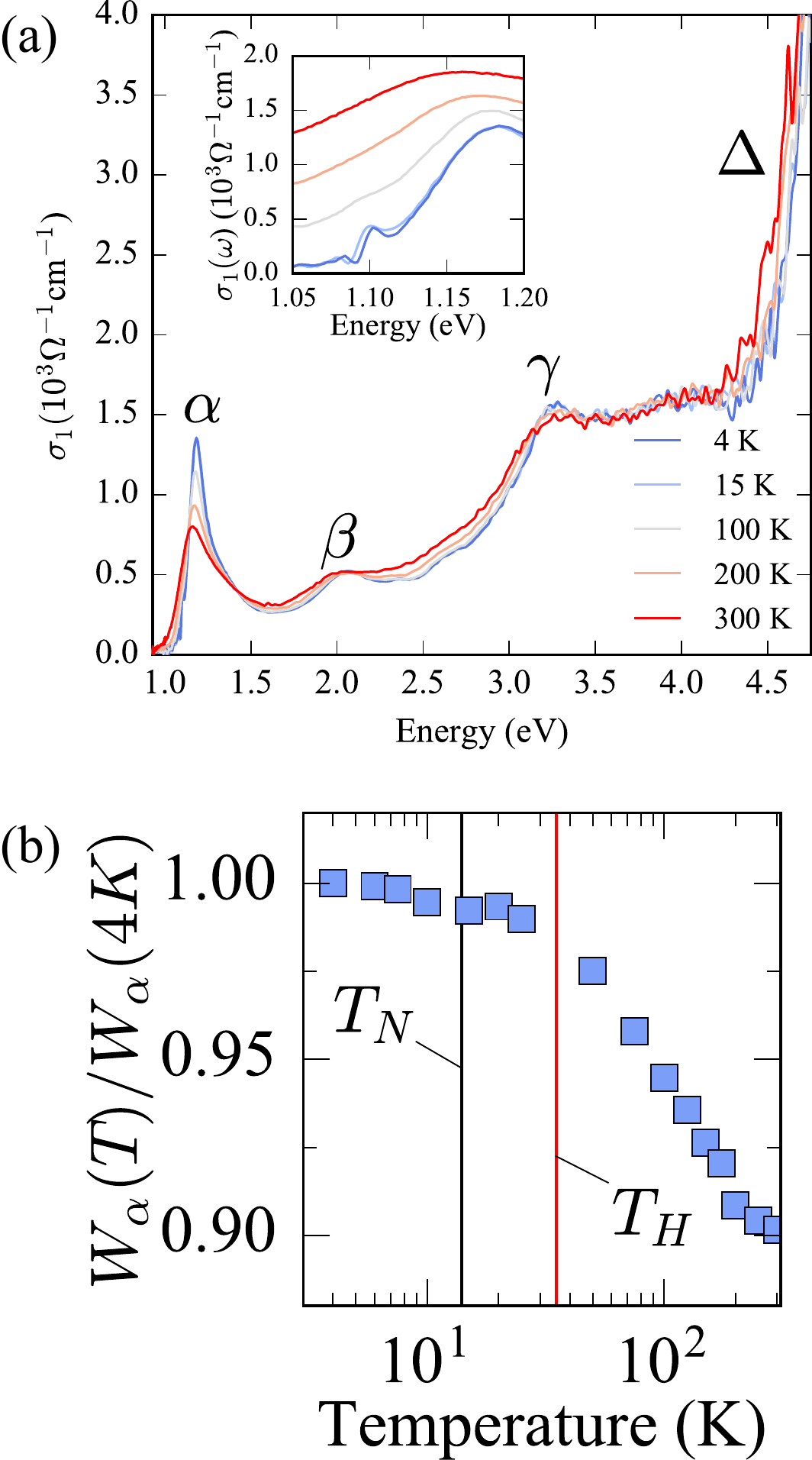}
\caption{
(Color online) 
(a) Real part of the optical conductivity obtained for $\alpha$-RuCl$_3$ at several $T$.
The inset shows the low-energy detail around the peak $\alpha$; the data for $100$, $200$, and $300$~K are offset for clarity.
(b) $T$ dependence of the spectral weight of the peak $\alpha$ in (a) integrated in the energy range between 0.9 to 1.4~eV.
The data are normalized by that at 4~K.
In (b), $T_H$ is shown by assuming $J= 8$~meV, and $T_N$ denotes the critical temperature for the magnetic ordering of this sample.
Reprinted with permission from Ref.~\citen{Sandilands2016} $\copyright$ (2016) the American Physical Society. 
}
\label{fig:SS_RuCl3}
\end{center}
\end{figure}

The equal-time spin correlation was indirectly obtained for $\alpha$-RuCl$_3$ by an optical measurement~\cite{Sandilands2016}.
In this experiment, several peaks were identified in the optical conductivity above the Mott gap $\sim 1$~eV, as shown in Fig.~\ref{fig:SS_RuCl3}(a).
Among them, the lowest-energy peak just above the Mott gap, denoted as $\alpha$ in Fig.~\ref{fig:SS_RuCl3}(a), shows considerable $T$ dependence.
As this excitation reflects virtual motions of electrons beyond the Mott gap, the $T$ dependence is considered to contain the information on the development of spin correlations originating from the virtual exchange processes.
The $T$ dependence of the spectral weight of the peak $\alpha$ is shown in Fig.~\ref{fig:SS_RuCl3}(b).
The data show that, while decreasing $T$, the spectral weight grows down to $\sim 40$~K, whereas it almost saturates in the lower-$T$ region, even below the critical temperature $T_N$.
This behavior resembles the $T$ dependence of the NN spin correlations in the Kitaev model plotted in Fig.~\ref{fig:corr}(a), where $T_H\sim 35~$~K by assuming $J\sim 8$~meV.
Nevertheless, the change of the weight in Fig.~\ref{fig:SS_RuCl3}(b) is rather small ($\sim 10$~\%), which might be due to contributions from other excitations in the optical spectrum and the $T$ independent contributions in the virtual exchange processes.
It is desired to make further quantitative comparison and also to perform more direct measurement of the spin correlations.

\subsection{Magnetic susceptibility}
\label{sec:chi}

\begin{figure}[t]
\begin{center}
\includegraphics[width=0.9\columnwidth,clip]{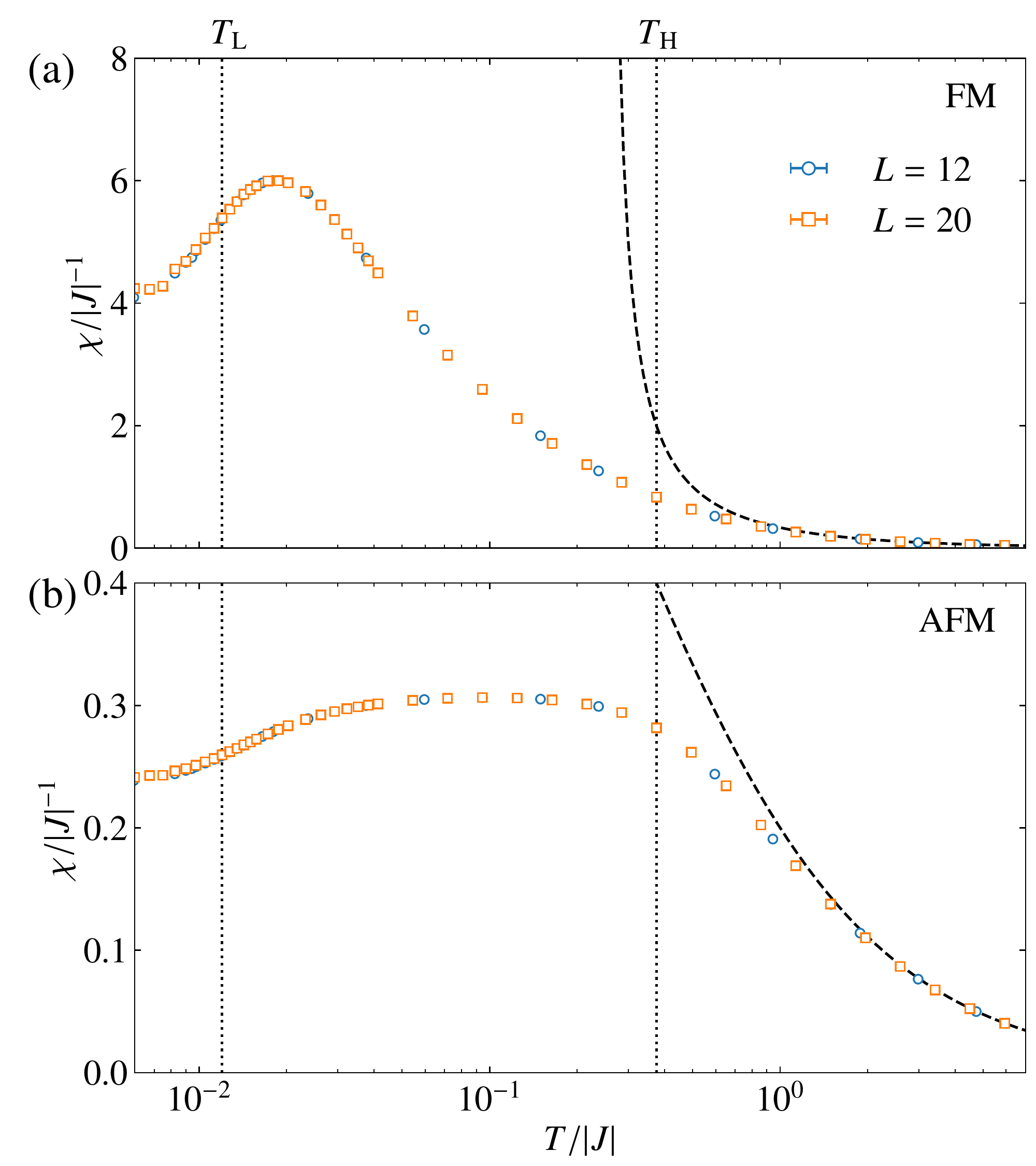}
\caption{
(Color online) 
$T$ dependence of the magnetic susceptibility for the honeycomb Kitaev model with isotropic coupling $J_x=J_y=J_z=J$ obtained by combining the Majorana-based QMC and CTQMC methods.
(a) and (b) correspond to the cases with FM and AFM Kitaev coupling, respectively.
The dashed curves represent the Curie-Weiss behaviors.
The data are taken from Ref.~\citen{Yoshitake2017b}.
}
\label{fig:chi}
\end{center}
\end{figure}

The magnetic susceptibility for the Kitaev model was calculated by Majorana-based numerical techniques~\cite{Yoshitake2016,Yoshitake2017a,Yoshitake2017b}, by using the formula
\begin{align}
\chi^{\mu\nu} = \frac1N \sum_{i,j} \int_0^\beta \langle S_i^\mu(\tau) S_j^\nu \rangle d\tau,
\end{align}
where $\beta=1/(k_{\rm B}T)$ is the inverse temperature (we set the Boltzmann constant $k_{\rm B}=1$), and $\langle S_i^z(\tau) S_j^z \rangle$ is the dynamical spin correlation in the ($2+1$)-dimensional space, where $S_i^\mu(\tau)=e^{\tau {\cal H}}S_i^\mu e^{-\tau {\cal H}}$ ($\tau$ is the imaginary-time).
Figure~\ref{fig:chi} shows the results obtained by the combined technique between the Majorana-based QMC and CTQMC methods~\cite{Yoshitake2017b} (see Appendix~\ref{sec:quantum-monte-carlo} and \ref{sec:cont-time-quant}).
Note that all the off-diagonal components $\chi^{\mu\nu}$ with $\mu\neq \nu$ vanish in the Kitaev model~\cite{Baskaran2007}, and $\chi^{xx}=\chi^{yy}=\chi^{zz}=\chi$ in the isotropic case. 

As shown in Fig.~\ref{fig:chi}, although the $T$ dependence as well as the overall magnitude is different between the cases with FM and AFM Kitaev coupling, the two cases share the following features.
(i) At sufficiently high $T$, $\chi$ obeys the Curie-Weiss law as in other magnets, but it starts to deviate below $T\sim J$.
The Curie-Weiss behavior is given by $\chi = 1/(4T-J)$ for the FM case and $\chi=1/(4T+J)$ for the AFM case.
(ii) $\chi$ exhibits a peak in the fractional PM region between $T_L$ and $T_H$.
The peak temperature is at $T\simeq 0.02J$ in the FM case and $T\simeq 0.1J$ in the AFM case.
(iii) $\chi$ decreases rapidly around $T_L$ with showing an inflection point.
This suppression is attributed to the freezing of $Z_2$ flux excitations by the gap opening.
(iv) In the low-$T$ limit, $\chi$ approaches a nonzero value.
Similar asymptotic behavior is commonly seen in the magnetic systems in which the total spin is not conserved.
In the present case, owing to the Dirac-like linear dispersion in the fermionic excitations, the asymptotic behavior is expected to be proportional to $T^3$ up to a constant, but it is hard to extract such behavior from the numerical results~\cite{Yoshitake_thesis}.

\begin{figure}[t]
\begin{center}
\includegraphics[width=0.8\columnwidth,clip]{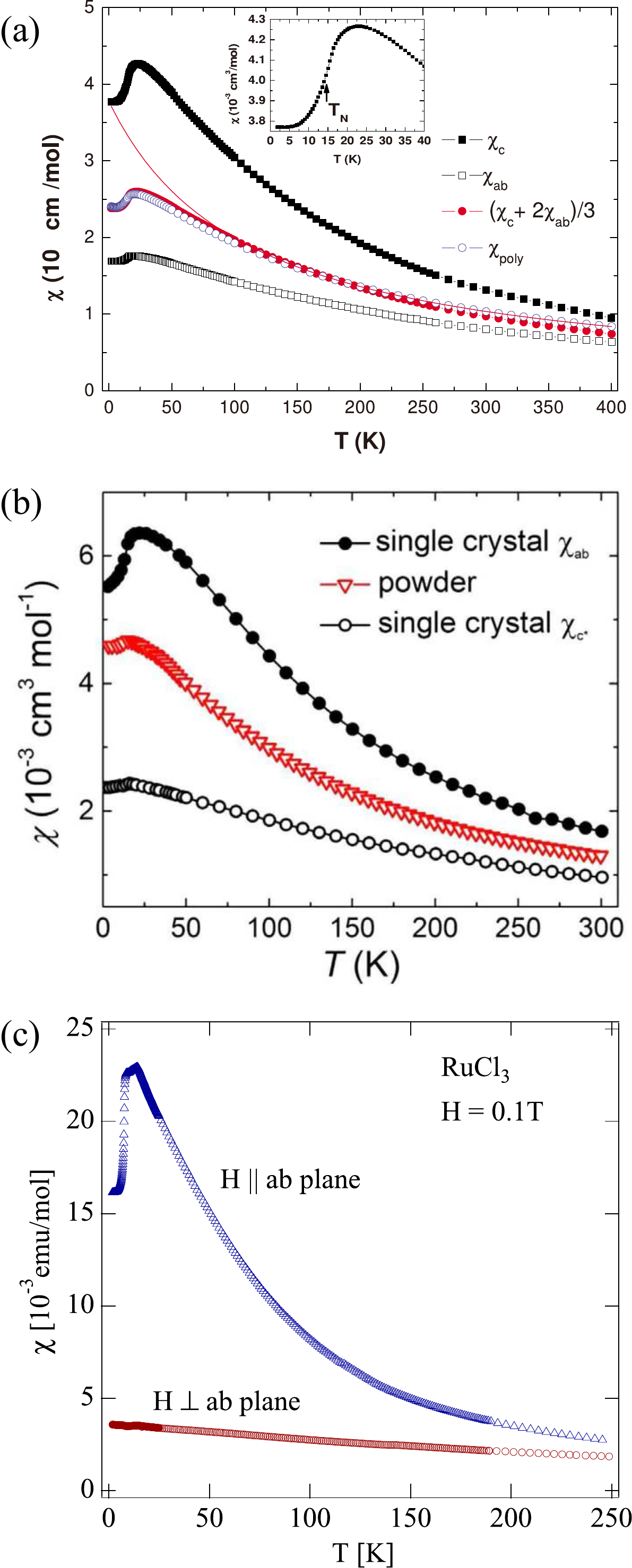}
\caption{
(Color online) 
$T$ dependences of the magnetic susceptibility for single crystals of (a) Na$_2$IrO$_3$, (b) $\alpha$-Li$_2$IrO$_3$, and (c) $\alpha$-RuCl$_3$.
In (a), the red curve represents the fitting by the Curie-Weiss law for the high-$T$ data.
The inset displays the enlarged plot for the low-$T$ part around the critical temperature $T_N$.
The data for powder samples are also plotted in (a) and (b).
The figure (a) is reprinted with permission from Ref.~\citen{Singh2010} $\copyright$ (2010) the American Physical Society. 
The figure (b) is reprinted from Ref.~\citen{Freund2016}. 
The figure (c) is reprinted with permission from Ref.~\citen{Kubota2015} $\copyright$ (2015) the American Physical Society. 
}
\label{fig:chi_exp}
\end{center}
\end{figure}

For comparison, we showcase the experimental data for the candidate materials in Fig.~\ref{fig:chi_exp}.
The data for Na$_2$IrO$_3$ in Fig.~\ref{fig:chi_exp}(a) shows that the susceptibility obeys the Curie-Weiss law above $\sim 150$~K, but starts to deviate at lower $T$~\cite{Singh2010}.
Similar behavior was observed also for $\alpha$-Li$_2$IrO$_3$~\cite{Singh2012} [see also Fig.~\ref{fig:chi_exp}(b)~\cite{Freund2016}].
In both cases, a peak appears at a slightly higher $T$ than the critical temperature $T_N \simeq 15$~K.
Below the peak, the susceptibility turns to decrease and exhibits an inflection point around $T_N$, and finally approaches a nonzero constant at the lowest $T$.
These behaviors appear to be at least qualitatively similar to the theoretical results in Fig.~\ref{fig:chi}, although one cannot compare the data below $T_N$.
Similar behaviors were observed for $\alpha$-RuCl$_3$~\cite{Kubota2015,Sears2015} [see Fig.~\ref{fig:chi_exp}(c)].

Meanwhile, a readily-seen discrepancy between theory and experiment is the magnetic anisotropy in the experimental data, as shown in Fig.~\ref{fig:chi_exp}.
The theoretical results are isotropic for the isotropic case, and it is also difficult to explain the magnetic anisotropy by the anisotropy in the Kitaev coupling~\cite{Yoshitake2017a}.
The importance of additional non-Kitaev interactions as well as the anisotropy of the $g$ factor was pointed out for the magnetic anisotropy~\cite{Chaloupka2016,Yadav2016,Janssen2017,Lampen-Kelley2018}.
It remains as a future issue to quantitatively explain the $T$ dependence of the anisotropic susceptibility and to determine the magnitude and sign of the Kitaev coupling.
We will comment on the sign of the Kitaev coupling in Sec.~\ref{sec:Sqw}.

Can we say that the comparison between theory and experiment for the magnetic susceptibility provides evidence for the proximity to the Kitaev QSL?
As in the case of the specific heat and entropy in Sec.~\ref{sec:Cv_S}, the similarity found in the $T$ dependence is suggestive but not sufficient to draw conclusions.
This is because the deviation from the Curie-Weiss behavior and the broad peak structure at a lower $T$ are commonly observed in a wide class of frustrated magnets as a consequence of the growth of short-range spin correlations under the frustration.
A more decisive feature would be an experimental observation of the rapid decrease around $T_L$ with the inflection point originating from the freezing of the localized $Z_2$ fluxes.
This is, however, hindered again by the magnetic ordering in the real compounds.

\subsection{Inelastic neutron scattering}
\label{sec:Sqw}

\begin{figure*}[t]
\begin{center}
\includegraphics[width=0.8\textwidth,clip]{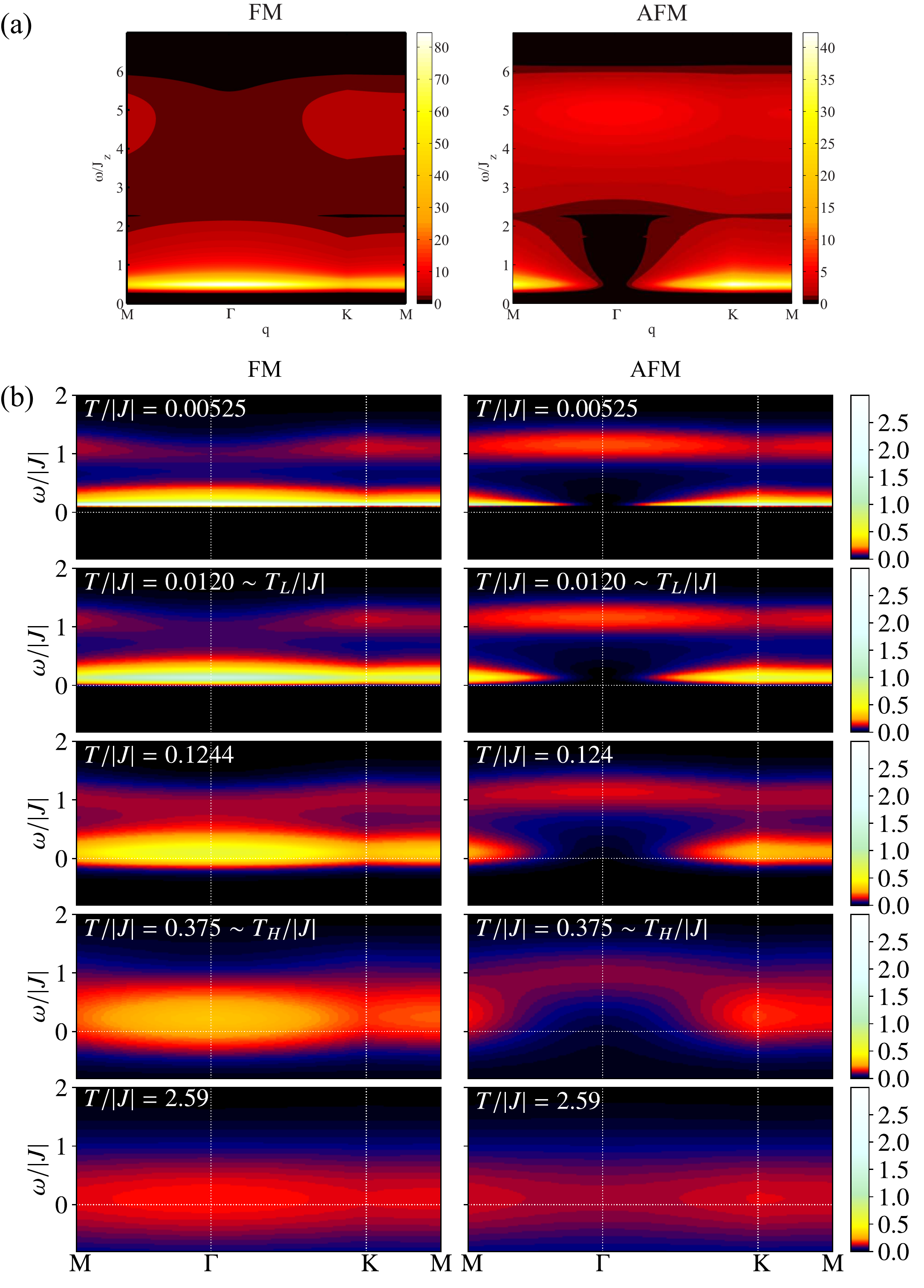}
\caption{
(Color online) 
Dynamical spin structure factor $S(\mathbf{q},\omega)$ calculated for the honeycomb Kitaev model with isotropic coupling $J_x=J_y=J_z=J$ for both cases with FM and AFM Kitaev coupling: (a) at $T=0$ and (b) for finite $T$.
The finite-$T$ results are obtained by combining the Majorana-based QMC and CTQMC methods.
Note that the energy scale in (a) is four-times different from the definition in this article used in (b):
$\omega/J_z = 4$ in (a) corresponds to $\omega/J=1$ in (b).
The figure (a) is reprinted with permission from Ref.~\citen{Knolle2015} $\copyright$ (2015) the American Physical Society. 
The data in (b) are taken from Ref.~\citen{Yoshitake2017b}. 
}
\label{fig:Sqw}
\end{center}
\end{figure*}

Inelastic neutron scattering is a powerful experimental tool to probe the spin dynamics.
The scattering intensity is proportional to the dynamical spin structure factor which includes the information on the spin dynamics as a function of wave vector $\mathbf{q}$ and frequency $\omega$.
Figure~\ref{fig:Sqw}(a) displays the theoretical results for the QSL ground state of the Kitaev model for both FM and AFM cases~\cite{Knolle2014,Knolle2015}.
Here, the dynamical spin structure factor is calculated by
\begin{align}
S^{\mu\nu}(\mathbf{q},\omega) &= \frac1N \sum_{i,j} \int_{-\infty}^\infty\frac{dt}{2\pi}\langle S_i^\mu(t) S_j^\nu \rangle e^{i (\omega t -\mathbf{q} \cdot \mathbf{r}_{ij})},
\end{align}
where $\mathbf{r}_{ij}$ is the vector connecting the sites $i$ and $j$, and $S_i^\mu(t)$ is the Heisenberg representation of $S_i^\mu$.
In the following, we mainly discuss the sum of the diagonal components, $S(\mathbf{q},\omega) = S^{xx}(\mathbf{q},\omega)+S^{yy}(\mathbf{q},\omega)+S^{zz}(\mathbf{q},\omega)$.

There are several interesting features in the results shown in Fig.~\ref{fig:Sqw}(a).
(i) The $\mathbf{q}$ dependence is weak.
This is due to the fact that the Kitaev model possesses extremely short-ranged spin correlations, as discussed in Sec.~\ref{sec:groundstate}.
(ii) The intensity vanishes below the rather strong response at low energy $\omega\sim 0.4J_z$ which corresponds to $\omega\sim 0.1J$ in our definition (see the figure caption).
This is due to the gap opening in the flux excitations.
As discussed in Sec.~\ref{sec:frac}, the spins are fractionalized into the Majorana fermions and the $Z_2$ fluxes, meaning that the elementary spin-flip excitation is given by a composite of the Majorana fermion excitation and the $Z_2$ flux excitation.
Hence, the spin excitation spectrum in $S(\mathbf{q},\omega)$ reflects the gap opening in the fractional excitations of the $Z_2$ fluxes.
At the same time, this suggests that the strong intensity above the gap predominantly originates from the $Z_2$ flux excitations.
(iii) In addition to the low-energy response, the spectrum has a broad incoherent intensity in the high-energy region extending up to $\omega\sim 6J_z$ corresponding to $\omega\sim 1.5J$ in our definition.
This reflects mainly the itinerant Majorana fermion excitations, which has the bandwidth $\sim 1.5J$ as shown in Sec.~\ref{sec:frac}.

\begin{figure}[t]
\begin{center}
\includegraphics[width=0.85\columnwidth,clip]{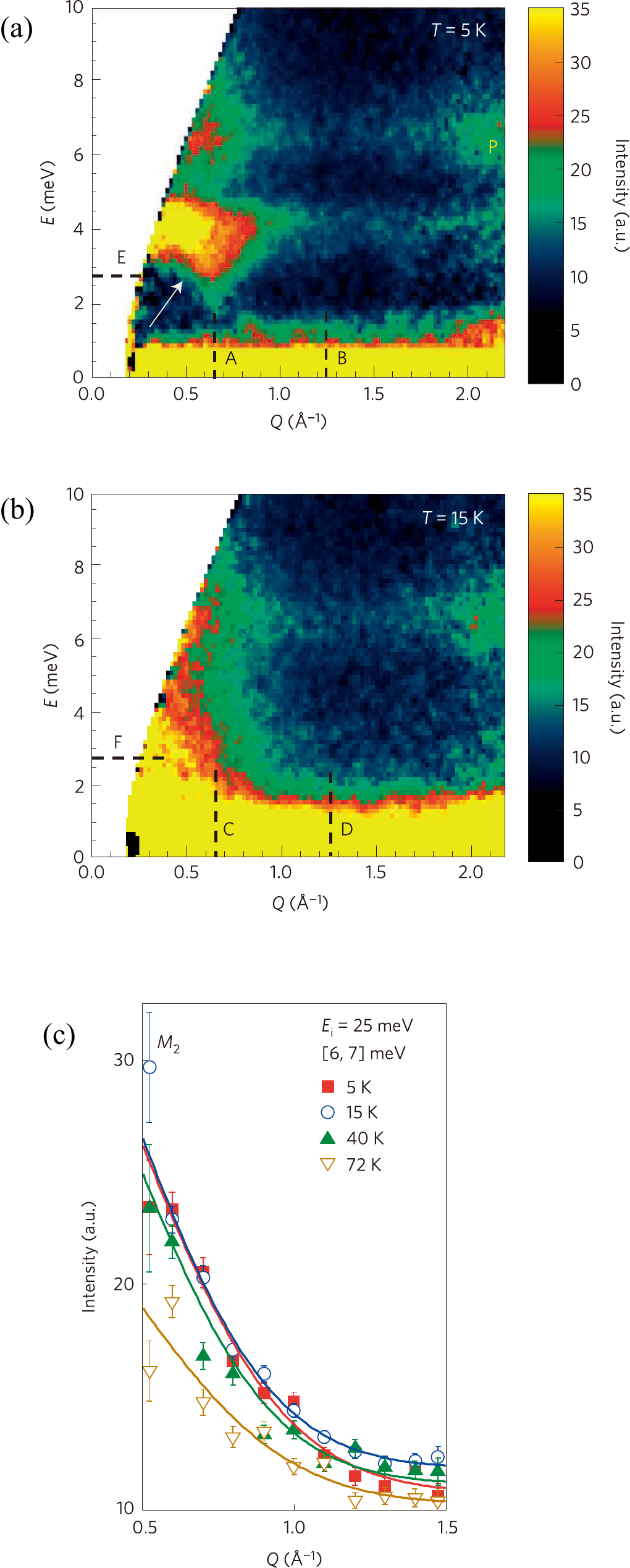}
\caption{
(Color online) 
Inelastic neutron scattering data measured for a powder sample of $\alpha$-RuCl$_3$:
the spectra at (a) $T=5$~K below $T_N$ and (b) $T=15$~K just above $T_N$ ($T_N\simeq 14$~K in this sample), and (c) the spectral weight integrated between 6 and 7~meV for several $T$.
Reprinted with permission from Ref.~~\citen{Banerjee2016} $\copyright$ (2016) Springer Nature.
}
\label{fig:INS_RuCl3-1}
\end{center}
\end{figure}

Recently, the inelastic neutron scattering measurements have been intensively performed for $\alpha$-RuCl$_3$.
In an early experiment for powder samples, an unusual incoherent intensity was observed in the energy range of $\omega=6$-8~meV, in both below and above the critical temperature $T_N$, as shown in Figs.~\ref{fig:INS_RuCl3-1}(a) and \ref{fig:INS_RuCl3-1}(b), respectively~\cite{Banerjee2016}.
This is clearly distinguished from the strong response at a lower energy only appearing below $T_N$ [indicated by the white arrow in Fig.~\ref{fig:INS_RuCl3-1}(a)] which is regarded as the spin-wave excitations in the ordered phase.
The incoherent response at high energy has a resemblance to that in the theoretical result at $T=0$ shown in Fig.~\ref{fig:Sqw}(a).
More interestingly, it remains visible up to $\sim 70$~K, which is much higher than $T_N$, as shown in Fig.~\ref{fig:INS_RuCl3-1}(c)~\cite{Banerjee2016}.
Nevertheless, at this stage, there was no theory for the $T$ dependence for comparison.

The $T$ dependence of the dynamical spin structure factor $S(\mathbf{q},\omega)$ was computed by using the combined techniques between the Majorana-based CDMFT and CTQMC~\cite{Yoshitake2016,Yoshitake2017a}, and the Majorana-based QMC and CTQMC methods~\cite{Yoshitake2017b} (see Appendix).
The calculations were done by
\begin{align}
S^{\mu\nu}(\mathbf{q},\omega) =\frac1N \sum_{i,j} S^{\mu\nu}_{i,j}(\omega) e^{-i \mathbf{q} \cdot \mathbf{r}_{ij}},
\label{eq:Sqw_finiteT}
\end{align}
where $S^{\mu\nu}_{i,j}(\omega)$ is obtained by solving
\begin{align}
\langle S_i^\mu(\tau) S_j^\nu \rangle = \int S_{i,j}^{\mu\nu}(\omega) e^{-\omega\tau} d\omega,
\label{eq:Stau_Sw}
\end{align}
by using the maximum entropy method with the Legendre polynomial expansion (see Ref.~\citen{Yoshitake2017a} for the details).

The results obtained by the Majorana-based QMC and CTQMC method are shown in Fig.~\ref{fig:Sqw}(b).
There are several interesting features.
(i) At sufficiently high $T$ in the conventional PM region above $T_H$  [lowest two panels in Fig.~\ref{fig:Sqw}(b)], $S(\mathbf{q},\omega)$ has an almost $\mathbf{q}$-independent broad peak centered around $\omega=0$.
(ii) While approaching $T_H$ with a decrease of $T$, however, an incoherent response gradually grows at high energy centered at $\omega \sim J$ [middle-lower two panels in Fig.~\ref{fig:Sqw}(b)].
This high-energy incoherent response persists down to the lowest $T$, gradually developing a weak $\mathbf{q}$ dependence.
(iii) With a further decrease of $T$ toward $T_L$, a quasi-elastic response grows in the low-energy region [center and middle-upper panels in Fig.~\ref{fig:Sqw}(b)].
(iv) Below $T_L$, this quasi-elastic response is shifted to the $\omega>0$ region with opening a small gap [upper two panels in Fig.~\ref{fig:Sqw}(b)], and the entire spectrum smoothly converges to the $T=0$ results shown in Fig.~\ref{fig:Sqw}(a).

The contrasting $T$ dependence between the high-energy incoherent response and the low-energy quasi-elastic response reflects the distinct nature between the two types of fractional excitations arising from the thermal fractionalization.
As discussed in Sec.~\ref{sec:2D_honeycomb}, the crossovers at $T=T_H$ and $T_L$ are caused by the itinerant Majorana fermions and the localized $Z_2$ fluxes, respectively.
Therefore, the growth of the high-energy incoherent response in $S(\mathbf{q},\omega)$ below $T\simeq T_H$ is considered to be dominated by the itinerant Majorana fermions, while that of the quasi-elastic response toward $T\simeq T_L$ as well as the gap opening below $T_L$ is by the localized $Z_2$ fluxes.
This is consistent with the assignment discussed above for the $T=0$ results in Fig.~\ref{fig:Sqw}(a).

\begin{figure*}[t]
\begin{center}
\includegraphics[width=\textwidth,clip]{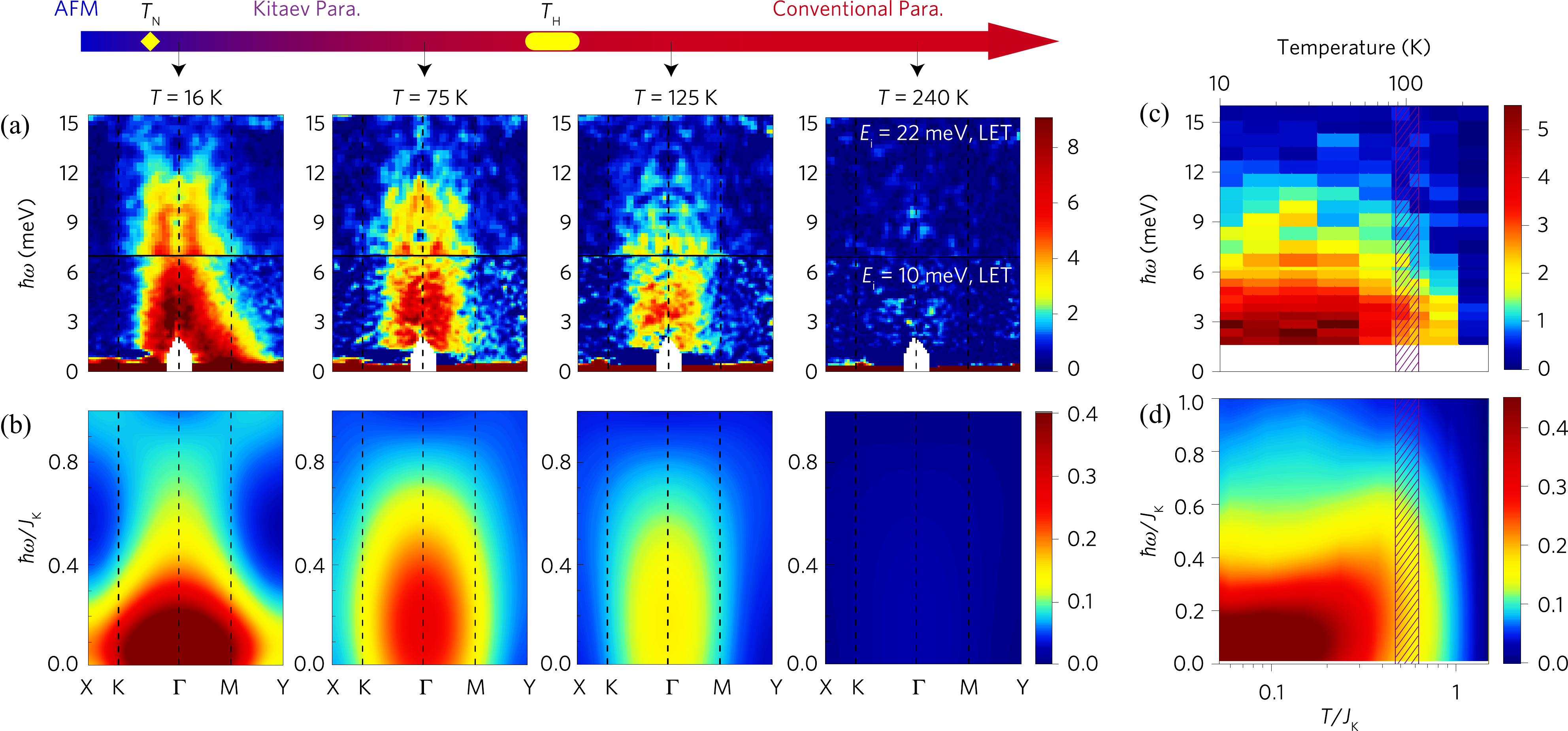}
\caption{
(Color online) 
Comparison of the dynamical spin structure factors between experiment and theory.
(a) and (c) show the experimental data for single crystals of $\alpha$-RuCl$_3$, and (b) and (d) are the theoretical results for the honeycomb Kitaev model with isotropic FM coupling calculated by using the combined technique between the Majorana-based CDMFT and CTQMC methods.
(c) and (d) display the $T$ and $\omega$ dependences of the spectra at the $\Gamma$ point $\mathbf{q}=0$.
The Bose factor correction is applied to both experimental and theoretical results.
The figure is reprinted from Ref.~\citen{Do2017}.
}
\label{fig:INS_RuCl3-2}
\end{center}
\end{figure*}

After the theoretical studies for $T>0$, inelastic neutron scattering experiments were performed for single crystals of $\alpha$-RuCl$_3$~\cite{Banerjee2017,Do2017}.
An example is shown in Fig.~\ref{fig:INS_RuCl3-2}.
As observed in the theoretical results in Fig.~\ref{fig:Sqw}(b), an unconventional incoherent response appears in a wide energy range up to $\sim 12$~meV below $\sim 100$~K.
(The differences in the energy and $T$ scales from Fig.~\ref{fig:INS_RuCl3-1} might be ascribed to the sample difference~\cite{Do2017,Park2016preprint}.)
The detailed comparison with theory in Fig.~\ref{fig:INS_RuCl3-2} indicates that, in the wide-$T$ range from the conventional PM region to just above $T_N$, the overall $\mathbf{q}$ and $\omega$ dependences of the spectra can be accounted for by the theoretical results for the Kitaev model with isotropic FM coupling.
Although the growth of the quasi-elastic response toward $T_L$ as well as the gap opening below $T_L$ predicted by theory was not observed in experiments because of the magnetic ordering at $T_N$, the agreement strongly suggests that the candidate material $\alpha$-RuCl$_3$ is in proximity to the Kitaev QSL~\cite{note3}.

Despite the overall good agreement, there remain some discrepancies between theory and experiment, especially at low $T$ and low $\omega$.
A representative feature is a star shape in the $\mathbf {q}$ dependence of the scattering intensity at low $\omega$ above $T_N$~\cite{Banerjee2017,Do2017}.
The numerical results for the Kitaev model show a round shape, unlike the star one~\cite{Do2017}.
The coexistence of such a low-energy feature and the high-energy incoherent response was discussed by considering the role of additional non-Kitaev interactions~\cite{Song2016,Banerjee2017,Gohlke2017,Winter2017b,Knolle2018}.

Let us briefly comment on the sign of the Kitaev coupling.
The comparison in Fig.~\ref{fig:INS_RuCl3-2} indicates that the FM Kitaev coupling well accounts for the weak $\mathbf{q}$ dependence in the experimental data.
In the earlier studies~\cite{Banerjee2016,Banerjee2017}, however, the AFM Kitaev coupling was deduced from the comparison between experiment and theory for the weak $\mathbf{q}$ dependence of the high-energy continuum.
The AFM Kitaev coupling was also suggested by theory based on first-principles calculations~\cite{Kim2015b}.
On the other hand, other theoretical studies based on quantum chemistry electronic-structure calculations~\cite{Yadav2016} and first-principles calculations~\cite{Winter2016} suggest the FM Kitaev coupling.
We note that, in a later experimental study by the same group~\cite{Banerjee2018}, the FM Kitaev coupling was deduced from the careful analyses of the spectral weights.

\begin{figure}[t]
\begin{center}
\includegraphics[width=\columnwidth,clip]{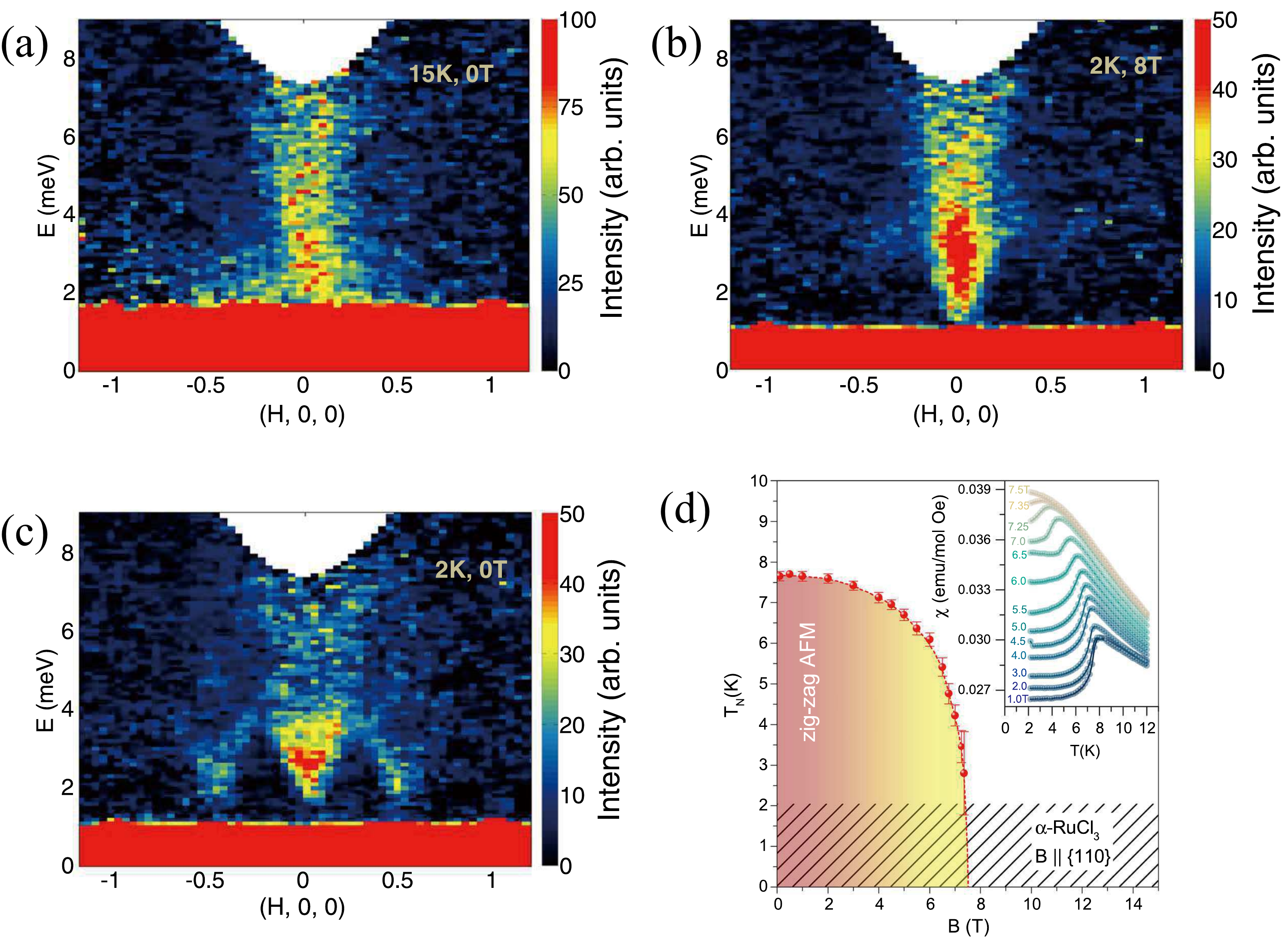}
\caption{
(Color online) 
Inelastic neutron scattering spectra measured for a powder sample of $\alpha$-RuCl$_3$ at (a) 15~K and 0~T, (b) 2~K and 8~T, and (c) 2~K and 0~T.
(d) displays the magnetic phase diagram determined by the $T$ dependence of the magnetic susceptibility shown in the inset.
The figures are reprinted with permission from Ref.~\citen{Banerjee2018} $\copyright$ (2018) by Springer Nature. 
}
\label{fig:INS_RuCl3-3}
\end{center}
\end{figure}

More recently, inelastic neutron scattering experiments have been done in a magnetic field~\cite{Banerjee2018,Balz2019}.
The experimental data for a powder sample of $\alpha$-RuCl$_3$ are shown in Fig.~\ref{fig:INS_RuCl3-3}.
The results show that the spin excitation spectrum in the region where the magnetic order is suppressed by the magnetic field [Fig.~\ref{fig:INS_RuCl3-3}(b)] is qualitatively similar to that above $T_N$ at zero field [Fig.~\ref{fig:INS_RuCl3-3}(a)];
the low-energy contribution from magnon excitations observed in the ordered phase [Fig.~\ref{fig:INS_RuCl3-3}(c)] is absent, and the high-energy incoherent response is commonly observed in Figs.~\ref{fig:INS_RuCl3-3}(a) and \ref{fig:INS_RuCl3-3}(b).
This suggests that an unconventional state potentially described by the Kitaev QSL is realized in the field-induced PM region [see the phase diagram in Fig.~\ref{fig:INS_RuCl3-3}(d); see also Fig.~\ref{fig:sketch_phase}(a) in Sec.~\ref{sec:schematic_phase_diagram}].

Theoretically, it is hard to obtain reliable results in a magnetic field since the exact solvability is lost and the Majorana-based numerical techniques cannot be applied straightforwardly, as described in Sec.~\ref{sec:field}.
However, the spin dynamics was recently obtained by a CTQMC method in a wide range of field and $T$~\cite{Yoshitake2019preprint}.
It was shown that $S(\mathbf{q},\omega)$ preserves the unconventional features reflecting the fractional excitations in the wide-field range before entering the forced FM region in the high field and low $T$.
This may explain the unconventional spectrum in the field-induced PM state discovered in the experiment above.
Moreover, the theoretical result unveiled a crossover behavior from the fractional quasiparticle picture to the conventional magnon picture while increasing the magnetic field, which is one of the confinement-deconfinement phenomena~\cite{Yoshitake2019preprint}.
Similar issue was studied by the exact diagonalization of a 24-site cluster for a model including non-Kitaev interactions~\cite{Winter2018}. 
While an experiment was performed recently~\cite{Balz2019}, further detailed comparison between theory and experiment is highly desired for these interesting issues.

\subsection{Nuclear magnetic resonance}
\label{sec:1/T1}

In addition to the inelastic neutron scattering, the NMR is an important probe of the spin dynamics.
The NMR relaxation rate is a measure of the dynamical spin susceptibility through the formula~\cite{Moriya1962}
\begin{align}
\frac{1}{T_1} \propto T \sum_\mathbf{q} | A_\mathbf{q} |^2 \frac{{\rm{Im}} \chi^\perp(\mathbf{q},\omega_0)}{\omega_0},
\label{eq:1/T1_def}
\end{align}
where $A_\mathbf{q}$ is the hyperfine coupling constant, $\chi^\perp(\mathbf{q},\omega_0)$ is the dynamical susceptibility for the spin component perpendicular to the field direction, and $\omega_0$ is the resonance frequency in the NMR measurement.
Note that the dynamical susceptibility $\chi(\textbf{q},\omega)$ is related with the dynamical spin structure factor discussed in the previous section through the fluctuation-dissipation theorem, as
\begin{align}
S(\mathbf{q}, \omega) = \frac{1}{\pi (1-e^{-\beta \omega})}{\rm{Im}}\chi(\mathbf{q},\omega).
\label{eq:S-chi}
\end{align}
In the NMR measurements, $\omega_0$ is in general negligibly small compared to the typical energy scale of the system, $J$ in the present case.
Thus, by taking the limit of $\omega_0 \rightarrow 0$ in Eq.~(\ref{eq:1/T1_def}) and using Eq.~(\ref{eq:S-chi}), one can obtain
\begin{align}
\frac{1}{T_1} \propto \sum_\mathbf{q} | A_\mathbf{q} |^2 S^\perp(\mathbf{q}, \omega=0),
\label{eq:1/T1_def2}
\end{align}
where $S^\perp(\mathbf{q}, \omega)$ is the dynamical spin structure factor for the spin components perpendicular to the field.

The NMR relaxation rate $1/T_1$ was calculated for the Kitaev model by using the Majorana-based numerical techniques~\cite{Yoshitake2016,Yoshitake2017a,Yoshitake2017b}.
The calculations were done in the limit of zero field, which correspond to the nuclear quadrupole resonance (NQR) in experiments.
Considering the fact that the Kitaev model has nonzero spin correlations only for the same site and between the NN sites (see Sec.~\ref{sec:groundstate}), $1/T_1$ for the magnetic field along the $z$ direction is computed by Eq.~(\ref{eq:1/T1_def2}) as
\begin{align}
1/T^z_1 &= a_{0,x} S^{xx}_{i,i}(\omega=0) + a_{0,y} S^{yy}_{i,i}(\omega=0) \nonumber \\
&+ a_{1,x} S^{xx}_{\rm NN}(\omega=0) + a_{1,y} S^{yy}_{\rm NN}(\omega=0),
\label{eq:1/T1z}
\end{align}
where $S^{\mu\mu}_{\rm NN}(\omega)$ represents the NN component on the $\mu$ bond [see also Eqs.~(\ref{eq:Sqw_finiteT}) amd (\ref{eq:Stau_Sw})].
In Eq.~(\ref{eq:1/T1z}), the coefficients $a_{0,x}$, $a_{0,y}$, $a_{1,x}$, and $a_{1,y}$ are determined by the hyperfine coupling constant $A_\mathbf{q}$ depending on the details of the actual compounds.

\begin{figure}[t]
\begin{center}
\includegraphics[width=0.9\columnwidth,clip]{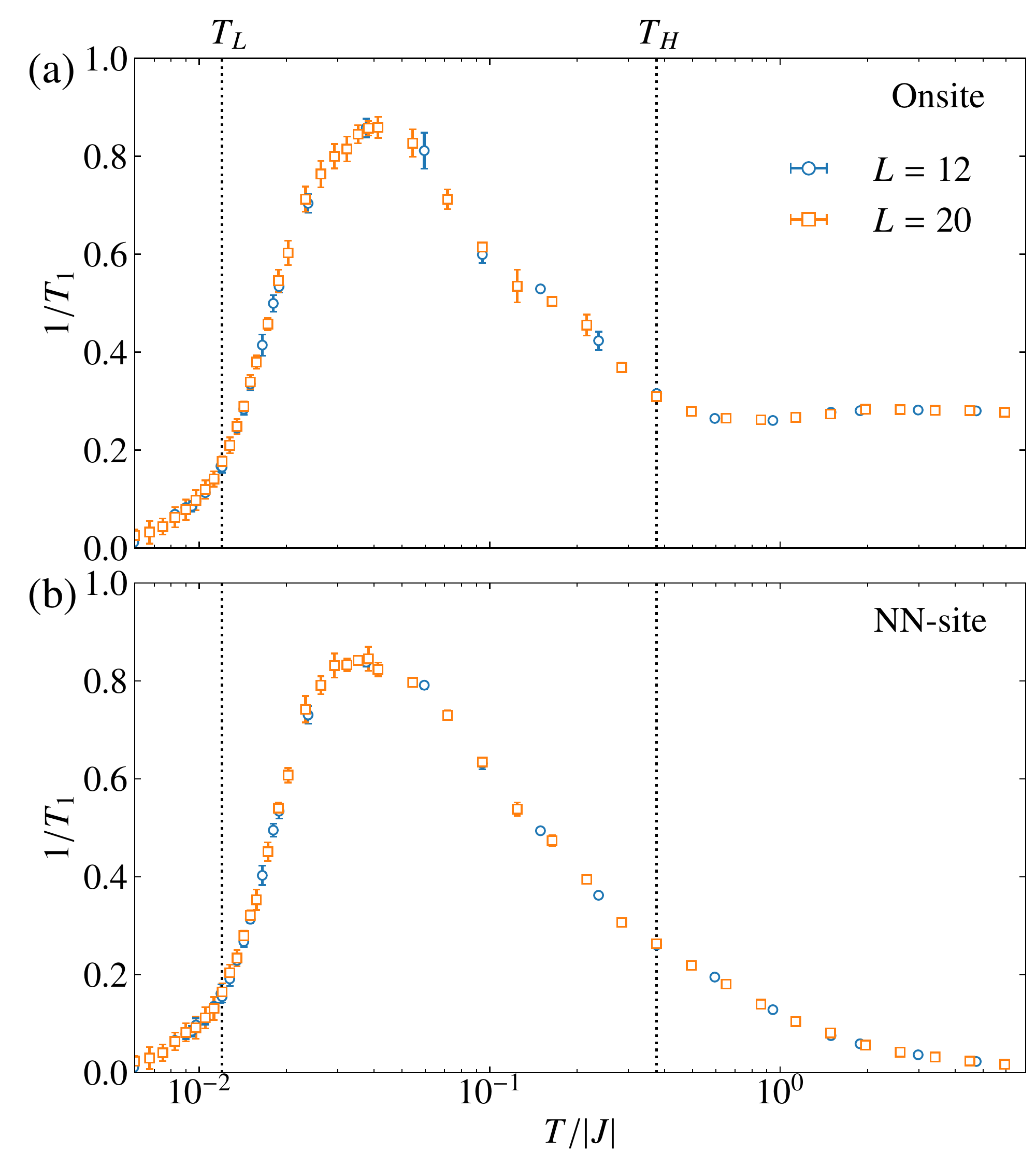}
\caption{
(Color online) 
$T$ dependence of the NMR relaxation rate $1/T_1$ for the honeycomb Kitaev model with isotropic coupling $J_x=J_y=J_z=J$ in the zero-field limit.
The results are obtained by a combined technique of the Majorana-based QMC and CTQMC methods~\cite{Yoshitake2017b}.
(a) and (b) display the contributions from onsite and NN sites [Eqs.~(\ref{eq:T1inv_onsite}) and (\ref{eq:T1inv_NNsite})], respectively.
The data are taken from Ref.~\citen{Yoshitake2017b}.
}
\label{fig:T1}
\end{center}
\end{figure}

Figure~\ref{fig:T1} shows the results for (a) the onsite and (b) NN-site components separately, defined as
\begin{align}
\label{eq:T1inv_onsite}
& 1/T_1^z = S^{xx}_{j,j}(\omega = 0) + S^{yy}_{j,j}(\omega = 0), \\
\label{eq:T1inv_NNsite}
& 1/T_1^z = \pm\{S^{xx}_{\rm{NN}}(\omega = 0) + S^{yy}_{\rm{NN}}(\omega = 0)\},
\end{align}
respectively; here we omit the coefficients $a_{0,x}$, $a_{0,y}$, $a_{1,x}$, and $a_{1,y}$.
In Eq.~(\ref{eq:T1inv_NNsite}), the sign is $+$($-$) for the FM (AFM) case [$S^{\mu\mu}_{\rm{NN}}(\omega = 0)$ changes sign but the absolute value is the same for both cases]. 
Note that $1/T_1^x = 1/T_1^y = 1/T_1^z = 1/T_1$ for the isotropic case.
Comparison to experiments can be made for the superpositions with appropriate coefficients determined by $A_\mathbf{q}$.
The results in Fig.~\ref{fig:T1} unveil the following characteristic behaviors.
(i) In the conventional PM region above $T_H$, the onsite component is almost independent of $T$, while the NN-site one decreases to zero while increasing $T$.
The almost constant behavior of the onsite component is consistent with PM spin fluctuations governed by $J$~\cite{Moriya1956}.
(ii) Below $T_H$, the onsite and NN-site components show almost the same $T$ dependence. 
This indicates that the dynamical spin correlations are almost the same for the two components after the fractionalization sets in.
(iii) While decreasing $T$ below $T_H$, both components grow in the fractional PM region and show a broad peak at $T\simeq 0.04J$.
(iv) Both components are rapidly suppressed around $T_L$.
This is ascribed to the gap opening in the $Z_2$ flux excitations, as observed in $S(\mathbf{q},\omega)$ in Sec.~\ref{sec:Sqw}.
Indeed, the low-$T$ behaviors are well fitted by the activation-type function proportional to $\exp\{-a\Delta_f/(k_{\rm B}T)\}$, where $\Delta_f$ is the flux gap and $a$ is a coefficient~\cite{Yoshitake_thesis}.

An interesting feature among these behaviors is the growth of $1/T_1$ in the fractional PM region below $T_H$.
This means that the dynamical spin correlations are developed in this $T$ region.
On the other hand, as discussed in Sec.~\ref{sec:crossovers} and \ref{sec:SS}, the equal-time spin correlations are almost saturated below $T_H$ and do not show significant $T$ dependence.
These observations indicate that the Kitaev model exhibits distinct $T$ dependences between the dynamical and static spin correlations below $T_H$ where the thermal fractionalization sets in.
Such dichotomy is hardly seen in conventional magnets, except for critical behaviors in magnetic ordering.
Thus, the strong enhancement with the broad peak in $1/T_1$ under the saturated static spin correlations would be an indication of the thermal fractionalization in the Kitaev QSL.

Related to this enhancement, let us make a remark on the Korringa law. 
As introduced in Sec.~\ref{sec:frac}, the system is described by noninteracting Majorana fermions coupled to the $Z_2$ fluxes. 
Indeed, the $T$-linear specific heat is observed in the fractional PM region, which we call the Majorana metal in Sec.~\ref{sec:Majorana_metal}. 
From this picture, one might expect the Korringa law, $1/(T_1T \chi^2) \sim$ constant, which holds for free fermion systems, in the same $T$ region. 
The numerical data, however, do not support this expectation~\cite{Yoshitake2017a}. 
This might be due to the fact that the spin-flip excitation is a composite of both itinerant Majorana fermions and localized $Z_2$ fluxes, as discussed in Sec.~\ref{sec:Sqw}.

\begin{figure}[t]
\begin{center}
\includegraphics[width=0.8\columnwidth,clip]{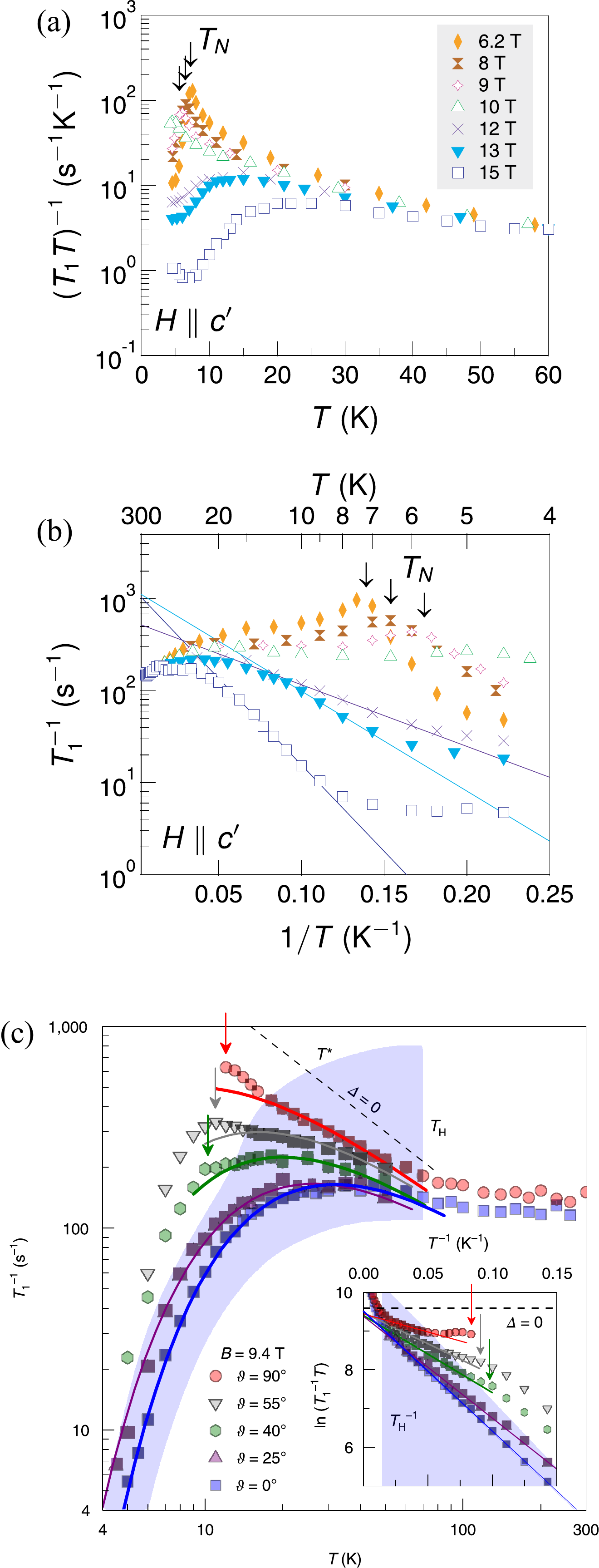}
\caption{
(Color online) 
$T$ dependences of the NMR relaxation rate $1/T_1$ for $\alpha$-RuCl$_3$.
In (a) and (b), the magnetic field is applied along the direction parallel to the electric field gradient at a Cl ion, while in (c), it is tilted from the $c$ axis by the angle $\theta$ with fixed magnitude at 9.4~T.
The arrows indicate the critical temperatures $T_N$ for magnetic ordering.
The lines in (b) represent the fitting by $1/T_1\propto \exp(-\Delta/T)$.
The colored curves in (c) are the fittings by an empirical function $1/T_1 \propto (1/T)\exp\{-0.67\Delta/(k_{\rm B}T)\}$ for the data in the blue hatched area.
The figures (a) and (b) are reprinted with permission from Ref.~\citen{Baek2017} $\copyright$ (2017) by the American Physical Society. 
The figure (c) is reprinted with permission from Ref.~\citen{Jansa2018} $\copyright$ (2018) Springer Nature. 
}
\label{fig:T1_RuCl3}
\end{center}
\end{figure}

NMR measurements have been done for $\alpha$-RuCl$_3$ by several groups~\cite{Baek2017,Zheng2017,Jansa2018,Nagai2019preprint}.
The representative data are shown in Fig.~\ref{fig:T1_RuCl3}.
In the low-field region for $\lesssim 9$~T where the magnetic ordering takes place at low $T$, $1/T_1$ grows gradually while decreasing $T$, and shows a sharp anomaly at the critical temperature $T_N$, followed by a rapid decrease below $T_N$.
On the other hand, in the higher-field region where the magnetic order is suppressed, $1/T_1$ grows gradually but turns to decrease after showing a broad peak, as shown in Fig.~\ref{fig:T1_RuCl3}(a).
While increasing the magnetic field, the peak height is gradually decreased and the peak temperature is shifted to higher $T$.
The low-$T$ decrease is well fitted by the activation-type function, as shown in Fig.~\ref{fig:T1_RuCl3}(b), while $1/T_1$ appears to approach a nonzero constant or show a slight increase at the lowest $T$ measured in this experiment.
We note, however, that the low-$T$ behaviors of $1/T_1$ are scattered among the data from different groups.
For instance, in Ref.~\citen{Zheng2017}, the power-law $T$ dependence was observed in some range of the magnetic field, from which the existence of gapless excitations was concluded.
Meanwhile, from the measurement down to 0.4~K in Ref.~\citen{Nagai2019preprint}, another exponential decrease was found at the lower-$T$ region than measured in the previous studies, from which two gap structure was identified.

As mentioned above, the theoretical results in Fig.~\ref{fig:T1} are obtained in the zero-field limit, which correspond to NQR, and hence, the direct comparison with the experimental data is not straightforward~\cite{note4}.
Nevertheless, it is interesting to point out that the experimental data in the high-field PM region look similar to the theoretical results in the points (iii) and (iv) raised above, while the low-$T$ asymptotic behaviors are controversial in experiments.
This suggests the possibility that the fractional PM state is realized in the magnetic field.
Indeed, in Ref.~\citen{Jansa2018}, the authors proposed an empirical function for the $T$ dependence of $1/T_1$ by analyzing the theoretical results at zero field, and showed that it fits well the experimental data in a wide range of the magnetic field, as presented in Fig.~\ref{fig:T1_RuCl3}(c).
Interestingly, the estimates of the gap by this fitting procedure appear to be consistent with the prediction from the perturbation theory in Sec.~\ref{sec:field}: The gap is proportional to $h^3$ up to a constant.

Theoretical analysis was recently extended to nonzero-field regions by a CTQMC method~\cite{Yoshitake2019preprint}.
The results indicate that the overall behavior of $1/T_1$ is retained in a wide range of $T$ and field; in particular, the broad peak structure is preserved with a decrease of the peak height and a shift of the peak temperature to a higher $T$ while increasing the magnetic field.
These behaviors are apparently consistent with the experimental data shown in Fig.~\ref{fig:T1_RuCl3}.
The agreement suggests that the Kitaev model qualitatively explain the behavior of $1/T_1$ in the field-induced PM region, and furthermore, that the fractional PM state appears to extend to a wide-field region in the candidate material $\alpha$-RuCl$_3$.

\subsection{Thermal conductivity}
\label{sec:kappa_xx}

In Sec.~\ref{sec:Sqw} and \ref{sec:1/T1}, we have discussed the signatures of the thermal fractionalization in spin dynamics.
As mentioned above, however, a spin-flip excitation is a composite excitation of the itinerant Majorana fermion and the localized $Z_2$ flux.
It is therefore not straightforward to observe the two types of fractional quasiparticles in a well-separated manner in spin dynamics, despite their signatures in the characteristic $T$, $\mathbf{q}$, and $\omega$ dependences.
Then, what kinds of physical quantities are suitable for such a separate observation?

One suitable probe is thermal transport.
This is because in the Kitaev QSL heat is carried solely by the itinerant Majorana fermions, as the $Z_2$ fluxes are completely localized.
The thermal response is measured as the thermal conductivity $\kappa_{\alpha\beta}$, which is defined by $J_Q^{\alpha}=\kappa_{\alpha\beta}\nabla_{\beta} T$, where $J_Q^{\alpha}$ is the thermal current flowing in the $\alpha$ direction induced by the thermal gradient applied to the $\beta$ direction, $\nabla_{\beta} T$.
Here, $\alpha,\beta=x,y$, which correspond to the $a$ and $b$ directions of the Cartesian coordinate shown in Fig.~\ref{fig:lattice}.

\begin{figure}[t]
\begin{center}
\includegraphics[width=0.9\columnwidth,clip]{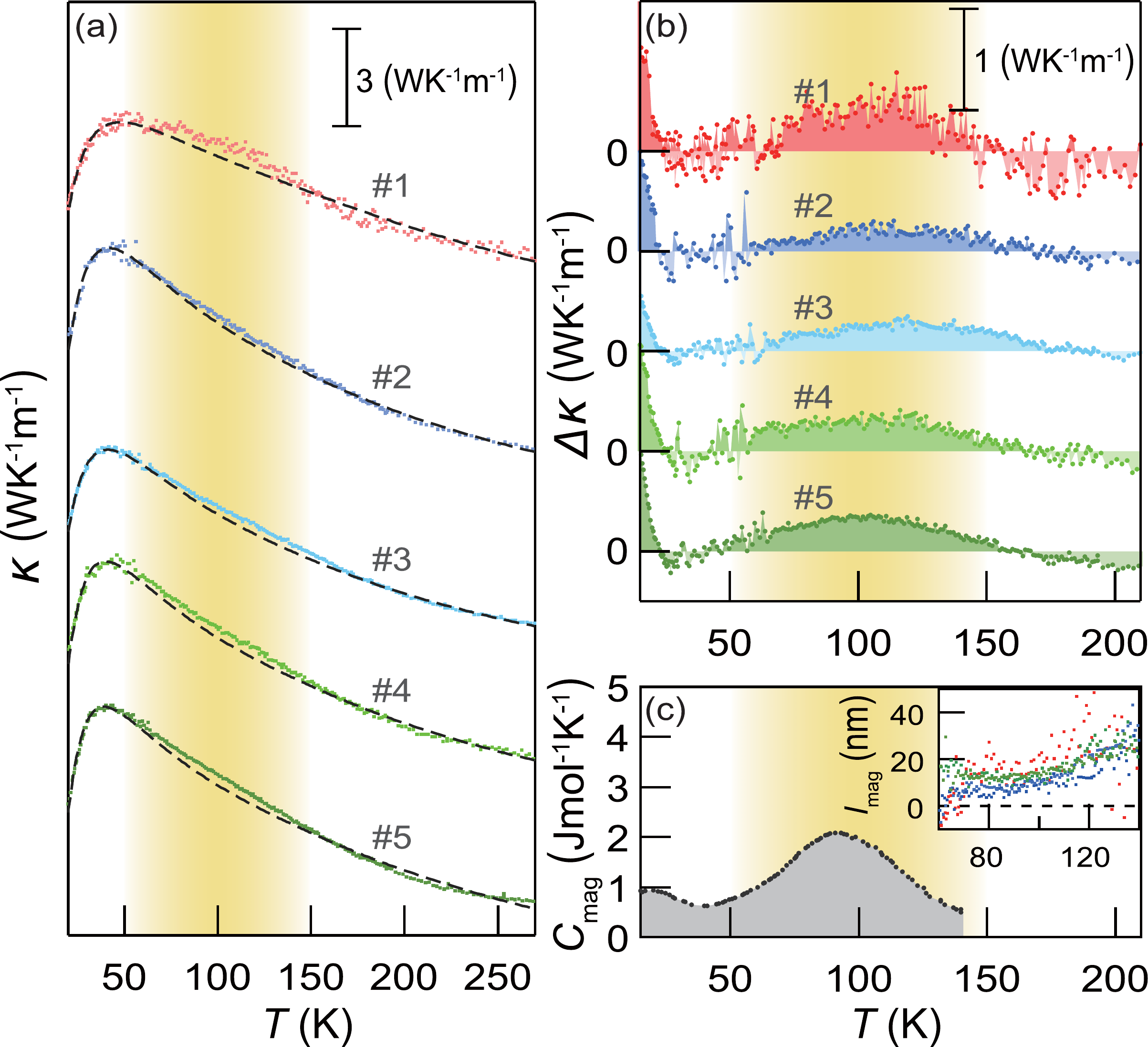}
\caption{
(Color online) 
(a) $T$ dependence of the thermal conductivity $\kappa$ for $\alpha$-RuCl$_3$ and (b) the data after the subtraction of the contributions from phonons.
\#1-\#5 denote different samples.
(c) $T$ dependence of the magnetic specific heat $C_{\rm mag}$.
The inset of (c) shows the $T$ dependence of the mean free path estimated from the analysis of $\kappa$ and $C_{\rm mag}$.
Reprinted with permission from Ref.~\citen{Hirobe2017} $\copyright$ (2017) the American Physical Society. 
}
\label{fig:kxx_RuCl3}
\end{center}
\end{figure}

In order to capture the itinerant nature of Majorana fermions, the longitudinal component of the thermal conductivity, $\kappa = \kappa_{\alpha\alpha}$, was measured for $\alpha$-RuCl$_3$~\cite{Hirobe2017}.
The results are shown in Fig.~\ref{fig:kxx_RuCl3}(a) for several samples.
Figure~\ref{fig:kxx_RuCl3}(b) shows the results after careful subtraction of the contributions from phonons.
The data indicate that there are additional contributions in a wide-$T$ range centered at $\sim 100$~K.

Theoretical results were obtained for the Kitaev model almost at the same time by using the Majorana-based QMC method~\cite{Nasu2017a}.
In the calculations, the thermal current $\textbf{J}_Q$ is defined by the time derivative of the energy polarization $\textbf{P}_E$ as
\begin{align}
 \textbf{J}_Q=\frac{\partial \textbf{P}_E}{\partial t}=i[{\cal H},\textbf{P}_E],
\end{align}
where $\textbf{P}_E$ is introduced from the Hamiltonian by replacing the exchange constant $J_\mu$ on the bond $\langle ij\rangle$ to $J_\mu \textbf{R}_{ij}$ with $\textbf{R}_{ij}=\frac12(\textbf{r}_i+\textbf{r}_j)$.
Using the above definitions, the longitudinal thermal conductivity was computed by the Kubo formula given as
\begin{align}
 \kappa_{\alpha\alpha}=\frac{1}{TV}\int_0^\infty dt e^{i(\omega+i\delta) t}\int_0^\beta d\lambda\means{J_Q^{\alpha}(-i\lambda)J_Q^{\alpha}(t)}\Bigg|_{\omega,\delta\to 0},
\label{eq:thermal}
\end{align}
where $J_Q^{\alpha}(t)$ is the Heisenberg representation of $J_Q^{\alpha}$, and $V$ is the volume of the system.
Note that the thermal current operator $\textbf{J}_Q$ commutes with all the $Z_2$ bond variables $\eta_r$ in the Hamiltonian in Eq.~(\ref{eq:H_Maj}), and therefore, Eq.~(\ref{eq:thermal}) can be calculated by using the sign-free Majorana-based QMC technique in Appendix~\ref{sec:quantum-monte-carlo}.

\begin{figure}[t]
\begin{center}
\includegraphics[width=0.9\columnwidth,clip]{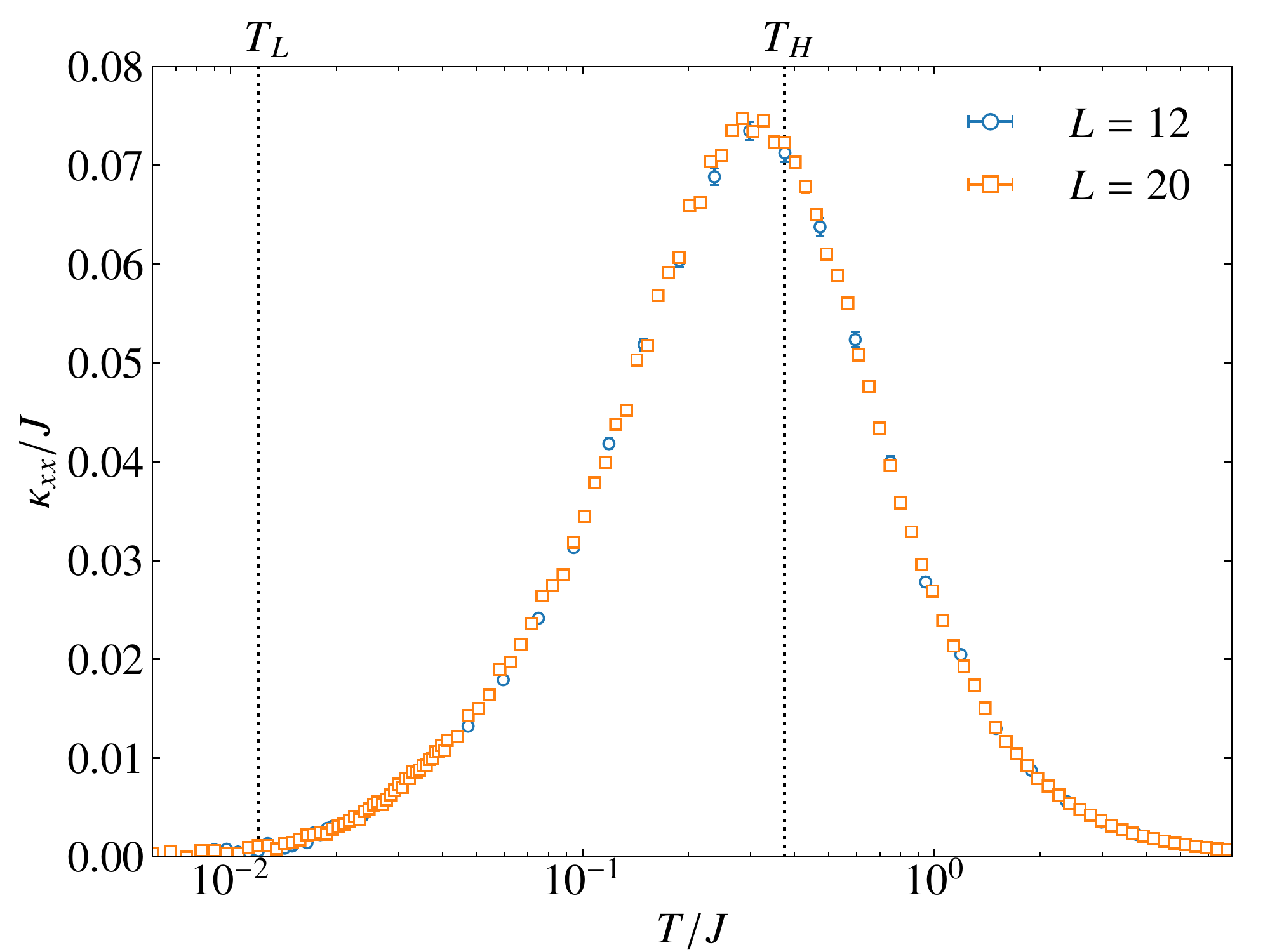}
\caption{
(Color online) 
$T$ dependence of the longitudinal thermal conductivity $\kappa_{xx}$ for the honeycomb Kitaev model with isotropic coupling $J_x=J_y=J_z=J$, obtained by the Majorana-based QMC method.
The result is common to the FM and AFM cases. 
The data are taken from Ref.~\citen{Nasu2017a}.
}
\label{fig:kxx}
\end{center}
\end{figure}

Figure~\ref{fig:kxx} shows the longitudinal thermal conductivity in the isotropic case of the honeycomb Kitaev model.
Note that $\kappa_{xx} = \kappa_{yy}$ and the result is common to the FM and AFM cases. 
The result indicates that the thermal conductivity exhibits a broad peak around $T_H$.
This is a direct consequence of the thermal fractionalization; the itinerant Majorana fermions appear in the system when the thermal fractionalization sets in by approaching $T_H$ from high $T$, but their thermally-activated population decreases with a further decrease of $T$ because of the Fermi degeneracy.
The theoretical result resembles qualitatively the experimental data in Fig.~\ref{fig:kxx_RuCl3}.

\subsection{Raman scattering}
\label{sec:Raman}

The comparison of the thermal conductivity in the previous section suggests the existence of heat carriers in the insulating material besides phonons.
However, it is not straightforward to conclude that the carriers are Majorana fermions.
In this section, we discuss another measurement, the Raman scattering, which could probe the Majorana fermions more directly.

The Raman scattering is a powerful tool to identify the magnetic excitations by using light.
Theoretically, the intensity of the Raman scattering spectrum is calculated as~\cite{Knolle2014b}
\begin{align}
 I(\omega)=\frac{1}{N}\int_{-\infty}^{\infty}dt e^{i\omega t}\means{{\cal R}(t){\cal R}}.
\label{eq:raman}
\end{align}
Here, ${\cal R}$ is the Loudon-Fleury operator~\cite{Fleury1968} given by
\begin{align}
  {\cal R}=\sum_{\means{ij}_\mu}(\boldsymbol{\epsilon}_{\rm in}\cdot \textbf{d}^\mu)(\boldsymbol{\epsilon}_{\rm out}\cdot \textbf{d}^\mu)J_\mu S_i^\mu S_j^\mu,
\label{eq:LF}
\end{align}
where $\boldsymbol{\epsilon}_{\rm in}$ and $\boldsymbol{\epsilon}_{\rm out}$ are the polarization vectors of the incoming and outgoing lights, and $\textbf{d}^\mu$ is the vector connecting a NN $\mu$ bond for the sites $i$ and $j$.
Note that, in the isotropic case with $J_x=J_y=J_z=J$ assumed here, there is no polarization dependence~\cite{Knolle2014b}.

\begin{figure}[t]
\begin{center}
\includegraphics[width=0.9\columnwidth,clip]{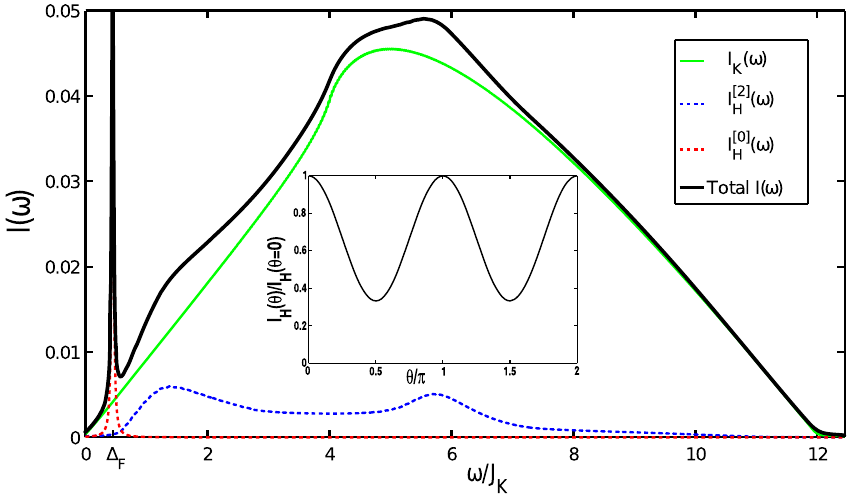}
\caption{
(Color online) 
Intensity of the Raman scattering spectrum calculated for the exact QSL ground state of the honeycomb Kitaev model with isotropic coupling.
Note that the energy scale is four-times different, as in Fig.~\ref{fig:Sqw}(a).
The result is common to the FM and AFM cases. 
The green curve represent the result for the Kitaev model, while the red and blue dashed curves show the contributions from additional exchange interactions; the black curve displays the summation of the three contributions.
Reprinted with permission from Ref.~\citen{Knolle2014b} $\copyright$ (2014) the American Physical Society.
}
\label{fig:Raman_zeroT}
\end{center}
\end{figure}

Figure~\ref{fig:Raman_zeroT} shows the Raman scattering intensity $I(\omega)$ calculated for the exact QSL ground state of the Kitaev model~\cite{Knolle2014b}.
The spectrum includes a broad incoherent response in a wide-energy range up to about $3J$ [note that the energy scale in Fig.~\ref{fig:Raman_zeroT} is four-times larger than the present definition, as in Fig.~\ref{fig:Sqw}(a)].
Equations~(\ref{eq:raman}) and (\ref{eq:LF}) indicate that on the basis of the Loudon-Fleury approach~\cite{Fleury1968} the Raman response originates solely from the itinerant Majorana fermions in the Kitaev QSL, as $S_i^\mu S_j^\mu$ are written by the Majorana operators $\gamma_i$ and $\gamma_j$ and do not affect the $Z_2$ variable configurations $\{\eta_r\}$~\cite{Knolle2014b}.
Hence, the broad response in Fig.~\ref{fig:Raman_zeroT} is a direct consequence of the fermionic excitations with the wide bandwidth shown in Sec.~\ref{sec:frac}.

\begin{figure}[t]
\begin{center}
\includegraphics[width=\columnwidth,clip]{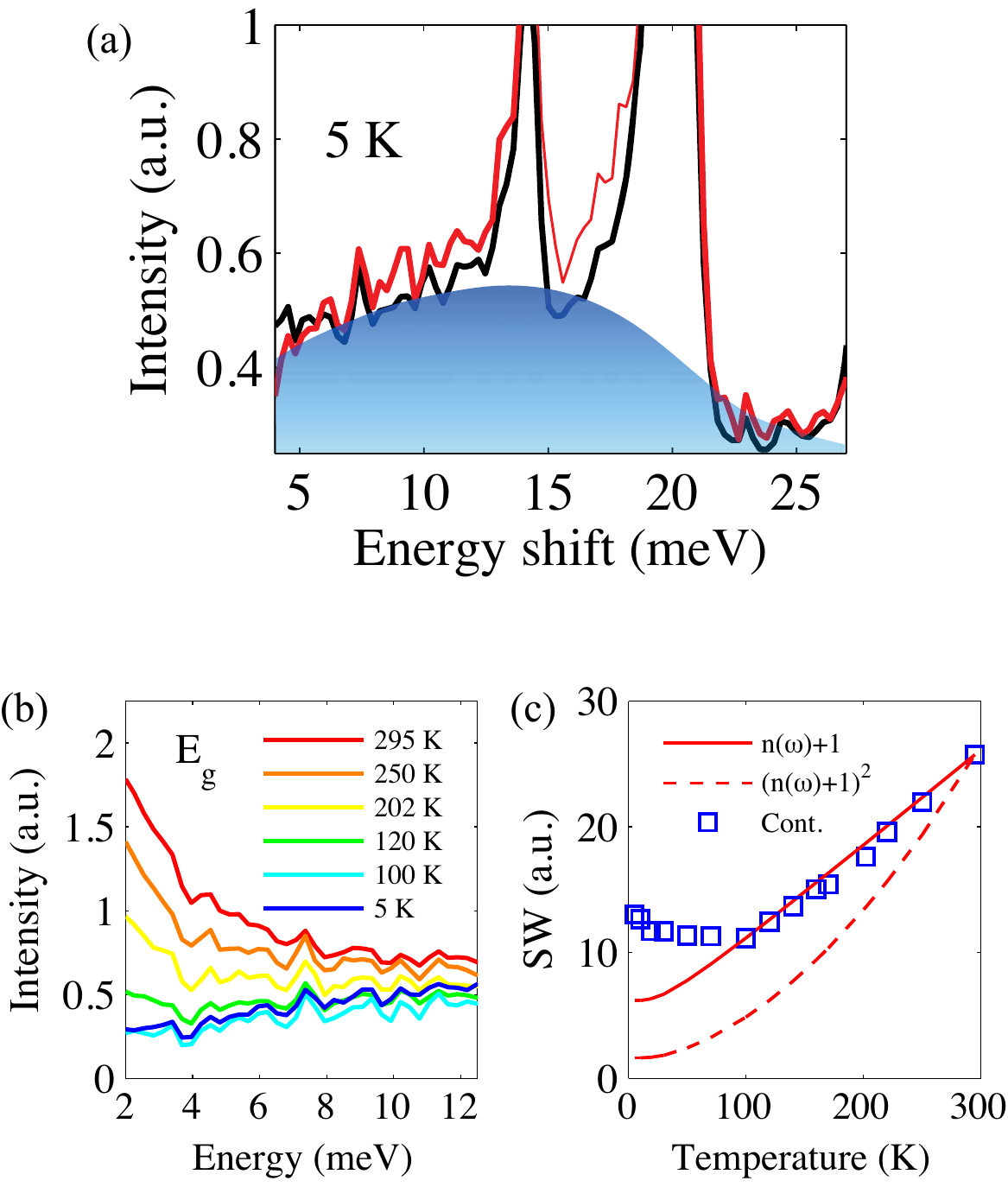}
\caption{
(Color online) 
(a) Raman scattering intensity measured for $\alpha$-RuCl$_3$ at 5~K.
(b) Magnetic contributions of the Raman intensity for several $T$.
(c) $T$ dependence of the magnetic Raman intensity integrated between 2.5 and 12.5~meV.
The solid and dashed lines represent the fittings by using the Bose-Einstein distribution function $n(\omega)$.
Reprinted with permission from Ref.~\citen{Sandilands2015} $\copyright$ (2015) the American Physical Society.
}
\label{fig:Raman_RuCl3}
\end{center}
\end{figure}

Figure~\ref{fig:Raman_RuCl3}(a) displays the experimental result measured for $\alpha$-RuCl$_3$ at 5~K~\cite{Sandilands2015}.
In addition to the sharp peaks around 14~meV and 20~meV, which are presumably from phonon excitations through the spin-lattice coupling, the spectrum exhibits a broad incoherent response ranging up to $\sim 25$~meV, as indicated by the blue shade in Fig.~\ref{fig:Raman_RuCl3}(a).
This incoherent response is similar to that found in the theoretical result at $T=0$ in Fig.~\ref{fig:Raman_zeroT}, suggesting the existence of the itinerant Majorana fermions.

Figure~\ref{fig:Raman_RuCl3}(b) displays the magnetic contributions for several $T$,
and Fig.~\ref{fig:Raman_RuCl3}(c) plots the $T$ dependence of the intensity integrated between 2.5~meV and 12.5~meV~\cite{Sandilands2015}.
In conventional magnets, $T$ dependence of the intensity is usually well fitted by using the Bose-Einstein distribution function $n(\omega)$, since the excitations are given by magnons and phonons, both of which obey the Bose-Einstein statistics.
The result plotted in Fig.~\ref{fig:Raman_RuCl3}(c) shows that this is not the case for $\alpha$-RuCl$_3$: There are additional contributions that cannot be fitted by using $n(\omega)$ in the wide-$T$ range. 
This peculiar $T$ dependence could be evidence of the Majorana fermions, but there was no theoretical result at finite $T$ at this stage.

\begin{figure}[t]
\begin{center}
\includegraphics[width=\columnwidth,clip]{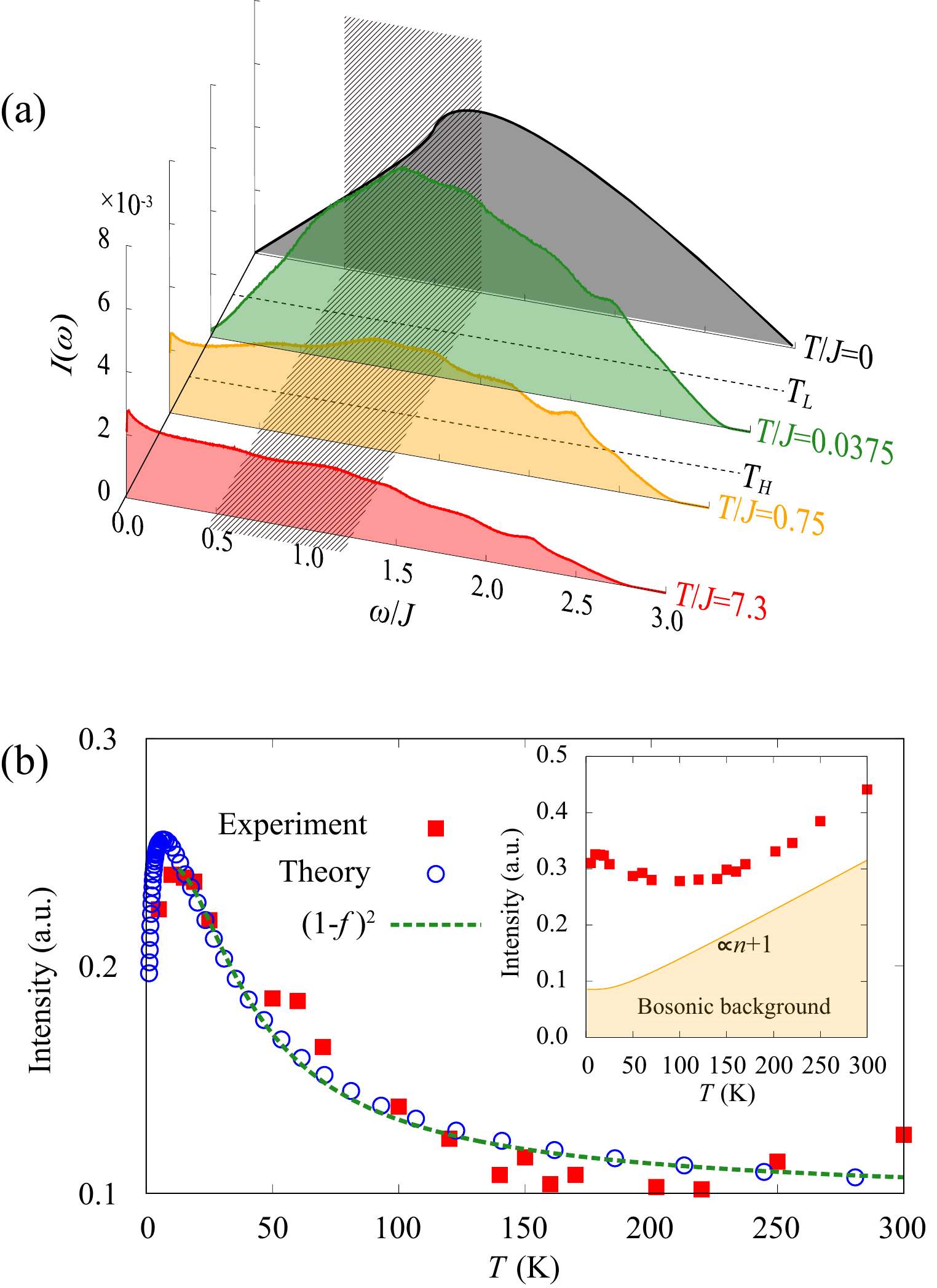}
\caption{
(Color online) 
(a) $T$ and $\omega$ dependence of the Raman scattering intensity for the honeycomb Kitaev model with isotropic coupling at several $T$, obtained by the Majorana-based QMC method.
The result is common to the FM and AFM cases. 
(b) Comparison between the theoretical result and the experimental data in Fig.~\ref{fig:Raman_RuCl3}.
The experimental data are obtained by integrating the intensity between 5~meV and 12.5~meV (shown in the inset), and correspondingly, the theoretical results are integrated in the hatched energy range in (a) by assuming the isotropic Kitaev coupling $J$ as 10~meV.
The orange shaded area in the inset of (b) represents the bosonic contribution subtracted for comparison.
The green dashed curve in the main panel represents the fitting by $(1-f)^2$, where $f$ is the Fermi-Dirac distribution function.
The figures are reprinted from Ref.~\citen{Nasu2016}.
}
\label{fig:Raman}
\end{center}
\end{figure}

Finite-$T$ behaviors of the Raman scattering intensity for the Kitaev model were obtained by using the Majorana-based QMC method~\cite{Nasu2016}.
Note that this dynamical quantity can also be calculated by the sign-free QMC technique in Appendix ~\ref{sec:quantum-monte-carlo}, since the Loudon-Fleury operator ${\cal R}$ commutes with all $\eta_r$ as the thermal current operator $\textbf{J}_Q$ in Sec.~\ref{sec:kappa_xx}.
Figure~\ref{fig:Raman}(a) displays the results for the $T$ and $\omega$ dependence.
While increasing $T$ from the ground state, the incoherent nature of the spectrum is retained, but the weight distribution changes gradually;
the low-$\omega$ weight increases continuously up to $T\simeq T_H$ and saturates above $T_H$, while the weight around $\omega=J$ shows a slight increase up to $T\simeq 0.05J$, which is slightly above $T_L$, but turns to decreases at higher $T$~\cite{Nasu2016}.

Figure~\ref{fig:Raman}(b) presents the comparison between theory and experiment.
In this comparison, the experimental spectrum is assumed to be a simple summation of the magnetic contribution from the Kitaev model and that from (unidentified) bosonic excitations.
The result in Fig.~\ref{fig:Raman}(b) shows that the $T$ dependence of the Raman intensity integrated in the middle-energy range from 5~meV to 12.5~meV is well reproduced by the theoretical result in a wide-$T$ range by assuming $J=10$~meV.
Furthermore, the theoretical calculations showed that the dominant contribution in this $T$ range comes from pair creations and annihilations of the emergent fermions composed of the Majorana fermions~\cite{Nasu2016}.
This is indeed seen from the fact that the theoretical result is well reproduced by a simple function $(1-f)^2$, where $f$ is the Femi-Dirac distribution function, as indicated by the green dashed curve in Fig.~\ref{fig:Raman}(b).
The good agreement between theory and experiment strongly suggests the existence of additional fermionic excitations in the experimental data in the wide PM region above $T_N$, which are absent in conventional magnets.

After this surprising result, similar analyses were performed for another candidates, iridium oxides $\beta$- and $\gamma$-Li$_2$IrO$_3$~\cite{Glamazda2016} (see Sec.~\ref{sec:3D_iridates}).
Despite the 3D honeycomblike structures in these compounds, the Raman scattering intensity exhibits similar $T$ dependence, which is well fitted by $(1-f)^2$.
This suggests that the fermionic excitations are commonly present in the candidate materials for the Kitaev QSL.
Also, we note that contributions of non-Kitaev interactions~\cite{Rousochatzakis2019,Wang2018preprint} and an external magnetic field~\cite{Wulferding2019preprint} were recently discussed.

\subsection{Thermal Hall conductivity}
\label{sec:kappa_xy}

The unconventional contribution in the Raman intensity strongly suggests the existence of fermionic excitations, but it is still difficult to conclude that the excitations are nothing but the Majorana fermions, especially solely from the experimental data.
To prove the existence of the Majorana fermions, one needs to explicitly identify the consequence from their peculiar nature, e.g., the equivalence between the particle and its anti-particle.
In this section, we discuss one of such direct consequences discovered in the recent measurements of the thermal Hall transport.

As discussed in Sec.~\ref{sec:field}, Kitaev showed by using the perturbation theory that a weak magnetic field induces a gapped topologically-nontrivial state showing the half-quantized thermal Hall effect due to the chiral Majorana edge mode~\cite{Kitaev2006} (see Fig.~\ref{fig:hall}). 
As the half quantization is a direct consequence of the fact that the Majorana fermions carry half degrees of freedom of the electrons, its measurement provides a smoking gun for the Majorana nature. 

\begin{figure}[t]
\begin{center}
\includegraphics[width=0.9\columnwidth,clip]{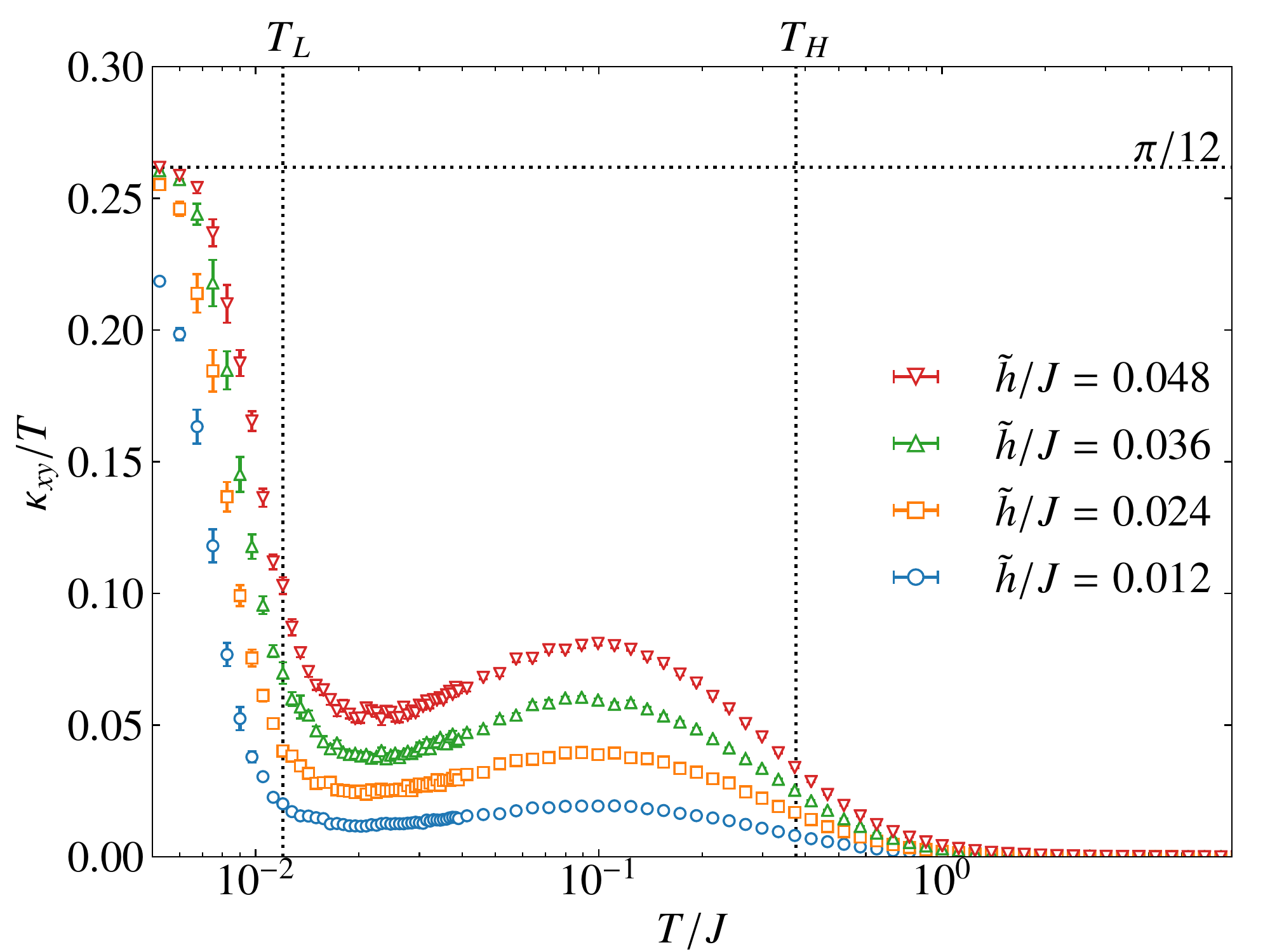}
\caption{
(Color online) 
$T$ dependence of the thermal Hall conductivity $\kappa_{xy}$ divided by $T$ calculated for the effective model for the Kitaev model in a magnetic field derived by the perturbation theory [Eqs~(\ref{eq:H_Maj}) and (\ref{eq:H'_Maj})].
$\tilde{h}$ represents the magnitude of the effective magnetic field, which is proportional to the cube of the actual field strength $h$.
The results are obtained by using the Majorana-based QMC method, and common to the FM and AFM cases.
The horizontal dotted line represents the half quantizated value $\pi/12$. 
The data are taken from Ref.~\citen{Nasu2017a}.
}
\label{fig:kxy}
\end{center}
\end{figure}

Prior to experiments, $T$ dependence of $\kappa_{xy}$ was numerically calculated by using the Majorana-based QMC method for the effective model derived by the perturbation theory given by Eqs.~(\ref{eq:H_Maj}) and (\ref{eq:H'_Maj})~\cite{Nasu2017a}.
In the calculations, a contribution from ``the gravitational magnetization'' was taken into account in addition to the Kobo formula similar to the longitudinal case given in Eq.~(\ref{eq:thermal})~\cite{Nomura2012,Sumiyoshi2013}.
Figure~\ref{fig:kxy} shows the results.
While decreasing $T$, $\kappa_{xy}/T$ increases gradually below $T\sim J$, and approaches rapidly the half quantized value $\pi/12$ below $T_L$.
The low-$T$ asymptotic behavior is fitted by $\propto \exp(-\Delta_f/T)$, where $\Delta_f$ is the flux gap.
Interestingly, $\kappa_{xy}/T$ shows nonmonotonic $T$ dependence in the intermediate-$T$ region, originating from thermal fluctuations of the localized $Z_2$ fluxes which scatter the itinerant Majorana fermions~\cite{Nasu2017a}.

\begin{figure}[t]
\begin{center}
\includegraphics[width=\columnwidth,clip]{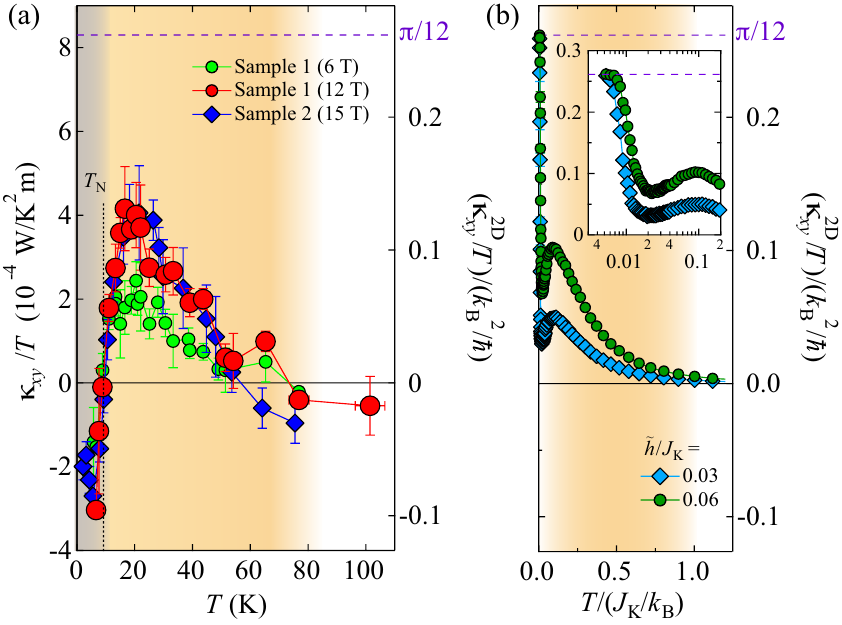}
\caption{
(Color online) 
(a) $T$ dependence of the thermal Hall conductivity $\kappa_{xy}$ divided by $T$ measured for $\alpha$-RuCl$_3$ in a magnetic field applied perpendicular to the $ab$ plane, and (b) theoretical results for comparison. 
The inset in (b) shows the enlarged plot for the low-$T$ part. 
Reprinted with permission from Ref.~\citen{Kasahara2018a} $\copyright$ (2018) the American Physical Society. 
}
\label{fig:kxy_RuCl3-1}
\end{center}
\end{figure}

The corresponding experiment was performed for $\alpha$-RuCl$_3$~\cite{Kasahara2018a}.
The results are shown in Fig.~\ref{fig:kxy_RuCl3-1}(a).
The $T$ dependence above $T_N$ is qualitatively similar to that in the theoretical results replotted in Fig.~\ref{fig:kxy_RuCl3-1}(b); $\kappa_{xy}/T$ becomes nonzero below $\sim 80$~K and shows a broad peak above $T_N$.
While further decreasing $T$, however, the experimental data decrease and change the sign to negative below $T_N$.
In this experiment, the magnetic field was applied along the $c$ axis, which cannot suppress the magnetic order in the field range measured, and hence, the half quantization, if any, is hindered by the magnetic ordering.

\begin{figure}[t]
\begin{center}
\includegraphics[width=0.9\columnwidth,clip]{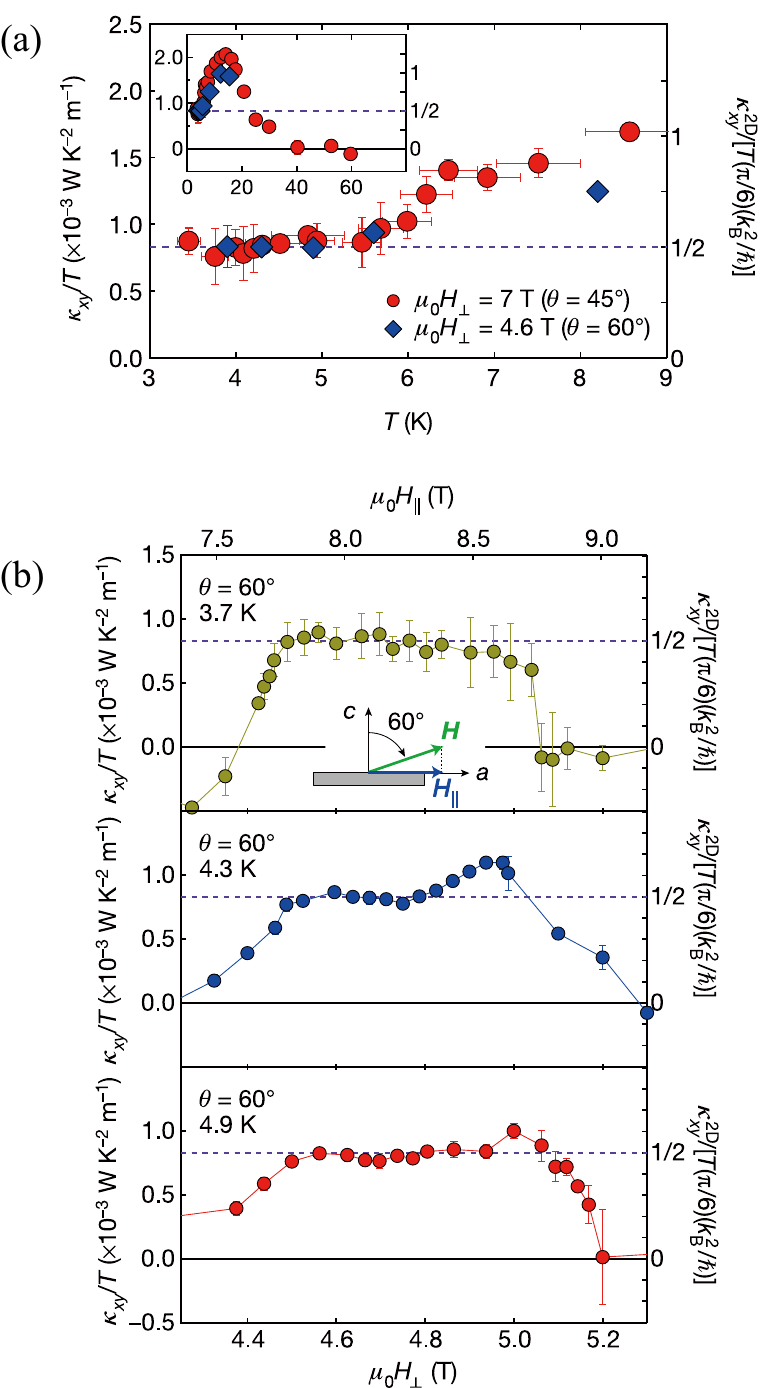}
\caption{
(Color online) 
(a) $T$ dependence of $\kappa_{xy}/T$ for $\alpha$-RuCl$_3$ in a magnetic field tilted from the $c$ axis to the $a$ axis by the angle $\theta$.
The horizontal dashed line represents the half quantization value.
The inset displays the data in a wider-$T$ range.
(b) Field dependences of $\kappa_{xy}/T$ for three temperatures at $\theta=60^\circ$.
The figures are reprinted from Ref.~\citen{Kasahara2018b}.
}
\label{fig:kxy_RuCl3-2}
\end{center}
\end{figure}

Recently, $\kappa_{xy}/T$ was measured in the magnetic field tilted from the $c$ axis, which can suppress the magnetic ordering~\cite{Kasahara2018b}.
Note that $\alpha$-RuCl$_3$ has strong easy-plane anisotropy as shown in Fig.~\ref{fig:chi_exp}(c).
The typical experimental data for the $T$ dependence are shown in Fig.~\ref{fig:kxy_RuCl3-2}(a).
As shown in the inset, $\kappa_{xy}/T$ increases from zero below $\sim 60$~K and once overshoots the half quantized value below $\sim 20$~K.
With a further decrease of $T$, $\kappa_{xy}/T$ turns to decrease, and below $\sim 5.5$~K, it becomes almost $T$ independent as shown in the main panel; the asymptotic constant value indeed coincides with the half quantized value within the experimental errors.
The field dependences at low $T$ are presented in Fig.~\ref{fig:kxy_RuCl3-2}(b).
The results clearly show that the half quantization appears in a narrow but finite range of the magnetic field.
These results strongly suggest the existence of the chiral Majorana edge mode in the topologically-nontrivial state in the field-induced PM region.

We note, however, that the $T$ dependence of $\kappa_{xy}/T$ is different from the theoretical results in Fig.~\ref{fig:kxy} both quantitatively and qualitatively.
The experimental data exhibits the overshoot above the half quantization value, which is not obtained in the theoretical results.
Moreover, the set-in temperature of the half quantization is considerably high compared to the theoretical prediction: The asymptotic convergence in theory appears well below $T_L$, which roughly corresponds to $\sim 1$~K, as shown in Fig.~\ref{fig:kxy}.
One of the reasons for such discrepanscies is that the theoretical results were obtained for the effective model which is justified in the weak-field limit.
More sophisticated theory beyond the perturbation is highly desired.
Furthermore, non-Kitaev interactions may play an important role in the topological phenomena.
Indeed, it was pointed out that a symmetric off-diagonal interaction contributes to the stabilization of the gapped topological state~\cite{Takikawa2019}.
Another caveat is the contribution from phonons.
The large value of the longitudinal thermal conductivity $\kappa_{xx}$ at low $T$ suggests the dominant phonon contribution~\cite{Hentrich2018}.
The possibility of the observation of quantized $\kappa_{xy}$ even in such a situation was theoretically discussed~\cite{Ye2018,Vinker-Aviv2018}.

\begin{figure}[t]
\begin{center}
\includegraphics[width=0.9\columnwidth,clip]{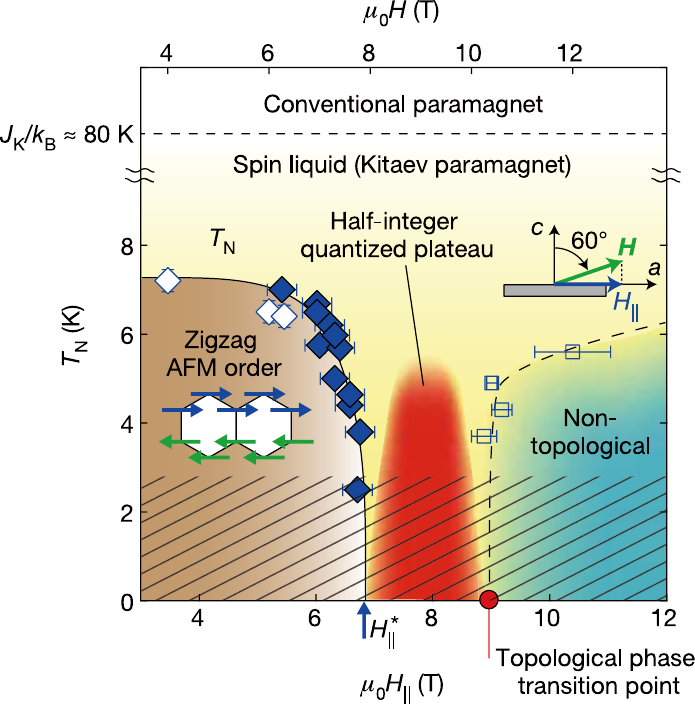}
\caption{
(Color online) 
Phase diagram of $\alpha$-RuCl$_3$ in a magnetic field.
The red area indicates the region where the half quantization of $\kappa_{xy}/T$ is observed, while the brown area shows the magnetically-ordered phase.
The yellow and green areas represent the fractional PM region and the topologically-trivial PM state at high fields.
The figure is reprinted from Ref.~\citen{Kasahara2018b}.
}
\label{fig:kxy_RuCl3-phase}
\end{center}
\end{figure}

Figure~\ref{fig:kxy_RuCl3-phase} summarizes the field-$T$ phase diagram elaborated by the experiments.
In the field region between $\sim 7$~T and $\sim 9$~T after the magnetic order is suppressed (red area in Fig.~\ref{fig:kxy_RuCl3-phase}), the half quantization of $\kappa_{xy}/T$ is observed below $\sim 5$~K.
This is the region where the Majorana topological state is suggested to be realized.
Thus, the results offer strong evidence of the Kitaev-type QSL with a gapped excitation in the field-induced PM state [see also the schematic phase diagram in Fig.~\ref{fig:sketch_phase}(a) in Sec.~\ref{sec:schematic_phase_diagram}].

The half quantization of $\kappa_{xy}/T$ has attracted great attention since it can be regarded as the direct evidence of the spin fractionalization in the Kitaev system, especially the Majorana fermionic nature.
Moreover, it is intriguing as the set-in temperature is rather high compared to other topological phenomena like the quantized anomalous Hall effect in magnetic topological insulators~\cite{Chang2013,Chang2015}.
Theoretically, it was pointed out that nonabelian anyons emergent in the topological state can be utilized for fault-tolerant quantum computation~\cite{Kitaev2003,Freedman2003}.
Thus, the experimental finding of the half quantization may offer a first step toward topological quantum computing based on peculiar quasiparticles in magnets.

\section{Summary and perspectives}
\label{sec:summary}

In this article, we have overviewed the recent development in the research of the Kitaev quantum spin liquids and their experimental realization.
We have reviewed finite-$T$ properties of the Kitaev model, including the spin dynamics, which have been revealed by the Majorana-based numerical techniques developed by the authors and their collaborators.
In the Kitaev model, the spin degree of freedom is fractionalized into two different types of quasiparticles: itinerant Majorana fermions and localized $Z_2$ fluxes.
They have largely different energy scales and affect the thermodynamics and spin dynamics in a peculiar manner, which we call thermal fractionalization.
We have discussed a number of fingerprints of the thermal fractionalization in experimentally observable quantities, and compare them with available experimental data for the candidate materials.
Let us summarize the main points, focusing on the 2D honeycomb case:
\begin{itemize}
\item
The Kitaev model exhibits two characteristic temperatures corresponding to the two types of quasiparticle excitations.
They define two crossovers at largely different temperatures $T_H$ and $T_L$ ($T_H\gg T_L$), signaled by two broad peaks in the specific heat and corresponding successive releases of the entropy by half $\ln 2$. (Sec.~\ref{sec:crossovers}) 
Similar behavior corresponding to the high-$T$ crossover was observed experimentally (Sec.~\ref{sec:Cv_S}). 
\item
The high-$T$ crossover at $T=T_H$ is caused by the itinerant Majorana fermions, while the low-$T$ one at $T=T_L$ is by the localized $Z_2$ fluxes.
The temperature scales $T_H$ and $T_L$ are set by the center of mass of the density of states for the complex fermion band and the $Z_2$ flux gap, respectively (Sec.~\ref{sec:Tscales}).
\item
The two crossovers define three distinct regimes: the conventional paramagnetic state for $T\gtrsim T_H$, the fractional paramagnetic state for $T_L\lesssim T\lesssim T_H$, and the asymptotic quantum spin liquid state for $T\lesssim T_L$ [see Fig.~\ref{fig:2d3d-phase}(a)].
\item
In the intermediate-$T$ range in the fractional paramagnetic state, the specific heat shows $T$-linear behavior, because the fermion density of states becomes nonzero at zero energy by thermally-fluctuating $Z_2$ fluxes.
We call this state the Majorana metal. (Sec.~\ref{sec:Majorana_metal})
\item
While decreasing $T$, the static spin correlations grow rapidly around $T=T_H$, and almost saturate at lower $T$ (Sec.~\ref{sec:crossovers}).
Similar behavior was inferred from the optical measurement for a candidate material $\alpha$-RuCl$_3$ (Sec.~\ref{sec:SS}).
\item
The magnetic susceptibility deviates from the Curie-Weiss law below $T\sim J$ ($J$ is the Kitaev coupling), shows a peak in the intermediate-$T$ region between $T_L$ and $T_H$.
Similar behaviors were observed for the candidate materials, Na$_2$IrO$_3$, $\alpha$-Li$_2$IrO$_3$, and $\alpha$-RuCl$_3$.
Theoretically, the susceptibility shows a rapid decrease around $T=T_L$ and approaches a nonzero value in the low-$T$ limit, but these behaviors are hindered by the magnetic ordering in the real compounds. (Sec.~\ref{sec:chi})
\item
The dynamical spin structure factor $S(\mathbf{q},\omega)$ shows a characteristic $T$ dependence.
Below $T\simeq T_H$, $S(\mathbf{q},\omega)$ develops an incoherent response at $\omega\simeq J$ with less $\mathbf{q}$ dependence, which persists down to the lowest $T$.
The overall $T$, $\mathbf{q}$, and $\omega$ dependences agree well with the experimental data by inelastic neutron scattering for $\alpha$-RuCl$_3$.
Theoretically, while approaching $T_L$, an additional quasielastic response grows rapidly, and it is gapped out below $T_L$ reflecting the gap opening in the flux excitations.
But, these behaviors are not observed in the experiments due to the magnetic ordering. (Sec.~\ref{sec:Sqw})
\item
The NMR relaxation rate $1/T_1$ increases below $T_H$, where the static spin correlations are almost saturated.
This dichotomy between dynamical and static spin correlations is a possible indication of the thermal fractionalization.
$1/T_1$ exhibits a broad peak above $T_L$, and decreases exponentially below $T_L$ reflecting the flux gap opening.
Similar behaviors were observed in experiments under a magnetic field, suggesting the potential realization of the Kitaev quantum spin liquid in the field-induced paramagnetic state. (Sec.~\ref{sec:1/T1})
\item
The itinerant Majorana fermions can contribute to heat transport.
Indeed, in the Kitaev model, the longitudinal thermal conductivity shows a broad peak around $T=T_H$.
Similar behavior was observed in experiments. (Sec.~\ref{sec:kappa_xx})
\item
The Kitaev model predicts an incoherent Raman response because of the Majorana fermions.
This was indeed observed in experiments for $\alpha$-RuCl$_3$.
Furthermore, an unconventional $T$ dependence of the scattering weight was unveiled in the experiments, and well explained by the theoretical results for the Kitaev model.
This provides strong evidence for the existence of unconventional fermionic excitations in $\alpha$-RuCl$_3$.
Similar behaviors were observed also for the 3D candidates $\beta$- and $\gamma$-Li$_2$IrO$_3$. (Sec.~\ref{sec:Raman})
\item
An external magnetic field opens a gap in the quasiparticle band and makes it topologically nontrivial (Sec.~\ref{sec:field}).
Reflecting the topological nature, the thermal Hall conductivity is asymptotically quantized at low $T$ below $T_L$.
The quantization value is half of that in the integer quantum Hall state reflecting that the heat carriers are Majorana fermions.
Such a half quantization of the thermal Hall conductivity was observed in $\alpha$-RuCl$_3$, which has recently gathering tremendous attention as direct evidence of the Majorana fermions and their topological state. (Sec.~\ref{sec:kappa_xy})
\end{itemize}
We have also discussed interesting signatures of the thermal fractionalization for the Kitaev models with some extensions from the original honeycomb one.
There appear a variety of phase transitions and crossovers, as schematically summarized in Fig.~\ref{fig:2d3d-phase} in Sec.~\ref{sec:phase_diagram}.
We list the key aspects in the following, which await for the experimental confirmation:
\begin{itemize}
\item
In the 3D Kitaev model, the nature of the $Z_2$ flux excitations is qualitatively different from that in two dimensions.
Because of the local constraint on the $Z_2$ fluxes, the excitations are allowed only in the form of closed loops in three dimensions.
This changes the low-$T$ crossover in the 2D cases into a phase transition.
This transition takes place between the high-$T$ paramagnet and the low-$T$ quantum spin liquid, which can be regarded as a gas-liquid transition in terms of the spin degree of freedom of insulating magnets. (Sec.~\ref{sec:loop_proliferation})
\item
When extending the Kitaev model by adding non-Kitaev interactions, the system may undergo phase transitions among three states of matter --- gas, liquid, and solid.
The phase diagram is distinct between the 2D and 3D cases, reflecting the different nature of the $Z_2$ flux excitations. (Sec.~\ref{sec:gas-liq-solid})
\item
When the Kitaev model is defined on the lattice structures with odd-site loops, the ground state can be a chiral spin liquid.
In this case, the low-$T$ crossover is replaced by a finite-$T$ phase transition with breaking of time-reversal symmetry caused by $Z_2$ flux ordering. (Sec.~\ref{sec:2D_CSL} and \ref{sec:3D_CSL})
\end{itemize}

Despite the clarification of many intriguing aspects of the thermal fractionalization in the Kitaev model and the successful comparison with experimental data, there remain a number of open issues in this rapidly growing field.
We hope that the present review will be helpful for studying the following issues in future studies.
\begin{itemize}

\item {\it Further theoretical understanding of the Kitaev model and its extensions:}
\begin{itemize}
\item
It is highly desired to clarify the effect of the external magnetic field on the phase diagram, the topological properties of the elementary excitations, and the excitation spectra.
This is crucially important for comparison with experimental data, especially the remarkable properties discovered in the field-induced paramagnetic state in $\alpha$-RuCl$_3$.
\item
In the magnetic field, the case of the antiferromagnetic Kitaev coupling is also intriguing, since an additional topological phase was predicted theoretically, as discussed in Sec.~\ref{sec:field}.
It is also important to find the candidate materials for the antiferromagnetic Kitaev coupling, by pushing forward the recent efforts introduced in Sec.~\ref{sec:field}.
\item
It is also important to clarify the effects of non-Kitaev interactions which exist in real compounds, as mentioned in Sec.~\ref{sec:other_exchanges}.
In particular, it is crucial to study such effects in sufficiently large system sizes with high resolution in both energy and momentum, since the subdominant interactions can lead to keen competitions between different phases and fine structures in the excitation spectra.
\item
It is worth extending the analyses to other lattices, especially in three dimensions. 
Besides the hyperhoneycomb, hyperoctagon, and hypernonagon structures discussed in Sec.~\ref{sec:3D}, a variety of extensions were discussed for other lattices~\cite{O'Brien2016}. 
Interestingly, depending on the underlying lattice structures, the itinerant Majorana fermions form Majorana Fermi surfaces, nodal lines, or topologically-protected Weyl nodes. 
In addition, the $Z_2$ flux configurations can be suffered from frustration~\cite{Eschmann2019}. 
A comprehensive study of finite-$T$ properties for such extensions will deepen our understanding of the Kitaev quantum spin liquids and fractionalization. 
\item
It would also be interesting to consider extensions of the Kitaev model to larger spins.
The local conserved quantity on each plaquette exists also in the larger spin cases~\cite{Baskaran2008}.
Recently, thermodynamic properties were studied numerically~\cite{Oitmaa2018,SuzukiYamaji2018,Koga2018,Dwivedi2018}.
While the realization of such systems was theoretically proposed~\cite{Stavropoulos2019}, the search for the candidate materials has just begun~\cite{Xu2018,Lee2019preprint}.
\item
Development of new theoretical techniques is a key to breakthrough in understanding of the effects of the magnetic field and non-Kitaev interactions listed above.
\end{itemize}

\item {\it Further quantitative comparison with experiments:}
\begin{itemize}
\item
Further experimental identification of fractional quasiparticles is an important issue.
In particular, the $Z_2$ flux excitations have not been identified clearly thus far.
It would be helpful to further study the field-induced paramagnetic state in $\alpha$-RuCl$_3$ at lower $T$.
\item
Regarding the potential topological quantum spin liquid in the field-induced paramagnetic state, the crucial questions are what kind of the gap protects the topological state, how large the gap is, and how it depends on the field. 
Extensive experiments have been done for the excitation gap in the magnetic field, for instance, the specific heat~\cite{Sears2017,Wolter2017,Widmann2019},  NMR~\cite{Baek2017,Zheng2017,Jansa2018,Nagai2019preprint}, electron spin resonance~\cite{Ponomaryov2017}, terahertz spectroscopy~\cite{Wang2017,Little2017,Reschke2019}, inelastic neutron scattering~\cite{Banerjee2018,Balz2019}, and thermal conductivity measurements~\cite{Leahy2017,Yu2018,Hentrich2018}, but there still remains controversy, even among the results obtained by the same experimental probes. 
Although the theoretical study in the field is also very difficult, close comparison between experiment and theory on this gap issue will be crucial to deeper understanding of the field-induced state. 
\item
It is important to precisely estimate the additional non-Kitaev interactions for each candidate material by further comparison between theory and experiment. 
This issue has been addressed by the analyses of, e.g., the magnon spectra in the ordered phases~\cite{Choi2012,Ran2017}. 
The gap problem above would also be helpful to this issue. 
Also, further detailed analysis on the magnetic anisotropy would play an important role, as stated in Sec.~\ref{sec:chi}.
\item
It is also important to discuss the effect of disorder, which is inevitably present in real compounds, on the physical observables at finite $T$.
This includes nonmagnetic/magnetic impurities~\cite{Willans2010,Dhochak2010,Vojta2016}, dislocations~\cite{Petrova2014}, chemical inhomogeneity, and so on.
\end{itemize}

\item
{\it Coupling to other degrees of freedom:}
\begin{itemize}
\item
Given the fractional quasiparticles, it will be very interesting to consider the coupling to other degrees of freedom, for instance, the electric charge. 
The dynamics of a single hole doped into the Kitaev quantum spin liquid was studied~\cite{Halasz2014,Halasz2016}. 
It was also predicted that carrier doping to the Kitaev model and its extensions may lead to topological superconductivity, reflecting the exotic nature of the Kitaev quantum spin liquid~\cite{You2012,Okamoto2013,Schmidt2018}. 
Theoretical studies beyond the mean-field calculations as well as the experimental realizations are highly desired. 
\item
It will also be intriguing to study the proximity effect to other magnets, metals, and superconductors.
Recent development in the heterostructure of $\alpha$-RuCl$_3$ and graphene, which was introduced in Sec.~\ref{sec:RuCl3}, is a good example in this direction.
The coupling between the fractional quasiparticles and other degrees of freedom, such as mobile electrons, Cooper pairs, magnons, and phonons, may lead to unprecedented physics. 
Indeed, the coupling to mobile electrons was discussed for the Kitaev-Kondo model, and topological superconductivity was predicted~\cite{Seifert2018,Choi2018}. 
In addition, effects of lattice strain are also worth investigating as a source of exotic states~\cite{Rachel2016,Perreault2017}. 
\end{itemize}

\item {\it Further materialization of Kitaev quantum spin liquids:}
\begin{itemize}
\item
As partly reviewed in this article, the candidate materials for the Kitaev spin liquids are still limited.
Further exploration is needed.
In particular, highly desired are candidates which show the Kitaev spin liquid nature at zero or weaker magnetic field.
Materials with the AFM Kitaev coupling are also desired, as mentioned above.
\item
Material design for new lattice structures is important.
In particular, 3D materials are desired for studying the intriguing physics listed above.
Interesting proposals were made by using metal organic frameworks~\cite{Yamada2017a,Yamada2017b}.
In addition, quasi-one-dimensional candidates, e.g., with a ladder structure, are also interesting to further clarify the nature of fractional quasiparticles.
\item
It would also be important to explore candidates in the form of thin films and heterostructures, especially for studies of the proximity effects mentioned above.
\end{itemize}

\item
{\it Control of fractional quasiparticles:}
\begin{itemize}
\item
In the topologically nontrivial phase under the magnetic field, each excited flux in the bulk accompanies a Majorana zero mode, which obeys nonabelian statistics [see Fig.~\ref{fig:hall}(b)].
Toward topological quantum computation by using the nonabelian anyons, it is a crucial task to invent a way for controlling them, e.g., braiding and fusion.
A potential way will be to use local geometry of the system, such as defects, dislocations, edges, and interfaces.
Another way would be local perturbations, e.g., by using the scanning tunneling microscope.
\item
Along this direction, it will be quite important to clarify nonequilibrium dynamics of the fractional quasiparticles, as the topological quantum computing will be implemented by the time evolution of the quasiparticles.
Although there were several attempts for clarifying the nonequilibrium dynamics by theory~\cite{Sengupta2008,Mondal2008,Hikichi2010,Patel2012,Sato2014preprint,Bhattacharya2016,Rademaker2017preprint,Sameti2019,Nasu2019} and also in experiments~\cite{Alpichshev2015,Hinton2015,Nembrini2016,Zhang2019preprint}, but further studies are desired. 
\end{itemize}

\end{itemize}

\begin{acknowledgment}
The authors thank
T. A. Bojesen, K. S. Burch, K.-Y. Choi, T. Eschmann, M. Hermanns, K. Ido, S.-H. Jang, S. Ji, Y. Kamiya, Y. Kasahara, T. Kaji, Y. Kato, J. Knolle, A. Koga, D. L. Kovrizhin, Y. Matsuda, K. Matsuura, T. Misawa, P. A. Mishchenko, R. Moessner, K. O'Brien, J.-H. Park, R. Sano, T. Shibauchi, Y. Shimizu, M. Shimozawa, K. Sugii, Y. Sugita, R. Takashima, H. Tanaka, H. Tomishige, S. Trebst, M. Udagawa, M. Yamashita, J. Yoshitake, and Y. Wang for fruitful collaborations and constructive discussions.
They also thank
T. Arima, L. Balents, A. Banerjee, C. D. Batista, W. Brenig, G. Chen, Y. P. Chen, G.-W. Chern, S. Fujimoto, P. Gegenwart, M. Gohlke, E. A. Henriksen, C. Hickey, Z. Hiroi, K. Hukushima, M. Imada, H. Ishizuka, H. Izduchi, G. Jackeli, G. Khaliullin, H.-Y. Kee, Y. B. Kim, M. Klanj\v{s}ek, I. Kimchi, K. Kitagawa, P. Lemmens, Y. Matsumoto, Y. Maeno, K. Matsuhira, J. Matsuno, T. Minakawa, T. Miyake, E.-G. Moon, T. Moriyama, N. Nagaosa, S. Nagler, S. Nakauchi, K. Nomura, J. Ohara, K. Ohgushi, T. Ono, N. B. Perkins, S. Rachel, A. Rosch L. J. Sandilands, M. Sato, H. Shinaoka, K. Shiozaki, S. Suzuki, T. Suzuki, H. Takagi, S. Takagi, T. Takayama, O. Tchernyshyov, R. Thomale, Y. Tokura, T. Tohyama, N. Trivedi, A. Tsukazaki, R. Valent\'i, Y. Yamaji, H. Yao, and M. G. Yamada for their helpful discussions.
J.N. acknowledges the support of Leading Initiative for Excellent Young Researchers in MEXT.
This work was supported by JSPS KAKENHI Grant Nos. JP24340076, JP15K13533, JP16H00987, JP16H02206, JP16K17747, JP18H04223, and JP19K03742, and by JST CREST (JP-MJCR18T2). Parts of the numerical calculations were performed in the supercomputing systems in ISSP, the University of Tokyo.
\end{acknowledgment}

\appendix
\section{Majorana-based numerical techniques}
\label{sec:App}

\subsection{Quantum Monte Carlo method}
\label{sec:quantum-monte-carlo}

In this section, we show the framework of the Majorana-based QMC technique for the Kitaev model which has been developed in Ref.~\citen{Nasu2014}.
The Majorana representation of the Kitaev Hamiltonian in Eq.~(\ref{eq:H_Maj}) for a given configuration of $\{\eta_r\}$ is written by
\begin{align}
 {\cal H}^{\{\eta_r\}}=\sum_{i<j} A_{ij}^{\{\eta_r\}} \gamma_i \gamma_j=\frac{1}{2}\sum_{ij} A_{ij}^{\{\eta_r\}} \gamma_i \gamma_j,
\label{eq-supp:hamil}
\end{align}
where $A^{\{\eta_r\}}$ is an $N\times N$ Hermite matrix with pure imaginary matrix elements, and therefore, $A_{ij}^{\{\eta_r\}}=-A_{ji}^{\{\eta_r\}}$.
This is diagonalized as
\begin{align}
 {\cal H}^{\{\eta_r\}}=E_0^{\{\eta_r\}} + \sum_{\lambda}E_\lambda^{\{\eta_r\}}f_\lambda^\dagger f_\lambda,
\end{align}
where $f_\lambda^\dagger$ and  $f_\lambda$ are the creation and annihilation operators of the complex fermion corresponding to the energy $E_\lambda^{\{\eta_r\}}$ ($>0$), and $E_0^{\{\eta_r\}}=-\frac{1}{2}\sum_{\lambda}E_\lambda^{\{\eta_r\}}$ is the ground-state energy.
Here and hereafter, the sum $\sum_{\lambda}$ is taken for positive energies ($\lambda=1,2,\cdots, N/2$) although both the eigenvalues of $A^{\{\eta_r\}}$ appear in pairs as $\pm \frac12 E_\lambda$.
The complex fermions $\{f_\lambda\}$ are introduced such that
\begin{align}
 \gamma_j=\sqrt{2}\sum_{\lambda}\left(U_{j\lambda}^{\{\eta_r\}}f_{\lambda}+U_{j\lambda}^{\{\eta_r\}*}f_{\lambda}^\dagger\right),
\end{align}
where $U_{j\lambda}^{\{\eta_r\}}$ is the $j$th component of the eigenvector associated with the eigenvalue $\frac12 E_\lambda^{\{\eta_r\}}$ of the matrix $A^{\{\eta_r\}}$.

To calculate thermodynamic quantities, we introduce the partition function by
\begin{align}
 Z=\sum_{\{\eta_r=\pm1\}}{\rm Tr}_{\{\gamma_i\}} e^{-\beta {\cal H}^{\{\eta_r\}}}.
\end{align}
This is rewritten as
\begin{align}
 Z=\sum_{\{\eta_r=\pm1\}}e^{-\beta F_\gamma^{\{\eta_r\}}},\label{eq-supp:2}
\end{align}
where $F_\gamma^{\{\eta_r\}}$ is the free energy of the Majorana fermion system for the configuration of $\{\eta_r\}$, which is given by
\begin{align}
 F^{\{\eta_r\}}= -\frac{1}{\beta} \ln Z_\gamma^{\{\eta_r\}}=-\frac{1}{\beta} \ln \left[{\rm Tr}_{\{\gamma_i\}} e^{-\beta {\cal H}^{\{\eta_r\}}}\right].
\end{align}
Using the eigenvalues of the matrix $A^{\{\eta_r\}}$, the partition function of the Majorana fermion system is evaluated as
\begin{align}
 Z_\gamma^{\{\eta_r\}}=\prod_\lambda2\cosh\frac{\beta E_\lambda^{\{\eta_r\}}}{2}.
\end{align}

Similar to the Hamiltonian, an operator commuting with all $\{\eta_r\}$ can be labeled by $\{\eta_r\}$ as ${\cal O}^{\{\eta_r\}}$.
The thermal average of such an operator can be calculated by
\begin{align}
 \means{{\cal O}}=\frac{1}{Z}\sum_{\{\eta_r=\pm1\}}{\rm Tr}_{\{\gamma_i\}}\left[{\cal O} e^{-\beta {\cal H}^{\{\eta_r\}}}\right]=\means{\bar{\cal O}^{\{\eta_r\}}}_\eta,
\end{align}
where we introduce the expectation value of ${\cal O}$ for the configuration of $\{\eta_r\}$ as
\begin{align}
 \bar{\cal O}^{\{\eta_r\}}=\frac{1}{Z_\gamma^{\{\eta_r\}}} {\rm Tr}_{\{\gamma_i\}}\left[{\cal O}^{\{\eta_r\}} e^{-\beta {\cal H}^{\{\eta_r\}}}\right],\label{eq-supp:mean-eta}
\end{align}
and
\begin{align}
 \means{\cdots}_\eta=\frac{1}{Z}\sum_{\{\eta_r=\pm1\}}[\cdots]e^{-\beta F_\gamma^{\{\eta_r\}}}.\label{eq-supp:1}
\end{align}
On the other hand, one cannot straightforwardly calculate thermal averages of the operators not commuting with $\{\eta_r\}$, such as dynamical spin correlations.
We will introduce a way to calculate such quantities in Appendix~\ref{sec:cont-time-quant}.

Using Eqs.~(\ref{eq-supp:2}) and (\ref{eq-supp:1}), finite-$T$ properties of the Kitaev model can be calculated by using the MC sampling on the configurations of $\{\eta_r\}$.
At a certain temperature, we calculate the free energy $F^{\{\eta_r\}}$ and $\bar{\cal O}^{\{\eta_r\}}$ for a given configuration $\{\eta_r\}$ in a finite-size cluster by exact diagonalization of the Hermite matrix $A^{\{\eta_r\}}$.
Using the Markov-chain MC simulation, the sequence $(\{\eta_r \}_1, \{\eta_r \}_2, \{\eta_r \}_3, \cdots \{\eta_r \}_{N_{\rm MC}})$ is successively generated so as to reproduce the probability distribution $e^{-\beta F_{f}^{\{\eta_r\}}}/Z$.
In the sequence of $\{\eta_r \}$, the thermal average of an operator ${\cal O}$ is evaluated by replacing $\means{\cdots}_\eta$ by $\means{\cdots}_{\rm MC}$ as
\begin{align}
 \means{{\cal O}}=\means{\bar{\cal O}^{\{\eta_r\}}}_{\rm MC}=\frac{1}{N_{\rm MC}} \sum_{\ell=1}^{N_{\rm MC}} \bar{\cal O}^{\{\eta_r\}_\ell}.\label{eq-supp:expectation}
\end{align}

In Sec.~\ref{sec:crossovers} and \ref{sec:loop_proliferation}, this technique is applied to calculate the internal energy, specific heat, entropy per site, and the DOS for the complex fermion band.
The internal energy per site is calculated as
\begin{align}
 E=-\frac{1}{N}\frac{\partial}{\partial \beta} \ln Z = \frac{1}{N}\means{\bar{E}^{\{\eta_r\}}}_\eta=\frac{1}{N N_{\rm MC}} \sum_{\ell=1}^{N_{\rm MC}} \bar{E}^{\{\eta_r\}_\ell},
\end{align}
where
\begin{align}
 \bar{E}^{\{\eta_r\}}=-\frac{\partial}{\partial \beta} \ln Z_\gamma^{\{\eta_r\}}=-\sum_{\lambda} \frac{E_{\lambda}^{\{\eta_r\}}}{2}\tanh\frac{\beta E_{\lambda}^{\{\eta_r\}}}{2}.
\end{align}
The specific heat per site can also be calculated as
\begin{align}
 C_v=\frac{d E}{dT}=\frac{1}{NT^2}
\left(
\mean{\left(\bar{E}^{\{\eta_r\}}\right)^2}_{\rm MC}-\mean{\bar{E}^{\{\eta_r\}}}_{\rm MC}^2-\mean{\frac{\partial \bar{E}^{\{\eta_r\}}}{\partial \beta}}_{\rm MC}
\right).
\label{eq-supp:Cv}
\end{align}
From the specific heat, the entropy per site is obtained as
\begin{align}
 S=\ln 2 - \int_T^{\infty} dT' \frac{C_v}{T'}.
\label{eq-supp:entropy}
\end{align}
In addition, the contributions from itinerant Majorana fermions and localized $Z_2$ fluxes are separately calculated as
\begin{align}
\label{eq:Cv_Maj}
 C_v^\gamma &=-\frac{1}{NT^2}
\mean{\frac{\partial \bar{E}^{\{\eta_r\}}}{\partial \beta}}_{\rm MC},\\
\label{eq:Cv_flux}
 C_v^f &=\frac{1}{NT^2}
\left(
\mean{\left(\bar{E}^{\{\eta_r\}} \right)^2}_{\rm MC}-\mean{\bar{E}^{\{\eta_r\}} }_{\rm MC}^2\right),
\end{align}
respectively.
The corresponding contributions to the entropy are calculated in a similar manner to Eq.~(\ref{eq-supp:entropy}).
The fermion DOS is computed by
\begin{align}
 D(\omega) = \mean{\frac{2}{N}\sum_\lambda\delta(\omega-E_{\lambda}^{\{\eta_r\}})}_{\rm MC},
\end{align}
which depends on temperature $T$.
Using this expression, $E$ and $C_v^\gamma$ are written as
\begin{align}
 E&=-\int d\omega D(\omega)\frac{\omega}{4}\tanh\frac{\beta \omega}{2},\\
C_v^\gamma&=\int d\omega D(\omega)\frac{\beta^2\omega^2}{4}\frac{1}{1+\cosh\beta \omega}.
\end{align}

The same method is applied to compute the thermal conductivity and the Raman scattering intensity in Sec.~\ref{sec:kappa_xx} and \ref{sec:Raman}, respectively.
These are feasible as the thermal current operator and the Raman operator commute with all $\{\eta_r \}$.
In Sec.~\ref{sec:kappa_xy}, the thermal Hall conductivity is calculated in the same manner, but in this case, for the Hamiltonian including the effect of the Zeeman coupling effectively in Eq.~(\ref{eq:H'_Maj}).
For this effective Hamiltonian, $\{\eta_r \}$ are still conserved and the thermal current operator commutes with $\{\eta_r \}$.

\subsection{Cluster dynamical mean-field theory}
\label{sec:clust-dynam-mean}

In the Majorana representation, one can also apply the CDMFT, which has been developed for interacting fermion systems~\cite{Kotliar2001}.
In the case of the Kitaev model, the system can be regarded as a noninteracting fermion system coupled with localized classical variables, similar to the Falicov-Kimball and the double-exchange models, as mentioned in Sec.~\ref{sec:finiteT}.
For this category of the models, the impurity problem in the CDMFT calculations can be solved exactly~\cite{Furukawa1994,Freericks1998,Freericks2003}.
In the following, we present the framework of the CDMFT for the Kitaev model in the Majorana representation which has been developed in Refs.~\citen{Yoshitake2016} and \citen{Yoshitake2017a}.

In the CDMFT, we assume that the system is composed of a periodic array of clusters with several lattice sites.
Accordingly, the Hamiltonian given in Eq.~(\ref{eq-supp:hamil}) is written in the form of
\begin{align}
{\cal H}^{\{\eta_r\}} ={\cal H}_0 + V^{\{\eta_r\}},
\label{eq-supp:cdmft}
\end{align}
with
\begin{align}
 {\cal H}_0&=\sum_{ll'ss'} \frac{1}{2}A^0_{(ls)(l's')}\gamma_{ls} \gamma_{l's'},\\
V^{\{\eta_r\}}&=\sum_{ll'ss'} \frac{1}{2}B^{\{\eta_r\}}_{(ls)(l's')}\gamma_{ls} \gamma_{l's'},
\end{align}
where $A^{\{\eta_r\}}=A^0+B^{\{\eta_r\}}$ with $A^0$ ($B^{\{\eta_r\}}$) being the $\eta$-independent (dependent) part of $A^{\{\eta_r\}}$, namely, $A^0$ ($B^{\{\eta_r\}}$) originates from the interactions on the $x$ and $y$ bonds (the $z$ bonds) in the original spin Hamiltonian in Eq.~(\ref{eq:H}).
The indices $l$ and $s$ label clusters and sites in the cluster, respectively.

Green's function is introduced as
\begin{align}
 G_{ss'}(\textbf{k}, i \omega_n)=-\frac{1}{2}\sum_{l}\int_0^\beta d\tau  \means{\textrm{T}_\tau \gamma_{ls}(\tau)\gamma_{0s'}}e^{i (\omega_n\tau -\textbf{k}\cdot \textbf{r}_l)},\label{eq-supp:green}
\end{align}
where $\textbf{k}$ is the wave number for the periodic array of the clusters, $\omega_n = (2n+1)\pi T$ is the Matsubara frequency with $n$ being an integer; $\textrm{T}_\tau$ is the time-ordering operator with respect to imaginary time $\tau$, and $\textbf{r}_l$ denotes the position of the cluster $l$.
In a similar manner to Eq.~(\ref{eq-supp:green}), Green's function for ${\cal H}_0$ is calculated as
\begin{align}
 G_{ss'}^0(\textbf{k}, i \omega_n)&=-\frac{1}{2}\sum_{l}\int_0^\beta d\tau  \means{\textrm{T}_\tau \gamma_{ls}(\tau)\gamma_{0s'}}_0e^{i (\omega_n\tau -\textbf{k}\cdot \textbf{r}_l)}\nonumber\\
&=\left[\left(i \omega_n-2A^0(\mathbf{k})\right)^{-1}\right]_{ss'},\label{eq-supp:green0}
\end{align}
where $\means{\cdots}_0$ is the expectation value for ${\cal H}_0$ and
\begin{align}
A^{0}_{ss'}(\mathbf{k}) =
\sum_{l} A^{0}_{(ls)(0s')} e^{-i \mathbf{k}\cdot\mathbf{r}_l}.
\end{align}
Using the above relations, the matrix form of Eq.~(\ref{eq-supp:green}) is formally given as
\begin{align}
G(\mathbf{k}, i \omega_n)=\left(G^0(\textbf{k}, i \omega_n)^{-1} - \Sigma(\mathbf{k}, i \omega_n) \right)^{-1},
\label{eq-supp:green2}
\end{align}
where $\Sigma(\mathbf{k}, i \omega_n)$ is the self-energy.

In the CDMFT, the $\textbf{k}$ dependence of the self-energy is omitted as $\Sigma(i \omega_n)$ and local Green's function within a cluster is given as
\begin{align}
G_{ss'}^{\rm cl}(i \omega_n)= \frac{1}{N_c}\sum_{\mathbf{k}} \left[\left(i \omega_n-2A(\mathbf{k}) - \Sigma(i \omega_n) \right)^{-1}\right]_{ss'},
\label{eq-supp:green-local}
\end{align}
where $N_c$ is the number of the clusters.
Cavity Green's function is introduced as
\begin{align}
 {\cal G} (i \omega_n)^{-1}=G^{\rm cl}(i \omega_n)^{-1}+\Sigma(i \omega_n).
\end{align}
This is obtained in the path integral formalism in the Majorana fermion representation introduced in Ref.~\citen{Nilsson2013} (see Refs.~\citen{Yoshitake2016} and \citen{Yoshitake2017a} for more details).

In the DMFT scheme~\cite{Metzner1989,Georges1996}, the original lattice problem is reduced to an impurity problem embedded in a dynamical medium.
In general, the impurity problem is still difficult to solve because of quantum many-body interactions. In the present case, however, the impurity problem can be solved exactly, as the Majorana fermions are noninteracting.
Green's function for the impurity, which is in this case for a cluster, is calculated as
\begin{align}
G_{ss'}^{\rm{imp}}(i \omega_n) = \sum_{ {\{\eta_r\}} } p(\{\eta_r\})G_{ss'}^{\{\eta_r\}}(i \omega_n),
\label{eq-supp:Gimp}
\end{align}
with
\begin{align}
 G^{\{\eta_r\}}(i \omega_n)=\left({\cal G}(i \omega_n)^{-1}- 2B^{\{\eta_r\}}\right)^{-1},
\end{align}
where $p(\{\eta_r\})$ is the weight of the configuration of $\{\eta_r\}$, which is given by
\begin{align}
p(\{\eta_r\}) = \frac{Z_\gamma^{\{\eta_r\}}}{\sum_{\{\eta_r\}}Z_\gamma^{\{\eta_r\}}}.
\end{align}
$Z_\gamma^{\{\eta_r\}}$ is calculated from Green's functions as
\begin{align}
Z_\gamma^{\{\eta_r\}}
=\prod_{n \geq 0} \textrm{det}\left[-G^{\{\eta_r\}}(i \omega_n)\right].
\end{align}

Finally, the self-energy for the impurity problem is obtained as
\begin{align}
\Sigma(i \omega_n) = {\cal G}(i \omega_n)^{-1}- G^{\rm{imp}}(i \omega_n)^{-1}.
\label{eq-supp:self-energy}
\end{align}
Using this self-energy, local Green's function is recalculated by Eq.~(\ref{eq-supp:green-local}).
These procedures are iterated until the following self-consistent condition is satisfied:
\begin{align}
G^{\rm cl}(i \omega_n) = G^{\rm{imp}}(i \omega_n).
\end{align}

In the calculation of $G_{ss'}^{\rm{imp}}(i \omega_n)$ in Eq.~(\ref{eq-supp:Gimp}), $p(\{\eta_r\})$ and $G^{\{\eta_r\}}(i \omega_n)$ are exactly enumerated for all the $2^{N_s/2}$ configurations of ${\{\eta_r\}}$ within the cluster ($N_s$ is the number of sites in the cluster, namely, $N=N_sN_c$).
Thus, the Majorana-based CDMFT technique provides a concise method without statistical errors.
It is also free from and any biased approximation except for the cluster approximation.
Although the mean-field treatment under the cluster approximation leads to a fictitious phase transition at low $T$, the cluster-size dependence is sufficiently small in the entire range of $T$ above the critical temperature~\cite{Yoshitake2016,Yoshitake2017a}.

\subsection{Continuous-time quantum Monte Carlo method}
\label{sec:cont-time-quant}

Although the Majorana-based QMC and CDMFT techniques enable to compute thermodynamic quantities, they cannot be applied to computation of spin dynamics since the dynamical spin correlations do not commute with the local conserved quantities $\{\eta_r\}$.
To overcome this difficulty, the Majorana-based CTQMC technique was developed in Refs.~\citen{Yoshitake2016,Yoshitake2017a,Yoshitake2017b}.
We introduce the framework in the following.

Let us focus on the dynamical spin correlation $\means{S^z_i(\tau)S^z_j}$, where the sites $i$ and $j$ belong to the $z$ bond $r=r_0$.
As the spin operator $S_i^z(\tau)$ is given by $S^z_i(\tau) = \pm i \gamma_i(\tau)\bar{\gamma}_i(\tau)$ (the sign depends on the sublattice of the honeycomb structure), we need to track the time evolution of $\gamma_i$ and $\bar{\gamma}_i$ on the bond $r_0$.
In the Kitaev model, the dynamical spin correlations of the $\mu$ component are nonzero only on NN $\mu$ bonds, similar to the static correlations in the ground state.
Hence, all other $\eta_r$ for $r\neq r_0$ remain static in the time evolution of $S^{\mu}_i(\tau)$. 
The situation is similar to the impurity Anderson model, to which the CTQMC technique has been applied, particularly as an impurity solver in the DMFT.

The procedure for the calculation of the dynamical spin correlations is as follows.
First, we prepare the configurations of $\{\eta_r\}$ by using the Majorana-based QMC or CDMFT technique in the previous sections.
Then, the dynamical spin correlation is calculated as
\begin{align}
 \means{S^z_i(\tau)S^z_j}=\frac{1}{N_\eta} \sum_{\{\eta_r\}} \left[\overline{S^{\mu}_i(\tau)S^{\mu}_j}\right]^{\{\eta_r\}'},
\label{eq-supp:thermal-ave}
\end{align}
where $N_\eta$ is the number of the $\{\eta_r\}$ configurations; $\bar{\cal O}^{\{\eta_r\}'}$ is calculated in a similar manner to Eq.~(\ref{eq-supp:mean-eta}) by taking the trace over the configuration $\{\eta_r\}$ except for $\eta_{r_0}$.
The CTQMC technique is applied to the numerical calculation of $\left[\overline{S^z_i(\tau)S^z_j}\right]^{\{\eta_r\}'}$ for each configuration of $\{\eta_r\}'$.

In the calculation of the dynamical spin correlations, following the CTQMC technique for the impurity Anderson model used in the DMFT scheme, the Hamiltonian Eq.~(\ref{eq-supp:hamil}) is divided into three parts:
\begin{align}
 {\cal H}^{\{\eta_r\}'}= {\cal H}_{\rm loc} + {\cal H}_{\rm hyb} + {\cal H}_{\rm bath}^{\{\eta_r\}'},
\end{align}
where
\begin{align}
 {\cal H}_{\rm loc}&=\frac{1}{2}\sum_{ij\in r_0} A_{ij}^{{\rm loc}} \gamma_i \gamma_j,\\
 {\cal H}_{\rm hyb}&=\frac{1}{2}\sum_{i\in r_0,j\notin r_0\ {\rm and}\ i\notin r_0,j\in r_0} A_{ij}^{{\rm hyb}} \gamma_i \gamma_j,\\
 {\cal H}_{\rm bath}^{\{\eta_r\}'}&=\frac{1}{2}\sum_{ij\notin r_0} A_{ij}^{\{\eta_r\}'} \gamma_i \gamma_j.\label{eq-supp:bath}
\end{align}
Here, $A_{ij}^{{\rm loc}}=\frac{J_z}{4}\bar{\gamma}_i\bar{\gamma}_j$ for the bond $r_0$, $A_{ij}^{{\rm hyb}}$ stands for a matrix element connecting between a site on $r_0$ and another one not on $r_0$, and $A_{ij}^{\{\eta_r\}'}$ in Eq.~(\ref{eq-supp:bath}) represents a matrix element between sites not on $r_0$.
Note that $A_{ij}^{{\rm loc}}$ and $A_{ij}^{{\rm hyb}}$ do not depend on $\{\eta_r\}'$.
Thus, the problem corresponds to the two-site impurity problem for ${\cal H}_{\rm loc}$ in the CTQMC calculations based on the strong-coupling (hybridization) expansion~\cite{Werner2006}.
Tracing out the bath Hamiltonian ${\cal H}_{\rm bath}^{\{\eta_r\}'}$ by using the path integral approach, the effective action for the two sites on the bond $r_0$ is given by
\begin{align}
\mathcal{S}_{\text{eff}}^{\{\eta\}'} =
\mathcal{S}_{\rm{hyb}}^{\{\eta\}'} + \mathcal{S}_{\rm{loc}},
\label{eq-supp:S_eff}
\end{align}
where
\begin{align}
\mathcal{S}_{\rm{hyb}}^{\{\eta\}'} =&-\frac{1}{2}\sum_{ij\in r_0} \int_0^\beta d\tau \int_0^\beta d\tau' \chi_{i}(\tau)\Delta_{ij}^{\{\eta_r\}'}(\tau-\tau')\chi_{j}(\tau'),
\label{eq-supp:ImTimeActionHyb}\\
\mathcal{S}_{\rm{loc}} =& \sum_{ij\in r_0} \int_0^\beta d\tau \chi_{i}(\tau) \left(\frac{\delta_{ij}}{2}\frac{\partial}{\partial\tau} + A_{ij}^{\rm loc} \right) \chi_{j}(\tau).
\label{eq-supp:ImTimeActionLoc}
\end{align}
Here, $\chi_{i}$ stands for the Grassmann number corresponding to $\gamma_i/\sqrt{2}$.
The hybridization function is given as
\begin{align}
\Delta_{ij}^{\{\eta_r\}'}(\tau) = T\sum_n e^{-i \omega_n \tau} \Delta_{ij}^{\{\eta_r\}'}(i \omega_n),
\end{align}
with
\begin{align}
\Delta_{ij}^{\{\eta_r\}'}(i \omega_n) =
-4\sum_{ll'\notin r_0}A_{il}^{\rm hyb}\left[\left(i \omega_n - 2A^{\{\eta_r\}'}\right)^{-1}\right]_{ll'}A_{l'j}^{\rm hyb}.
\end{align}

In this formalism, the partition function for the two sites is given by
\begin{align}
 Z_{\rm{ loc}} = \int \mathcal{D}\chi e^{ - \mathcal{S}_{{\rm loc}}},
\end{align}
where $\mathcal{D}\chi=\prod_{i,n}d\chi_{i,\omega_n}$.
Using this, the expectation value of ${\cal O}$ in the two-site problem is obtained as
\begin{align}
\langle{\cal O}\rangle_{{\rm loc}} = \frac{\int \mathcal{D}\chi
{\cal O}e^{ - \mathcal{S}_{{\rm loc}}}}{\int \mathcal{D}\chi e^{ - \mathcal{S}_{{\rm loc}}}}.
\end{align}
The partition function of the whole system is written by using the above expression
as
\begin{align}
\frac{Z_\gamma^{\{\eta_r\}'}}{Z_{{\rm loc}}} = \frac{\int \mathcal{D}\chi e^{-\mathcal{S}_{\rm{hyb}}^{\{\eta_r\}'}}e^{- \mathcal{S}_{{\rm loc}}} }
{ \int \mathcal{D}\chi e^{ - \mathcal{S}_{{\rm loc}}}}
= \langle e^{-\mathcal{S}_{{\rm hyb}}^{\{\eta_r\}'}} \rangle_{{\rm loc}}.
\end{align}
Then, the dynamical spin correlation for a given configuration of $\{\eta_r\}'$ is given by
\begin{align}
\left[\overline{S^z_i(\tau)S^z_j}\right]^{\{\eta_r\}'}
&=\frac{1}{Z_\gamma^{\{\eta_r\}'}}\int \mathcal{D}\chi
S^z_i(\tau)S^z_j e^{-\mathcal{S}_{\rm{hyb}}^{\{\eta_r\}'}} e^{ - \mathcal{S}_{{\rm loc}}}\notag\\
&=\frac{\langle e^{-\mathcal{S}_{\rm{hyb}}^{\{\eta_r\}'}} S^z_i(\tau)S^z_j\rangle_{{\rm loc}}}{\langle e^{-\mathcal{S}_{\rm{hyb}}^{\{\eta\}'}} \rangle_{{\rm loc}}}.\label{eq-supp:ss}
\end{align}
This is calculated by expanding the hybridization $e^{-\mathcal{S}_{\rm{hyb}}^{\{\eta_r\}'}}$ in the expectation values on the bond $r_0$ as
\begin{align}
\means{ e^{-\mathcal{S}_{\rm{hyb}}^{\{\eta_r\}'}} {\cal O}}_{{\rm loc}}
=&\sum_{d}\sum_{i_1, \cdots,i_{2d}\in r_0} \int_{0}^{\beta} d\tau_1 \cdots \int_{0}^{\beta} d\tau_{2d}\nonumber\\
&\times\frac{1}{(2d)!} \langle {\textrm{T}_\tau \chi_{i_1}(\tau_1) \cdots \chi_{i_{2d}}(\tau_{2d}) {\cal O}} \rangle_{{\rm loc}}\nonumber\\
&\times{\rm Pf}(\hat{\Delta}^{\{\eta_r\}'}(d, i_1, \cdots, i_{2d}, \tau_1, \cdots, \tau_{2d})),
\label{eq-supp:integral}
\end{align}
where $d$ is the order of $\mathcal{S}_{\rm{hyb}}^{\{\eta_r\}'}$ in the expansion of $e^{-\mathcal{S}_{\rm{hyb}}^{\{\eta_r\}'}}$, Pf($M$) is the Pfaffian of a skew-symmetric matrix $M$, and $\hat{\Delta}^{\{\eta_r\}'}(d, i_1, \tau_1, \cdots, i_{2d}, \tau_{2d})$ is a $2d \times 2d$ matrix, whose matrix element is given by
\begin{align}
\hat{\Delta}^{\{\eta_r\}'}(d, i_1, \cdots, i_{2d}, \tau_1, ..., \tau_{2d})_{mn} = \Delta_{i_m i_n}^{\{\eta_r\}'}(\tau_m-\tau_n).
\end{align}
Note that the coefficient $1/(2d)!$ in Eq.~(\ref{eq-supp:integral}) comes from the product of $1/(2^dd!)$ yielded from Eq.~(\ref{eq-supp:ImTimeActionHyb}) and $1/(2d-1)!!$ whose denominator corresponds to the number of terms in the Pfaffian.

To calculate Eq.~(\ref{eq-supp:integral}), the configurations of $(d, i_1,\tau_1,\cdots, i_{2d},\tau_{2d})$ in Eq.~(\ref{eq-supp:integral}) are generated using the Markov-chain MC method by regarding the integral as the statistical weight for each configuration. 
In each MC step, the configuration is updated by, for example, an increase of the order of expansion $d$ as $(d, i_1, \cdots, i_{2d}, \tau_1, \cdots, \tau_{2d})$ to
$(d+1, i_1, \cdots, i_{2d}, i_{2d+1}, i_{2d+2}, \tau_1, \cdots, \tau_{2d}, \tau_{2d+1}, \tau_{2d+2})$
 by adding $(i_{2d+1},\tau_{2d+1}),(i_{2d+2},\tau_{2d+2})$.
To carry out the update of the configuration, one needs to calculate the ratio of the Pfaffians obtained by adding two rows and columns in the matrix $\hat{\Delta}^{\{\eta_r \}'}$:
\begin{align}
\frac{{\rm Pf}\left[\hat{\Delta}^{\{\eta_r\}'}(d, i_1, \tau_1,\cdots, i_{2d}, \tau_{2d})\right]}
{{\rm Pf}\left[\hat{\Delta}^{\{\eta_r\}'}(d+1, i_1, \tau_1,\cdots, i_{2d+2}, \tau_{2d+2})\right]}.
\end{align}
This can be evaluated by the fast update algorithm, which has been applied for interacting fermion problems (for example, see Ref.~\citen{PhysRevB.72.035122});
the calculation cost is in the order of $d^2$.
On the other hand, $\langle {\textrm{T}_\tau \chi_{i_1}(\tau_1) \cdots \chi_{i_{2d}}(\tau_{2d}) } \rangle_{{\rm loc}}$ and $\langle {\textrm{T}_\tau \chi_{i_1}(\tau_1) \cdots \chi_{i_{2d}}(\tau_{2d}) S^z_i(\tau)S^z_j} \rangle_{{\rm loc}}$ in Eqs.~(\ref{eq-supp:ss}) and (\ref{eq-supp:integral}) are calculated by considering the imaginary-time evolution of all the four states in the two-site problem on the bond $r_0$.

The Majorana-based CTQMC technique is applied to compute the dynamical quantities: the magnetic susceptibility in Sec.~\ref{sec:chi}, the dynamical spin structure factor in Sec.~\ref{sec:Sqw}, and the NMR relaxation rate in Sec.~\ref{sec:1/T1}.
Although it give essentially the same results with the use of either the Majorana-based QMC or CDMFT technique, the combination with the QMC technique can provides the results at lower $T$, as the CDMFT results suffer from the fictitious phase transition at low $T$ as mentioned in Appendix~\ref{sec:clust-dynam-mean}.


\begin{thebibliography}{999}
\bibitem{Majorana1937} E. Majorana, Il Nuovo Cimento {\bf 14}, 171 (1937).
\bibitem{Doi1985} M. Doi,  T. Kotani,  E. Takasugi, Prog. Theor. Phys. Supp. {\bf 83}, 1 (1985). 
\bibitem{Mohapatra2006} R. N. Mohapatra and A. Y. Smirnov, Annu. Rev. Nucl. Part. Sci. {\bf 56}, 569 (2006).
\bibitem{Akhmedov2015} E. Akhmedov, {\it Majorana neutrinos and other Majorana particles: Theory and experiment}, Chap. 15 in {\it The Physics of Ettore Majorana} by Salvatore Esposito (Cambridge Univ. Press, 2015). 
\bibitem{Wilczek2009} F. Wilczek, Nat. Phys. {\bf 5}, 614 (2009).
\bibitem{Moore1991} G. Moore and N. Read, Nucl. Phys. B {\bf 360}, 362 (1991).
\bibitem{Stomer1999} H. L. Stomer, D. C. Tsui, and A. C. Gossard, Rev. Mod. Phys. {\bf 71}, S298 (1999).
\bibitem{Read2000} N. Read and D. Green, Phys. Rev. B {\bf 61}, 10267 (2000).
\bibitem{Nayak2008} C. Nayak, S. H. Simon, A. Stern, M. Freedman, and S. D. Sarma, Rev. Mod. Phys. {\bf 80}, 1083 (2008).
\bibitem{Jain2015} J. K. Jain, Annu. Rev. Condens. Matter Phys. {\bf 6}, 39 (2015).
\bibitem{DasSarma2006} S. Das Sarma, C. Nayak, and S. Tewari, Phys. Rev. B {\bf 73}, 220502 (2006).
\bibitem{Jackiw1981} R. Jackiw, and P. Rossi, Nucl. Phys. B {\bf 190}, 681 (1981).
\bibitem{Fu2008} L. Fu and C. Kane, Phys. Rev. Lett. {\bf 100}, 096407 (2008).
\bibitem{Sato2016} M. Sato and S. Fujimoto, J. Phys. Soc. Jpn. {\bf 85}, 072001 (2016). 
\bibitem{Sato2017} M. Sato and Y. Ando, Rep. Prog. Phys. {\bf 80}, 076501 (2017).
\bibitem{Kitaev2003} A. Kitaev: Ann. Phys. {\bf 303}, 2 (2003).
\bibitem{Freedman2003} M. H. Freedman, A. Kitaev, M. J. Larsen, and Z. Wang, Bull. Amer. Math. Soc. {\bf 40} 31 (2003). 
\bibitem{Anderson1973} P. W. Anderson, Mater. Res. Bull. {\bf 8}, 153 (1973).
\bibitem{Wen2004} X.-G. Wen, {\it Quantum Field Theory of Many-Body Systems} (Oxford Univ. Press, 2004.
\bibitem{Savary2017} L. Savary and L. Balents, Rep. Prog. Phys. {\bf 80}, 016502 (2017).
\bibitem{Read1989} N. Read and B. Chakraborty, Phys. Rev. B {\bf 40}, 7133 (1989).
\bibitem{Wen1991} X. G. Wen, Phys. Rev. B {\bf 44}, 2664 (1991).
\bibitem{Balents2010} L. Balents, Nature. {\bf 464} 199 (2010).
\bibitem{Lacroix2011} C. Lacroix, P. Mendels, and F. Mila, {\it Introduction to Frustrated Magnetism} (Springer, 2011). 
\bibitem{Diep2013} H. T. Diep, {\it Frustrated Spin Systems} (World Scientific, 2013) 2nd ed. 
\bibitem{Zhou2017} Y. Zhou, K. Kanoda, and T.-K. Ng, Rev. Mod. Phys. {\bf 89}, 025003 (2017). 
\bibitem{Kitaev2006} A. Kitaev, Ann. Phys. {\bf 321}, 2 (2006).
\bibitem{Khaliullin2005} G. Khaliullin, Prog. Theor. Phys. Suppl. {\bf 160}, 155 (2005).
\bibitem{Jackeli2009} G. Jackeli and G. Khaliullin, Phys. Rev. Lett. {\bf 102}, 017205 (2009).
\bibitem{Nussinov2015} Z. Nussinov and J. van den Brink, Rev. Mod. Phys. {\bf 87}, 1 (2015).
\bibitem{Trebst2017preprint} S. Trebst, preprint (arXiv:1701.07056).
\bibitem{Winter2017a} S. M. Winter, A. A. Tsirlin , M. Daghofer, J. van den Brink, Y. Singh, P. Gegenwart, and R. Valent\'i, J. Phys.: Condens. Matter {\bf 29}, 493002 (2017).
\bibitem{Hermanns2018} M. Hermanns, I. Kimchi, and J. Knolle, Annu. Rev. Condens. Matter Phys. {\bf 9}, 17 (2018).
\bibitem{Knolle2019a} J. Knolle and R. Moessner, Annu. Rev. Condens. Matter Phys. {\bf 10}, 451 (2019).
\bibitem{Takagi2019} H. Takagi, T. Takayama, G. Jackelli, G. Khaliullin, and S. E. Nagler, Nat. Rev. Phys. {\bf 1}, 264 (2019).
\bibitem{Janssen2019} L. Janssen and M.Vojta, J. Phys.: Condens. Matter {\bf 31} 423002 (2019).
\bibitem{Nasu2014} J. Nasu, M. Udagawa, and Y. Motome, Phys. Rev. Lett. {\bf 113}, 197205 (2014).
\bibitem{Nasu2015} J. Nasu, M. Udagawa, and Y. Motome, Phys. Rev. B {\bf 92}, 115122 (2015).
\bibitem{Yoshitake2016} J. Yoshitake, J. Nasu, and Y. Motome, Phys. Rev. Lett. {\bf 117}, 157203 (2016).
\bibitem{Yoshitake2017a} J. Yoshitake, J. Nasu, Y. Kato, and Y. Motome, Phys. Rev. B {\bf 96}, 024438 (2017).
\bibitem{Yoshitake2017b} J. Yoshitake, J. Nasu, and Y. Motome, Phys. Rev. B {\bf 96}, 064433 (2017).
\bibitem{Mishchenko2017} P. A. Mishchenko, Y. Kato, and Y. Motome, Phys. Rev. B {\bf 96}, 125124 (2017).
\bibitem{Nasu2016} J. Nasu, J. Knolle, D. L. Kovrizhin, Y. Motome, and R. Moessner, Nat. Phys. {\bf 12}, 912 (2016).
\bibitem{Nasu2017a} J. Nasu, J. Yoshitake, and Y. Motome, Phys. Rev. Lett. {\bf 119}, 127204 (2017).
\bibitem{Baskaran2008} G. Baskaran, D. Sen, and R. Shankar, Phys. Rev. B {\bf 78}, 115116 (2008).
\bibitem{Abragam1970} A. Abragam and B. Bleaney, {\it Electron Paramagnetic Resonance of Transition Ions} (Clarendon, Oxford, 1970). 
\bibitem{Liu2018} H. Liu and G. Khaliullin, Phys. Rev. B {\bf 97}, 014407 (2018).
\bibitem{Sano2018} R. Sano, Y. Kato, and Y. Motome, Phys. Rev. B {\bf 97}, 014408 (2018).
\bibitem{Yan2019} J.-Q. Yan, S. Okamoto, Y. Wu, Q. Zheng, H. D. Zhou, H. B. Cao, and M. A. McGuire
Phys. Rev. Materials {\bf 3}, 074405 (2019).
\bibitem{Yao2019preprint} W. Yao and Y. Li, preprint (arXiv:1908.09427).
\bibitem{Zhong2019preprint} R. Zhong, T. Gao, N. P. Ong, and R. J. Cava, preprint (arXiv:1910.08577).
\bibitem{Li2017} F.-Y. Li, Y.-D. Li, Y. Yu, A. Paramekanti, and G. Chen, Phys. Rev. B {\bf 95}, 085132 (2017). 
\bibitem{Rau2018} J. G. Rau and M. J. P. Gingras, Phys. Rev. B {\bf 98}, 054408 (2018).
\bibitem{Jang2019} S.-H. Jang, R. Sano, Y. Kato, and Y. Motome, Phys. Rev. B {\bf 99}, 241106(R) (2019).
\bibitem{Luo2019preprint} Z.-X. Luo and G. Chen, preprint (arXiv:1903.02530). 
\bibitem{Xing2019preprint} J. Xing, H. Cao, E. Emmanouilidou, C. Hu, J. Liu, D. Graf, A. P. Ramirez, G. Chen, and N. Ni, preprint (arXiv:1903.03615). 
\bibitem{Chen2007} H.-D. Chen and J. Hu, Phys. Rev. B {\bf 76}, 193101 (2007).
\bibitem{Feng2007} X.-Y. Feng, G.-M. Zhang, and T. Xiang, Phys. Rev. Lett. {\bf 98}, 087204 (2007).
\bibitem{Chen2008} H.-D. Chen and Z. Nussinov, J. Phys. A {\bf 41}, 075001 (2008).
\bibitem{note1} A numerical method was recently developed for the former Majorana representation in M. Udagawa, Phys. Rev. B {\bf 98}, 220404(R) (2018) and C. N. Self, J. Knolle, S. Iblisdir, and J. K. Pachos, Phys. Rev. B {\bf 99}, 045142 (2019).
\bibitem{Lieb1994} E. H. Lieb, Phys. Rev. Lett. {\bf 73}, 2158 (2994).
\bibitem{Baskaran2007} G. Baskaran, S. Mandal, and R. Shankar, Phys. Rev. Lett. {\bf 98}, 247201 (2007).
\bibitem{Falicov1969} L. M. Falicov and J. C. Kimball, Phys. Rev. Lett. {\bf 22}, 997 (1969).
\bibitem{Zener1951} C. Zener, Phys. Rev. B {\bf 82}, 403 (1951).
\bibitem{Motome2001} Y. Motome and N. Furukawa, J. Phys. Soc. Jpn. {\bf 70}, 1487 (2001).
\bibitem{Haldane1988} F. D. M. Haldane, Phys. Rev. Lett. {\bf 61}, 2015 (1988).
\bibitem{Jiang2011} H.-C. Jiang, Z.-C. Gu, X.-L. Qi, and S. Trebst, Phys. Rev. B {\bf 83}, 245104 (2011).
\bibitem{Yadav2016} R. Yadav, N. A. Bogdanov, V. M. Katukuri, S. Nishimoto, J. van den Brink, and L. Hozoi, Sci. Rep. {\bf 6}, 37925 (2016).
\bibitem{Zhu2018} Z. Zhu, I. Kimchi, D. N. Sheng, and L. Fu, Phys. Rev. B {\bf 97}, 241110(R) (2018).
\bibitem{Gohlke2018} M. Gohlke, R. Moessner, and F. Pollmann, Phys. Rev. B {\bf 98}, 014418 (2018).
\bibitem{Hickey2019} C. Hickey and S. Trebst, Nat. Commun. {\bf 10}, 530 (2019).
\bibitem{Gordon2019} J. S. Gordon, A. Catuneanu, E. S. Sorensen, and H.-Y. Kee, Nat. Commun. {\bf 10}, 2470 (2019).
\bibitem{Nasu2018} J. Nasu, Y. Kato, Y. Kamiya, and Y. Motome, Phys. Rev. B {\bf 98}, 060416(R) (2018).
\bibitem{Liang2018} S. Liang, M.-H. Jiang, W. Chen, J.-X. Li, and Q.-Hua Wang, Phys. Rev. B {\bf 98}, 054433 (2018). 
\bibitem{Ronquillo2019} D. C. Ronquillo, A. Vengal, and N. Trivedi, Phys. Rev. B {\bf 99}, 140413(R) (2019). 
\bibitem{Patel2019} N. D. Patel and N. Trivedi, PNAS {\bf 116}, 12199 (2019). 
\bibitem{Sugita2019preprint} Y. Sugita, Y. Kato, and Y. Motome, preprint (arXiv:1905.12139).
\bibitem{Yoshitake2019preprint} J. Yoshitake, J. Nasu, Y. Kato, and Y. Motome, preprint (arXiv:1907.07299).
\bibitem{Chaloupka2010} J. Chaloupka, G. Jackeli, and G. Khaliullin, Phys. Rev. Lett. {\bf 105}, 027204 (2010).
\bibitem{Chaloupka2013} J. Chaloupka, G. Jackeli, and G. Khaliullin, Phys. Rev. Lett. {\bf 110}, 097204 (2013).
\bibitem{Rau2014} J. Rau, E. K.-H. Lee, and H.-Y. Kee, Phys. Rev. Lett. {\bf 112}, 077204 (2014).
\bibitem{Rusnacko2019} J. Rusna\v{c}ko, D. Gotfryd, and J. Chaloupka, Phys. Rev. B {\bf 99}, 064425 (2019).
\bibitem{Kitagawa2018} K. Kitagawa, T. Takayama, Y. Matsumoto, A. Kato, R. Takano, Y. Kishimoto, S. Bette, R. Dinnebier, G. Jackeli, and H. Takagi, Nature {\bf 554}, 341 (2018).
\bibitem{Slagle2018} K. Slagle, W. Choi, L. E. Chern, and Y. B. Kim, Phys. Rev. B {\bf 97}, 115159 (2018).
\bibitem{Yadav2018} R. Yadav, R. Ray, M. S. Eldeeb, S. Nishimoto, L. Hozoi, and J. van den Brink, Phys. Rev. Lett. {\bf 121}, 197203 (2018).
\bibitem{Li2018} Y. Li, S. M. Winter, and R. Valent\'i, Phys. Rev. Lett. {\bf 121}, 247202 (2018).
\bibitem{Knolle2019b} J. Knolle, R. Moessner, and N. B. Perkins, Phys. Rev. Lett. {\bf 122}, 047202 (2019).
\bibitem{Foyevtsova2013} K. Foyevtsova, H. O. Jeschke, I. I. Mazin, D. I. Khomskii, and R. Valent\'i, Phys. Rev. B {\bf 88}, 035107 (2013).
\bibitem{Katukuri2014} V. M. Katukuri, S Nishimoto, V. Yushankhai, A. Stoyanova, H. Kandpal, S. Choi, R. Coldea, I. Rousochatzakis, L. Hozoi, and J. van den Brink, New J. Phys. {\bf 16}, 013056 (2014).
\bibitem{Yamaji2014} Y. Yamaji, Y. Nomura, M. Kurita, R. Arita, and M. Imada, Phys. Rev. Lett. {\bf 113}, 107201 (2014).
\bibitem{Winter2016} S. M. Winter, Y. Li, H. O. Jeschke, and Roser Valent\'i, Phys. Rev. B {\bf 93}, 214431 (2016).
\bibitem{Sandilands2015} L. J. Sandilands, Y. Tian, K. W. Plumb, Y.-J. Kim, and K. S. Burch, Phys. Rev. Lett. {\bf 114}, 147201 (2015).
\bibitem{Banerjee2016} A. Banerjee, C. A. Bridges, J.-Q. Yan, A. A. Aczel, L. Li, M. B. Stone, G. E. Granroth, M. D. Lumsden, Y. Yiu, J. Knolle, S. Bhattacharjee, D. L. Kovrizhin, R. Moessner, D. A. Tennant, D. G. Mandrus, and S. E. Nagler, Nat. Mat. {\bf 15}, 733 (2016).
\bibitem{Do2017} S.-H. Do, S.-Y. Park, J. Yoshitake, J. Nasu, Y. Motome, Y. S. Kwon, D. T. Adroja, D. J. Voneshen, K. Kim, T.-H. Jang, J.-H. Park, K.-Y. Choi, and S. Ji, Nat. Phys. {\bf 13}, 1079 (2017).
\bibitem{note2} It was pointed out that the DOS exhibits logarithmic divergence in the low-$\omega$ limit in C. N. Self, J. Knolle, S. Iblisdir, and J. K. Pachos, Phys. Rev. B {\bf 99}, 045142 (2019).
\bibitem{Yao2007} H. Yao and S. A. Kivelson, Phys. Rev. Lett. {\bf 99}, 247203 (2007).
\bibitem{Dusuel2008} S. Dusuel, K. P. Schmidt, J. Vidal, and R. L. Zaffino, Phys. Rev. B {\bf 78}, 125102 (2008).
\bibitem{Nasu2015b} J. Nasu and Y. Motome, Phys. Rev. Lett. {\bf 115}, 087203 (2015).
\bibitem{Mandal2009} S. Mandal and N. Surendran, Phys. Rev. B {\bf 79}, 024426 (2009).
\bibitem{Modic2014} K.A. Modic, T. E. Smidt, I. Kimchi, N. P. Breznay, A. Biffin, S. Choi, R. D. Johnson, R. Coldea, P. Watkins-Curry, G. T. McCandless, J. Y. Chan, F. Gandara, Z. Islam, A. Vishwanath, A. Shekhter, R. D. McDonald, and J. G. Analytis, Nat. Commun. {\bf 5}, 4203 (2014).
\bibitem{Takayama2015} T. Takayama, A. Kato, R. Dinnebier, J. Nuss, H. Kono, L. S. I. Veiga, G. Fabbris, D. Haskel, and H. Takagi, Phys. Rev. Lett. {\bf 114}, 077202 (2015).
\bibitem{Nasu2014b} J. Nasu, T. Kaji, K. Matsuura, M. Udagawa, and Y. Motome, Phys. Rev. B {\bf 89}, 115125 (2014).
\bibitem{Kimchi2014} I. Kimchi, J. G. Analytis, and A. Vishwanath, Phys. Rev. B {\bf 90}, 205126 (2014). 
\bibitem{Kamiya2015} Y. Kamiya, Y. Kato, J. Nasu, and Y. Motome, Phys. Rev. B {\bf 92}, 100403(R) (2015).
\bibitem{Mandal2011} S. Mandal, S. Bhattacharjee, K. Sengupta, R. Shankar, and G. Baskaran, Phys. Rev. B {\bf 84}, 155121 (2011). 
\bibitem{Nasu2017b} J. Nasu, Y. Kato, J. Yoshitake, Y. Kamiya, and Y. Motome, Phys. Rev. Lett. {\bf 118}, 137203 (2017). 
\bibitem{Kato2017} Y. Kato, Y. Kamiya, J. Nasu, and Y. Motome, Phys. Rev. B {\bf 96}, 174409 (2017).
\bibitem{Mishchenko2019preprint} P. A. Mishchenko, Y. Kato, K. O'Brien, T. A. Bojesen, T. Eschmann, M. Hermanns, S. Trebst, and Y. Motome, preprint (arXiv:1907.10241).
\bibitem{Kalmeyer1987} V. Kalmeyer, and R. B. Laughlin, Phys. Rev. Lett. {\bf 59}, 2095 (1987).
\bibitem{Singh2010} Y. Singh and P. Gegenwart, Phys. Rev. B {\bf 82}, 064412 (2010).
\bibitem{Singh2012} Y. Singh, S. Manni, J. Reuther, T. Berlijn, R. Thomale, W. Ku, S. Trebst, and P. Gegenwart, Phys. Rev. Lett. {\bf 108}, 127203 (2012).
\bibitem{Freund2016} F. Freund, S. C. Williams, R. D. Johnson, R. Coldea, P. Gegenwart, and A. Jesche, Sci. Rep. {\bf 6}, 35362 (2016).
\bibitem{Comin2012} R. Comin, G. Levy, B. Ludbrook, Z.-H. Zhu, C. N. Veenstra, J. A. Rosen, Y. Singh, P. Gegenwart, D. Stricker, J. N. Hancock, D. van der Marel, I. S. Elfimov, and A. Damascelli, Phys. Rev. Lett. {\bf 109}, 266406 (2012). 
\bibitem{Sohn2013} C. H. Sohn, H.-S. Kim, T. F. Qi, D. W. Jeong, H. J. Park, H. K. Yoo, H. H. Kim, J.-Y. Kim, T. D. Kang, Deok-Yong Cho, G. Cao, J. Yu, S. J. Moon, and T. W. Noh, Phys. Rev. B {\bf 88}, 085125 (2013). 
\bibitem{Chun2015} S. H. Chun, J.-W. Kim, J. Kim, H. Zheng, C. C. Stoumpos, C. D. Malliakas, J. F. Mitchell, K. Mehlawat, Y. Singh, Y. Choi, T. Gog, A. Al-Zein, M. M. Sala, M. Krisch, J. Chaloupka, G. Jackeli, G. Khaliullin, and B. J. Kim, Nat. Phys. {\bf 11}, 462 (2015).
\bibitem{Das2019} S. D. Das, S. Kundu, Z. Zhu, E. Mun, R. D. McDonald, G. Li, L. Balicas, A. McCollam, G. Cao, J. G. Rau, H.-Y. Kee, V. Tripathi, and S. E. Sebastian, Phys. Rev. B {\bf 99}, 081101(R) (2019). 
\bibitem{Liu2011} X. Liu, T. Berlijn, W.-G. Yin, W. Ku, A. Tsvelik, Y.-J. Kim, H. Gretarsson, Y. Singh, P. Gegenwart, and J. P. Hill, Phys. Rev. B {\bf 83}, 220403(R) (2011).
\bibitem{Ye2012} F. Ye, S. Chi, H. Cao, B. C. Chakoumakos, J. A. Fernandez-Baca, R. Custelcean,
T. F. Qi, O. B. Korneta, and G. Cao, Phys. Rev. B {\bf 85}, 180403(R) (2012).
\bibitem{Williams2016} S. C. Williams, R. D. Johnson, F. Freund, S. Choi, A. Jesche, I. Kimchi, S. Manni, A. Bombardi, P. Manuel, P. Gegenwart, and R. Coldea, Phys. Rev. B {\bf 93}, 195158 (2016).
\bibitem{Todorova2011} V. Todorova, A. Leineweber, L. Kienle, V. Duppel, and M. Jansen, J. Solid State Chem. {\bf 184}, 1112 (2011).
\bibitem{Roudebush2016} J. H. Roudebush, K. A. Ross, and R. J. Cava, Dalton Trans. {\bf 45}, 8783 (2016).
\bibitem{Abramchuk2017} M. Abramchuk, C. Ozsoy-Keskinbora, J. W. Krizan, K. R. Metz, D. C. Bell, and F. Tafti, J. Am. Chem. Soc. {\bf 139}, 15371 (2017). 
\bibitem{Choi2019} Y. S. Choi, C. H. Lee, S. Lee, S. Yoon, W.-J. Lee, J. Park, A. Ali, Y. Singh, J.-C. Orain, G. Kim, J.-S. Rhyee, W.-T. Chen, F. Chou, and K.-Y. Choi, Phys. Rev. Lett. {\bf 122}, 167202 (2019). 
\bibitem{Kenny2019} E. M. Kenney, C. U. Segre, W. Lafargue-Dit-Hauret, O. I. Lebedev, M. Abramchuk, A. Berlie, S. P. Cottrell, G. Simutis, F. Bahrami, N. E. Mordvinova, G. Fabbris, J. L. McChesney, D. Haskel, X. Rocquefelte, M. J. Graf, and F. Tafti, Phys. Rev. B {\bf 100}, 094418 (2019). 
\bibitem{Takahashi2019} S. K. Takahashi, J. Wang, A. Arsenault, T. Imai, M. Abramchuk, F. Tafti, and P.p M. Singer, Phys. Rev. X {\bf 9}, 031047 (2019). 
\bibitem{Plumb2014} K. W. Plumb, J. P. Clancy, L. J. Sandilands, V. Vijay Shankar, Y. F. Hu, and K. S. Burch, Phys. Rev. B {\bf 90}, 041112(R) (2014).
\bibitem{Fletcher1963} J. M. Fletcher, W. E. Gardner, E. W. Hooper, K. R. Hyde, F. H. Moore, and J. L. Woodhead, Nature (London) {\bf 199}, 1089 (1963).
\bibitem{Fletcher1967} J. M. Fletcher, W. E. Gardner, A. C. Fox, and G. Topping, J. Chem. Soc. A, 1038 (1967).
\bibitem{Brodersen1968} K. Brodersen, G. Thiele, H. Ohnsorge, I. Recke, and F. Moers,
J. Less-Common Met. {\bf 15}, 347 (1968).
\bibitem{Kubota2015} Y. Kubota, H. Tanaka, T. Ono, Y. Narumi, and K. Kindo, Phys. Rev. B {\bf 91}, 094422 (2015).
\bibitem{Johnson2015} R. D. Johnson, S. C. Williams, A. A. Haghighirad, J. Singleton, V. Zapf, P. Manuel, I. I. Mazin, Y. Li, H. O. Jeschke, R. Valent\'i, and R. Coldea, Phys. Rev. B {\bf 92}, 235119 (2015).
\bibitem{Cao2016} H. B. Cao, A. Banerjee, J.-Q. Yan, C. A. Bridges, M. D. Lumsden, D. G. Mandrus, D. A. Tennant, B. C. Chakoumakos, and S. E. Nagler, Phys. Rev. B {\bf 93}, 134423 (2016).
\bibitem{Kim2016} H.-S. Kim and H.-Y. Kee, Phys. Rev. B {\bf 93}, 155143 (2016). 
\bibitem{Koitzsch2016} A. Koitzsch, C. Habenicht, E. M\"uller, M. Knupfer, B. B\"uchner, H. C. Kandpal, J. van den Brink, D. Nowak, A. Isaeva, and Th. Doert, Phys. Rev. Lett. {\bf 117}, 126403 (2016).
\bibitem{Sinn2016} S. Sinn, C. H. Kim, B. H. Kim, K. D. Lee, C. J. Won, J. S. Oh, M. Han, Y. J. Chang, N. Hur, H. Sato, B.-G. Park, C. Kim, H.-D. Kim, and T. W. Noh, Sci. Rep. {\bf 6}, 39544 (2016).
\bibitem{Sears2015} J. A. Sears, M. Songvilay, K. W. Plumb, J. P. Clancy, Y. Qiu, Y. Zhao, D. Parshall, and Y.-J. Kim, Phys. Rev. B {\bf 91}, 144420 (2015).
\bibitem{Weber2016} D. Weber, L. M. Schoop, V. Duppel, J. M. Lippmann, J. Nuss, and B. V. Lotsch, Nano Lett. {\bf 16}, 3578 (2016). 
\bibitem{Ziatdinov2016} M. Ziatdinov, A. Banerjee, A. Maksov, T. Berlijn, W. Zhou, H. B. Cao, J.-Q. Yan, C. A. Bridges, D. G. Mandrus, S. E. Nagler, A. P. Baddorf, and S. V. Kalinin, Nat. Commun. {\bf 7}, 13774 (2016).
\bibitem{Gronke2018} M Gr\"onke, P. Schmidt, M. Valldor, S. Oswald, D. Wolf, A. Lubk, B. B\"uchner, and S. Hampel, Nanoscale {\bf 10}, 19014 (2018). 
\bibitem{Zhou2019a} B. Zhou, Y. Wang, G. B. Osterhoudt, P. Lampen-Kelley, D. Mandrus, R. Hee, K. S. Burch, and E. A. Henriksen, J. Phys. Chem. Solid. {\bf 128}, 291 (2019).
\bibitem{Zhou2019b} B. Zhou, J. Balgley, P. Lampen-Kelley, J.-Q. Yan, D. G. Mandrus, and E. A. Henriksen, Phys. Rev. B {\bf 100}, 165426 (2019).
\bibitem{Mashhadi2019} S. Mashhadi, Y. Kim, J. Kim, D. Weber, T. Taniguchi, K. Watanabe, N. Park, B. Lotsch, J. H. Smet, M. Burghard, and K. Kern, Nano Lett. {\bf 19}, 4659 (2019).
\bibitem{Biswas2019preprint} S. Biswas, Y. Li, S. M. Winter, J. Knolle, and R. Valent\'i, preprint (arXiv:1908.04793).
\bibitem{Gerber2019preprint} E. Gerber, Y. Yao, T. A. Arias, and E.-A. Kim, preprint (arXiv:1902.09550). 
\bibitem{Kim2015a} H.-S. Kim, E. K.-H. Lee, and Y. B. Kim, Eur. Phys. Lett. {\bf 112}, 67004 (2015). 
\bibitem{Katukuri2016} V. M. Katukuri, R. Yadav, L. Hozoi, S. Nishimoto, and J. van den Brink, Sci. Rep. {\bf 6}, 29585 (2016). 
\bibitem{Biffin2014a} A. Biffin, R. D. Johnson, S. Choi, F. Freund, S. Manni, A. Bombardi, P. Manuel, P. Gegenwart, and R. Coldea, Phys. Rev. B {\bf 90}, 205116 (2014).
\bibitem{Biffin2014b} A. Biffin, R. D. Johnson, I. Kimchi, R. Morris, A. Bombardi, J. G. Analytis, A. Vishwanath, and R. Coldea, Phys. Rev. Lett. {\bf 113}, 197201 (2014).
\bibitem{Ruiz2017} A. Ruiz, A. Frano, N. P. Breznay, I. Kimchi, T. Helm, I. Oswald, J. Y. Chan, R.J. Birgeneau, Z. Islam, and J. G. Analytis, Nat. Commun. {\bf 8}, 961 (2017).
\bibitem{Modic2017} K. A. Modic, B. J. Ramshaw, J. B. Betts, N. P. Breznay, J. G. Analytis, R. D. McDonald, and A. Shekhter, Nat. Commun. {\bf 8}, 180 (2017). 
\bibitem{Breznay2017} N. P. Breznay, A. Ruiz, A. Frano, W. Bi, R. J. Birgeneau, D. Haskel, and J. G. Analytis, Phys. Rev. B {\bf 96}, 020402(R) (2017). 
\bibitem{Takayama2019} T. Takayama, A. Krajewska, A. S. Gibbs, A. N. Yaresko, H. Ishii, H. Yamaoka, K. Ishii, N. Hiraoka, N. P. Funnell, C. L. Bull, and H. Takagi, Phys. Rev. B {\bf 99}, 125127 (2019). 
\bibitem{Mehlawat2017} K. Mehlawat, A. Thamizhavel, and Y. Singh, Phys. Rev. B {\bf 95}, 144406 (2017).
\bibitem{Yamaji2016} Y. Yamaji, T. Suzuki, T. Yamada, S. Suga, N. Kawashima, and M. Imada, Phys. Rev. B {\bf 93}, 174425 (2016).
\bibitem{Suzuki2018} T. Suzuki and S. Suga, Phys. Rev. B {\bf 97}, 134424 (2018).
\bibitem{Sears2017} J. A. Sears, Y. Zhao, Z. Xu, J. W. Lynn, and Y.-J. Kim, Phys. Rev. B {\bf 95}, 180411(R) (2017). 
\bibitem{Wolter2017} A. U. B. Wolter, L. T. Corredor, L. Janssen, K. Nenkov, S. Sch\"onecker, S.-H. Do, K.-Y. Choi, R. Albrecht, J. Hunger, T. Doert, M. Vojta, and B. B\"uchner, Phys. Rev. B {\bf 96}, 041405(R) (2017). 
\bibitem{Widmann2019} S. Widmann, V. Tsurkan, D. A. Prishchenko, V. G. Mazurenko, A. A. Tsirlin, and A. Loidl, Phys. Rev. B {\bf 99}, 094415 (2019).
\bibitem{Sandilands2016} L. J. Sandilands, C. H. Sohn, H. J. Park, S. Y. Kim, K. W. Kim, J. A. Sears, Y.-J. Kim, and T. W. Noh, Phys. Rev. B {\bf 94}, 195156 (2016).
\bibitem{Yoshitake_thesis} J. Yoshitake, PhD thesis (The University of Tokyo, 2019).
\bibitem{Chaloupka2016} J. Chaloupka and G. Khaliullin, Phys. Rev. B {\bf 94}, 064435 (2016).
\bibitem{Janssen2017} L. Janssen, E. C. Andrade, and M. Vojta, Phys. Rev. B {\bf 96}, 064430 (2017). 
\bibitem{Lampen-Kelley2018} P. Lampen-Kelley, S. Rachel, J. Reuther, J.-Q. Yan, A. Banerjee, C. A. Bridges, H. B. Cao, S. E. Nagler, and D. Mandrus, Phys. Rev. B {\bf 98}, 100403(R) (2018). 
\bibitem{Knolle2014} J. Knolle, D. L. Kovrizhin, J. T. Chalker, and R. Moessner, Phys. Rev. Lett. {\bf 112}, 207203 (2014).
\bibitem{Knolle2015} J. Knolle, D. L. Kovrizhin, J. T. Chalker, and R. Moessner, Phys. Rev. B {\bf 92}, 115127 (2015).
\bibitem{Banerjee2017} A. Banerjee, J. Yan, J. Knolle, C. A. Bridges, M. B. Stone, M. D. Lumsden, D. G. Mandrus, D. A. Tennant, R. Moessner, and S. E. Nagler, Science {\bf 356}, 1055 (2017).
\bibitem{Park2016preprint} S.-Y. Park, S.-H. Do, K.-Y. Choi, D. Jang, T.-H. Jang, J. Schefer, C.-M. Wu, J. S. Gardner, J. M. S. Park, J.-H. Park, and S. Ji, preprint (arXiv:1609.05690). 
\bibitem{note3} The high-energy incoherent feature is known to be reproducible at the level of a classical approximation~\cite{Samarakoon2017}.
\bibitem{Samarakoon2017} A. M. Samarakoon, A. Banerjee, S.-S. Zhang, Y. Kamiya, S. E. Nagler, D. A. Tennant, S.-H. Lee, and C. D. Batista, Phys. Rev. B {\bf 96}, 134408 (2017).
\bibitem{Song2016} X.-Y. Song, Y.-Z. You, and L. Balents, Phys. Rev. Lett. {\bf 117}, 037209 (2016). 
\bibitem{Gohlke2017} M. Gohlke, R. Verresen, R. Moessner, and F. Pollmann, Phys. Rev. Lett. {\bf 119}, 157203 (2017). 
\bibitem{Winter2017b} S. M. Winter, K. Riedl, P. A. Maksimov, A. L. Chernyshev, A. Honecker, and R. Valent\'i, Nat. Commun. {\bf 8}, 1152 (2017). 
\bibitem{Knolle2018} J. Knolle, S. Bhattacharjee, and R. Moessner, Phys. Rev. B {\bf 97}, 134432 (2018). 
\bibitem{Kim2015b} H.-S. Kim, V. Shankar V., A. Catuneanu, and H.-Y. Kee, Phys. Rev. B {\bf 91}, 241110(R) (2015).
\bibitem{Banerjee2018} A. Banerjee, P. Lampen-Kelley, J. Knolle, C. Balz, A. A. Aczel, B. Winn, Y. Liu, D. Pajerowski, J. Yan, C. A. Bridges, A. T. Savici, B. C. Chakoumakos, M. D. Lumsden, D. A. Tennant, R. Moessner, D. G. Mandrus, and S. E. Nagler, npj Quantum Materials {\bf 3}, 8 (2018).
\bibitem{Balz2019} C. Balz, P. Lampen-Kelley, A. Banerjee, J. Yan, Z. Lu, X. Hu, S. M. Yadav, Y. Takano, Y. Liu, D. A. Tennant, M. D. Lumsden, D. Mandrus, and S. E. Nagler, Phys. Rev. B {\bf 100}, 060405(R) (2019). 
\bibitem{Winter2018} S. M. Winter, K. Riedl, D. Kaib, R. Coldea, and R. Valent\'i, Phys. Rev. Lett. {\bf 120}, 077203 (2018). 
\bibitem{Moriya1962} T. Moriya, Prog. Theor. Phys. {\bf 28}, 371 (1962).
\bibitem{Moriya1956} T. Moriya, Prog. Theor. Phys. {\bf 16}, 23 (1956).
\bibitem{Baek2017} S.-H. Baek, S.-H. Do, K.-Y. Choi, Y. S. Kwon, A. U. B. Wolter, S. Nishimoto, J. van den Brink, and B. B\"uchner, Phys. Rev. Lett. {\bf 119}, 037201 (2017).
\bibitem{Zheng2017} J. Zheng, K. Ran, T. Li, J. Wang, P. Wang, B. Liu, Z.-X. Liu, B. Normand, J. Wen, and W. Yu, Phys. Rev. Lett. {\bf 119}, 227208 (2017).
\bibitem{Jansa2018} N. Jan\v{s}a, A. Zorko, M. Gomil\v{s}ek, M Pregelj, K. W. Kr\"amer, D. Biner,
A. Biffin, C. R\"uegg, and M. Klanj\v{s}ek, Nat. Phys. {\bf 14}, 786 (2018).
\bibitem{Nagai2019preprint} Y. Nagai, T. Jinno, Y. Yoshitake, J. Nasu, Y. Motome, M. Itoh, and Y. Shimizu, preprint (arXiv:1810.05379).
\bibitem{note4} The NQR measurement was done in Ref.~\citen{Nagai2019preprint}, but the magnetic ordering and associated fluctuations make the comparison difficult.
\bibitem{Hirobe2017} D. Hirobe, M. Sato, Y. Shiomi, H. Tanaka, and E. Saitoh, Phys. Rev. B {\bf 95}, 241112(R) (2017).
\bibitem{Knolle2014b} J. Knolle, G.-W. Chern, D. L. Kovrizhin, R. Moessner, and N. B. Perkins, Phys. Rev. Lett. {\bf 113}, 187201 (2014).
\bibitem{Fleury1968} P. A. Fleury and R. Loudon, Phys. Rev. {\bf 166}, 514 (1968).
\bibitem{Glamazda2016} A. Glamazda, P. Lemmens, S.-H. Do, Y. S. Choi, and K.-Y. Choi, Nat. Commun. {\bf 7}, 12286 (2016).
\bibitem{Rousochatzakis2019} I. Rousochatzakis, S. Kourtis, J. Knolle, R. Moessner, and N. B. Perkins, Phys. Rev. B {\bf 100}, 045117 (2019). 
\bibitem{Wang2018preprint} Y. Wang, G. B. Osterhoudt, Y. Tian, P. Lampen-Kelley, A. Banerjee, T. Goldstein, J. Yan, J. Knolle, J. Nasu, Y. Motome, S. Nagler, D. Mandrus, and K. S. Burch, preprint (arXiv:1809.07782). 
\bibitem{Wulferding2019preprint} D. Wulferding, Y. Choi, S.-H. Do, C. H. Lee, P. Lemmens, C. Faugeras, Y. Gallais, and K.-Y. Choi, preprint (arXiv:1910.00800). 
\bibitem{Nomura2012}  K. Nomura, S. Ryu, A. Furusaki, and N. Nagaosa, Phys. Rev. Lett. {\bf 108}, 026802 (2012).
\bibitem{Sumiyoshi2013}  H. Sumiyoshi and S. Fujimoto, J. Phys. Soc. Jpn. {\bf 82}, 023602 (2013).
\bibitem{Kasahara2018a} Y. Kasahara, K. Sugii, T. Ohnishi, M. Shimozawa, M. Yamashita, N. Kurita, H. Tanaka, J. Nasu, Y. Motome, T. Shibauchi, and Y. Matsuda, Phys. Rev. Lett. {\bf 120}, 217205 (2018).
\bibitem{Kasahara2018b} Y. Kasahara, T. Ohnishi, Y. Mizukami, O. Tanaka, S. Ma, K. Sugii, N. Kurita, H. Tanaka, J. Nasu, Y. Motome, T. Shibauchi, and Y. Matsuda, Nature {\bf 559}, 227 (2018).
\bibitem{Takikawa2019} D. Takikawa and S. Fujimoto, Phys. Rev. B {\bf 99}, 224409 (2019).
\bibitem{Hentrich2018} R. Hentrich, A. U. B. Wolter, X. Zotos, W. Brenig, D. Nowak, A. Isaeva, T. Doert, A. Banerjee, P. Lampen-Kelley, D. G. Mandrus, S. E. Nagler, J. Sears, Y.-J. Kim, B. B\"uchner, and C. Hess, Phys. Rev. Lett. {\bf 120}, 117204 (2018).
\bibitem{Ye2018} M. Ye, G. B. Hal\'asz, L. Savary, and L. Balents, Phys. Rev. Lett. {\bf 121}, 147201 (2018).
\bibitem{Vinker-Aviv2018} Y. Vinkler-Aviv and A. Rosch, Phys. Rev. X {\bf 8}, 031032 (2018).
\bibitem{Chang2013} C.-Z. Chang, J. Zhang, X. Feng, J. Shen, Z. Zhang, M. Guo, K. Li, Y. Ou, P. Wei, L.-L. Wang, Z.-Q. Ji, Y. Feng, S. Ji, X. Chen, J. Jia, X. Dai, Z. Fang, S.-C. Zhang, K. He, Y. Wang, L. Lu, X.-C. Ma, and Q.-K. Xue, Science {\bf 340}, 167 (2013).
\bibitem{Chang2015} C.-Z. Chang, W. W. Zhao, D. Y. Kim, H. J. Zhang, B. A. Assaf, D. Heiman, S.-C. Zhang, C. X. Liu, M. H. W. Chan, and J. S. Moodera, Nat. Mater. {\bf 14}, 473 (2015).
\bibitem{O'Brien2016} K. O'Brien, M. Hermanns, and S. Trebst, Phys. Rev. B {\bf 93}, 085101 (2016). 
\bibitem{Eschmann2019} T. Eschmann, P. A. Mishchenko, T. A. Bojesen, Y. Kato, M. Hermanns, Y. Motome, and S. Trebst, Phys. Rev. Research {\bf 1}, 032011(R) (2019).  
\bibitem{Oitmaa2018} J. Oitmaa, A. Koga, and R. R. P. Singh, Phys. Rev. B {\bf 98}, 214404 (2018).
\bibitem{SuzukiYamaji2018} T. Suzuki and Y. Yamaji, Physica B: Condens. Matter {\bf 536}, 637 (2018).
\bibitem{Koga2018} A. Koga, H. Tomishige, and J. Nasu, J. Phys. Soc. Jpn. {\bf 87}, 063703 (2018).
\bibitem{Dwivedi2018} V. Dwivedi, C. Hickey, T. Eschmann, and S. Trebst, Phys. Rev. B {\bf 98}, 054432 (2018).
\bibitem{Stavropoulos2019} P. P. Stavropoulos, D. Pereira, and H.-Y. Kee, Phys. Rev. Lett. {\bf 123}, 037203 (2019).
\bibitem{Xu2018} C. Xu, J. Feng, H. Xiang, and L. Bellaiche, npj Comput. Mater. {\bf 4}, 57 (2018). 
\bibitem{Lee2019preprint} I. Lee, F. G. Utermohlen, K. Hwang, D. Weber, C. Zhang, J. van Tol, J. E. Goldberger, N. Trivedi, and P. C. Hammel, preprint (arXiv:1902.00077).
\bibitem{Ponomaryov2017} A. N. Ponomaryov, E. Schulze, J. Wosnitza, P. Lampen-Kelley, A. Banerjee, J.-Q. Yan, C. A. Bridges, D. G. Mandrus, S. E. Nagler, A. K. Kolezhuk, and S. A. Zvyagin, Phys. Rev. B {\bf 96}, 241107(R) (2017). 
\bibitem{Wang2017} Z. Wang, S. Reschke, D. H\"uvonen, S.-H. Do, K.-Y. Choi, M. Gensch, U. Nagel, T. R\~{o}\~{o}m, and A. Loidl, Phys. Rev. Lett. {\bf 119}, 227202 (2017). 
\bibitem{Little2017} A. Little, L. Wu, P. Lampen-Kelley, A. Banerjee, S. Patankar, D. Rees, C. A. Bridges, J.-Q. Yan, D. Mandrus, S. E. Nagler, and J. Orenstein, Phys. Rev. Lett. {\bf 119}, 227201 (2017). 
\bibitem{Reschke2019} S. Reschke, V. Tsurkan, S.-H. Do, K.-Y. Choi, P. Lunkenheimer, Zhe Wang, and A. Loidl, Phys. Rev. B {\bf 100}, 100403(R) (2019). 
\bibitem{Leahy2017} I. A. Leahy, C. A. Pocs, P. E. Siegfried, D. Graf, S.-H. Do, K.-Y. Choi, B. Normand, and M. Lee, Phys. Rev. Lett. {\bf 118}, 187203 (2017). 
\bibitem{Yu2018} Y. J. Yu, Y. Xu, K. J. Ran, J. M. Ni, Y. Y. Huang, J. H. Wang, J. S. Wen, and S. Y. Li, Phys. Rev. Lett. {\bf 120}, 067202 (2018). 
\bibitem{Choi2012} S. K. Choi, R. Coldea, A. N. Kolmogorov, T. Lancaster, I. I. Mazin, S. J. Blundell, P. G. Radaelli, Y. Singh, P. Gegenwart, K. R. Choi, S.-W. Cheong, P. J. Baker, C. Stock, and J. Taylor, Phys. Rev. Lett. {\bf 108}, 127204 (2012). 
\bibitem{Ran2017} K. Ran, J. Wang, W. Wang, Z.-Y. Dong, X. Ren, S. Bao, S. Li, Z. Ma, Y. Gan, Y. Zhang, J. T. Park, G. Deng, S. Danilkin, S.-L. Yu, J.-X. Li, and J. Wen, Phys. Rev. Lett. {\bf 118}, 107203 (2017). 
\bibitem{Willans2010} A. J. Willans, J. T. Chalker, and R. Moessner, Phys. Rev. Lett. {\bf 104}, 237203 (2010). 
\bibitem{Dhochak2010} K. Dhochak, R. Shankar, and V. Tripathi, Phys. Rev. Lett. {\bf 105}, 117201 (2010). 
\bibitem{Vojta2016} M. Vojta, A. K. Mitchell, and F. Zschocke, Phys. Rev. Lett. {\bf 117}, 037202 (2016). 
\bibitem{Petrova2014} O. Petrova, P. Mellado, and O. Tchernyshyov, Phys. Rev. B {\bf 90}, 134404 (2014). 
\bibitem{Halasz2014} G. B. Hala\'sz, J. T. Chalker, and R. Moessner, Phys. Rev B {\bf 90}, 035145 (2014). 
\bibitem{Halasz2016} G. B. Hala\'sz and J. T. Chalker, Phys. Rev B {\bf 94}, 235105 (2016). 
\bibitem{You2012} Y.-Z. You, I. Kimchi, and A. Vishwanath, Phys. Rev. B {\bf 86}, 085145 (2012). 
\bibitem{Okamoto2013} S. Okamoto, Phys. Rev. B {\bf 87}, 064508 (2013). 
\bibitem{Schmidt2018} J. Schmidt, D. D. Scherer, and A. M. Black-Schaffer, Phys. Rev. B {\bf 97}, 014504 (2018). 
\bibitem{Seifert2018} U. F. P. Seifert, T. Meng, and M. Vojta, Phys. Rev. B {\bf 97}, 085118 (2018). 
\bibitem{Choi2018} W. Choi, P. W. Klein, A. Rosch, and Y. B. Kim, Phys. Rev. B {\bf 98}, 155123 (2018). 
\bibitem{Rachel2016} S. Rachel, L. Fritz, and M. Vojta, Phys. Rev. Lett. {\bf 116}, 167201 (2016).
\bibitem{Perreault2017} B. Perreault, S. Rachel, F. J. Burnell, and J. Knolle, Phys. Rev. B {\bf 95}, 184429 (2017). 
\bibitem{Yamada2017a} M. G. Yamada, H. Fujita, and M. Oshikawa, Phys. Rev. Lett. {\bf 119}, 057202 (2017).
\bibitem{Yamada2017b} M. G. Yamada, V. Dwivedi, and M. Hermanns, Phys. Rev. B {\bf 96}, 155107 (2017).
\bibitem{Sengupta2008} K. Sengupta, D. Sen, and S. Mondal, Phys. Rev. Lett. {\bf 100}, 077204 (2008).
\bibitem{Mondal2008} S. Mondal, D. Sen, and K. Sengupta, Phys. Rev. B {\bf 78}, 045101 (2008).
\bibitem{Hikichi2010} T. Hikichi, S. Suzuki, and K. Sengupta, Phys. Rev. B {\bf 82}, 174305 (2010).
\bibitem{Patel2012} A. A. Patel and A. Dutta, Phys. Rev. B {\bf 86}, 174306 (2012). 
\bibitem{Sato2014preprint} M. Sato, Y. Sasaki, and T. Oka, preprint (arXiv:1404.2010). 
\bibitem{Bhattacharya2016} U. Bhattacharya, S. Dasgupta, and A. Dutta, Eur. Phys. J. B {\bf 89}, 216 (2016). 
\bibitem{Rademaker2017preprint} L. Rademaker, preprint (arXiv:1710.09761). 
\bibitem{Sameti2019} M. Sameti and M. J. Hartmann, Phys. Rev. A {\bf 99}, 012333 (2019).
\bibitem{Nasu2019} J. Nasu and Y. Motome, Phys. Rev. Research {\bf 1}, 033007 (2019). 
\bibitem{Alpichshev2015} Z. Alpichshev, F. Mahmood, G. Cao, and N. Gedik Phys. Rev. Lett. {\bf 114}, 017203 (2015). 
\bibitem{Hinton2015} J. P. Hinton, S. Patankar, E. Thewalt, A. Ruiz, G. Lopez, N. Breznay, A. Vishwanath, J. Analytis, J. Orenstein, J. D. Koralek, and I. Kimchi, Phys. Rev. B {\bf 92}, 115154 (2015). 
\bibitem{Nembrini2016} N. Nembrini, S. Peli, F. Banfi, G. Ferrini, Y. Singh, P. Gegenwart, R. Comin, K. Foyevtsova, A. Damascelli, A. Avella, and C. Giannetti, Phys. Rev. B {\bf 94}, 201119(R) (2016). 
\bibitem{Zhang2019preprint} H. Zhang, S. Kim, Y.-J. Kim, H.-Y. Kee, and L. Yang, preprint (arXiv:1908.04807). 
\bibitem{Kotliar2001} G. Kotliar, S. Y. Savrasov, G. P\'alsson, and G. Biroli, Phys. Rev. Lett. \textbf{87}, 186401 (2001).
\bibitem{Freericks1998}J. K. Freericks, V. Zlati\'c, Phys. Rev. B {\bf 58}, 322 (1998)
\bibitem{Freericks2003}J. K. Freericks, V. Zlati\'c, Rev. Mod. Phys. {\bf 75}, 1333 (2003)
\bibitem{Furukawa1994} N. Furukawa, J. Phys. Soc. Jpn. \textbf{63}, 3214 (1994).
\bibitem{Nilsson2013} J. Nilsson and M. Bazzanella, Phys. Rev. B \textbf{88}, 045112 (2013).
\bibitem{Metzner1989} W. Metzner and D. Vollhardt, Phys. Rev. Lett. \textbf{62}, 324 (1989).
\bibitem{Georges1996} A. Georges, G. Kotliar, W. Krauth, and M. J. Rozenberg, Rev. Mod. Phys. \textbf{68}, 13 (1996).
\bibitem{Werner2006} P. Werner, A. Comanac, L. de' Medici, M. Troyer, and A. J. Mills, Phys. Rev. Lett. \textbf{97}, 076405 (2006).
\bibitem{PhysRevB.72.035122} A. N. Rubtsov, V. V. Savkin, and A. I. Lichtenstein, Phys. Rev. B \textbf{72}, 035122 (2005).
\end{thebibliography}
\end{document}